# "Strukturelle Charakterisierung und Optimierung der Beugungseigenschaften von $Si_{1-x}Ge_x$ Gradientenkristallen, die aus der Gasphase gezogen wurden"

Von der Mathematisch-Naturwissenschaftlichen Fakultät der Rheinisch-Westfälischen Technischen Hochschule Aachen zur Erlangung des akademischen Grades eines Doktors der Naturwissenschaften genehmigte Dissertation

vorgelegt von

**Diplom-Physiker Klaus-Dieter Liß**

aus

Idar-Oberstein, Rheinland-Pfalz.

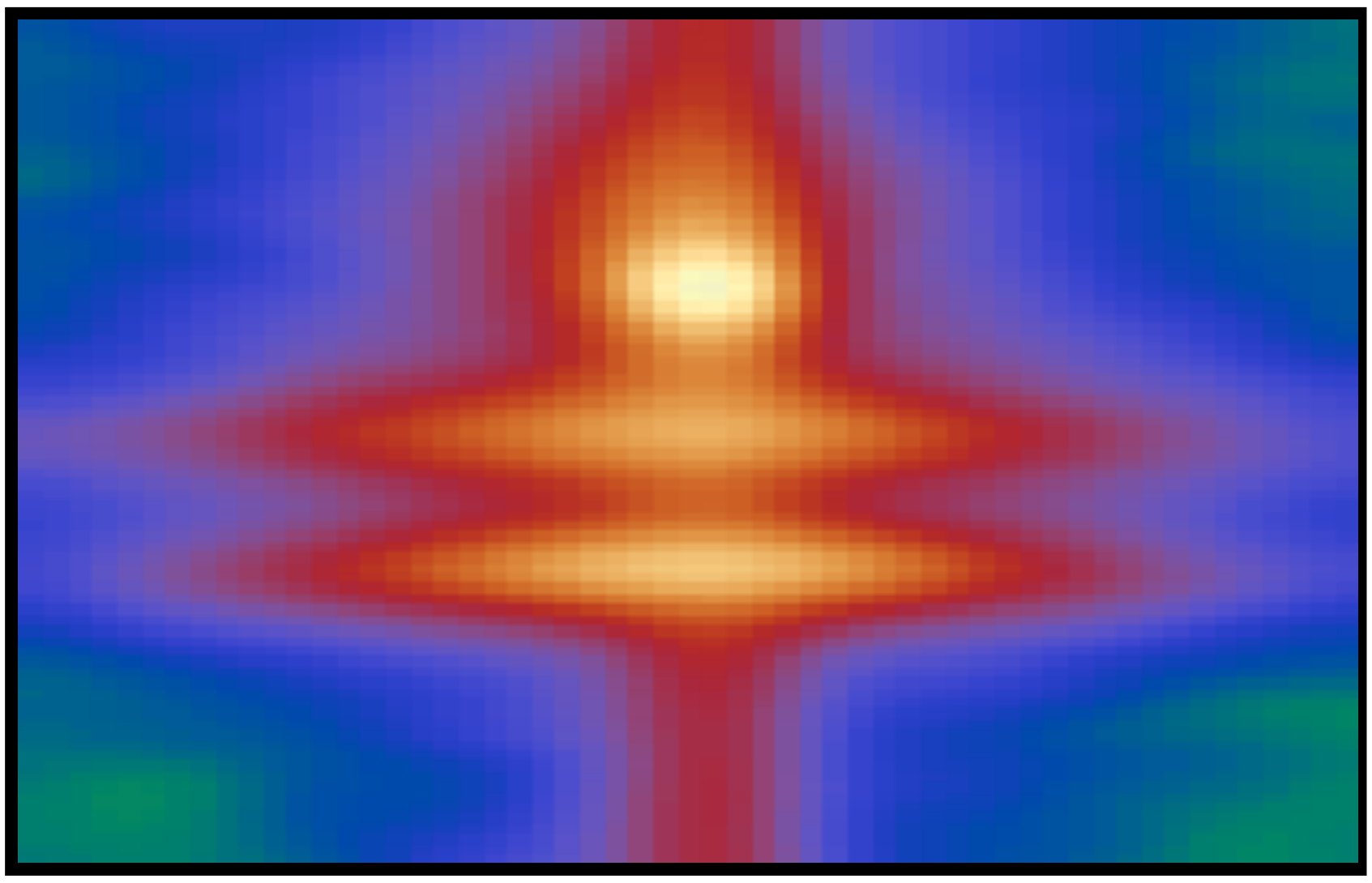

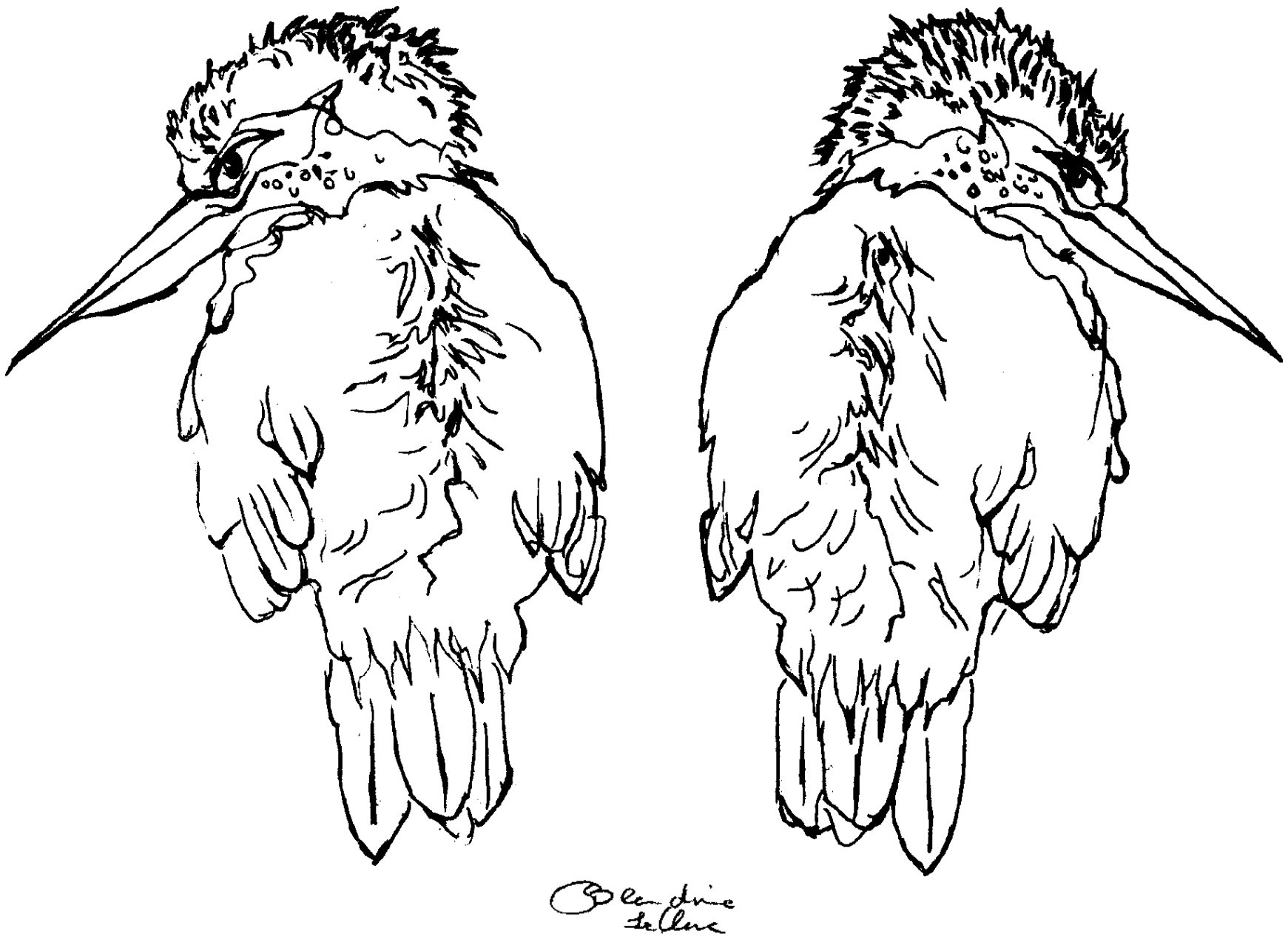

für meine liebe Frau, Laure

# Inhaltsübersicht

# 1. Einleitung

Die heute bestehenden Neutronenquellen liegen womöglich nahe an ihrer technischen Grenze und ähneln in ihren Neutronenflüssen denen vor 20 Jahren. Sie lassen sich mit dem Photonenfluß einer Kerze vergleichen [1]. Vor diesem Hintergrund ist es von fundamentaler Bedeutung die Auflösungselemente der Problemstellung eines Experimentes optimal anzupassen. Dies bedeutet, daß die interessierenden Komponenten des Auflösungselementes genau definiert sein sollen, während die dazu orthogonalen Komponenten zugunsten eines Intensitätsgewinns um zwei oder drei Größenordnungen relaxieren können. Ein Beispiel hierfür ist ein Mosaikmonochromator, also ein Kristall, der aus vielen, um die Vorzugsrichtung orientierten, Blöckchen zusammengesetzt ist. Durch diese transversale Streuvektorverteilung wird die sonst sehr scharf definierte Braggbedingung für viele, benachbarte Wellenvektoren erfüllt. Hierdurch wird dessen Bandbreite, und damit die reflektierte Intensität gegenüber dem Idealkristall erheblich erhöht. Die Mosaikverteilung $\eta$ führt allerdings eine Strahldivergenzaufweitung um $2\,\eta$ mit sich. Sie ist häufig, wenn nicht gerade die wohlgeformte Wellenzahl-Richtungs-Beziehung einer dispersionsfreien Doppelkristallanordnung ausgenutzt werden kann, der auflösungsbegrenzende Faktor.

Ein Gradientenkristall, der durch eine longitudinale Gitteraufweitung charakterisiert ist, reflektiert hingegen die innerhalb seiner Akzeptanz einfallenden Wellenvektoren ohne Aufweitung, wie an einem optischen Spiegel. Die Intensitätsgewinne sind mit denen eines Mosaikkristalls vergleichbar. Eine ausführliche Diskussion dieser beiden entgegengesetzten optischen Eigenschaften wurde vor kurzem im Rahmen eines Arbeitskreises über fokussierende Braggoptik geführt [2]. Als ein Beispiel sei hier erwähnt, daß zur Messung großer Objekte, also kleiner Streuvektoren, wie dies bei magnetischen Überstrukturen, Schichtsystemen oder großen Molekülen auftritt, die Monochromatisierung durch den Gradientenkristall vorteilhaft erscheint. Insbesonders die derzeit aufkommende Neutronenreflektometrie würde durch einen derartigen Monochromator begünstigt. Auch für die hochauflösende Rückstreuspektroskopie, die spezifisch auf eine longitudinale Variation des Gittervektors anspricht, verspricht ein Gradientenkristall direkte Vorteile, wenn Auflösung und Fluß optimiert werden müssen.

Die Idee der Verwendung von Gradientenkristallen ist so alt wie die Neutronenstreuung selbst. Leider kommen diese Kristalle in der Natur nicht vor. Sie müssen künstlich hergestellt werden, was mit einem erheblichen Entwicklungsaufwand verbunden ist. Dennoch mag die Qualität der Kristalle ungenügend erscheinen, und deshalb kommen sie nur relativ selten zum Einsatz. Bei unterschiedlichen Versuchen wurden bislang sowohl Temperaturgradienten [3], als auch Cu-Ge Legierungsgradienten [4] und in jüngster Zeit auch die Gradienten akustischer Wellenfelder im Monochromator [5] verwendet. Bereits in den sechziger Jahren wurde von

Maier-Leibnitz das Legierungssystem $Si_{1-x}Ge_x$ vorgeschlagen. Dabei spielt der durch die Halbleiterindustrie begünstigte technische Aspekt weltweiter Erfahrungen auf dem Gebiet der Ausgangsstoffe eine wesentliche Rolle. Allerdings stellte es sich sehr bald heraus, daß diese Kristalle nicht aus der Schmelze gezogen werden können.

Mittlerweile hat die Methode der chemischen Gasphasenabscheidung ihren Platz in der Kristallzucht behauptet. Hier können Legierungskonzentrationen in einfacher Weise durch die Gaszusammensetzung geregelt werden. Gegenüber anderen Epitaxiemethoden zeichnet sich die Gasphasenabscheidung durch erhebliche Wachstumsgeschwindigkeiten aus. Sie sind in günstigen Fällen ausreichend, um millimeterdicke Kristalle innerhalb weniger Tage zu ziehen. So wurden in einer Machbarkeitsstudie der Monochromatorgruppe am Institut Laue-Langevin zusammen mit einer erfahrenen Epitaxiegruppe der Universität in Grenoble die Grundlagen dieser Kristallzucht für das System $Si_{1-x}Ge_x$ ausgearbeitet.

Die Aufgabe der vorliegenden Arbeit war die Ausweitung des Zuchtverfahrens auf großflächige Proben. Dazu mußten Teile der Kristallzuchtanlage neu konzipiert und aufgebaut werden. Nachdem die komplette Anlage eingefahren war, wurde zunächst auf die Bestimmung der optimalen Zuchtparameter mit Blick auf Homogenität, Kristallinität und Wachstumsrate hingearbeitet, um schließlich für die Beugungsexperimente anwendungsnahe Proben zu ziehen. Neben der strukturellen Charakterisierung, insbesonders aus dem Bereich der Materialwissenschaften, spielen Beugungsexperimente an Gradientenkristallen mit Röntgen- und Neutronenstrahlen eine wesentliche Rolle. Zur theoretischen Beschreibung der gemessenen Reflektionskurven wurden verschiedene Ansätze im Rahmen der kinematischen und der dynamischen Theorie mit kohärenter oder inkohärenter Überlagerung gemacht. Die beiden am weitesten vorangetriebenen, nämlich die kinematische Theorie sowie eine exakte Transfermatrizenmethode im Rahmen der dynamischen Beugungstheorie, werden in dieser Schrift hergeleitet und, über die Anwendung auf Gradientenkristalle hinaus, in vielen Ergebnissen diskutiert.

# 2. Wirkungsquerschnitte und Streulängen

In der Streuphysik hängt die Intensität einer gestreuten Welle im Verhältnis zur einfallenden Welle vom Wirkungsquerschnitt der streuenden Teilchen ab. Betrachtet man die Amplitude einer kohärent abgebeugten Welle, so ist ihr Verhältnis zur einfallenden durch die sogenannte Streuamplitude $f(\Omega)$ im Raumwinkel $\Omega$ gegeben. Diese steht durch die Beziehung

$$\sigma_c = \int f(\Omega)\, d\Omega \tag{1}$$

mit dem kohärenten Wirkungsquerschnitt $\sigma_c$ in Verbindung [6]. Für thermische und kältere Neutronen ist die Streuamplitude aufgrund der geringen Ausdehnung des Kerns gegenüber der Wellenlänge nicht vom Streuwinkel abhängig und als kohärente Streulänge

$$f(\Omega) = b_c \tag{2}$$

bekannt. Sie ist von den streuenden Kernen und deren Spinzuständen abhängig und kann nicht nur von Element zu Element, sondern auch von Isotop zu Isotop erheblich schwanken. Die kohärente Streuung beschreibt die Interferenz eines Körpers bekannter Spinzustände und Isotopenverteilung. Neben ihr wird das Nichtwissen aufgrund statistischer Fluktuationen, z. B. einer Anhäufung nichtpolarisierter Kerne oder einer statistischen Isotopenverteilung im inkohärenten Wirkungsquerschnitt $\sigma_i$ verpackt. Diese Streuung ist isotrop und trägt daher für die Braggoptik zu einem unerwünschten Untergrund bei. Neben Streuung kann auch Absorption der Welle stattfinden, die durch den Absorptionsquerschnitt $\sigma_a$ beschrieben wird. Er ist umgekehrt proportional zur Geschwindigkeit des Neutrons und wird üblicherweise für eine bestimmte Neutronengeschwindigkeit $v_0$ tabelliert, kann also mittels

$$\sigma_a = \frac{v_0}{v}\,\sigma_a(v_0) = \frac{k_0}{k}\,\sigma_a(k_0) = \frac{\lambda}{\lambda_0}\,\sigma_a(\lambda_0) \tag{3}$$

der betrachteten Neutronenwelle angepaßt werden.
In Tabelle (1) sind die Streulängen und Wirkungsquerschnitte für die Isotopen von Silizium und Germanium zusammengestellt. Wie man sieht, unterscheiden sich die kohärenten Streulängen bei den Germaniumisotopen mehr als bei Silizium, was demnach bei natürlicher Isotopenverteilung einen größeren inkohärenten Beitrag mit sich führt.

Betrachtet man die Streuung von Röntgenstrahlen, so kann man die Streuamplitude durch das Produkt aus der Streulänge eines freien Elektrons, also dem klassischem Elektronenradius $r_e$, und dem Atomformfaktor $f(\vec{Q})$ ausdrücken:

$$f(\Omega) = r_e\, f(\vec{Q}) \qquad (4)$$

Aufgrund der, im Vergleich zur Wellenlänge ausgedehnten Elektronenverteilung des Atoms hängt die Streuamplitude nun vom betrachteten Impulsübertrag $\vec{Q}$ zwischen einfallender und gestreuter Welle ab und geht aus der Fouriertransformation dieser räumlichen Verteilung hervor. Der atomare Formfaktor $f(\vec{Q})$ ist somit eine dimensionslose Zahl, die bei Vorwärtsstreuung gleich der Ordnungszahl Z des Elements ist:

$$f(0) = Z \qquad (5)$$

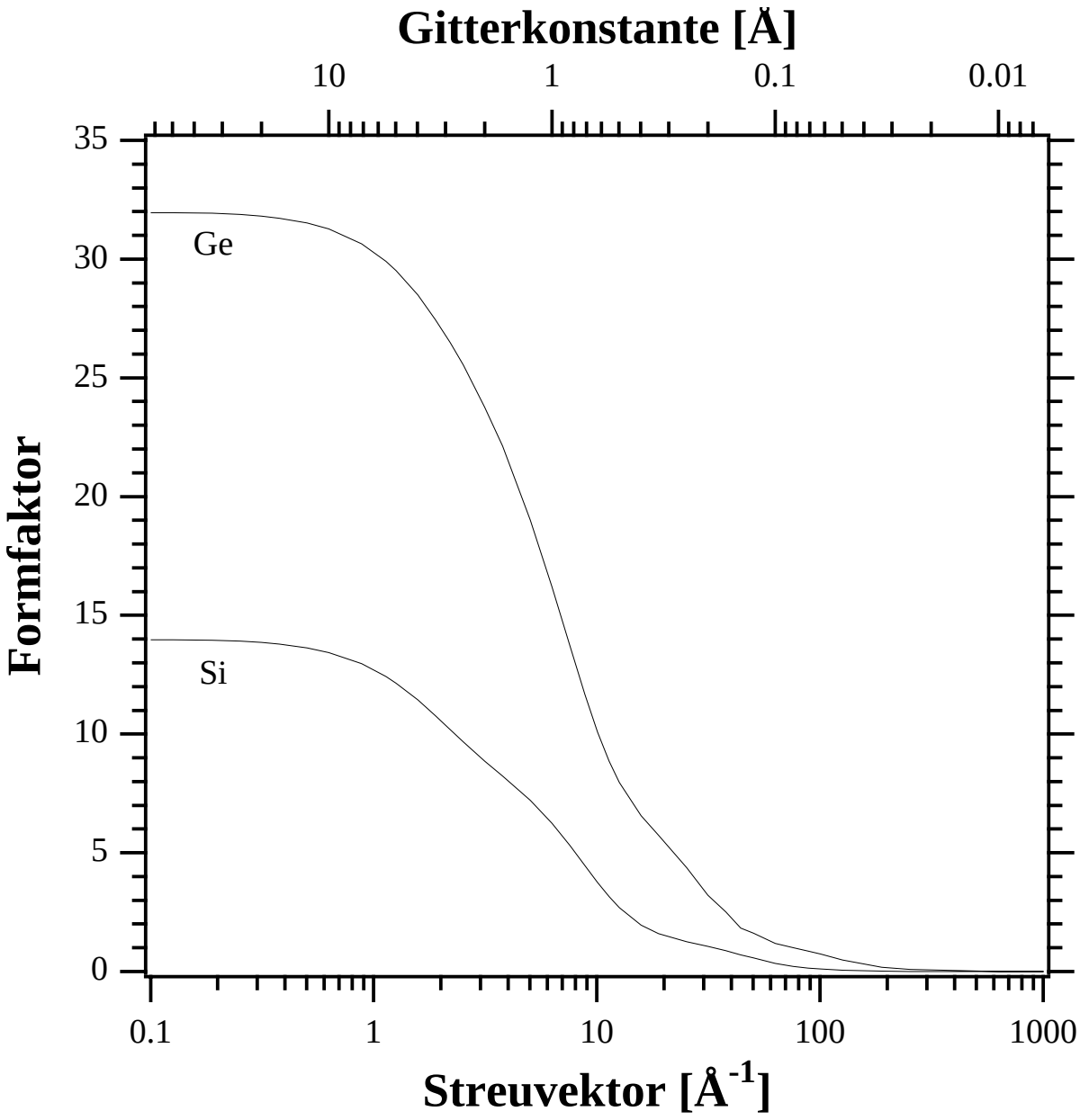

*Abbildung (1):*
*Atomare Formfaktoren für Silizium und Germanium als Funktion des Streuvektors und der Gitterkonstanten bei Braggreflektion an einem Kristall. Nach [7].*

In Abbildung (1) sind die atomaren Formfaktoren für Silizium und Germanium wiedergegeben [7]. Die Absorption von Röntgenstrahlung geschieht hauptsächlich durch Photoabsorption und fällt, außer an Absorptionskanten, gemäß eines Potenzgesetzes mit zunehmender Photonenenergie ab. Comptonstreuung und Paarbildungsprozesse spielen in unseren Energiebereichen keine Rolle. Eine Übersicht der einzelnen Beiträge ist in Abbildung (2) dargestellt [8, 9].

Die Wechselwirkung der Welle mit Materie bewirkt im allgemeinen mit zunehmender Tiefe t eine Abschwächung der eingestrahlten Intensität. Abgesehen von kohärenter Streuung, bei der z. B. auch oszillierende Interferenzeffekte auftreten können, folgt diese Abschwächung einem Exponentialgesetz

$$I(t) = I_0\, e^{-\mu t} \qquad (6)$$

mit dem linearen Abschwächungskoeffizienten

$$\mu = \frac{N_A\, \sigma\, \rho}{A}, \qquad (7)$$

der Avogadrokonstanten $N_A$, der relativen Atommasse A und der Dichte $\rho$. $\mu$ kann gemäß $\sigma$ in seine verschiedenen Anteile für Absorption oder Extinktion zerlegt werden. Die Abschwächungslänge $t_0$ wird als die Tiefe definiert, in der die Strahlung auf $e^{-1}$ abgefallen ist, also

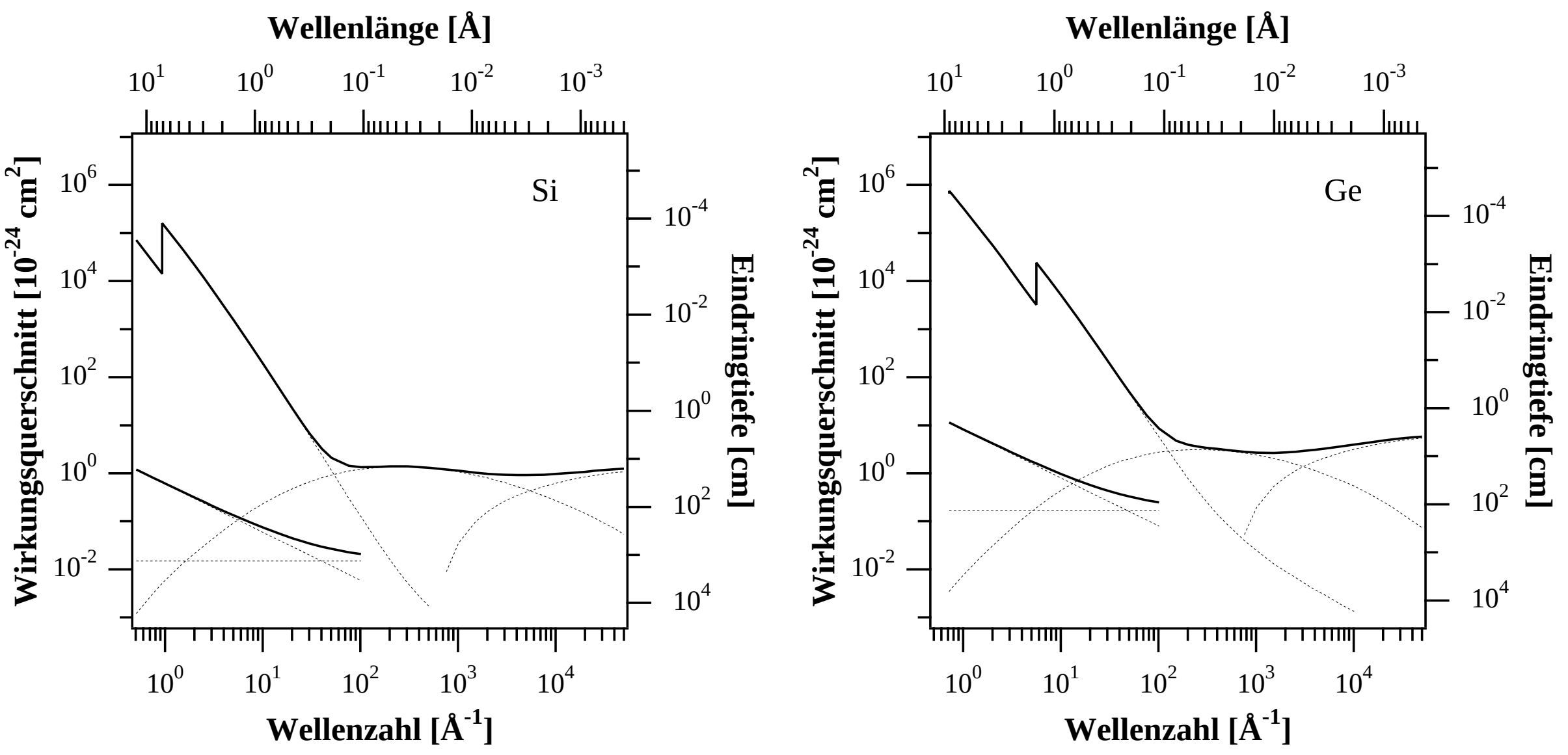

*Abbildung (2):*
*Wirkungsquerschnitte und Eindringtiefen ohne kohärente Beiträge in Abhängigkeit von Wellenzahl und Wellenlänge für Silizium (links) und Germanium (rechts). Die totalen Wirkungsquerschnitte für Röntgenstrahlen (obere, durchgezogene Linien) setzen sich aus den Anteilen für Photoabsorption, Comptonstreuung und Paarbildung (von links oben nach rechts unten), die für Neutronenstrahlen (untere, linke, durchgezogene Linien) aus Absorption und inkohärenter Streuung zusammen. Nach [8, 9].*

$$t_0 = \frac{1}{\mu} = \frac{A}{N_A \, \sigma \, \rho} \, . \tag{8}$$

In Abbildung (2) sind diese Tiefen für Neutronen und Photonen bei der Wechselwirkung mit Silizium und Germanium auf der rechten Skala aufgetragen.

Bei der Ausbildung eines Braggreflexes tritt ebenfalls Extinktion durch kohärente Streuung auf. Die dafür notwendigen Tiefen müssen mit obigen Abschwächungstiefen verglichen werden, um Kriterien für die optimale Strahlung zur Untersuchung eines Körpers zu finden.

Ein $Si_{1-x}Ge_x$ Legierungskristall, bei dem die Elemente statistisch auf die Gitterplätze verteilt sind, kann, die Wirkungsquerschnitte betrachtend, wie ein Isotopengemisch angesehen werden. Somit ergibt sich für die resultierende Streulänge der mit der Konzentration $C_\alpha$ gewichtete Mittelwert aus den Einzelisotopen $\alpha$

$$b_c = \sum_\alpha C_\alpha \, b_c^\alpha \, . \tag{9}$$

Der inkohärente Wirkungsquerschnitt aufgrund des Isotopengemischs ergibt sich aus der quadratischen Summe aller paarweisen Differenzen der kohärenten Einzelstreulängen, wobei der Index $\alpha$ über alle Silizium- als auch Germaniumisotope zu laufen hat:

$$\sigma_i = 4\pi \sum_{\alpha} \sum_{\beta < \alpha} C_\alpha C_\beta \left| b_c^\alpha - b_c^\beta \right|^2 . \qquad (10)$$

Abbildung (3) zeigt den, aufgrund der Isotopenmischung auftretenden inkohärenten Wirkungsquerschnitt für $Si_{1-x}Ge_x$ als Funktion der Germaniumkonzentration x bei natürlicher Isotopenzusammensetzung. Da die inkohärente Streuung mit der kohärenten verglichen werden muß, gibt die gestrichelte Linie zur rechten Skala das Verhältnis zwischen inkohärentem und kohärentem Wirkungsquerschnitt wieder. Wie man sieht, ist die inkohärente Streuung maximal bei x = 0,58 und beträgt 0,15 barn. Die kohärente Streuung steigt von 2,16 barn bei Silizium auf 8,44 barn bei Germanium. Dadurch ist das Maximum im Verhältnis von inkohärentem und kohärentem Querschnitt zu kleineren Germaniumkonzentrationen, also nach x = 0,37 und einem Wert von 0,15 verschoben.

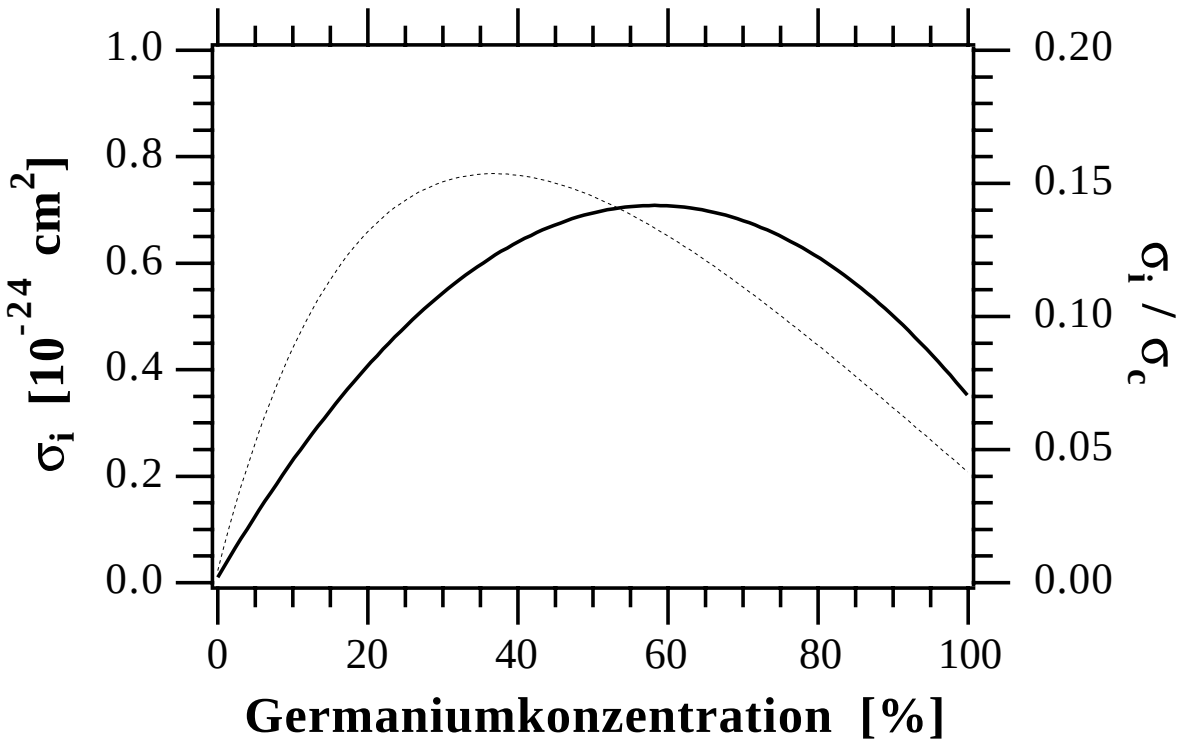

*Abbildung (3):*
*Berechneter, inkohärenter Wirkungsquerschnitt $\sigma_i$ (durchgezogene Linie) in Abhängigkeit der Germaniumkonzentration einer $Si_{1-x}Ge_x$ Legierung und dessen Verhältnis zum kohärenten Anteil $\sigma_c$ (gestrichelte Linie).*

| Element | Z | A | I(π) | C [%] | $b_c$ [fm] | $b_i$ [fm] | $\sigma_c$ [barn] | $\sigma_i$ [barn] | $\sigma_s$ [barn] | $\sigma_a$ [barn] |
|---|---|---|---|---|---|---|---|---|---|---|
| **Si** | **14** | | | | **4,149** | | **2,163** | **0,015** | **2,178** | **0,171** |
| Si | 14 | 28 | 0 | 92,23 | 4,106 | 0 | 2,119 | 0 | 2,119 | 0,177 |
| Si | 14 | 29 | 1/2 | 4,67 | 4,7 | -1,1 | 2,78 | 0,15 | 2,93 | 0,101 |
| Si | 14 | 30 | 0 | 3,10 | 4,58 | 0 | 2,64 | 0 | 2,64 | 0,107 |
| **Ge** | **32** | | | | **8,1929** | | **8,435** | **0,17** | **8,60** | **2,3** |
| Ge | 32 | 70 | 0 | 20,5 | 9,5 | 0 | 11,3 | 0 | 11,3 | 3,43 |
| Ge | 32 | 72 | 0 | 27,4 | 8,8 | 0 | 9,7 | 0 | 9,7 | 0,98 |
| Ge | 32 | 73 | 9/2 | 7,8 | 3,2 | 2 | 1,3 | 0,5 | 1,8 | 15 |
| Ge | 32 | 74 | 0 | 36,5 | 7,9 | 0 | 7,8 | 0 | 7,8 | 0,51 |
| Ge | 32 | 76 | 0 | 7,8 | 9 | 0 | 10 | 0 | 10 | 0,15 |

*Tabelle (1):*
*Neutronenwirkungsquerschnitte für Silizium und Germanium. Nach [6].*
*Z: Ordnungszahl, A: Nukleonenzahl, I(π): Spin (Parität) des nuklearen Grundzustandes, C: natürliches Isotopenverhältnis, $b_c$: kohärente, $b_i$: inkohärente Streulänge, $\sigma_c$: kohärenter, $\sigma_i$: inkohärenter, $\sigma_s$: totaler Wirkungsquerschnitt, $\sigma_a$: Absorptionsquerschnitt für $v_0$ = 2200 m/s.*

# 3. Grundlagen zum Kristallgitter

Silizium und Germanium kristallisieren beide in der Diamantstruktur, einem kubischen Punktgitter der Basis

$$\mathfrak{A} = [\vec{a}_1\ \vec{a}_2\ \vec{a}_3] = a_0 \begin{bmatrix} 1 & 0 & 0 \\ 0 & 1 & 0 \\ 0 & 0 & 1 \end{bmatrix} \tag{11}$$

und der achtatomigen Einheitszelle mit den Ortsvektoren

$$\vec{\rho}_\alpha \in \left\{ \begin{bmatrix} 0 \\ 0 \\ 0 \end{bmatrix}, \begin{bmatrix} 1/4 \\ 1/4 \\ 1/4 \end{bmatrix} \right\} \oplus \left\{ \begin{bmatrix} 0 \\ 0 \\ 0 \end{bmatrix}, \begin{bmatrix} 0 \\ 1/2 \\ 1/2 \end{bmatrix}, \begin{bmatrix} 1/2 \\ 0 \\ 1/2 \end{bmatrix}, \begin{bmatrix} 1/2 \\ 1/2 \\ 0 \end{bmatrix} \right\}, \tag{12}$$

wobei der ⊕-Operator die paarweise Addition der Elemente beider Mengen bezeichnet. Die Struktur besteht also aus zwei kubisch flächenzentrierten Untergittern (rechte Menge), die um ein Viertel der elementaren Raumdiagonalen gegeneinander versetzt sind (linke Menge). Jedes Atom ist dabei tetraederförmig von vier nächsten Nachbarn umgeben.

Die Basis des reziproken Gitters geht mit

$$\vec{G}_i = 2\pi \frac{\vec{a}_j \times \vec{a}_k}{\vec{a}_i \cdot (\vec{a}_j \times \vec{a}_k)} \ ; \ (i, j, k) = (1, 2, 3) \text{ zykisch} \tag{13}$$

aus $\mathfrak{A}$ hervor und ergibt sich zu

$$\mathfrak{G} = [\vec{G}_1\ \vec{G}_2\ \vec{G}_3] = G_0 \begin{bmatrix} 1 & 0 & 0 \\ 0 & 1 & 0 \\ 0 & 0 & 1 \end{bmatrix} \ ; G_0 = \frac{2\pi}{a_0}, \tag{14}$$

ist also parallel zu $\mathfrak{A}$. Die Punkte $\vec{G}$ des reziproken Gitters gehen mit dem aus ganzzahligen Komponenten bestehenden Millerindexvektor $\vec{h}$ aus der reziproken Basis hervor:

$$\vec{G} = \mathfrak{G}\,\vec{h}, \ \text{mit } \vec{h} = \begin{bmatrix} h \\ k \\ l \end{bmatrix}. \tag{15}$$

Aufgrund der Struktur (12) der Einheitszelle ist nicht jeder reziproke Gitterpunkt besetzt, und es ergibt sich allgemein für den Strukturfaktor:

$$
\begin{aligned}
F(\vec{G}) = F(\vec{h}) = F_{hkl} &= \sum_{\alpha} \exp\left(i\, \vec{G}\, \vec{\rho}_{\alpha}\right) = \\
&= \left\{\exp(i\,0) + \exp\left(i\, \frac{\pi}{2}\,(h + k + l)\right)\right\} \left\{\exp(i\,0) + \exp\left(i\,\pi\,(k + l)\right) + \exp\left(i\,\pi\,(h + l)\right) + \exp\left(i\,\pi\,(h + k)\right)\right\} = \\
&= \left\{1 + i^{h + k + l}\right\} \left\{1 + (-1)^{k + l} + (-1)^{h + l} + (-1)^{h + k}\right\}
\end{aligned}
\tag{16}
$$

Die Werte von $F_{hkl}$ und deren Betrag $|F_{hkl}|$ in Abhängigkeit der Millerindizes sind in Tabelle (2) als Fallunterscheidung wiedergegeben.

| Bedingung der Millerindizes | $F_{hkl}$ | $\lvert F_{hkl} \rvert$ |
|---|---|---|
| (h + k + l) / 4 = ganze Zahl | 8 | 8 |
| h, k, l alle ungerade | $4\,(1 \pm i)$ | $4\sqrt{2}$ |
| ansonsten | 0 | 0 |

*Tabelle (2):*
*Strukturfaktoren des Diamantgitters*

Der Strukturfaktor spielt vor allem bei der Berechnung der Intensität eines Braggreflexes eine wesentliche Rolle.

Die Gitterebenen des Kristallgitters werden durch ihre Normalen, also die reziproken Gittervektoren $\vec{G}$ beschrieben, deren Betrag durch

$$
G = \frac{2\pi}{d} \tag{17}
$$

mit dem Netzebenenabstand d in Verbindung steht. Wegen der Diagonalität von $\mathfrak{G}$ und $\mathfrak{A}$ im kubischen Gitter hängen G und d gemäß Pythagoras von den Millerindizes ab:

$$
G = \sqrt{h^2 + k^2 + l^2}\; G_0 \quad \text{bzw.} \quad d = \frac{a_0}{\sqrt{h^2 + k^2 + l^2}} \tag{18}
$$

Für die experimentelle Ausrichtung der Kristalle sind die Winkel zwischen verschiedenen Gitterebenen von Interesse. Sie entsprechen den Winkeln zwischen den reziproken Gittervektoren und sind durch

$$
\sphericalangle\left(\vec{G}_1, \vec{G}_2\right) = \operatorname{acos}\left(\frac{\vec{G}_1 \cdot \vec{G}_2}{G_1\, G_2}\right) = \operatorname{acos}\left(\frac{\vec{h}_1 \cdot \vec{h}_2}{h_1\, h_2}\right) \tag{19}
$$

gegeben. Tabelle (3) stellt einige wichtige Winkel zusammen.

| Millerindizes | 1 0 0 | 1 1 0 | 1 1 1 |
|---|---|---|---|
| 0 0 1 | 90° | 90° | 54,74° |
| 0 1 1 | 45° | 60° | 35,26° |
| 1 $\bar{1}$ 1 | 54,74° | 90° | 70,53° |

*Tabelle (3):*
*Winkel zwischen verschiedenen Gitterebenen des Diamantgitters*

# 4. Intensitätsverteilungen bei der Beugung an Gradientenkristallen

Neben den zum Mosaikkristall komplementären optischen Eigenschaften liegt die grundsätzliche Idee von Gradientenkristallen bei ihrer Verwendung als Monochromator in der wesentlichen Eigenschaft des Intensitätsgewinns gegenüber dem Idealkristall [4, 10]. Dieser hängt nicht nur von der Aufweitung $\Delta d$ des Netzebenenabstands sondern auch, wie bei Idealkristallen, vom streuenden Kristallvolumen ab. Dabei muß man den Gradienten $\vec{g}$ mit der Pendellösungsperiode $\Delta$ vergleichen, innerhalb derer die Reflektivität eines Idealkristalls voll ausgebildet wird. Der ideale Gradient für maximale Reflektivität ist dann so bestimmt, daß sich die Gitterkonstante über $\Delta$ dermaßen gezielt ändert, um die Reflektionskurve eines Idealkristalls gerade um seine natürliche Linienbreite, die Darwinbreite $\delta d$ zu verschieben. Dieser Sachverhalt wird in Abbildung (4) für den Braggfall angedeutet. Hierbei werden aus den verschiedenen Kristallbereichen resultierende Darwinkurven aneinandergereiht. Die Reflektionskurve des gesamten Kristalls ist durch die Einhüllende gegeben. Ist der Gradient größer, das heißt, wird beim Fortschreiten um eine Pendellösungsperiode die Reflektionskurve weiter als die Darwinbreite verschoben, so fällt die Maximalintensität ab, und man kommt in den Gültigkeitsbereich der kinematischen Streutheorie, in der die über den gesamten Braggreflex integrierte Reflektivität dem Volumen proportional absinkt. Ist im Gegensatz dazu der Gradient zu klein, tritt primäre Extinktion auf, das heißt, es gibt Kristallbereiche, in denen die Intensität des Primärstrahls bereits durch davorliegende Kristallbereiche reflektiert und somit so weit abgeschwächt wurde, daß nur mehr ein kleiner Teil zur Reflektion übrigbleibt. Anders als beim Idealkristall trägt beim Gradientenkristall auch bei Extinktion das gesamte Kristallvolumen zur Reflektion bei, da die Braggbedingung als Funktion des Kristallortes für verschiedene, benachbarte Wellenlängen erfüllt wird. Somit muß eine Welle, die z. B. nahe der Kristallrückseite reflektiert wird, das gesamte vordere Kristallmaterial durchdringen, was dort zu Absorptionsverlusten führen kann. Abgesehen davon würde bei einem zu kleinen Gradienten wertvolles Kristallvolumen durch Extinktion verschenkt. Der ideale Gradient $\vec{g}_i$ bei dem die Reflektivität voll ausgebildet ist und der Kristall gerade nur so dick wie nötig ist, wird also durch

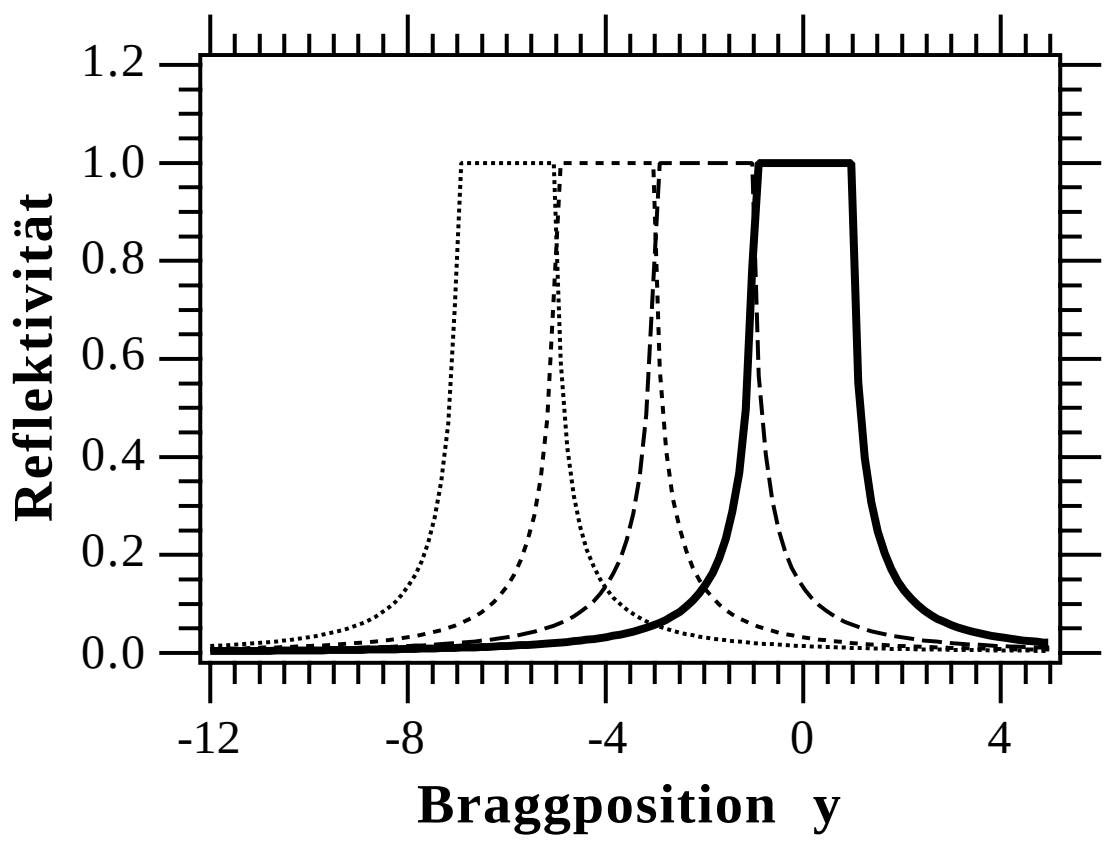

*Abbildung (4):*
*Interpretation der Reflektionskurve eines idealen Gradientenkristalls durch Aneinanderreihen von Darwinkurven, dargestellt als Funktion der halben Darwinbreite y.*

$$g_i = \frac{\delta d}{\Delta} \tag{20}$$

bestimmt.

Zur analytischen Betrachtung von Gradientenkristallen sind jedoch quantitative Aussagen notwendig, die im kommenden im Rahmen der kinematischen Theorie angeschnitten und im Anschluß durch einen Transfermatrizenalgorithmus innerhalb der dynamischen Streutheorie beschrieben werden sollen.

## 4.1. Kinematische Beugung an Gradientenkristallen

In der kinematischen Beugungstheorie wird von reflektierenden Gitterebenen ausgegangen, die miteinander interferieren. Dabei wird gemäß dem Huygens'schen Prinzip jede Ebene als eine Quelle der gestreuten Welle mit Amplitude $A_1$ aufgefaßt, deren Phasenunterschiede durch die einfallende Welle gegeben sind. Da die Wellengleichung bei einer Reflektion an einer Ebene gemäß Parallel- und Normalkomponenten separiert, kann die Reflektion eindimensional als Funktion der Koordinate x senkrecht zu den Ebenen berechnet werden. Die reflektierte Welle $\Xi$ ergibt sich durch phasengerechte Summation der Einzelwellen

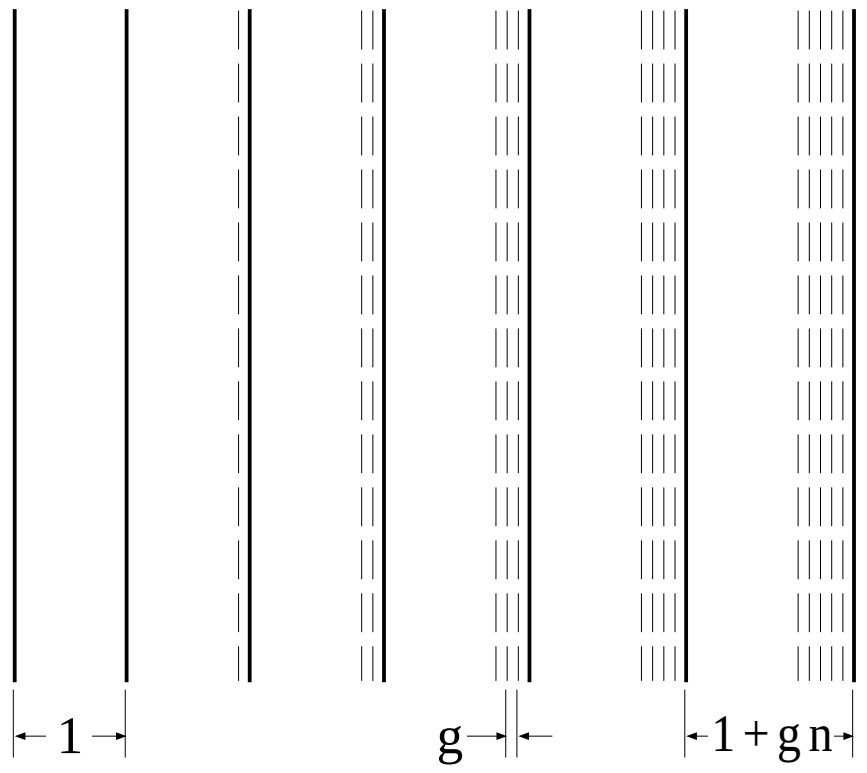

*Abbildung (5): Darstellung der Gitterebenen eines Gradientenkristalls. Die angegebenen Größen sind in Einheiten der Gitterkonstanten d.*

$$\Xi = A_1 \sum_{n=0}^{N} \exp(i\,2\,k\,x_n) \tag{21}$$

wobei n die N+1 Gitterebenen der Ortskoordinaten $x_n$ durchnumeriert. k ist die Wellenzahl und der Faktor 2 rührt von dem verdoppelten Weg der hin- und herlaufenden Welle her, wenn man um x in der Tiefe fortschreitet.

Während beim Idealkristall alle benachbarten Gitterebenen den gleichen Abstand d voneinander haben, unterscheidet sich dieser beim Gradientenkristall von Ebene zu Ebene relativ um

$$g = \frac{\partial d}{\partial x}, \tag{22}$$

liegen also mit den Abständen d, $d(1+g)$, $d(1+2g)$ usw. zueinander (siehe Abbildung (5)). Somit ist

$$x_n = \sum_{\nu=1}^{n} d_\nu = d \sum_{\nu=1}^{n} (1 + (\nu - 1)\, g) = d \left(1 - \frac{g}{2}\right) n + \frac{1}{2}\, d\, g\, n^2 \,. \tag{23}$$

Mit $|g| \ll 1$ erhalten wir

$$x_n = n\, d + \frac{1}{2}\, d\, g\, n^2 \quad . \tag{24}$$

Dies in (21) eingesetzt und die Anzahl der Gitterebenen durch die Kristalldicke D und den Netzebenenabstand d ausgedrückt ergibt für die reflektierte Welle

$$\Xi = A_1 \sum_{n=0}^{D/d} \exp(i\, 2\, k\, d\, n) \;\; \exp\left(i\, k\, d\, g\, n^2\right) . \tag{25}$$

In der Nähe des Braggmaximums kann die Phase k d durch ihre Abweichung $\kappa$ von dieser Position ausgedrückt werden

$$k\, d = \pi + \kappa\, \pi \,, \qquad\qquad |\kappa| \ll 1 \tag{26}$$

und wir erhalten

$$\Xi = A_1 \sum_{n=0}^{D/d} \exp(i\, 2\, \pi\, n) \;\; \exp(i\, 2\, \pi\, \kappa\, n) \;\; \exp\left(i\, \pi\, g\, n^2\right) \;\; \exp\left(i\, \pi\, \kappa\, g\, n^2\right) . \tag{27}$$

Der erste Faktor ergibt gerade 1, der letzte kann wegen $|\kappa| \ll 1$ und $|g| \ll 1$ in erster Ordnung vernachlässigt werden, so daß

$$\Xi = A_1 \sum_{n=0}^{D/d} \exp\left(i\, 2\, \pi \left(\kappa\, n + \frac{g}{2}\, n^2\right)\right) . \tag{28}$$

Beim Übergang von der Summe zum Integral erhält man

$$\Xi = A_1 \int_0^{D/d} \exp\left(i\, 2\, \pi \left(\kappa\, n + \frac{g}{2}\, n^2\right)\right) \; dn$$

$\Rightarrow$

$$\Xi = A_1 \int_0^{D/d} \exp\left(i\, \pi\, g \left(n + \frac{\kappa}{g}\right)^2\right) \;\; \exp\left(-i\, \pi\, \frac{\kappa^2}{g}\right) \; dn \,. \tag{29}$$

Hier kann ebenfalls der Faktor in $\kappa^2$ fortgelassen werden, also

$$\Xi = A_1 \int_0^{D/d} \exp\left(i\,\pi\,g\left(n + \frac{\kappa}{g}\right)^2\right)\,dn\,. \tag{30}$$

Mit der Substitution

$$\zeta = \sqrt{2\,g}\left(n + \frac{\kappa}{g}\right) \quad ; \quad dn = \frac{1}{\sqrt{2\,g}}\,d\zeta \tag{31}$$

kann das Integral in

$$\Xi = \frac{A_1}{\sqrt{2\,g}} \int_{\sqrt{2\,g}\,\frac{\kappa}{g}}^{\sqrt{2\,g}\left(\frac{D}{d}+\frac{\kappa}{g}\right)} \exp\left(i\,\frac{\pi}{2}\,\zeta^2\right)\,d\zeta \tag{32}$$

umgewandelt werden. Dieses Integral kann man durch die Fresnelschen Integrale

$$\begin{aligned} C(z) &= \int_0^z \cos\left(\frac{\pi}{2}\,t^2\right) dt \\ S(z) &= \int_0^z \sin\left(\frac{\pi}{2}\,t^2\right) dt \end{aligned} \quad , \tag{33}$$

ausdrücken, wie sie in mathematischen Formelsammlungen diskutiert werden [11]. Wir führen hier das komplexe Fresnelsche Integral

$$F(z) := C(z) + i\,S(z) \tag{34}$$

ein und erhalten

$$\Xi = \frac{A_1}{\sqrt{2\,g}}\left\{ F\left(\sqrt{2\,g}\left(\frac{D}{d} + \frac{\kappa}{g}\right)\right) - F\left(\sqrt{2\,g}\,\frac{\kappa}{g}\right) \right\}. \tag{35}$$

Um die anschließende Diskussion zu vereinfachen, soll als nächstes auf die Größen $\kappa \to y$, $D \to A$, $g \to c$ übergegangen werden, wie sie in der dynamischen Streutheorie üblich sind. Da bisher noch keine Angaben über die Streustärke gemacht wurden und in der dynamischen Theorie sämtliche Größen auf das Kristallpotential normiert sind, kann letzteres für die hiesige Herleitung gleich 1 gesetzt werden. Außerdem gelten hier wegen der Eindimensionalität die Bedingungen der Rückstreuung, also

$$d = \frac{\pi}{k}\,, \tag{36}$$

$$D = \frac{2}{k} A , \tag{37}$$

$$\kappa = \frac{y}{2} . \tag{38}$$

Mit der Definition

$$c := \frac{\partial y}{\partial A} \tag{39}$$

des Gradienten in den neuen Einheiten folgt mit $\Delta d/d = \Delta k/k = \kappa$ die Beziehung

$$g = \frac{\pi}{4} c . \tag{40}$$

Wie in Kapitel 4.2.1. gezeigt wird, ergibt sich für die Streuamplitude $A_1$ einer Gitterebene

$$A_1 = \frac{\pi}{2} . \tag{41}$$

Diese Beziehungen in (35) eingesetzt ergeben schließlich die endgültige Form unserer reflektierten Welle

$$\Xi = \sqrt{\frac{\pi}{2\,c}} \left\{ F\left(\sqrt{\frac{2}{\pi\,c}}\,(c\,A + y)\right) - F\left(\sqrt{\frac{2}{\pi\,c}}\;y\right) \right\}. \tag{42}$$

Für die Reflektivität R muß bekanntlich das Betragsquadrat von $\Xi$ gebildet werden:

$$R = |\Xi|^2 \tag{43}$$

In Abbildung (6) sind die Reflektionskurven für verschiedene c- und A-Werte dargestellt. Als erstes sticht dem Leser das, zwar mit Oszillationen besetzte, Plateau ins Auge, wie es dem Bild aneinandergereihter Reflektionskurven von Idealkristallen entsprechen muß. Die Anstiege finden an den Stellen statt, bei denen das Argument eines der Fresnelschen Integrale verschwindet, also bei

$$y = -c\,A \quad \text{und} \quad y = 0 . \tag{44}$$

Die Breite c A entspricht , wie erwartet, der Gesamtvariation der Gitterkonstanten $\Delta d/d$. Den Plateaumittelwert $\overline{R}_P$ des Plateaus, d. h. lokal über die Oszillationen gemittelt, erhalten wir, wenn wir das Fresnelsche Integral für sehr große Argumente betrachten:

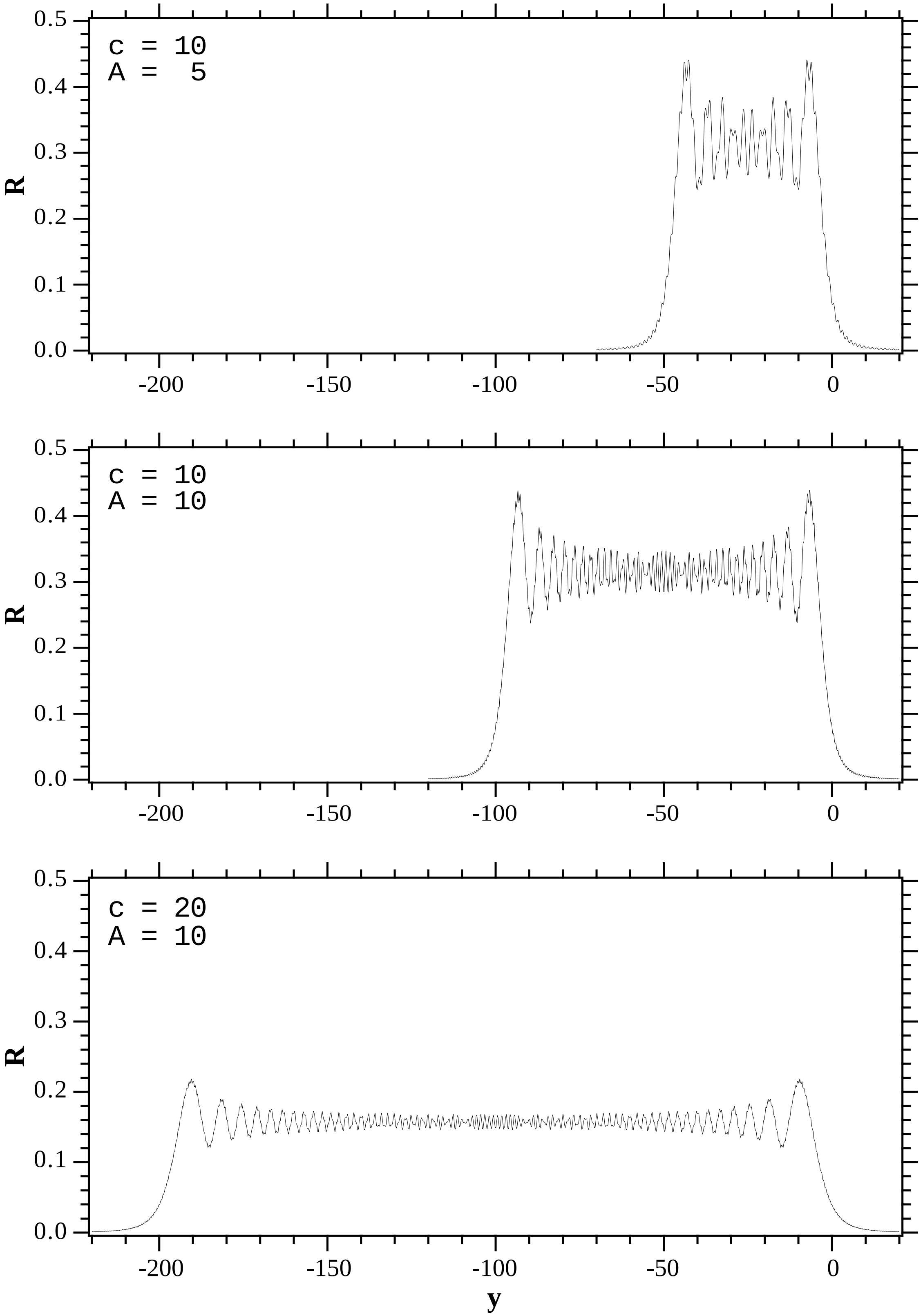

*Abbildung (6):*
*Kinematisch berechnete Reflektivität verschiedener Gradientenkristalle mit Gradienten c und Kristalldicken A als Funktion der Abweichung y von der Braggposition an der Kristalloberfläche. Vom ersten zum zweiten Bild verdoppelt sich A bei festgehaltenem c, was keinen Einfluß auf den Plateaumittelwert sowie auf die Oszillationen am Rand, sondern nur auf die Breite der Reflektionskurve hat. Vom zweiten zum dritten Bild wird A festgehalten und c verdoppelt, was zu einer halbierten Plateaureflektivität führt. Die Oszillationsperiode in der Plateaumitte bleibt jedoch unverändert.*

$$F(x) = \frac{1+i}{2} - \frac{i}{\pi x} \exp\left(i \frac{\pi}{2} x^2\right) + O\left(\frac{1}{x^2}\right) ; \quad x \gg 1 ; \tag{45}$$

$$F(x) = \frac{1+i}{2} + O\left(\frac{1}{x}\right) ; \quad x \to \infty . \tag{46}$$

Da F(x) um seinen Mittelwert (46) oszilliert, erhält man $\overline{R}_P$ für $A \to \infty$ bei $y = \frac{1}{2} c A$ unter der Berücksichtigung, daß $F(-x) = -F(x)$ eine ungerade Funktion ist

$$\overline{R}_P = \left| \sqrt{\frac{\pi}{2c}} \left( \frac{1+i}{2} + \frac{1+i}{2} \right) \right|^2 = \frac{\pi}{c} . \tag{47}$$

Dies entspricht der Tatsache, daß sich die Reflektionskurve proportional zu c verbreitert, die integrierte Gesamtreflektivität bei unveränderter Dicke jedoch gleich bleiben muß und folglich $\overline{R}_P$ umgekehrt proportional mit dem Gradienten abnimmt. Außerdem hängt der mittlere Plateauwert nicht von der Kristalldicke A ab. Dies wird so verstanden, daß bei festgehaltener Abweichung von der Braggposition y nur eine gewisse Anzahl Gitterebenen zur konstruktiven Interferenz beitragen, weil in anderen Kristallbereichen die Braggbedingung nicht mehr erfüllt ist. Diese Zahl, bzw. das dadurch definierte Kristallvolumen ist umgekehrt proportional zum Gradienten, und ebenso die Reflektivität $\overline{R}_P$ an dieser Stelle.
Proportional zur Kristalldicke verbreitert sich jedoch die Reflektionskurve, was der wohlbekannten Tatsache entspricht, daß im Rahmen der kinematischen Beugungstheorie die Gesamtreflektivität mit dem Kristallvolumen wächst.

Die Oszillationen, auch Taupin-Oszillationen genannt, sind am Plateaurand relativ langwellig in y, während ihre Oszillationsfrequenz $\omega$ zur Plateaumitte hin linear mit y zunimmt. Um $\omega_M$ in der Plateaumitte zu berechnen kann man für große Dicken Gleichung (45) benutzen und schreiben:

$$F\left(\sqrt{\frac{2}{\pi c}}\, y\right) = \frac{1+i}{2} - \sqrt{\frac{c}{2\pi}} \frac{1}{y} \exp\left(i \frac{y}{c} y\right) \tag{48}$$

Die Frequenz des oszillierenden Anteils ist durch $y/c$ gegeben, die sich bei der Betragsquadratbildung der gestreuten Intensität noch verdoppelt,

$$\omega = 2 \frac{y}{c} \tag{49}$$

und in der Plateaumitte für $y_M = A\,c/2$ ausgewertet

$$\omega_M = A \tag{50}$$

ergibt, also vom Gradienten c unabhängig ist! Demnach entsprechen die Oszillationen in der Plateaumitte der Transformation der Kristalldicke in den reziproken Raum.

Die Position $x_n$ des n-ten Maximums von $|F(x)|^2$ ist durch

$$x_n = \sqrt{4\,n - \frac{5}{2}} \tag{51}$$

gegeben, was unter Vernachlässigung von Schwebungen zwischen den beiden Fresnelschen Integralen für die Positionen $y_n$ der ersten Maxima zu

$$y_n = -\sqrt{\left(2\,n - \frac{5}{4}\right)\pi\, c} \tag{52}$$

führt. Im Gegensatz zur Plateaumitte hängen die Oszillationen am Plateaurand ausschließlich vom Gradienten, nicht aber von der Kristalldicke ab. Aus ihnen kann der Gradient direkt abgelesen werden.

## 4.2. Dynamische Beugung

Im Gegensatz zur kinematischen Beugungstheorie basiert die dynamische auf der Lösung der Wellengleichung, je nach Strahlung gemäß Schrödinger oder Maxwell im Kristallpotential und liefert außerhalb der kinematischen Näherung nicht nur quantitativ bessere, sondern auch neue Erkenntnisse. Das für die vorliegende Arbeit wohl wichtigste, grundlegende Ergebnis ist die Beschreibung der primären Extinktion: Wenn ein, in einem Kristall fortschreitender Primärstrahl braggreflektiert wird, verliert dieser zugunsten des Sekundärstrahls an Intensität. Dieser Abschwächung entspricht eine Extinktionstiefe $\Delta$, unterhalb derer quasi kein Primärstrahl mehr vorhanden ist. Dadurch wird das reflektierende Kristallvolumen beschränkt.
Dieser Sachverhalt ist für die Berechnung der reflektierten Intensitäten wichtig, die ja schon wegen der Teilchenzahlerhaltung nicht wie in der kinematischen Theorie dem Streuvolumen proportional ins Unermeßliche ansteigen dürfen. Durch die reziproke Extinktionslänge $2\pi/\Delta$ ist die natürliche Linienbreite des Braggreflexes am Idealkristall gegeben. Letztendlich wird, wie schon an früherer Stelle erwähnt, durch Extinktionslänge und natürliche Linienbreite ein optimaler Gradient definiert.

Im folgenden Kapitel sollen die für diese Arbeit wichtigsten Grundlagen und Ergebnisse der dynamischen Beugungstheorie zusammengestellt werden, um als Basis für den im Anschluß entwickelten Transfermatrizenalgorithmus zu dienen.

### 4.2.1. Grundlagen der dynamischen Beugung von Neutronen

Dieses Kapitel sowie die hier gebrauchten Bezeichnungen stützen sich stark auf die Arbeit von H. Rauch und D. Petrascheck [12, 13], die sich wiederum an die Nomenklatur des Buches über Röntgenstreuung von Zachariasen [14] lehnt.
Die dynamische Streutheorie behandelt die Lösung der stationären Schrödingergleichung

$$\left\{-\frac{\hbar^2}{2\,m}\vec{\partial}^2 + V(\vec{r}) - E\right\}\psi = 0 \qquad (53)$$

(bzw. Maxwellgleichungen für Röntgenstrahlen) im Kristallpotential, dem Fermi-Pseudopotential

$$V(\vec{r}) = \frac{2\,\pi\,\hbar^2}{m}\, b_c \sum_j \delta(\vec{r} - \vec{r}_j)\,. \qquad (54)$$

Wegen der Gitterperiodizität können Wellenfunktion $\psi$ und Potential V in Fourierreihen entwickelt werden. Die Wellenvektoren nichtverschwindender Entwicklungskomponenten unterscheiden sich nur durch reziproke Gittervektoren $\vec{G}$, wobei zunächst alle Vektoren des reziproken Gitters mitgeführt werden. Aus einem Blochwellenansatz

$$\psi(\vec{r}) = e^{i \vec{K} \vec{r}}\, u(\vec{r}) \qquad (55)$$

ergibt sich das algebraische Grundgleichungssystem der Beugungstheorie

$$\left\{ \frac{\hbar^2}{2\,m} |\vec{K} + \vec{G}|^2 - E \right\} u(\vec{G}) = - \sum_{\vec{G}'} V(\vec{G} - \vec{G}')\, u(\vec{G}') \qquad (56)$$

mit den Fourierkomponenten

$$V(\vec{G}) = \frac{2\,\pi\,\hbar^2\, b_c\, F_{hkl}}{m\, V_z} \quad \text{bzw.} \quad \frac{V(\vec{G})}{E} = \frac{4\,\pi\, b_c\, F_{hkl}}{k^2\, V_z} \qquad (57)$$

des Pseudopotentials und dem Volumen $V_z$ der Elementarzelle. Dieses Gleichungssystem beinhaltet unendlich viele, gleichwertige Wellen und ist nicht explizit lösbar. Deshalb werden Näherungen gemacht: Die Einstrahlnäherung behandelt nur eine angeregte Welle, die vorwärtsgebeugte. Sie gilt dann, wenn man weit von jeder Braggbedingung entfernt ist. Aus ihr folgt das Snell'sche Brechungsgesetz. Die Wellenzahlen $K_0$ im Medium und k im Vakuum unterscheiden sich im Verhältnis um ε, d. h.

$$K_0 = k\,(1 + \varepsilon) = k \left(1 - \frac{V(0)}{E}\right). \qquad (58)$$

Die Zweistrahlnäherung berücksichtigt zwei nichtverschwindende Kristallwellen, eine vorwärtsgebeugte (≈ einfallende) zum reziproken Gittervektor $\vec{0}$, sowie eine braggebeugte zum reziproken Gittervektor $\vec{G}$ gehörende. Von den Grundgleichungen bleiben nur noch zwei übrig, nämlich

$$\begin{aligned} \left\{ \frac{\hbar^2}{2\,m} \left|\vec{K}\right|^2 - E \right\} u(\vec{0}) &= - V(\vec{0})\, u(\vec{0}) - V(-\vec{G})\, u(\vec{G}) \quad ; \quad \text{für } \vec{G} = \vec{0} \\ \left\{ \frac{\hbar^2}{2\,m} \left|\vec{K} + \vec{G}\right|^2 - E \right\} u(\vec{G}) &= - V(\vec{G})\, u(\vec{0}) - V(\vec{0})\, u(\vec{G}) \quad ; \quad \text{für } \vec{G} \neq \vec{0} \quad . \end{aligned} \qquad (59)$$

Die Wellenzahlen im Kristall sind dann

$$K = k\,(1+\varepsilon)$$
$$\left|\vec{K}+\vec{G}\right| = k\,(1+\varepsilon_G) \tag{60}$$

mit den Anregungsfehlern $\varepsilon$ und $\varepsilon_G$. Setzt man diese Beziehungen in die Grundgleichungen ein und drückt durch Geometriebetrachtungen von Kristalloberfläche $\vec{n}$, einfallender Welle $\vec{K}$ und Lage $\vec{G}$ der Gitterebenen $\varepsilon_G$ durch $\varepsilon$ aus, so erhält man eine quadratische Gleichung in $\varepsilon$ mit den Lösungen $\varepsilon_1$ und $\varepsilon_2$. Die Schrödingergleichung erhält zu jedem $\varepsilon_n$, n = 1,2 eine Lösung, d. h. einen Eigenzustand:

$$\psi_n = \exp\left(i\,(\vec{k}+\kappa_n\,\vec{n})\,\vec{r}\right)\left\{u_n(0)+u_n(\vec{G})\exp\left(i\,\vec{G}\,\vec{r}\right)\right\}\;;\;n=1,2\,. \tag{61}$$

mit

$$\kappa_n = \frac{k\,\varepsilon_n}{\cos(\gamma_0)}\;;\;n=1,2\,. \tag{62}$$

Hier ist $\gamma$ der Winkel zwischen der Oberflächennormalen $\vec{n}$ und dem Wellenvektor $\vec{k}$, und die $u_n$ sind die zu den $\varepsilon_n$ gehörenden Blochwellenamplituden.

Aus den elementaren Grundgleichungen (59) können die Amplitudenverhältnisse

$$X_n := \frac{u_n(\vec{G})}{u_n(\vec{0})} = -\frac{2\,\varepsilon_n + \frac{V(\vec{0})}{E}}{\frac{V(-\vec{G})}{E}}\;;\;n=1,2 \tag{63}$$

zwischen den abgebeugten und vorwärtsgebeugten Anteilen abgeleitet werden. Die Wellen in Vorwärtsrichtung zu $\vec{G}=\vec{0}$ sowie für den reflektierten Strahl sind dann durch

$$\begin{aligned}\Psi_0 &= \exp\left(i\,\vec{k}\,\vec{r}\right)\left\{u_1(\vec{0})\exp\left(i\,\kappa_1\,\vec{n}\,\vec{r}\right)+u_2(\vec{0})\exp\left(i\,\kappa_2\,\vec{n}\,\vec{r}\right)\right\}\\ \Psi_G &= \exp\left(i\left(\vec{k}+\vec{G}\right)\vec{r}\right)\left\{X_1\,u_1(\vec{0})\exp\left(i\,\kappa_1\,\vec{n}\,\vec{r}\right)+X_2\,u_2(\vec{0})\exp\left(i\,\kappa_2\,\vec{n}\,\vec{r}\right)\right\}\end{aligned} \tag{64}$$

gegeben. Die Superposition der zu $\kappa_1$ und $\kappa_2$ gehörenden Eigenzustände führt zu den bekannten Pendellösungseigenschaften.

Die typische Anwendung der Zweistrahlnäherung sind Kristalle mit zwei planparallelen Oberflächen, von denen eine auch ins Unendliche rücken kann. Die Amplitudenfunktionen $u_n(\vec{0})$, $u_n(\vec{G})$ knüpfen phasengerecht durch die Randbedingungen der Oberflächen an die Amplituden im Vakuum $u_0(\vec{0})$, $u_0(\vec{G})$, deren Betragsquadrate unter Berücksichtigung von

Projektionsfaktoren die Intensitätsverteilungen R und T des reflektierten und des transmittierten Strahls geben.
Üblicherweise führt man hier die dimensionslosen Parameter

$$A := \frac{k}{2}\,\frac{1}{\sqrt{|\cos(\gamma_0)\cos(\gamma_G)|}}\left|\frac{V(\vec{G})}{E}\right| D \qquad (65)$$

für die Kristalldicke und

$$y := \frac{(b-1)\,\frac{V(\vec{0})}{E} + \alpha\, b}{2\sqrt{|b|}\left|\frac{V(\vec{G})}{E}\right|} \qquad (66)$$

für die Abweichung von der Braggposition ein. Hierin ist

$$b := \frac{\cos(\gamma_0)}{\cos(\gamma_G)} \qquad (67)$$

das Verhältnis der Richtungskosinusse der Strahlen zu den Oberflächen. Je nach Vorzeichen von b unterscheidet man zwischen Lauefall für b > 0, in dem der reflektierte Strahl durch die Kristallrückseite austritt und Braggfall für b < 0, in dem der reflektierte Strahl wieder durch die Eintrittsoberfläche das Kristallmedium verläßt. Sie sind in Abbildung (7) schematisch dargestellt. Speziell gibt es die symmetrischen Fälle, bei denen die streuenden Gitterebenen senkrecht (b = 1) oder parallel zur Oberfläche (b = -1) liegen.

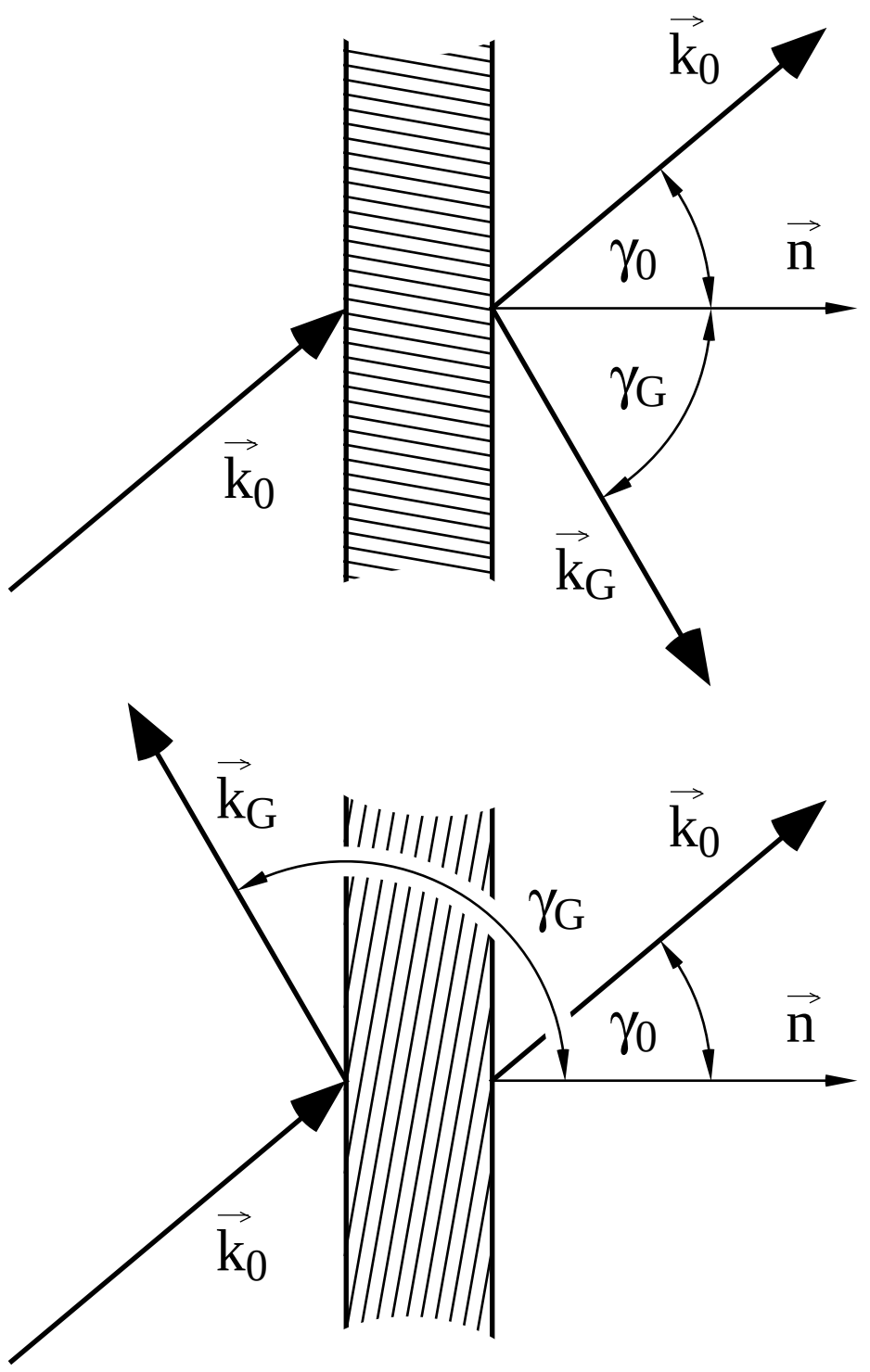

*Abbildung (7): Darstellung der Fallunterscheidung für Laue-, oben, und Braggeometrie, unten. Je nach Lage der Gitterebenen zu den Oberflächen kann der reflektierte Strahl mit $\vec{k}_G$ an der Kristallrückseite oder -Vorderseite aus dem Medium austreten.*

Der Parameter

$$\alpha := \left(\frac{\vec{G}}{k} + 2\,\frac{\vec{k}}{k}\right)\cdot\frac{\vec{G}}{k} \qquad (68)$$

beschreibt die Abweichung von der geometrischen Braggposition. Seine Konstruktion im reziproken Raum ist in Abbildung (8) wiedergegeben: α entspricht gerade der Projektion der Variation des Wellenvektors $\Delta\vec{k}/k$ auf den normierten reziproken Gittervektor $\vec{G}/k$. Ganz gleich, ob sich $\vec{k}$ in Länge oder Richtung ändert, beschreibt die Projektion immer eine zu $\vec{G}$ longitudinale Abweichung. Drücken wir

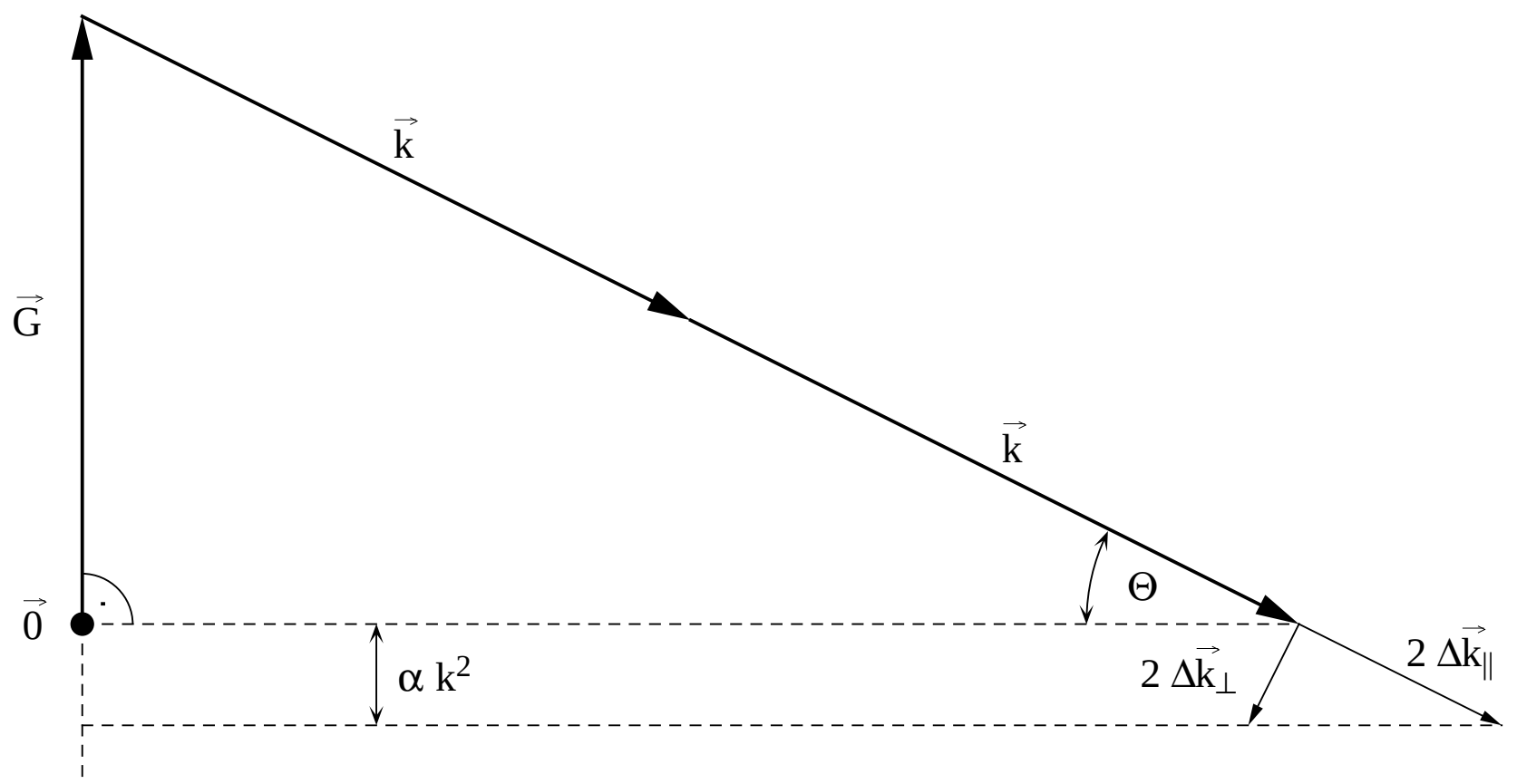

*Abbildung (8):*
*Konstruktion der Abweichung $\alpha$ von der geometrischen Braggposition. $\vec{k}$ kann in longitudinaler oder transversaler Richtung variiert werden, wichtig ist nur die Projektion von $\Delta\vec{k}$ auf den reziproken Gittervektor $\vec{G}$.*

$$\vec{k} = \vec{k}_B + \Delta\vec{k} \tag{69}$$

durch den, die geometrische Braggbedingung erfüllenden Vektor $\vec{k}_B$ und deren Abweichung $\Delta\vec{k}$ aus, so läßt sich Gleichung (68) in

$$\alpha = 2\,\frac{\Delta\vec{k}\,\vec{G}}{k^2} \tag{70}$$

umformen. Die Variation $\Delta\vec{k}$ kann im wesentlich auf zwei verschiedene Arten, nämlich einer transversalen Änderung, bei der die Länge von $\vec{k}$ festgehalten und die Einfallsrichtung um den Winkel $\Delta\Theta$ geändert wird, oder einer longitudinalen Änderung, bei der nicht die Richtung, sondern die Wellenzahl durchgestimmt wird, geschehen. Im ersten Fall wird

$$\Delta\vec{k} = \Delta\vec{k}_\perp\ ;\ \ \Delta k/k = \Delta\Theta \tag{71}$$

in die Gleichung (70) eingesetzt und, zusammen mit dem Braggesetz, $\Theta$ durch $\vec{G}$ und $\vec{k}$ ausgedrückt. Dies führt über die Rechenschritte

$$\alpha = 2\,\frac{\vec{G}}{k}\cdot\frac{\Delta\vec{k}_\perp}{k} = 2\,\frac{G\,\Delta k_\perp}{k^2}\cos(\Theta) = 4\sin(\Theta)\cos(\Theta)\,\Delta\Theta$$

zu dem Ergebnis

$$\alpha = 2\sin(2\,\Theta)\,\Delta\Theta\,. \qquad (\perp) \tag{72}$$

Auf die gleiche Weise erhält man bei Variation in longitudinaler Richtung

$$\Delta\vec{k} = \Delta k_{\parallel} = \vec{k}\,\frac{\Delta k}{k} \tag{73}$$

und

$$\alpha = -\frac{G^2}{k^2}\frac{\Delta k_{\parallel}}{k} = -4\,\frac{\Delta k_{\parallel}}{k}\sin^2(\Theta)\,. \qquad (\parallel) \tag{74}$$

Für die Beschreibung von Gradientenkristallen ist es auch wichtig, die Variation des reziproken Gittervektors mit $\alpha$ in Verbindung zu bringen, wodurch sich zu Gleichung (70) ein weiterer Term addiert:

$$\alpha = \frac{G^2}{k^2}\frac{\Delta G}{G} + 2\,\frac{\Delta\vec{k}\,\vec{G}}{k^2}\,. \tag{75}$$

Durch die Größen y und A werden die Reflektions- und Transmissionskurven R(y, A) und T(y, A) universell, das heißt unabhängig von den Materialeigenschaften und der Art der Variation. Es wird sich herausstellen, daß A in Einheiten von $\Delta/\pi$ der Pendellösungsperiode oder Extinktionslänge $\Delta$, und y in Einheiten der halben Darwinplateaubreite, also der halben natürlichen Linienbreite, mißt.

In Gleichung (66) setzt sich y aus zwei Summanden im Zähler zusammen. Der erste berücksichtigt die Snell'sche Brechung, also eine Verschiebung der geometrischen Braggposition, der zweite eben die Variation um die Braggeometrie.

Mit den so definierten Größen und

$$V(-\vec{G}) = V^*(\vec{G}) \tag{76}$$

für den Fall ohne Absorption lassen sich

$$\varepsilon_n = \frac{1}{2}\sqrt{|b|}\left|\frac{V(\vec{G})}{E}\right|\left(-y \pm \sqrt{y^2 \pm 1}\right) - \frac{1}{2}\frac{V(\vec{0})}{E}\;;\; n = 1, 2 \tag{77}$$

$$\underset{n=1,\,2}{\uparrow} \qquad \underset{\substack{+:\, b>0 \\ -:\, b<0}}{\uparrow}$$

$$X_n = \sqrt{|b|}\,\frac{V(\vec{G})}{\left|V(\vec{G})\right|}\left(-y \pm \sqrt{y^2 \pm 1}\right)\;;\; n = 1, 2 \tag{78}$$

$$X_1X_2 = -b\left(\frac{V(\vec{G})}{|V(\vec{G})|}\right)^2 = -b\,\frac{V(\vec{G})}{V(-\vec{G})} \qquad (79)$$

$$X_2 - X_1 = -2\,\sqrt{|b|}\,\frac{V(\vec{G})}{|V(\vec{G})|}\,\sqrt{y^2 \pm 1} \qquad (80)$$

ausdrücken.

Zur Berechnung der Reflektionskurven werden die Verhältnisse der reflektierten zur einfallenden Welle gebildet und quadriert. Die so erhaltenen Intensitäten

$$\frac{I_G}{I_e} = \left|\frac{\psi_G}{\psi_e}\right|^2 \qquad (81)$$

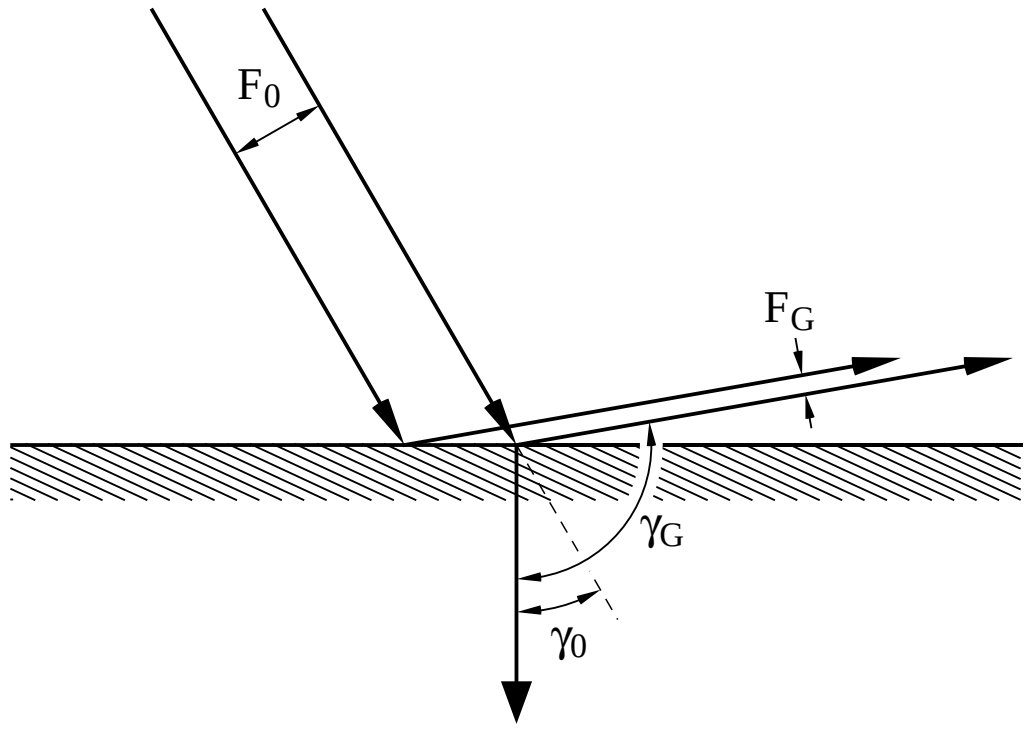

*Abbildung (9):*
*Bei asymmetrischer Reflektion ändert sich der Strahlquerschnitt um 1 / |b|, dem in der Intensitätsberechnung Aufmerksamkeit geschenkt werden muß.*

beziehen sich jedoch auf gleichbleibende Strahlquerschnitte. Weil dieser sich bei asymmetrischer Reflektion $|b| \neq 1$ ändert, skalieren die für die Intensitätsmessungen wichtigen Größen

$$R = \frac{P_G}{P_e} = \frac{1}{|b|}\frac{I_G}{I_e} \text{ und } T = \frac{P_0}{P_e} = \frac{I_0}{I_e} \qquad (82)$$

mit dem Verhältnis ihrer Querschnitte gemäß Abbildung (9).

Die Reflektions- und Transmissionskurven für Laue- und Braggeometrie ergeben sich zu

$$R_L(y, A) = \frac{\sin^2\left(A\sqrt{1+y^2}\right)}{1+y^2}; \qquad T_L(y, A) = 1 - R_L(y, A) \qquad (83)$$

und

$$R_B(y, A) = \frac{1}{y^2 + (y^2-1)\cot^2\left(A\sqrt{y^2-1}\right)}; \qquad T_B(y, A) = 1 - R_B(y, A)\,. \qquad (84)$$

Sie sind in Abbildung (10) graphisch dargestellt, während Abbildung (11) einige Querschnitte für feste A zeigt.

Bei sehr dünnen Kristallen findet man in beiden Fällen sehr flache, aber breite Kurven in y, die durch ein zentrales Hauptmaximum und seitliche Nebenmaxima ausgezeichnet sind. In beiden Fällen erhält man die gemeinsame Reflektionskurve

Kristall in Laueposition

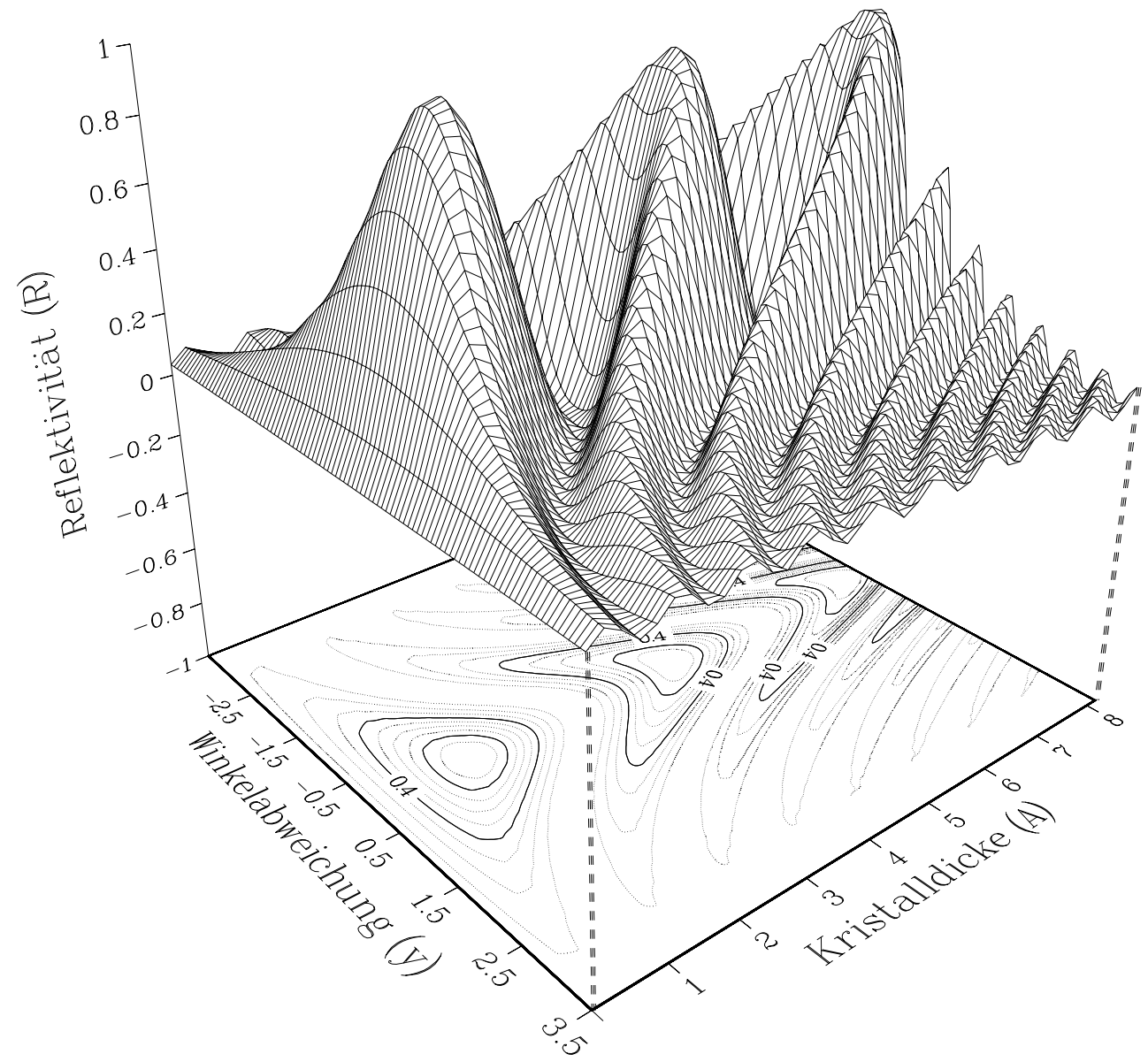

Kristall in Braggposition

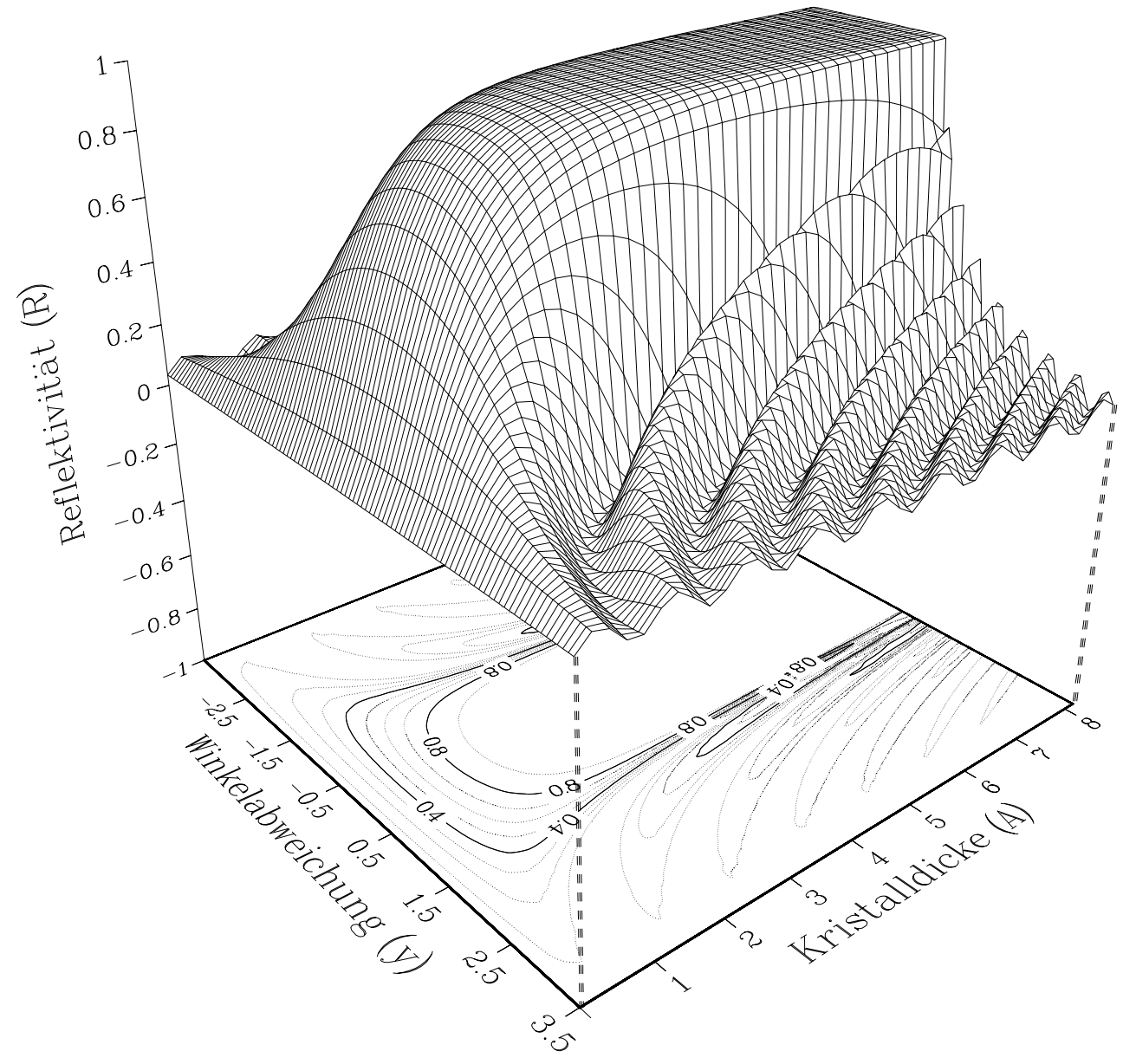

*Abbildung (10):*
*Reflektionskurven in Abhängigkeit der Kristalldicke A und der Abweichung von der geometrischen Braggposition y für Lauegeometrie, oben und Braggeometrie, unten.*

$$R(y, A) = \left(\frac{\sin(A\,y)}{y}\right)^2 \; ; A \ll 1 \,. \tag{85}$$

Das Hauptmaximum der Reflektionskurve verbreitert sich mit 1/A. Betrachten wir den Extremfall eines Kristalls der Dicke nur einer Gitterebene, so bleibt dessen Reflektivität auch noch in weit von der Braggposition entfernten Bereichen quasi konstant gleich $A^2$. Dies wurde, bis auf einen Phasenfaktor, in Kapitel 4.1. über die kinematische Beugung gebraucht, worin mittels $D = d$ in Gleichung (65), die Reflektionsamplitude

$$A_1 := \frac{k}{2} \frac{1}{\sqrt{|\cos(\gamma_0)\cos(\gamma_G)|}} \left|\frac{V(\vec{G})}{E}\right| d \tag{86}$$

einer einzelnen Gitterebene angesetzt wurde.

Mit anwachsender Kristalldicke rücken die Nebenmaxima immer enger zusammen, bis sie schließlich so nahe beieinander liegen, daß sie nicht mehr auflösbar sind. In diesem Fall kann über die schnell oszillierenden Anteile gemittelt werden, was zu den Reflektionskurven für unendlich dicke Kristalle, einer Lorentzkurve

$$R_L(y, \infty) = \frac{1}{2\left(1 + y^2\right)} \tag{87}$$

im Lauefall und der Darwinkurve

$$R_B(y, \infty) = \begin{cases} 1 & \text{für} \quad |y| < 1 \\ 1 - \sqrt{1 - \dfrac{1}{y^2}} & \text{für} \quad |y| \geq 1 \end{cases} \tag{88}$$

im Braggfall führt. Der frappierende Unterschied zwischen beiden Fällen sind die Maximal- und damit auch die Gesamtreflektivitäten. Während im Braggfall 100 % erreicht werden, steigt die Kurve im Lauefall auf nur 50 % an. Man erhält für die integrierten Reflektivitäten bei $A = \infty$

$$\Re_L(\infty) = \int_{-\infty}^{\infty} R_L(y, \infty)\, dy = \frac{\pi}{2} \tag{89}$$

und

$$\Re_B(\infty) = \int_{-\infty}^{\infty} R_B(y, \infty)\, dy = \pi \,. \tag{90}$$

Dieser Intensitätsverlust im Lauefall hängt damit zusammen, daß die abgebeugte Welle $\psi_G$ zunächst mit zunehmender Kristalldicke anwächst, in tieferen Schichten maximal wird, dabei mit der vorwärtsgebeugten Welle $\psi_0$ konkurriert und in diese wieder Intensität zurückstreut. Dem entspricht ein hin- und herpendeln der Intensität zwischen Primär- und Sekundärstrahl. Betrachtet man das Zentrum

$$R_L(0, A) = \sin^2(A) \tag{91}$$

der Lauekurve als Funktion der Kristalldicke A, so erhält man eine Periodizität von

$$A_\Delta = \pi \,. \tag{92}$$

Mittels (65) in metrische Einheiten umgerechnet bedeutet dies für die Pendellösungsperiode

$$\Delta = \frac{2\,\pi}{k} \sqrt{|\cos(\gamma_0)\cos(\gamma_G)|}\,\frac{1}{\left|\frac{V(\vec{G})}{E}\right|} \,. \tag{93}$$

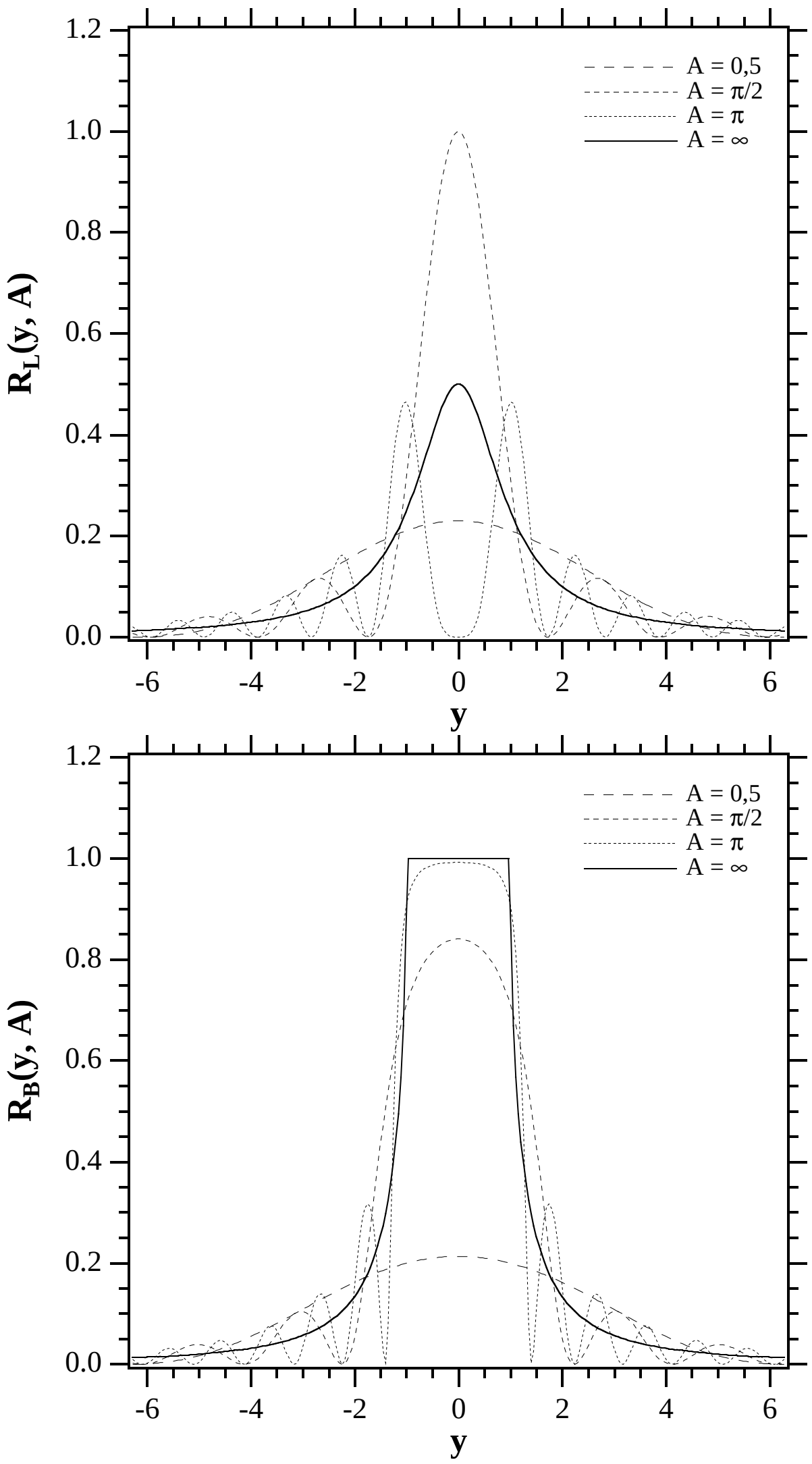

*Abbildung (11):*
*Laue- (oben) und Braggreflektionskurven (unten) für verschiedene Kristalldicken A. Während im Braggfall die Gesamtreflektivität stetig mit A ansteigt, oszilliert diese im Fall der Lauestellung.*

Im Braggfall gilt

$$R_B(0, A) = \tanh^2(A) \tag{94}$$

und

$$\Re_B(A) = \int_{-\infty}^{\infty} R_B(y, A)\, dy = \pi \tanh(A) \,. \tag{95}$$

Setzt man für A die Pendellösungsperiode $\pi$ in diese Formeln ein, so ist $\Re_B(\pi)$ gerade auf 100 % angestiegen, wodurch begründet wird, daß in der Dicke eine Pendellösungsperiode an Kristallmaterial gebraucht wird, um volle Reflektivität zu erhalten. Unterhalb von A = 1 fällt die Braggreflektivität quasi linear mit der Kristalldicke ab, was den Gültigkeitsbereich einer kinematischen Streutheorie characterisiert.

### 4.2.2. Eine Transfermatrizenmethode in der dynamischen Beugungstheorie

Zur Beschreibung der Reflektionseigenschaften geschichteter, optischer Medien werden seit Aufkommen leistungsfähiger Rechenanlagen mehr und mehr Transfermatrixalgorithmen herangezogen [15]. Sehr bequem erweist sich diese Methode für die Berechnung von Schichtsystemen, dessen Lagen sich durch unterschiedliche Brechungsindizes voneinander unterscheiden. Für Neutronen wurde dies zum Beispiel bei den Reflektionskurven von Superspiegeln angewandt [16].

Die Idee dieser Methode ist es, wie in Abbildung (12) dargestellt, die als Spaltenvektoren geschriebenen, unabhängigen Lösungen $\psi_0^{N+1}$ und $\psi_G^{N+1}$ hinter einem homogenen Medium durch ein Produkt aus einer $2 \times 2$ Matrix $\mathfrak{C}$ und den Lösungen $\psi_0^0$ und $\psi_G^0$ vor diesem Medium auszudrücken:

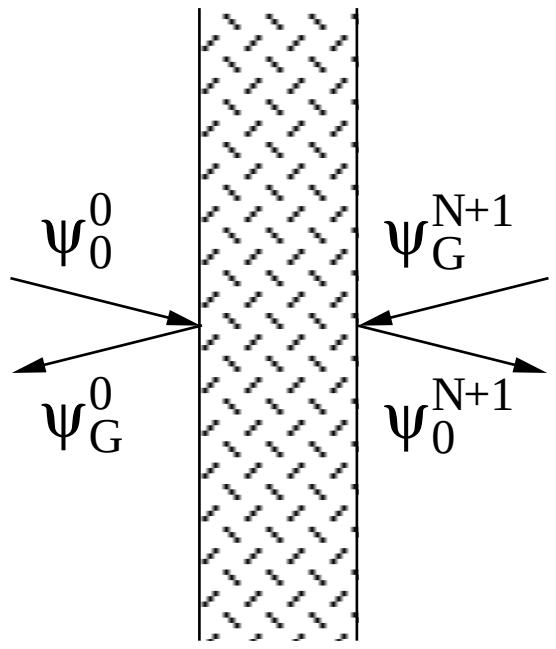

*Abbildung (12): Ebene, monochromatische Wellen vor und hinter einem planparallelen, homogenen Medium.*

$$\begin{bmatrix} \psi_0^{N+1} \\ \psi_G^{N+1} \end{bmatrix} = \mathfrak{C} \begin{bmatrix} \psi_0^0 \\ \psi_G^0 \end{bmatrix} \quad \text{mit} \quad \mathfrak{C} = \begin{bmatrix} C_{00} & C_{0G} \\ C_{G0} & C_{GG} \end{bmatrix} \tag{96}$$

Dies ist ganz allgemein möglich, da es sich bei den Wellengleichungen nach Schrödinger oder Maxwell um Differentialgleichungen zweiter Ordnung handelt, und diese überall, also vor, in und hinter dem betrachteten Medium zwei unabhängige Lösungen besitzen. Die auf einem Modell basierende Physik wird in die Transfermatrix hineingesteckt, die im allgemeinen eine Mischung der Eigenzustände durch die Stetigkeitsbedingungen an den Grenzflächen und die Propagation der Welle als Funktion des Ortes beschreibt. Aus globalen Randbedingungen, z. B. daß hinter dem Medium keine reflektierte Welle vorliegt, werden die speziellen Lösungen bestimmt, aus denen man die beobachtbaren Größen, wie Reflektion und Transmission erhält.

In diesem Kapitel soll zum ersten Mal ein Transfermatrizenalgorithmus im Rahmen der dynamischen Streutheorie erstellt werden.

Abbildung (13) skizziert ein geschichtetes Medium von N Lagen. Zählen wir das Vakuum vor und hinter diesem dazu, so werden die Schichten mit dem laufenden Index j der Reihe nach von 0 bis (N+1) durchnumeriert:

$$j \in \{0, N+1\} \,. \tag{97}$$

Größen, die sich auf die Lage j beziehen, werden rechts oben mit diesem Index verziert. Somit ist z. B. jeder Schicht eine Dicke $D^j$ zugewiesen.

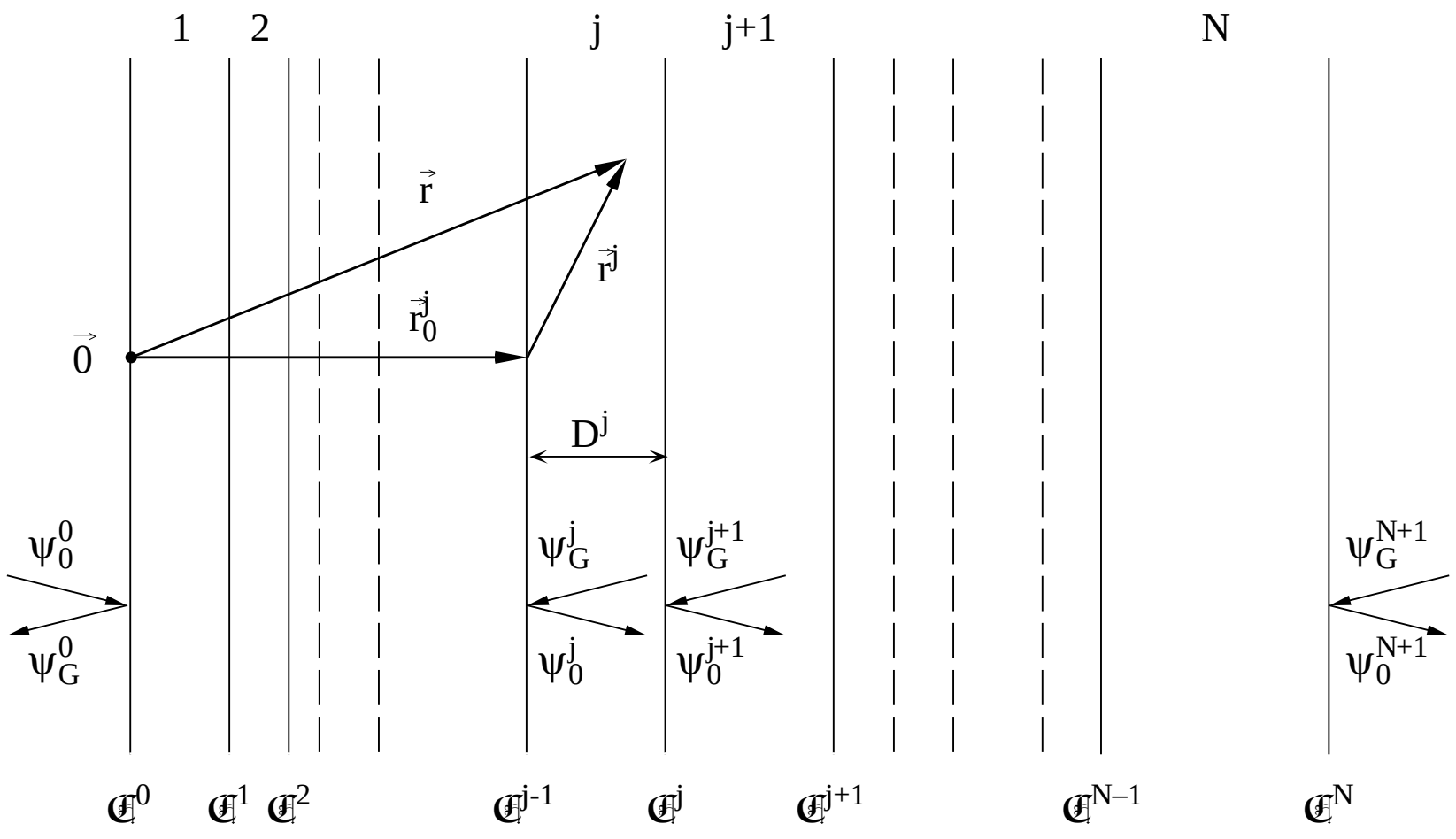

*Abbildung (13):*
*Geschichtetes Kristallmedium*

Der Ursprung $\vec{0}$ des Koordinatensystems kann ohne Beschränkung der Allgemeinheit auf der Eintrittsoberfläche des Mediums gewählt werden. Ortsvektoren $\vec{r}$ werden in ihre Anteile $\vec{r}_0^{\,j}$ entlang des gemeinsamen Normalenvektors $\vec{n}$ zur Grenzfläche der Schicht j und dem Anteil $\vec{r}^{\,j}$ innerhalb derselben Schicht zerlegt:

$$\vec{r} = \vec{r}_0^{\,j} + \vec{r}^{\,j} \quad \text{mit} \quad \vec{r}_0^{\,1} = \vec{0} \quad \text{und} \quad \vec{r}_0^{\,j} = \vec{n} \sum_{\nu = 1}^{j-1} D^{\nu} \,. \tag{98}$$

Die beiden linear unabhängigen Wellenfunktionen in jeder Schicht werden mit $\psi_0^j$ in vorwärtsgebeugter und $\psi_G^j$ in abgebeugter Richtung bezeichnet. Gemäß (96) beschreibt eine Matrix $\mathfrak{T}^j$ die Beziehung zwischen den Wellenfunktionen

$$\begin{bmatrix} \psi_0^{j+1} \\ \psi_G^{j+1} \end{bmatrix} = \mathfrak{T}^j \begin{bmatrix} \psi_0^{j} \\ \psi_G^{j} \end{bmatrix} . \tag{99}$$

Dabei werden Amplitude und Phase jeweils im Medium j und an der Grenzfläche zwischen j und (j-1) gemessen, oder in Formeln ausgedrückt:

$$\begin{bmatrix} \psi_0^{j} \\ \psi_G^{j} \end{bmatrix} = \begin{bmatrix} \psi_0^{j}(\vec{r}_0^{\,j}) \\ \psi_G^{j}(\vec{r}_0^{\,j}) \end{bmatrix} . \tag{100}$$

Für die Gesamtmatrix werden alle Einzelmatrizen aneinandermultipliziert, so daß wir

$$\begin{bmatrix} \psi_0^{N+1} \\ \psi_G^{N+1} \end{bmatrix} = \overrightarrow{\prod_{\nu=0}^{N}} \mathfrak{C}^{\nu} \begin{bmatrix} \psi_0^{0} \\ \psi_G^{0} \end{bmatrix} \qquad (101)$$

erhalten. Der Pfeil über dem Produktzeichen bedeutet eine sukzessive Multiplikation von links.

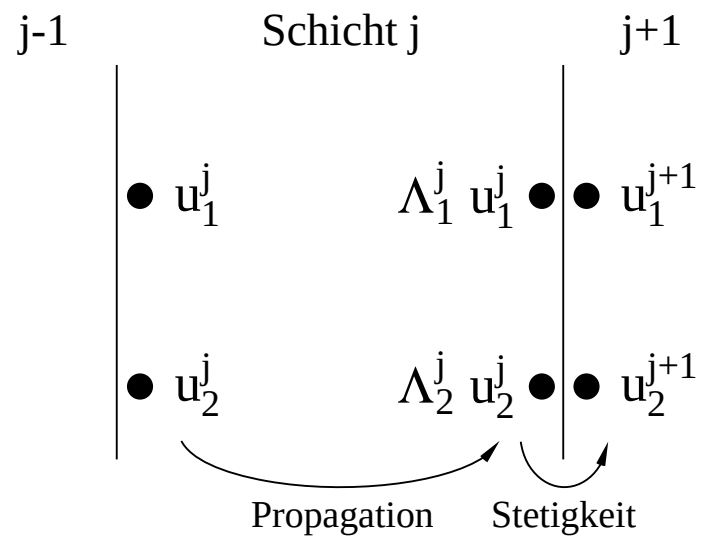

*Abbildung (14): Blochwellenamplituden $u_n^j$ beim Übergang von der Schicht j zur Schicht j+1. Die $u_n^j$ an der linken Grenzfläche propagieren durch das Medium und ergeben die $\Lambda_n^j u_n^j$ vor der rechten Grenzfläche. Dort wird die Stetigkeitsbedingung gefordert, um die $u_n^{j+1}$ hinter dieser Grenzfläche zu erhalten.*

Wie schon erwähnt, sollen die Medien in den einzelnen Lagen homogen sein. Für uns bedeutet dies, daß jede Schicht einen Idealkristall darstellt. Verschiedene Lagen können sich durch verschiedene Kristallpotentiale, das heißt, bei fester, einfallender Welle, durch unterschiedliche Abweichungen $y^j$ von der jeweils geometrischen Braggposition unterscheiden. Davon ausgehend, gibt es in jeder Lage einen individuellen, durch Gleichung (64) beschriebenen Satz von Lösungen

$$\begin{aligned} \Psi_0^j(\vec{r}) &= \exp\left(i\,\vec{k}\,\vec{r}\right) \left\{ u_1^j(\vec{0}) \exp\left(i\,\kappa_1^j\,\vec{n}\,\vec{r}^j\right) + u_2^j(\vec{0}) \exp\left(i\,\kappa_2^j\,\vec{n}\,\vec{r}^j\right)\right\} \\ \Psi_G^j(\vec{r}) &= \exp\left(i\left(\vec{k}+\vec{G}\right)\vec{r}\right) \left\{ X_1\, u_1^j(\vec{0}) \exp\left(i\,\kappa_1^j\,\vec{n}\,\vec{r}^j\right) + X_2\, u_2^j(\vec{0}) \exp\left(i\,\kappa_2^j\,\vec{n}\,\vec{r}^j\right)\right\} , \end{aligned} \qquad (102)$$

der an der Grenzfläche

$$\vec{n}\,\vec{r}^j = D^j \quad \text{und} \quad \vec{n}\,\vec{r}^{j+1} = 0 \qquad (103)$$

zwischen j und (j+1) die Stetigkeitsbedingung

$$\Psi_H^j\left(D^j\right) = \Psi_H^{j+1}(0) \ , \ \vec{H} = \vec{0}, \vec{G} \qquad (104)$$

erfüllen muß. Es kann allgemein gezeigt werden, daß hier die Stetigkeitsforderung der Ableitung vernachlässigbar ist. Diese führte zu den optischen Reflektionseigenschaften an der Grenzfläche, die wegen des kleinen Kristallpotentials und der gemäß (58) verbundenen geringen Abweichung des Brechungsindexes von 1 um viele Größenordnungen nicht ins Gewicht fällt.

Somit folgt mit der Abkürzung

$$\Lambda_n^j := \exp\left(i\,\kappa_n^j\,D^j\right) \qquad (105)$$

$$\begin{aligned} &\text{für } \psi_0: & u_1^j\,\Lambda_1^j + u_2^j\,\Lambda_2^j &= u_1^{j+1} + u_2^{j+1} \\ &\text{und } \psi_G: & X_1^j\,u_1^j\,\Lambda_1^j + X_2^j\,u_2^j\,\Lambda_2^j &= X_1^{j+1}\,u_1^{j+1} + X_2^{j+1}\,u_2^{j+1} , \end{aligned} \qquad (106)$$

Dieser Zusammenhang der Blochwellenamplituden $u_n^j$ soll nochmal in Abbildung (14) verdeutlicht werden. Obiges Gleichungssystem läßt sich nach $u_1^{j+1}$ und $u_2^{j+1}$ auflösen, so daß man eine Transfermatrizengleichung

$$\begin{bmatrix} u_1^{j+1} \\ u_2^{j+1} \end{bmatrix} = \frac{1}{\left(X_2^{j+1} - X_1^{j+1}\right)} \begin{bmatrix} \left(X_2^{j+1} - X_1^j\right) & \left(X_2^{j+1} - X_2^j\right) \\ -\left(X_1^{j+1} - X_1^j\right) & -\left(X_1^{j+1} - X_2^j\right) \end{bmatrix} \begin{bmatrix} \Lambda_1^j & 0 \\ 0 & \Lambda_2^j \end{bmatrix} \begin{bmatrix} u_1^j \\ u_2^j \end{bmatrix} \tag{107}$$

für die $u_n^j$ erhält. Die Matrix

$$\mathfrak{T}^j = \mathfrak{X}^j \, \mathfrak{L}^j \tag{108}$$

wurde als Produkt zweier Einzelmatrizen

$$\mathfrak{X}^j = \frac{1}{\left(X_2^{j+1} - X_1^{j+1}\right)} \begin{bmatrix} \left(X_2^{j+1} - X_1^j\right) & \left(X_2^{j+1} - X_2^j\right) \\ -\left(X_1^{j+1} - X_1^j\right) & -\left(X_1^{j+1} - X_2^j\right) \end{bmatrix} \tag{109}$$

und

$$\mathfrak{L}^j = \begin{bmatrix} \Lambda_1^j & 0 \\ 0 & \Lambda_2^j \end{bmatrix} \tag{110}$$

geschrieben. Die Diagonalmatrix $\mathfrak{L}^j$ beschreibt die Propagation der Phasen in der Lage j, wenn man von ihrer Eintrittsfläche um die Dicke $D^j$ zur Austrittsfläche fortschreitet. $\mathfrak{X}^j$ hingegen setzt sich aus den Differenzen der Amplitudenverhältnisse (63) zwischen den Medien (j+1) und j, normiert auf (j+1) zusammen. Sind beide Schichten gleich, so erhalten wir für $\mathfrak{X}^j$ die Einheitsmatrix, das heißt, die $u_n^j$ spüren keine Grenzfläche, sondern ändern sich nur durch $\mathfrak{L}$ in ihrer Phase. Schaltet man nun einen kleinen Potentialunterschied zwischen beiden Lagen ein, so bleiben die Diagonalmatrixelemente von $\mathfrak{X}^j$ in nullter Näherung unverändert, während Außerdiagonalelemente hinzukommen. Letztere beschreiben eine Mischung der zu $u_1^j$ und $u_2^j$ gehörenden Zustände.

Als nächstes sollen die Matrizen $\mathfrak{X}^j$ und $\mathfrak{L}^j$ vereinfacht und durch die wohlbekannten Größen $y^j$ und $A^j$ ausgedrückt werden. Für die Behandlung von $\mathfrak{X}^j$ zerlegen wir den Ausdruck (78) in

$$X_n^j = \sqrt{|b|} \, \frac{V^j\left(\vec{G}\right)}{\left|V^j\left(\vec{G}\right)\right|} \, \tilde{X}_n^j \tag{111}$$

mit

$$\tilde{X}_n^j = -y^j \pm \sqrt{y^{j2} \pm 1}\ . \tag{112}$$

Da der Vorfaktor $\sqrt{|b|}\, V^j(\vec{G}) \,/\, |V^j(\vec{G})|$ für alle n und j derselbe ist, kürzt sich dieser und wir können die $X_n^j$ in $\mathfrak{X}^j$ durch die $\tilde{X}_n^j$ ersetzen. Der in der Phasenmatrix vorkommende Term

$$\phi_n := \kappa_n^j\, D^j \tag{113}$$

kann mittels (62), (77) und der Definition (65) von A in

$$\phi_n^j = \left\{ -y^j \pm \sqrt{y^{j2} \pm 1} - \frac{1}{|b|} \frac{V^j(\vec{G})}{|V^j(\vec{G})|} \right\} A^j \tag{114}$$

umgeschrieben und in die beiden Anteile

$$\phi_n^j = \tilde{\phi}_n^j + \hat{\phi}_n^j \tag{115}$$

mit

$$\tilde{\phi}_n^j := \tilde{X}_n^j\, A^j \tag{116}$$

und

$$\hat{\phi}_n^j = -\frac{1}{\sqrt{|b|}} \frac{V^j(\vec{G})}{|V^j(\vec{G})|} A^j \tag{117}$$

zerlegt werden. Dabei hängt $\hat{\phi}_n^j$ nicht von n = 1, 2 ab und liefert im Exponentialausdruck (105) nur einen gemeinsamen Phasenfaktor, der aus der Matrix $\mathfrak{L}^j$ herausgezogen, und mit Hinblick auf eine spätere Betragsquadratbildung fortgelassen werden kann. Wir ersetzen also

$$\mathfrak{L}^j \;\rightarrow\; \tilde{\mathfrak{L}}^j = \exp\left(-i\, \hat{\phi}^j\right) \mathfrak{L}^j = \begin{bmatrix} \tilde{\Lambda}_1^j & 0 \\ 0 & \tilde{\Lambda}_2^j \end{bmatrix} \tag{118}$$

mit

$$\tilde{\Lambda}_n^j = \exp\left(\tilde{X}_n^j\, A^j\right). \tag{119}$$

Somit erhalten wir für die Transfermatrix

$$\mathfrak{T}^j \;\rightarrow\; \tilde{\mathfrak{T}}^j = \tilde{\mathfrak{X}}^j\, \tilde{\mathfrak{L}}^j \tag{120}$$

mit

$$\tilde{\mathfrak{X}}^j = \frac{1}{\left(\tilde{X}_2^{j+1} - \tilde{X}_1^{j+1}\right)} \begin{bmatrix} \left(\tilde{X}_2^{j+1} - \tilde{X}_1^j\right) & \left(\tilde{X}_2^{j+1} - \tilde{X}_2^j\right) \\ -\left(\tilde{X}_1^{j+1} - \tilde{X}_1^j\right) & -\left(\tilde{X}_1^{j+1} - \tilde{X}_2^j\right) \end{bmatrix} \tag{121}$$

und

$$\tilde{\mathfrak{L}}^j = \begin{bmatrix} \tilde{\Lambda}_1^j & 0 \\ 0 & \tilde{\Lambda}_2^j \end{bmatrix} . \tag{122}$$

Bisher haben wir die Transfermatrizen für die Beschreibung der Blochwellenamplituden $u_n^j$ im Kristallmedium erstellt. Beim Übergang zum Vakuum kombinieren diese Zustände gemäß Gleichung (102) zu den beobachtbaren Wellen. Diese Beziehung in Matrixdarstellung geschrieben bedeutet

$$\begin{bmatrix} \psi_0^\nu \\ \psi_G^\nu \end{bmatrix} = \exp\left(i\,\vec{k}\,\vec{r}\right) \mathfrak{M}^\nu \begin{bmatrix} u_1^\nu \\ u_2^\nu \end{bmatrix} \quad ; \;\nu = 0, (N+1) \tag{123}$$

mit der Vakuumtransfermatrix

$$\mathfrak{M}^\nu = \begin{bmatrix} \Lambda_1^\nu & \Lambda_2^\nu \\ \exp\left(i\,\vec{G}\,\vec{r}\right) X_1^\nu \Lambda_1^\nu & \exp\left(i\,\vec{G}\,\vec{r}\right) X_2^\nu \Lambda_2^\nu \end{bmatrix} . \tag{124}$$

Da ohne Beschränkung der Allgemeinheit die ein- und austretenden Wellen an den Kristall-oberflächen gemessen werden können, kann man

$$D^\nu \equiv 0 \quad \Rightarrow \quad \Lambda_n^\nu \equiv 1 \;\; ; \;\nu = 0, (N+1) \tag{125}$$

setzen. In diesen Grenzschichten verschwindender Dicke muß außerdem

$$u_n^0 \equiv u_n^1 \quad \text{und} \quad u_n^{N+1} \equiv u_n^N \tag{126}$$

gelten, was mit

$$\tilde{X}_n^0 \equiv \tilde{X}_n^1 \quad \text{bzw.} \quad \tilde{X}_n^{N+1} \equiv \tilde{X}_n^N \tag{127}$$

erfüllt wird. Für $\mathfrak{M}^\nu$ ergibt dies

$$\mathfrak{M}^\nu = \begin{bmatrix} 1 & 1 \\ X_1^\nu & X_2^\nu \end{bmatrix} \;\; ; \;\nu = 0, (N+1) . \tag{128}$$

Mit obigen Relationen kann nun die Beziehung zwischen den $\psi_H^\nu$, $\nu = 0, (N+1)$; $\vec{H} = \vec{0}, \vec{G}$ durch ein Produkt aus einer Eintrittsmatrix $(\mathfrak{M})^{-1}$ vom Vakuum ins Medium, einer sich aus dem Produkt der Einzelmatrizen zusammensetzenden Transfermatrix $\mathfrak{T}$ und einer Austrittsmatrix $\mathfrak{M}$ ausgedrückt werden:

$$\begin{bmatrix} \psi_0^{N+1} \\ \psi_G^{N+1} \end{bmatrix} = \mathfrak{C} \begin{bmatrix} \psi_0^0 \\ \psi_G^0 \end{bmatrix} \text{ mit } \mathfrak{C} = \begin{bmatrix} C_{00} & C_{0G} \\ C_{G0} & C_{GG} \end{bmatrix} = \mathfrak{M}^{N+1}\, \mathfrak{T}\, \left(\mathfrak{M}^0\right)^{-1} \tag{129}$$

Bevor nun weitere vereinfachende Annahmen gemacht werden können, müssen wir die Randbedingungen einführen:

Im Braggfall ($b < 0$) haben wir vor dem Medium sowohl Wellen in vorwärts- und abgebeugter Richtung, hinter diesem jedoch nur eine vorwärtsgebeugte, also

$$\begin{bmatrix} \psi_0^0 \\ \psi_G^0 \end{bmatrix} = \begin{bmatrix} \psi_e \\ \psi_G^0 \end{bmatrix} \qquad \text{bei } \vec{n}\,\vec{r} = 0 \tag{130}$$

und

$$\begin{bmatrix} \psi_0^{N+1} \\ \psi_G^{N+1} \end{bmatrix} = \begin{bmatrix} \psi_0^{N+1} \\ 0 \end{bmatrix} \qquad \text{bei } \vec{n}\,\vec{r} = \sum_{\nu=1}^{N} D^\nu . \tag{131}$$

Daraus folgt

$$\begin{bmatrix} \psi_e \\ \psi_G^0 \end{bmatrix} = \mathfrak{C}^{-1} \begin{bmatrix} \psi_0^{N+1} \\ 0 \end{bmatrix} \tag{132}$$

oder komponentenweise, wenn $C_{HI}^{-1}$ mit $H, I = 0, G$ die Elemente von $\mathfrak{C}^{-1}$ bezeichnet,

$$\psi_e = C_{00}^{-1}\, \psi_0^{N+1} \tag{133}$$

woraus für die Transmission, also die normierte Intensität in Vorwärtsrichtung

$$T_B = \frac{I_0^{N+1}}{I_e} = \left|\frac{\psi_0^{N+1}}{\psi_e}\right|^2 = \frac{1}{\left|C_{00}^{-1}\right|^2} \tag{134}$$

folgt und

$$\psi_G^0 = C_{G0}^{-1}\, \psi_0^{N+1}\,, \tag{135}$$

was zusammen mit (133) und (81) zur Reflektion

$$R_B = \frac{1}{|b|}\frac{I_G^0}{I_e} = \frac{1}{|b|}\left|\frac{\psi_G^0}{\psi_e}\right|^2 = \frac{1}{|b|}\left|\frac{C_{G0}^{-1}}{C_{00}^{-1}}\right|^2 \tag{136}$$

führt.

Der Lauefall zeichnet sich dadurch aus, daß die abgebeugte Welle vor dem Medium verschwindet, während sie an der Rückseite mit endlicher Amplitude selbiges verläßt:

$$\begin{bmatrix} \psi_0^0 \\ \psi_G^0 \end{bmatrix} = \begin{bmatrix} \psi_e \\ 0 \end{bmatrix} \qquad \text{bei } \vec{n}\,\vec{r} = 0\,, \tag{137}$$

woraus folgt

$$\begin{bmatrix} \psi_0^{N+1} \\ \psi_G^{N+1} \end{bmatrix} = \mathfrak{C} \begin{bmatrix} \psi_e \\ 0 \end{bmatrix} \tag{138}$$

und ähnlich wie im Braggfall durch komponentenweise Betrachtung

$$\psi_0^{N+1} = C_{00}\, \psi_e \tag{139}$$

und

$$\psi_G^{N+1} = C_{G0}\, \psi_e \tag{140}$$

ergibt. Damit erhalten wir

$$T_L = |C_{00}|^2 \tag{141}$$

und

$$R_L = \frac{1}{|b|}\, |C_{G0}|^2 \tag{142}$$

für die Meßgrößen der vorwärtsgebeugten und abgebeugten Strahlen.

Es läßt sich nun für eine weitere Vereinfachung zeigen, daß sich formell

$$\mathfrak{C} = \mathfrak{M}^{N+1}\,\mathfrak{T}\,\left(\mathfrak{M}^{0}\right)^{-1} = \begin{bmatrix} \alpha & \exp\left(-i\,\vec{G}\,\vec{r}\right)\frac{1}{\sqrt{|b|}}\,\Phi^{-1}\,\beta \\ \exp\left(i\,\vec{G}\,\vec{r}\right)\sqrt{|b|}\,\Phi\,\gamma & \delta \end{bmatrix} \qquad (143)$$

schreiben läßt, worin $\Phi = V(\vec{G}) \,/\, \left|V(\vec{G})\right|$ einen Phasenfaktor darstellt und $\alpha$, $\beta$, $\gamma$, $\delta$ nicht mehr von $\exp\left(-i\,\vec{G}\,\vec{r}\right)$, $|b|$ und $\Phi$ abhängen. Für $\mathfrak{C}^{-1}$ gilt dasselbe. Zusammen mit den Randbedingungen (130) und (131) oder (137) und der späteren Betragsquadratbildung können die Phasenfaktoren fortgelassen werden. Der Faktor $|b|$ kürzt sich jeweils in den Reflektivitätsformeln (136) und (142) heraus, weshalb wir die modifizierten, nur noch von $y^j$ und $A^j$ abhängigen Matrizen

$$\widetilde{\mathfrak{M}}^{\nu} = \begin{bmatrix} 1 & 1 \\ \widetilde{X}_1^{\nu} & \widetilde{X}_2^{\nu} \end{bmatrix} \text{ bzw. } \left(\widetilde{\mathfrak{M}}^{\nu}\right)^{-1} = \frac{1}{\widetilde{X}_1^{\nu} - \widetilde{X}_2^{\nu}} \begin{bmatrix} -\widetilde{X}_2^{\nu} & 1 \\ \widetilde{X}_1^{\nu} & -1 \end{bmatrix} ;\ \nu = 0, (N+1) \qquad (144)$$

für die Vakuumtransfermatrizen $\mathfrak{M}$ ersetzen können und die endgültigen Transfermatrizen

$$\widetilde{\mathfrak{C}} = \widetilde{\mathfrak{M}}^{N+1}\,\widetilde{\mathfrak{T}}\,\left(\widetilde{\mathfrak{M}}^{0}\right)^{-1} \qquad (145)$$

für unsere Berechnungen erhalten. Die Transmissionen und Reflektionen lassen sich dann folgendermaßen schreiben:

Bragg $b < 0$:

$$T_B\left(y^1, \cdots, y^N; A^1, \cdots, A^N\right) = \frac{1}{\left|\widetilde{C}_{00}^{-1}\right|^2} \qquad (146)$$

$$R_B\left(y^1, \cdots, y^N; A^1, \cdots, A^N\right) = \left|\frac{\widetilde{C}_{G0}^{-1}}{\widetilde{C}_{00}^{-1}}\right|^2 \qquad (147)$$

Laue $b > 0$:

$$T_L\left(y^1, \cdots, y^N; A^1, \cdots, A^N\right) = \left|\widetilde{C}_{00}\right|^2 \qquad (148)$$

$$R_L\left(y^1, \cdots, y^N; A^1, \cdots, A^N\right) = \left|\widetilde{C}_{G0}\right|^2 \qquad (149)$$

Sie hängen nur noch von den $y^j$ und $A^j$ aller Schichten ab.

### 4.2.3. Anwendungsbeispiele der Transfermatrizen

Im vorangegangenen Kapitel wurde eine Transfermatrizenmethode im Rahmen der dynamischen Beugungstheorie für geschichtete Medien hergeleitet, die nun an Beispielen angewendet werden soll. Das einfachste System ist natürlich die Behandlung eines Idealkristalls, worauf Einsätze der Methode auf Vielfachschichtsysteme und, als Grenzfall derer, Gradientenkristalle vorgeführt werden. Selbst die Beschreibung eines Kristallinterferometers wäre denkbar, doch würde sie den Rahmen dieser Arbeit sprengen.

Der Transfermatrizenalgorithmus an sich ist eine rein analytische Methode zur Beschreibung solcher Schichtsysteme, doch werden bei größeren Schichtzahlen die Matrixelemente so komplex, daß man die Matrizen nur noch auf numerischen Rechenmaschinen ausmultiplizieren kann. Wie wir sehen werden, wird im Fall der Gradientenkristalle noch ein numerischer Grenzübergang hinzukommen.

#### 4.2.3.1. Der planparallele Idealkristall

An diesem Beispiel soll die Analytik und die Konsistenz der Transfermatrixmethode gezeigt werden.

Bei einem Idealkristall herrscht in allen Lagen j das gleiche Kristallpotential und daher überall dieselbe Abweichung

$$y^j = y \tag{150}$$

von der Braggposition. Somit sind auch alle $\widetilde{X}_n^j = \widetilde{X}_n$ gleich. Die Gesamtdicke ist als Summe der einzelnen Lagendicken durch

$$A = \sum_{j=1}^{N} A^j \tag{151}$$

gegeben. Wie auch schon früher diskutiert, erhalten wir mit (150) und (121) für

$$\widetilde{\mathfrak{X}}^j = \begin{bmatrix} 1 & 0 \\ 0 & 1 \end{bmatrix} \tag{152}$$

die Einheitsmatrix, weshalb sich $\widetilde{\mathfrak{T}}$ zu

$$\widetilde{\mathfrak{T}} = \prod_{j=1}^{\overrightarrow{N}} \widetilde{\mathfrak{L}}^{j} = \prod_{j=1}^{\overrightarrow{N}} \begin{bmatrix} \exp\left(i\,\widetilde{X}_1\,A^j\right) & 0 \\ 0 & \exp\left(i\,\widetilde{X}_2\,A^j\right) \end{bmatrix} \tag{153}$$

oder mit

$$\widetilde{\Lambda}_n = \exp\left(i\,\widetilde{X}_n\,A\right) \tag{154}$$

zu

$$\widetilde{\mathfrak{T}} = \begin{bmatrix} \widetilde{\Lambda}_1 & 0 \\ 0 & \widetilde{\Lambda}_2 \end{bmatrix} \tag{155}$$

vereinfacht und nur die Phasenpropagation der Blochwellenamplituden durch den Gesamtkristall beschreibt. Es ist also gleichgültig, in wie viele Lagen ein Idealkristall mathematisch aufgeteilt wird. Physikalisch stellt er nur eine einzige dar, die durch die Matrix $\widetilde{\mathfrak{T}}$ beschrieben wird.

Um richtig an die Vakuumzustände anzuknüpfen, müssen wir das Produkt (145) mit der Vakuumtransfermatrix $\widetilde{M}^{N+1} \equiv \widetilde{M}^{0} \equiv \widetilde{M}$ ausführen und erhalten

$$\widetilde{\mathfrak{C}} = \frac{1}{\widetilde{X}_2 - \widetilde{X}_1} \begin{bmatrix} 1 & 1 \\ \widetilde{X}_1 & \widetilde{X}_2 \end{bmatrix} \begin{bmatrix} \widetilde{\Lambda}_1 & 0 \\ 0 & \widetilde{\Lambda}_2 \end{bmatrix} \begin{bmatrix} \widetilde{X}_2 & -1 \\ -\widetilde{X}_1 & 1 \end{bmatrix} \tag{156}$$

oder

$$\widetilde{\mathfrak{C}} = \frac{1}{\widetilde{X}_2 - \widetilde{X}_1} \begin{bmatrix} \widetilde{\Lambda}_1\,\widetilde{X}_2 - \widetilde{\Lambda}_2\,\widetilde{X}_1 & \widetilde{\Lambda}_2 - \widetilde{\Lambda}_1 \\ \left(\widetilde{\Lambda}_1 - \widetilde{\Lambda}_2\right)\widetilde{X}_1\,\widetilde{X}_2 & \widetilde{\Lambda}_2\,\widetilde{X}_2 - \widetilde{\Lambda}_1\,\widetilde{X}_1 \end{bmatrix}. \tag{157}$$

Damit wird zum Beispiel die durch (149) gegebene Reflektionskurve im Lauefall

$$R_L = \left|\widetilde{C}_{G0}\right|^2 = \left|\frac{\widetilde{X}_1\,\widetilde{X}_2}{\widetilde{X}_2 - \widetilde{X}_1}\right|^2 \left|\widetilde{\Lambda}_1 - \widetilde{\Lambda}_2\right|^2 \tag{158}$$

mittels den aus (79) und (80) erhaltenen Ausdrücken

$$\widetilde{X}_1\,\widetilde{X}_2 = \mp 1\,, \tag{159}$$

$$\widetilde{X}_2 - \widetilde{X}_1 = -2\,\sqrt{y^2 \pm 1} \tag{160}$$

und

$$\left|\widetilde{\Lambda}_1 - \widetilde{\Lambda}_2\right|^2 = \left|1 - \exp\left(-i\,2\,\sqrt{y^2+1}\,A\right)\right|^2 = 2 - 2\cos\left(2\,\sqrt{y^2+1}\,A\right) \tag{161}$$

sowie der Erinnerung, daß im Lauefall das + Zeichen unter der Wurzel gilt in

$$R_L = \frac{1 - \cos\left(2\sqrt{y^2 + 1}\ A\right)}{y^2 + 1} \tag{162}$$

umgeformt, was durch ein Additionstheorem der Winkelfunktionen in den Ausdruck

$$R_L = \frac{\sin^2\left(\sqrt{y^2 + 1}\ A\right)}{y^2 + 1} \tag{163}$$

übergeht. Wie wir sehen, ist diese Gleichung identisch mit (83). Genauso kann man auch die anderen Größen $T_L$ und für den Braggfall $R_B$ und $T_B$ erhalten.

### 4.2.3.2. Das kristalline Multilagensystem

In diesem Abschnitt sollen kristalline Doppelschichtsysteme betrachtet werden, die sich N-fach wiederholen. Eine schematische Darstellung ist in Abbildung (15) wiedergegeben. Numerieren wir die Lagen einer Doppelschicht mit j = 1, 2, so ergibt sich für die Transfermatrix des gesamten Mediums

$$\widetilde{\mathfrak{E}} = \widetilde{\mathfrak{M}}^3 \left(\widetilde{\mathfrak{T}}^2\ \widetilde{\mathfrak{T}}^1\right)^N \left(\widetilde{\mathfrak{M}}^0\right)^{-1} . \tag{164}$$

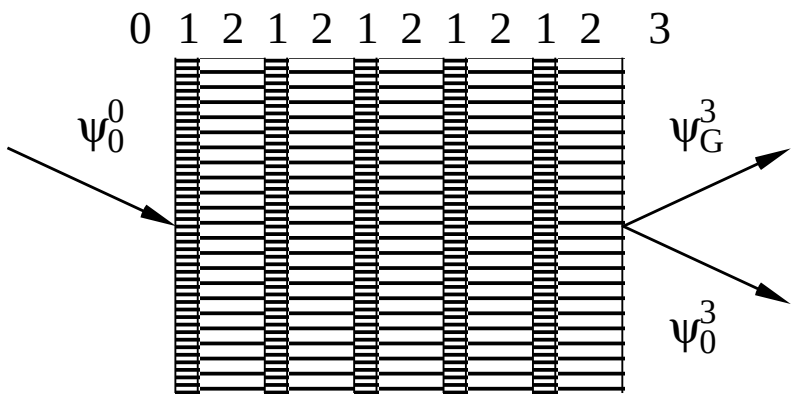

*Abbildung (15): Doppelschichtensystem mit 5-facher Wiederholung, im Lauefall dargestellt.*

In die $\widetilde{\mathfrak{T}}^j$ gehen gemäß (120) und (121) Differenzen aufgrund unterschiedlicher Abweichungen $y^j$ von der, in jedem Medium individuellen geometrischen Braggposition ein.
Gehen wir davon aus, daß beide Kristallagen in derselben Struktur kristallisieren, so kann y gemäß seiner Definition (66) auf zwei verschiedene Arten variiert werden, nämlich gemäß (75) als Abweichung der Gitterkonstanten d, bzw. des dazugehörigen Streuvektors G, und zum anderen als Variation der kohärenten Streulänge $b_c$. Um auch letztere Modifikation zu berücksichtigen, entwickeln wir (66) um $b_c$ mit $\Delta b_c = b_c' - b_c$ und erhalten

$$y = \frac{(b-1)\,\frac{V(\vec{0})}{E} + \alpha\, b \left(1 - \frac{\Delta b_c}{b_c}\right)}{2\sqrt{|b|}\,\left|\frac{V(\vec{G})}{E}\right|} . \tag{165}$$

Für α muß bei Variation der Gitterkonstanten die zu ΔG lineare Abhängigkeit (75) eingesetzt werden. Dabei verschiebt ΔG die Braggposition des Mediums 2 auf der Skala 1, wie es aus dem Braggesetz nicht anders zu erwarten ist. Eine Variation von $b_c$ bei festem G läßt jedoch die Braggbedingung an derselben Stelle, nämlich bei $y(\alpha = 0)$ erscheinen, staucht oder streckt aber die Skala des zweiten Mediums auf der des ersten. Dies soll nochmals in Abbildung (16) veranschaulicht werden.

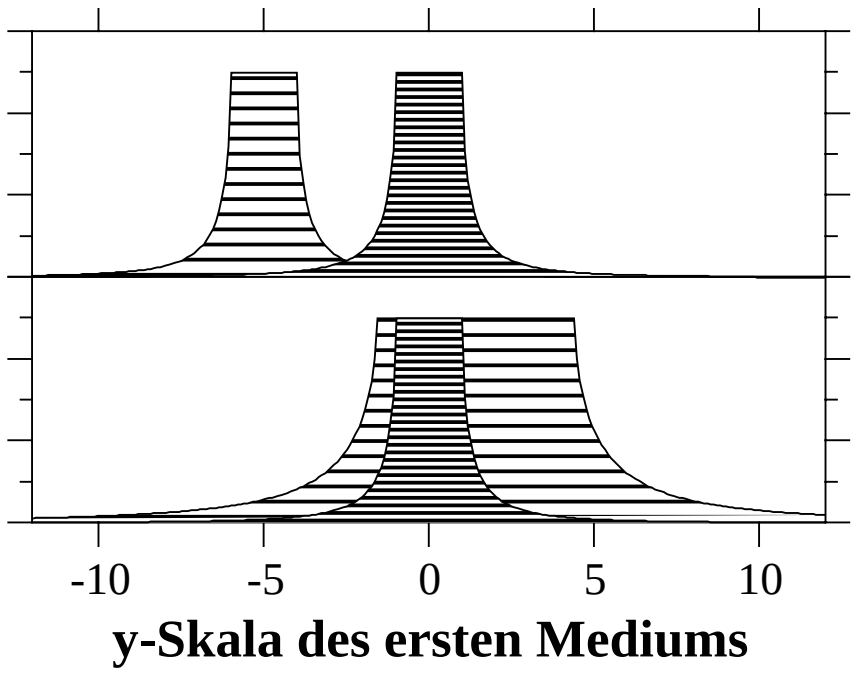

*Abbildung (16):*
*Variation des Braggreflexes in Abhängigkeit einer Änderung der Gitterkonstanten, oben, bzw. der kohärenten Streulänge, unten.*

Ein Beispiel für ein Lagensystem mit Gitterkonstantenvariation ist in erster Näherung das $Si_{1-x}Ge_x$ System, während eine Streulängenvariation durch ein Isotopensystem, z. B. ein Germaniumsystem betrachtet werden kann, bei dem sich die Lagen durch verschiedene Germaniumisotope unterscheiden. In Zahlen ausgedrückt, unterscheiden sich die Schichten von Si-$Si_{0,99}Ge_{0,01}$ in Rückstreuung und symmetrischer Braggeometrie um $\Delta y = 46$, während der Streckungsfaktor eines $^{70}$Ge-$^{73}$Ge Systems p = 1,5 betragen kann.

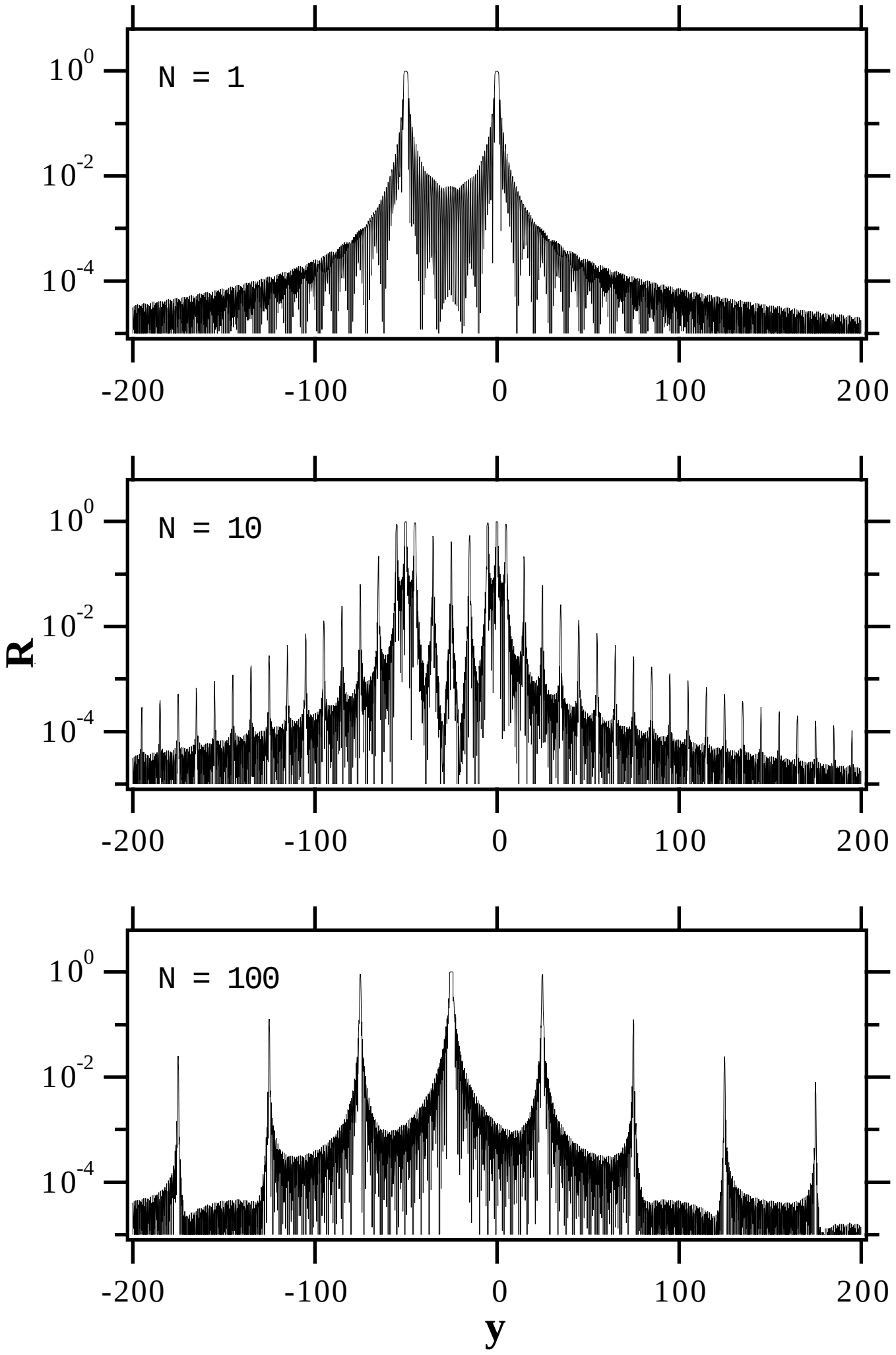

*Abbildung (17):*
*Berechnete Reflektionskurven in Braggeometrie bei Gitterkonstantenvariation von N-fach Doppelschichtsystemen bei festgehaltener Gesamtdicke $A = 2\pi$, also Einzelschichtdicken von $A^1 = A^2 = \pi / N$.*

Abbildung (17) zeigt einige Rechenergebnisse bei Variation der Gitterkonstanten mit $\Delta y = 50$ in Braggeometrie und logarithmischer Darstellung für die Reflektivität. Das erste Teilbild zeigt die Reflektionskurve einer Doppellage der Dicke $A^1 = A^2 = \pi$, also einer Pendellösungsperiode pro Schicht. Wir sehen die Überlagerung zweier, voll ausgeprägter Braggreflexe, den der ersten Schicht bei y = 0 sowie den der zweiten bei y = -50, also an den Braggpositionen jeder einzelnen Schicht. Die feinen Oszillationen sind uns bereits aus Kapitel 4.2.1. bekannt. Im zweiten Teilbild, das mit 10 Doppellagen der Dicken $A^1 = A^2 = \pi/10$, also gleichbleibender Gesamtdicke berechnet wurde, fallen dem Leser sofort eine Vielzahl ausgeprägter Nebenmaxima auf, die jeweils den Abstand $\delta y = \pi / A^1 = \pi / A^2$, also den der reziproken Einzelschichten voneinander haben. Die Hauptmaxima bei y = 0 und y = -50 sind zwar noch vorhanden, doch beginnen ihre benachbarten Nebenmaxima stark mit diesen zu konkurrieren. Erhöhen wir die Zahl der Doppellagen um einen weiteren Faktor 10, so hat, wie im untersten Teilbild dargestellt, bereits ein Grenzübergang stattgefunden: Die Hauptmaxima an den erwarteten Stellen sind vollkommen verschwunden. Stattdessen hat sich im Zentrum eine, dem Idealkristall sehr ähnliche Reflektionskurve ausgebildet, deren Position genau der mittleren Gitterkonstanten y = -25 entspricht. Jetzt sind die einzelnen Lagen so fein geworden, daß die Kristallwellen nur noch ihr mittleres Potential erkennen. Natürlich ist weiterhin eine Periodizität vorhanden, was wieder eine wohldefinierte Reihe von Nebenmaxima aufwirft. Im Gegensatz zum mittleren Teilbild sind diese doch nur um $\delta y = \pi / (A^1 + A^2)$, also um die reziproke Gesamtdicke voneinander separiert. Dies mag wieder mit obiger Grenzbetrachtung interpretiert werden, wobei für die Welle wohl noch die Periodizität vorhanden, in niederer Ordnung jedoch nicht mehr aufgelöst wird. Ein Beitrag höherer Ordnung mag dazu tragen, daß der Streuanteil zwischen den Maxima mal nach oben, mal nach unten „gekrümmt" ist.

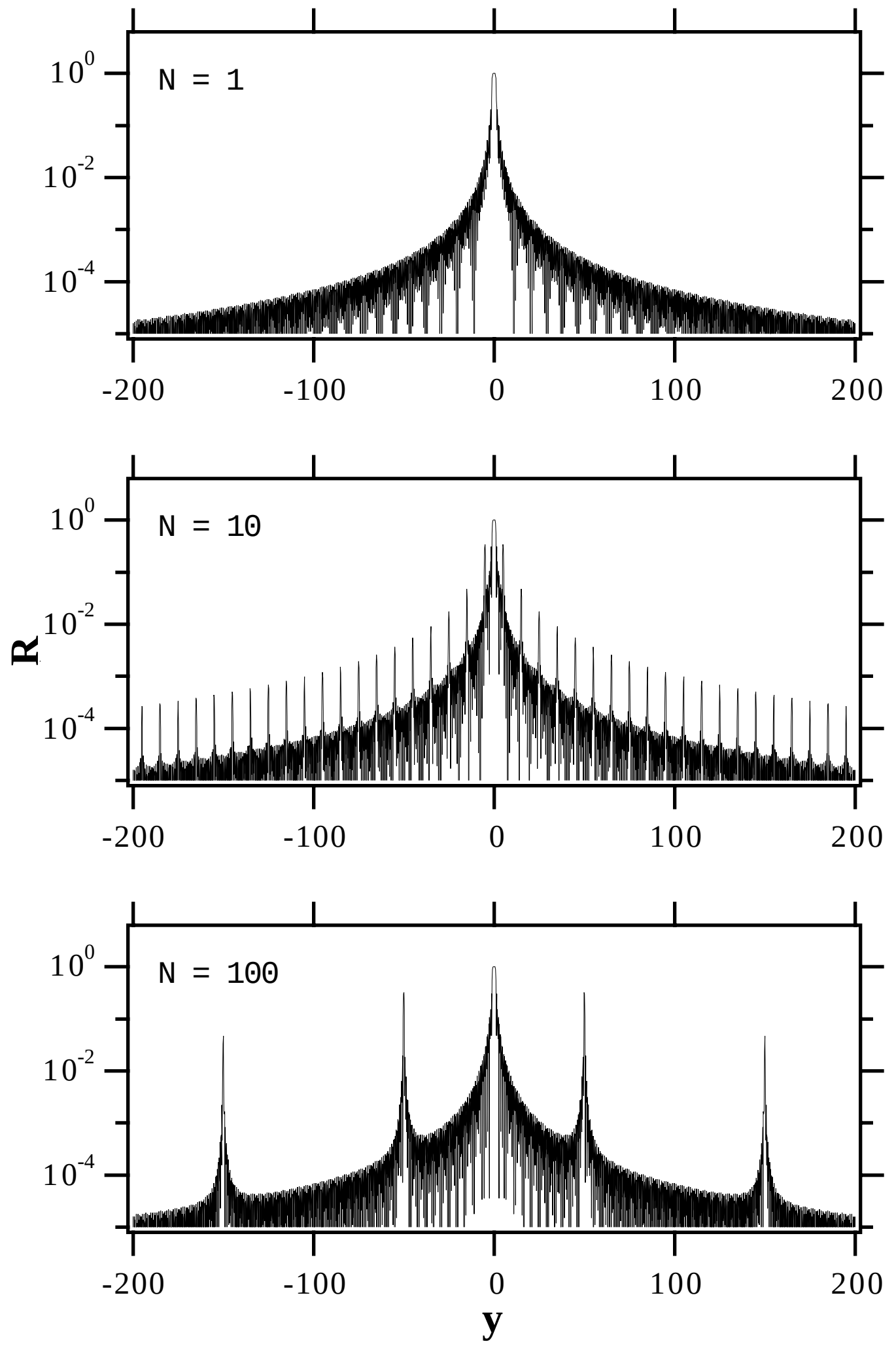

*Abbildung (18):*
*Berechnete Reflektionskurven in Braggeometrie bei Streulängenvariation um p = 1,5 von N-fach Doppelschichtsystemen bei festgehaltener Gesamtdicke. Die Einzelschichtdicken von  betragen $A^1 = \pi / N$ und $A^2 = \pi / (p\, N)$.*

Die Reflektionskurven bei Streulängenvariation mit p = 1,5 sind in Abbildung (18) wiedergegeben. Im oberen Teilbild sehen wir das Ergebnis einer Doppellage mit Schichtdicken $A^1 = \pi$ und $A^2 = \pi / p = 2{,}094$. Wenn wir in Betracht ziehen, daß die Größen im Medium 2 auf der Skala des ersten gemessen werden, entsprechen die Dicken jeweils einer Pendellösungsperiode. Die Braggreflexe beider Schichten überlagern sich nun aufgrund identischer Gitterkonstanten bei y = 0. Sieht man genauer hin als dargestellt, so erscheinen die feinen

Oszillationen der beiden Kurven um p gegeneinander gestreckt bzw. gestaucht und wirken sich in Schwebungserscheinungen derselben aus. Im mittleren Teilbild haben wir wieder bei fester Gesamtdicke die Schichtenzahl um 10 erhöht, wodurch sofort ausgeprägte Nebenmaxima des Abstands $\delta y = \pi / A^1$, also der reziproken Einzelschichtdicke entsprechend, hervorgerufen werden. Bei weiterer Erhöhung der Schichtenzahl auf $N = 100$ ändert sich qualitativ nichts mehr. Natürlich sind die Nebenmaxima wegen der kleineren Einzelschichtdicken weiter voneinander entfernt. Gegenüber der Gitterkonstantenvariation finden sich keine Maxima, die der Doppellagenperiodizität von $A^1 + p \cdot A^2$ entsprechen. Dies ist gerade auf unsere spezielle Wahl von $A^1$ und $A^2$ zurückzuführen, wobei die Phasenvorschübe durch Propagation der Welle durch die Lagen des einen oder anderen Mediums identisch sind und damit diese Reflexe verboten werden.

Bei genauerer Auswertung läßt sich in beiden Variationsfällen demonstrieren, daß die Nebenmaxima schmäler als die natürlichen Linienbreiten des Idealkristalls und, im kinematischen Bereich, umgekehrt proportional zur Schichtenzahl N sind. Letzteres Verhalten ist aus der Beugung am Strichgitter bekannt. Während bei einem Idealkristall das Streuvolumen durch die Pendellösungsperiode beschränkt ist, trägt der von den Multilagen herrührende Anteil nur relativ schwach zur Ausbildung der Nebenmaxima bei, wodurch das zugehörige, effektive Streuvolumen vervielfacht, und damit die Linienbreite im reziproken Raum verkleinert wird. Ein weiterer interessanter Gesichtspunkt erweist sich bei der Betrachtung dicker und dickerer Kristalle bei festgehaltener Einzelschichtdicke, dem ein Anstieg der Nebenreflexintensitäten entspricht, bis diese ihren Bereich primärer Extinktion erlangen und in Sättigung gehen. Die alsdann ausgeprägten Reflektionskurven ähneln, und dieses im mathematischen Sinn, denen von Idealkristallen, je nach Geometrie im Laue- oder Braggfall. Demnach strebt auch die Linienbreite einem, der reziproken Extinktionslänge entsprechenden, endlichen Wert zu, der sich umgekehrt proportional zur Einzelschichtdicke verhält.

Im Vorangegangenen wurde demonstriert, wie ein Vielfachdoppellagensystem auch Satellitenreflexe um die atomaren Braggreflexe aufwirft, wenn hinreichende Kohärenz zwischen den einzelnen, auch nicht benachbarten Lagen besteht. Im Gegensatz zur Reflektometrie kann bei Anwendung auf z. B. epitaktisch gewachsene Proben nicht nur auf die Lagenstruktur als ganzes, sondern auch auf die atomare Verteilung in diesen geschlossen werden. Es mag von Interesse sein, daß diese Methode mit Braggreflexen vergleichbare Streuvektoren benutzt. Damit können am Meßinstrument herkömmliche Strahlgeometrien bei großen Streuwinkeln verwendet werden. Insbesonders kann auch die Lauegeometrie Anwendung finden, da es bei der Ausbildung der Nebenmaxima qualitativ keinen Unterschied zur Braggeometrie gibt.

### 4.2.3.3. Stufen- und Gradientenkristalle

Dieses Kapitel soll dem eigentlichen Ziel dieser Arbeit, nämlich der Beschreibung von Reflektionskurven an Gradientenkristallen gewidmet werden. Dazu führen wir den Stufenkristall ein, um schließlich durch einen numerischen Grenzübergang die gewünschten Eigenschaften zu erhalten.

Ein Stufenkristall sei durch ein Schichtensystem definiert, dessen Lagen sich durch monoton wachsende oder fallende Kristallpotentiale unterscheiden und deren Variationsprofile, wie beispielsweise das der Gitterkonstanten in Abbildung (19), einer Treppenfunktion ähneln. Die Reflektionskurve läßt, wie im unteren Teil der Abbildung, für jede Lage ein individuelles Maximum erwarten.

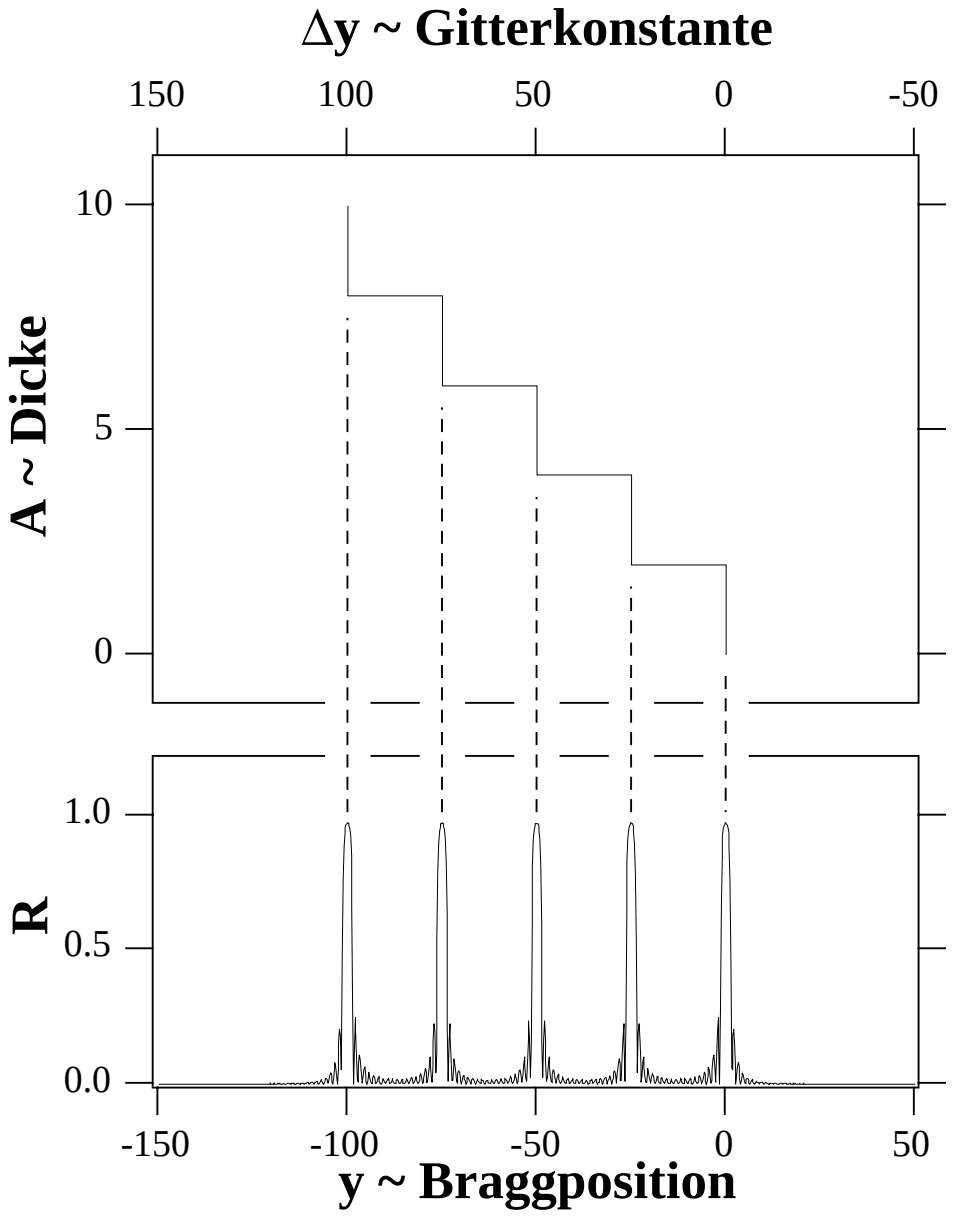

*Abbildung (19):*
*Variation Δy proportional zur Gitterkonstanten als Funktion der Kristalldicke A im oberen Diagramm, hier am Beispiel von 5 Schichten. Die Variation zwischen benachbarten Schichten ist so groß, daß jede für sich ihr eigenes Braggmaximum ausbildet*

Bei einem Gradientenkristall gehen wir von einer kontinuierlichen Variation des Kristallpotentials aus. Zur Berechnung der Reflektionskurve wird diese, Abbildung (20) entsprechend, in eine Vielzahl von Stufen unterteilt, die so fein sind, daß sich die resultierenden Reflektionskurven bei weiterer Unterteilung nicht mehr unterscheiden.

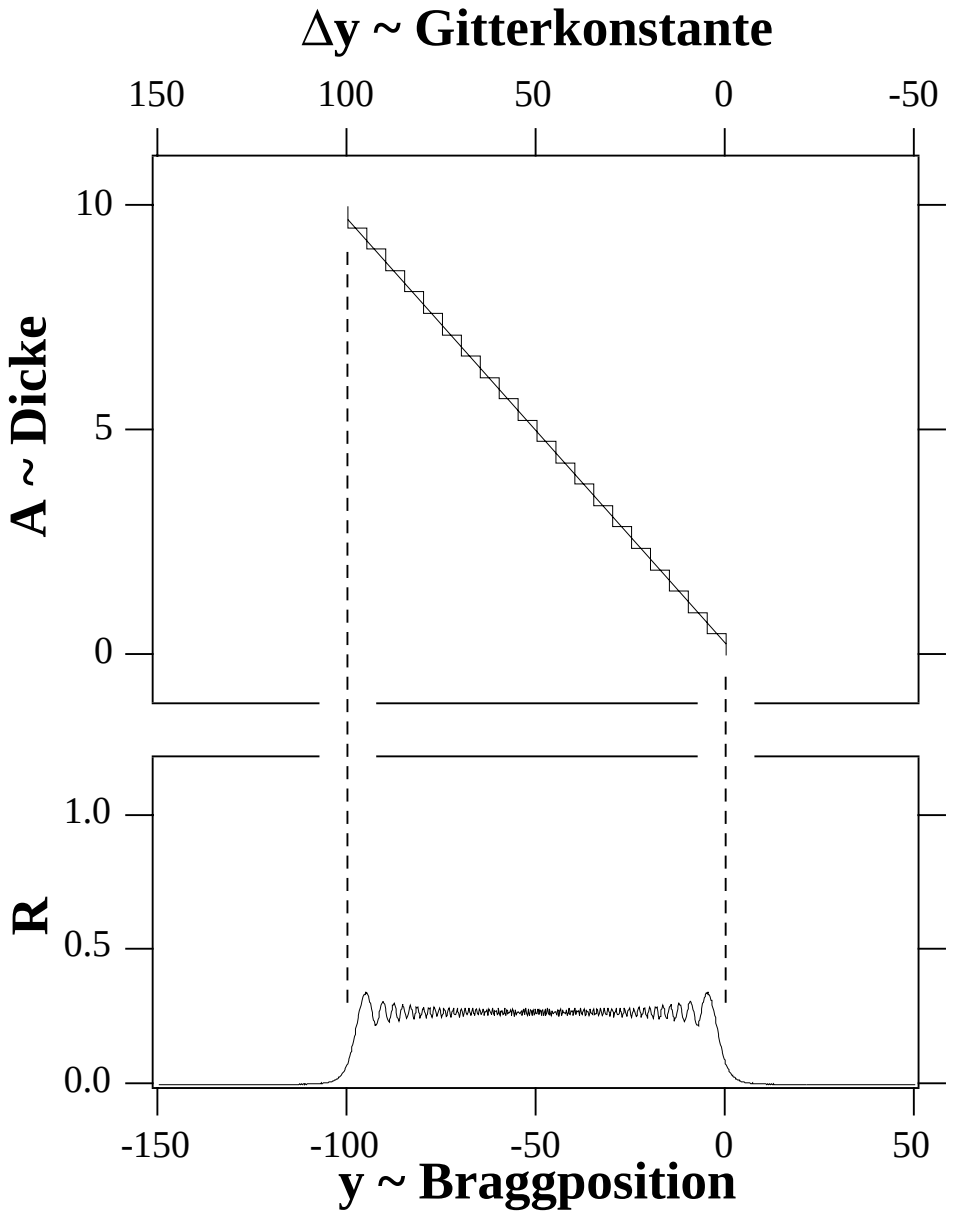

*Abbildung (20):*
*Die lineare Variation Δy eines Gradientenkristalls wird für die Berechnung mittels Transfermatrizenmethode in Stufen zerlegt, die fein genug sein müssen um numerisch das richtige Reflektionsprofil R(y) zu erzeugen.*

Zur Beschreibung eines Gradienten in den dimensionslosen Einheiten führen wir die Größe

$$c = \frac{\partial y}{\partial A} \qquad (166)$$

ein, die mittels (65) und (66) durch

$$c = \pm \sqrt{|b|}\, \frac{1}{2\pi}\, \frac{G^3}{k^3}\, \frac{\sqrt{|\cos(\gamma_0)\cos(\gamma_G)|}}{\left|\frac{V(\vec{G})}{E}\right|^2}\, g \qquad (167)$$

mit dem Gitterkonstantengradienten

$$g = \frac{\partial d}{\partial D} \tag{168}$$

in Verbindung steht. Für lineare Gradienten ist c eine Konstante. Damit ergibt sich z. B. mit der Darwinplateaubreite $\delta y = 2$ und der Pendellösungsperiode (92) für den in Gleichung (20) definierten, idealen Gradienten

$$c_i = \frac{2}{\pi}\,. \tag{169}$$

Sei

$$A = N\,A^1 \tag{170}$$

die Gesamtdicke des in N Stufen gleicher Dicken $A^1$ unterteilten Gradientenkristalls, so ergibt sich eine Variation von

$$\Delta y = \frac{c\,A}{N-1} \tag{171}$$

zwischen benachbarten Lagen. Genauso läßt sich bei einer Streulängenvariation um p für den gesamten Kristall der entsprechende Faktor

$$p_\Delta = \sqrt[N-1]{p} \tag{172}$$

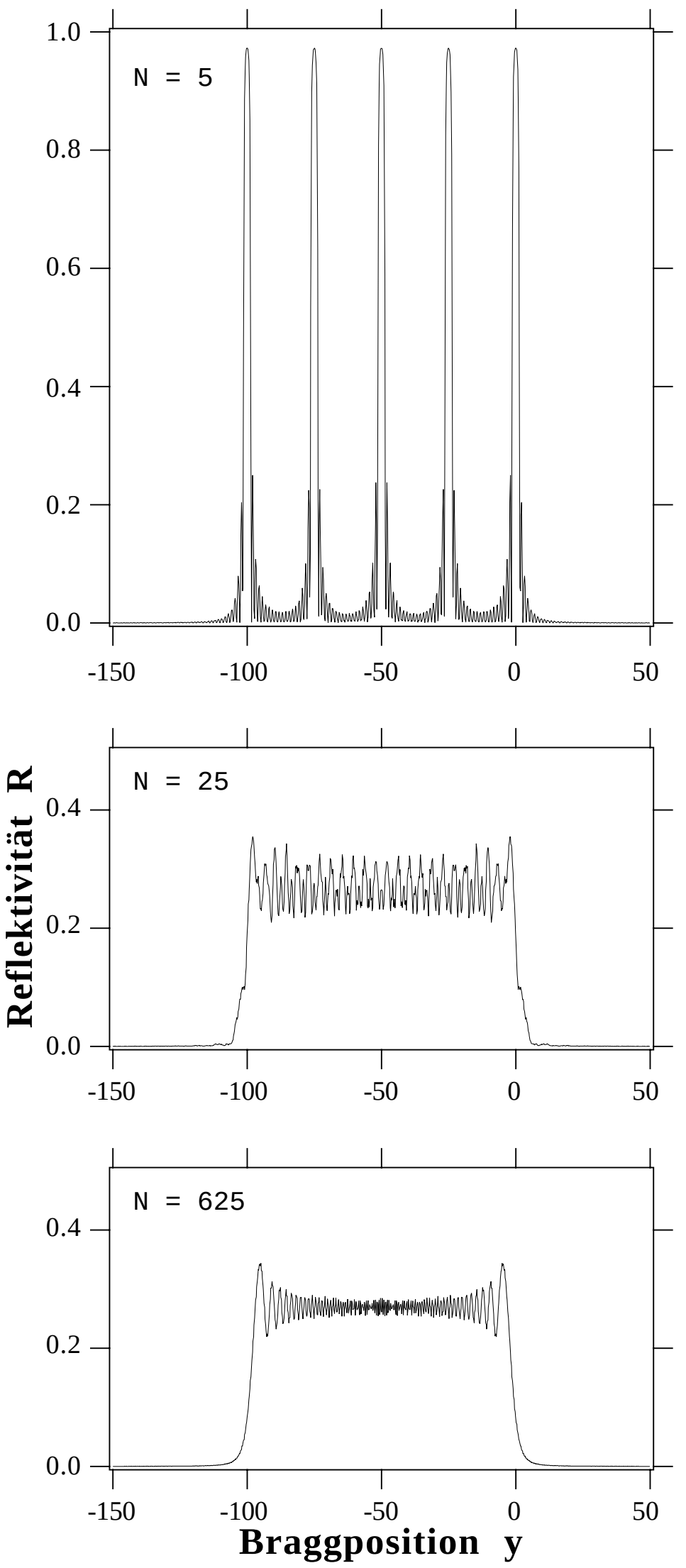

*Abbildung (21):*
*Reflektionskurven beim numerischen Übergang eines Stufenkristalls zum Gradientenkristall. Je feiner die Unterteilungen, d. h. je Größer N wird, desto besser gleicht die erhaltene Kurve der des Gradientenkristalls.*

zwischen benachbarten Schichten einführen, womit man mit den Werten $p^1 = 1$ und $y^1$ der ersten Lage

$$y^j = p_\Delta^{\,j-1}\left(y^1 + (j-1)\,c\,A^1\right) \tag{173}$$

für die j-te Lage erhält und in den Matrixformalismus einsetzt.

In Abbildung (21) sind somit erhaltene Rechenergebnisse im Braggfall für A = 10, c = 10 und p = 1 mit 5, 25 und 625 Unterteilungen dargestellt. Während das erste Teilbild die Reflektionskurve eines ausgeprägten Stufenkristalls zeigt, ähnelt das zweite, obwohl noch sehr eckig und kantig und noch mit 24 Hauptmaxima besetzt, schon eher der eines Gradientenkristalls wie im letzten Teilbild. Den relativen Fehler

$$E = \frac{\int_{-\infty}^{\infty} \left(R(y, A^1) - R(y, A^1 \to 0)\right)^2 dy}{\int_{-\infty}^{\infty} R(y, A^1 \to 0)\, dy} \qquad (174)$$

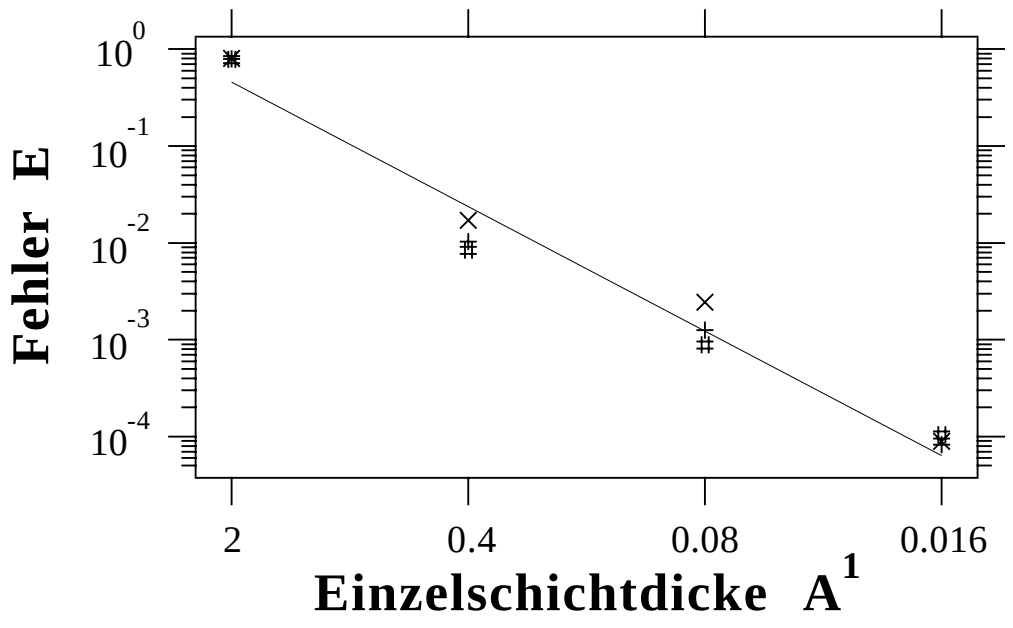

*Abbildung (22):*
*Die Rechenungenauigkeit E hängt bei breiten Reflektionskurven nur von der Einzelschichtdicke* $A^1$*, jedoch nicht vom Gradienten c ab.*
×: *A = 5, c = 10;*
+: *A = 10, c = 10;*
#: *A = 10, c = 20;*

aus numerischen Berechnungen für drei verschiedene Kristallsysteme relativ breiter Reflektionskurven und in Abhängigkeit der Einzelschichtdicke gibt Abbildung (22) wieder. Obwohl die Punkte wegen der Komplexität der Reflektionskurven nicht genau auf die angepaßte, durchgezogene Linie fallen, ist doch eine quadratische Gesetzmäßigkeit herauszulesen:

$$E \sim (A^1)^2 \qquad (175)$$

Interessanterweise hängt dessen Größenordnung weder von der Gesamtdicke A noch vom Gradienten c ab.

### 4.2.3.4. Ergebnisse für Gradientenkristalle

Als nächstes wollen wir untersuchen, wie sich die Reflektivitäten in Abhängigkeit der Gesamtdicke und des Gradienten verhalten. Dabei sei auch auf die Diskussion der wesentlichen Punkte im Braggfall von B. Klar und F. Rustichelli [17] hingewiesen.

Die Abbildung (23) zeigt die Evolution der Reflektionskurven in Bragg- und Lauegeometrie bei konstanter Gesamtvariation c A = 10 für verschiedene Kristalldicken bzw. Gradienten. Die Reflektionskurven dicker Idealkristalle sind gestrichelt eingetragen. Die Kurven zu $c = 10 / \pi$ erreichen noch nicht die Maximalreflektivität von 100 % und oszillieren außerdem mit großer Amplitude auf und ab. Erhöhen wir die Kristalldicke, so wächst auch die Reflektivität. Für den idealen Gradienten $c_i = 2 / \pi$, wie er durch die Beziehungen (20) und (169) für den Braggfall definiert wurde, wird bereits ein schönes Extinktionsplateau nahe der Maximalreflektivität ausgebildet, wobei bei weiterer Verdickung nur noch die Kanten desselben ausgeprägt werden. Bemerkenswert ist, daß auch im Lauefall das Reflektionsvermögen auf nahezu 100 % anwächst, obwohl beim Idealkristall nur die Hälfte erreicht wird. Dies kann durch einen Verlust der Pendellösungseigenschaften erklärt werden, wobei sich die Gitterkonstante beim Fortschreiten durch das Medium ändert und somit eine Rückreflektion des Sekundärstrahls zugunsten des primären zwar noch lokal, jedoch nicht mehr global stattfindet. Die Halbwertsbreiten für stark extinktionsbedingte Reflektionskurven ergeben sich zu

$$y_B^H = c\,A + \frac{4}{\sqrt{3}} \qquad \text{für } b < 0,\ A \rightarrow \infty,\ c\,A = \text{const.} \tag{176}$$

im Braggfall und

$$y_L^H = c\,A \qquad \text{für } b > 0,\ A \rightarrow \infty,\ c\,A = \text{const.} \tag{177}$$

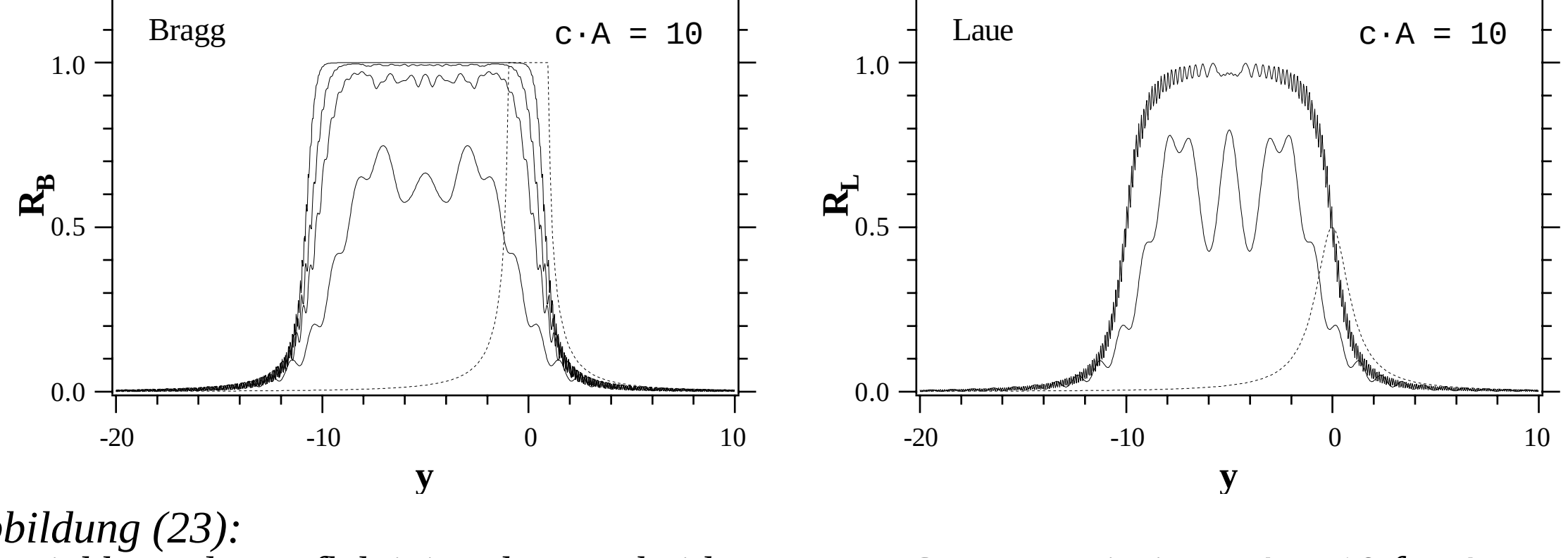

*Abbildung (23):*
*Entwicklung der Reflektivitätskurven bei konstanter Gesamtvariation c A = 10 für A = π, 3 π, 5 π und 10 π links im Braggfall sowie A = π und 10 π im Lauefall, rechts. Dickere Kristalle zeigen höhere Intensitäten. Zum Vergleich sind die Reflektivitätskurven dicker Idealkristalle punktiert eingezeichnet.*

im Lauefall. Im ersten Fall bilden sich die Plateaukanten bei y = 1 und y = -c A – 1 aus, so daß sich zur Variationsbreite des Streuvektors noch die natürliche Breite einer Darwinkurve hinzuaddiert. Damit können wir die integrierten Reflektivitäten $\Re^y$ dieser Grenzfälle abschätzen, die sich im Braggfall offensichtlich aus der Gesamtreflektivität $\pi$ eines Idealkristalls und der Verbreiterung aufgrund des Gradienten zusammensetzt

$$\Re_B^y = c\,A + \pi \qquad \text{für } b < 0,\ A \to \infty,\ c\,A = \text{const.} \tag{178}$$

Ersetzt man im Lauefall die schnell oszillierende Reflektionskurve durch einen über diese Oszillationen gemittelte, so läßt sich am Graphen abschätzen, daß ihr Integral mit dem einer Rechteckfunktion gleicher Breite identisch ist, also

$$\Re_L^y = c\,A \qquad \text{für } b > 0,\ A \to \infty,\ c\,A = \text{const.} \tag{179}$$

Numerisch berechnete Gesamtreflektivitäten für endliche Dicken sind in Abbildung (24) als Funktion dieser wiedergegeben. Bei kleinen Dicken gilt der punktiert eingezeichnete, kinematische Grenzfall

$$\Re^y = \pi\,A \qquad \text{für } A \to 0\,, \tag{180}$$

doch weichen die tatsächlichen Werte mit wachsendem A bald von dieser Proportionalität ab um im unendlichen ihre Sättigungswerte (178) und (179) zu erreichen.

Im folgenden wollen wir die Dickenabhängigkeit bei konstantem Gradienten c betrachten. Die

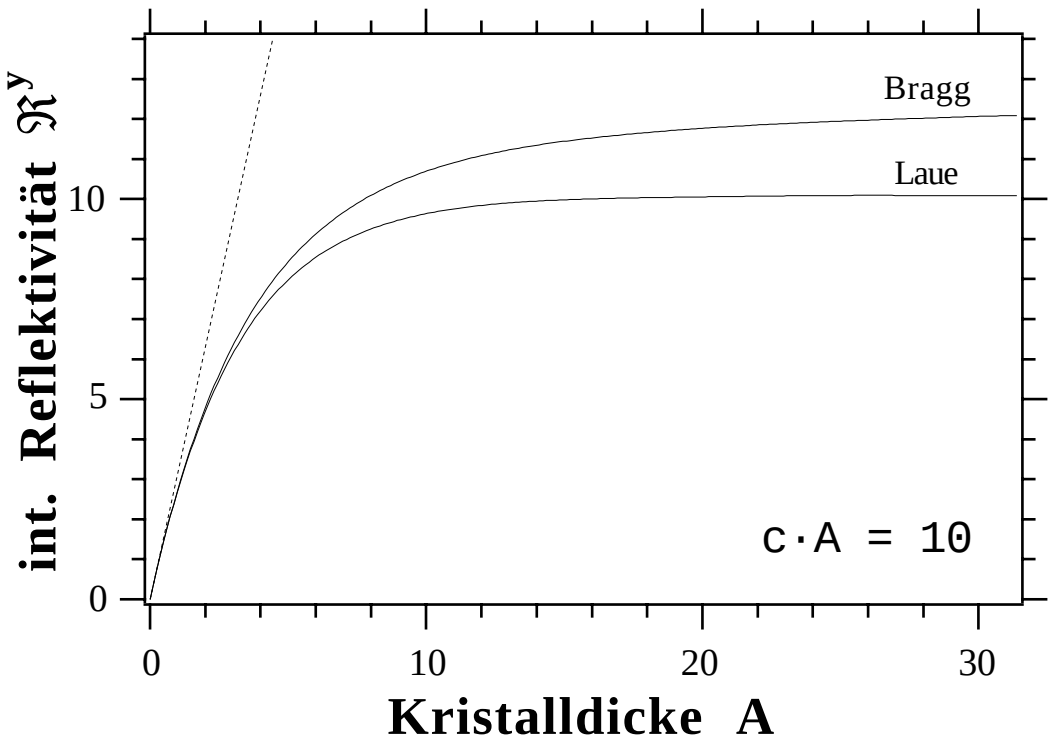

*Abbildung (24):*
*Integrierte Reflektivitäten im Bragg- und Lauefall bei festgehaltener Gesamtvariation c A = 10. Die kinematische Näherung ist durch die unterbrochene Linie wiedergegeben.*

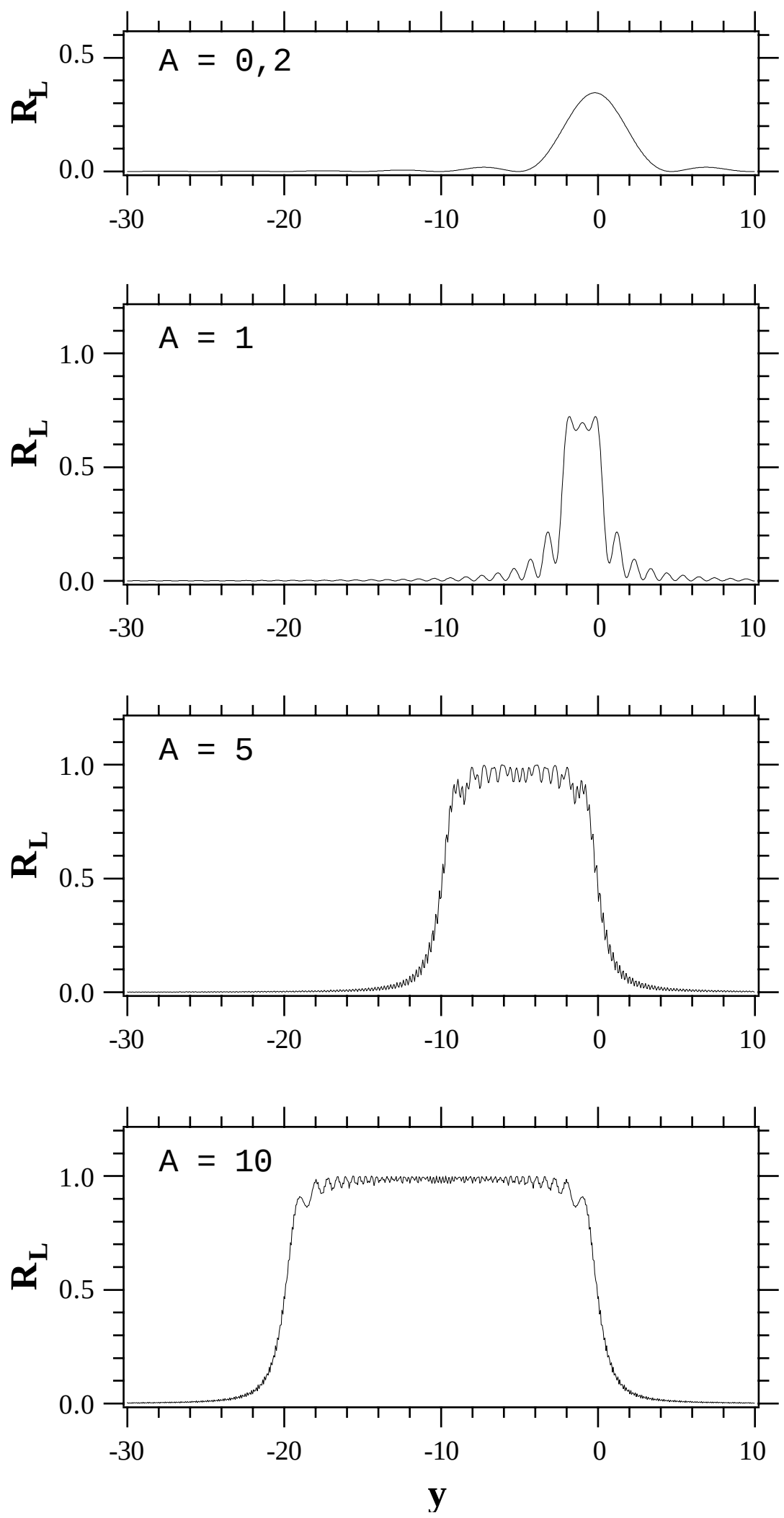

*Abbildung (25):*
*Evolution der Reflektionskurven im Lauefall mit der Gesamtdicke bei kontstantem $c = 2 / \pi$, dem idealen Gradienten.*

so berechneten Reflektionskurven ähneln für große Gradienten denen aus Abbildung (6) im Kapitel über kinematische Gradientenstreuung, obwohl bei kleinen Gradienten, wie hier für den idealen Gradienten in Abbildung (25) dargestellt, noch Extinktionseffekte hinzukommen. Während die Kurve für A = 0,2 im oberen Teilbild noch gut durch die Funktion (85) für dünne Kristalle beschrieben werden kann, bildet sich bei A = 1 schon ein kleines, auf den Gradienten zurückzuführendes Plateau aus, das bei weiterer Verdickung im dritten Teilbild bis zu seinem Plateauwert anwächst und schließlich, wie unten dargestellt, nur noch in die Breite gehen kann. Im Braggfall ergibt sich qualitativ derselbe Sachverhalt. Die damit zusammenhängenden integrierten Reflektivitäten sind in den Bildern (26) für Laue- und Braggeometrie in Abhängigkeit der Gesamtdicke aufgetragen. Für kleine A oder $c \to \infty$ gilt wieder der kinematische Grenzfall, bei dem $\mathfrak{R}^y$ mit dem Kristallvolumen gemäß (180) ansteigt. Bei Dicken A > 1 macht sich analog zu den Idealkristallen (c = 0) primäre Extinktion bemerkbar, so daß die integrierten Reflektivitäten abknicken und nur noch aufgrund einer Verbreiterung, aber nicht mehr durch Erhöhung des in Sättigung gegangenen Maximums weiterhin ansteigen. Der Anstieg der Kurven nach dem Einsatz dieser partiellen Extinktion ergibt sich offensichtlich zu

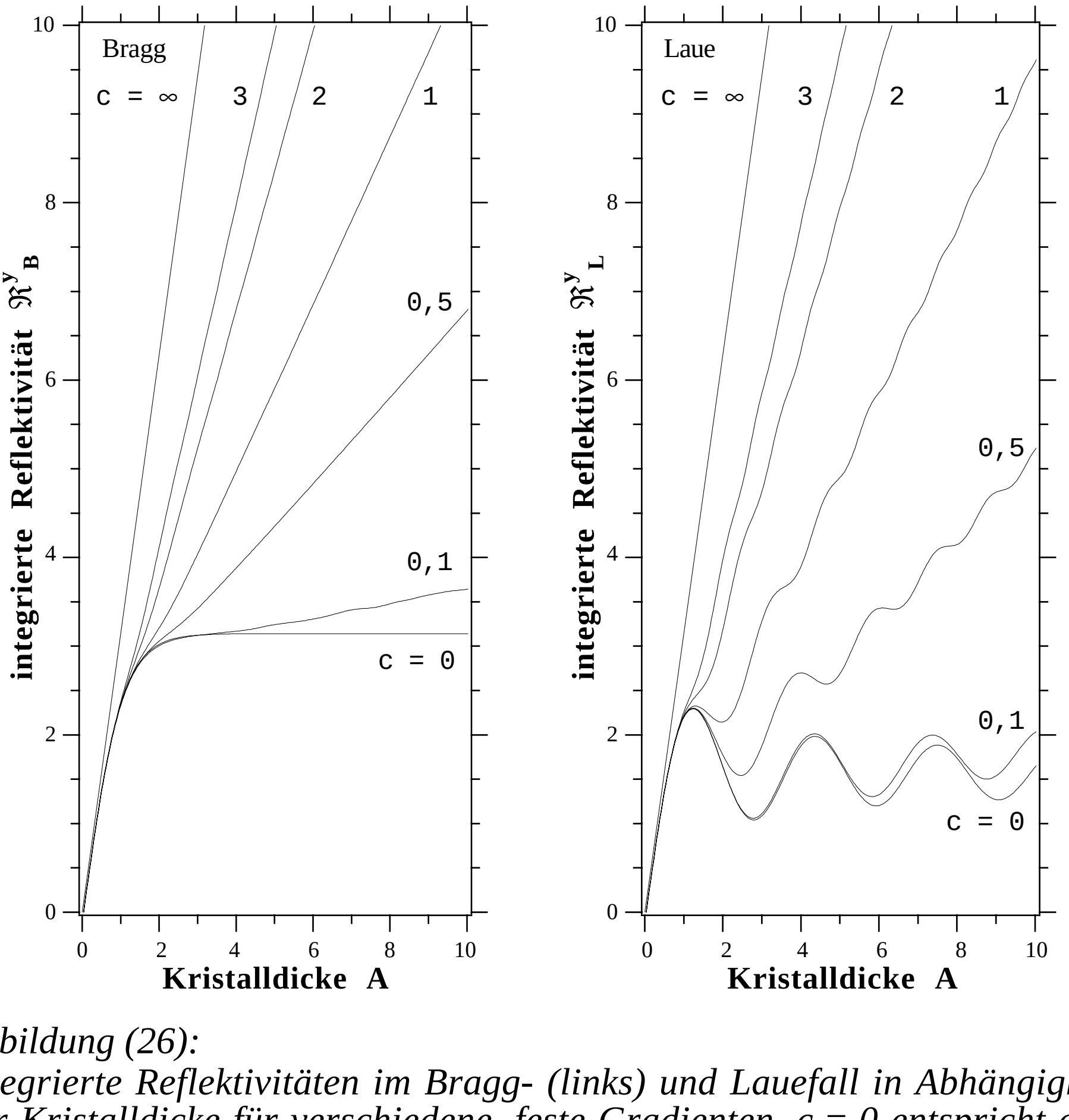

*Abbildung (26):*
*Integrierte Reflektivitäten im Bragg- (links) und Lauefall in Abhängigkeit der Kristalldicke für verschiedene, feste Gradienten. c = 0 entspricht dem Idealkristall und c = ∞ dem kinematischen Grenzfall.*

$$\frac{\partial \Re^y}{\partial A} = \pi \qquad \text{für } c \to \infty \tag{181}$$

und

$$\frac{\partial \Re^y}{\partial A} = c \qquad \text{für } c \to 0\ . \tag{182}$$

Im Lauefall erkennt man schön die Pendellösungsozillationen, die sogar noch bei relativ großen Gradienten vorhanden sind. Erst mit dem Verschwinden der partiellen Extinktion gehen auch diese gegen null.

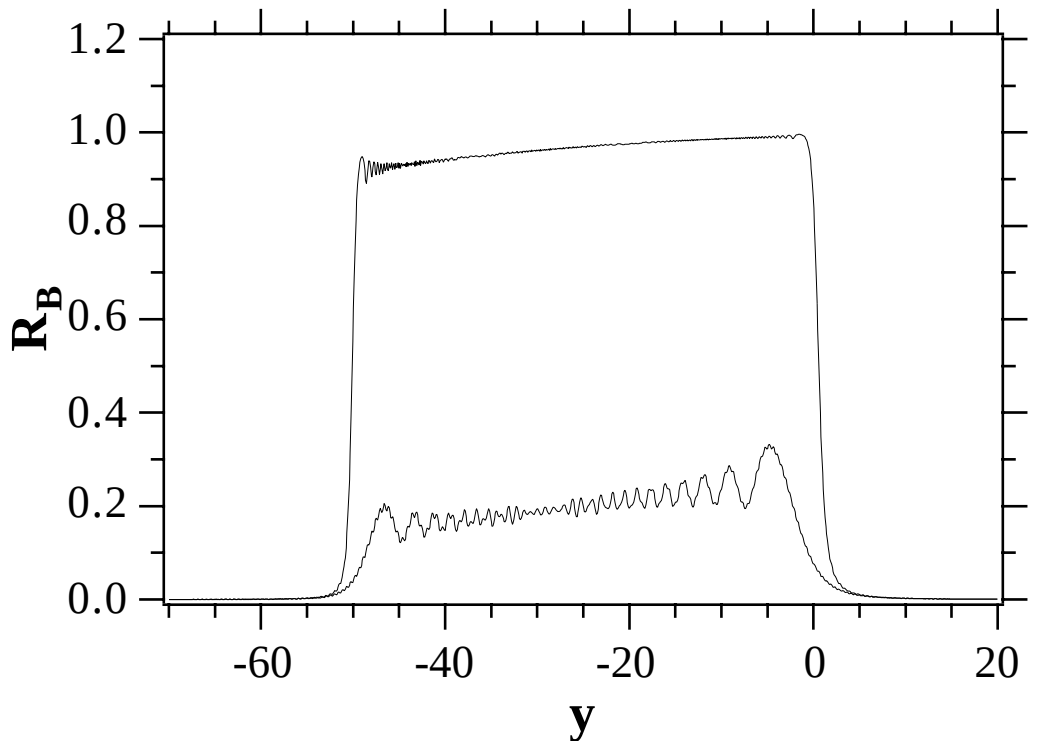

*Abbildung (27):*
*Reflektionsprofile in Braggeometrie mit Gitterkonstanten- und Streulängengradient c und p = 2, mit c·A = 50 und A = 5 für den unteren, nahezu kinematisch bestimmten und A = 25 π für den oberen, zumindest rechts schon fast die Maximalreflektivität erreichenden und damit extinktionsbestimmten Fall.*

Der Vollständigkeit halber soll hier noch der allgemeine Fall gleichzeitiger Gitterkonstanten- und Streulängenvariation angeführt werden, wie er strenggenommen auch beim System $Si_{1-x}Ge_x$ auftritt. Zwei Beispiele mit p = 2, c·A = 50 und A = 5 bzw. A = 25 π sind in Abbildung (27) für den Braggfall wiedergegeben. Interessant erscheint die, aufgrund der verschiedener Streulängendichten innerhalb unterschiedlicher Tiefen im Gradientenkristall hervorgerufene Neigung des Plateaus, solange die Sättigung noch nicht erreicht ist.

## 4.3. Vergleich der Theorien für Gradientenkristalle

In den vorangegangenen Kapiteln wurden zwei Methoden zur Beschreibung der Reflektionseigenschaften an Gradientenkristallen, eine analytische im kinematischen Grenzfall, die andere mittels Transfermatrizen im Rahmen der dynamischen Beugungstheorie neu hergeleitet. Zunächst wollen wir an Beispielen zeigen, daß die von diesen Theorien vorausgesagten Reflektionskurven in ihrem Gültigkeitsbereich übereinstimmen, bevor wir die Ergebnisse der dynamischen Lösung mit denen der weit etablierten, analytischen Methode nach Taupin vergleichen [17, 18].

### 4.3.1. Transfermatrizenergebnisse und kinematische Beugung

Die kinematische Näherung, wie sie in Kapitel 4.1. aufgestellt wurde, besitzt bekanntlich einen Gültigkeitsbereich für solche Kristalle, deren Dicken wesentlich dünner als die Pendellösungsperiode sind. Dann kann die Abschwächung des Primärstrahls durch Reflektion zugunsten des Sekundärstrahls beim Fortschreiten durch das Medium vernachlässigt werden, so daß der Superpositionsansatz an verschiedenen Gitterebenen gestreuter Wellen gleicher Amplituden gerechtfertigt ist.

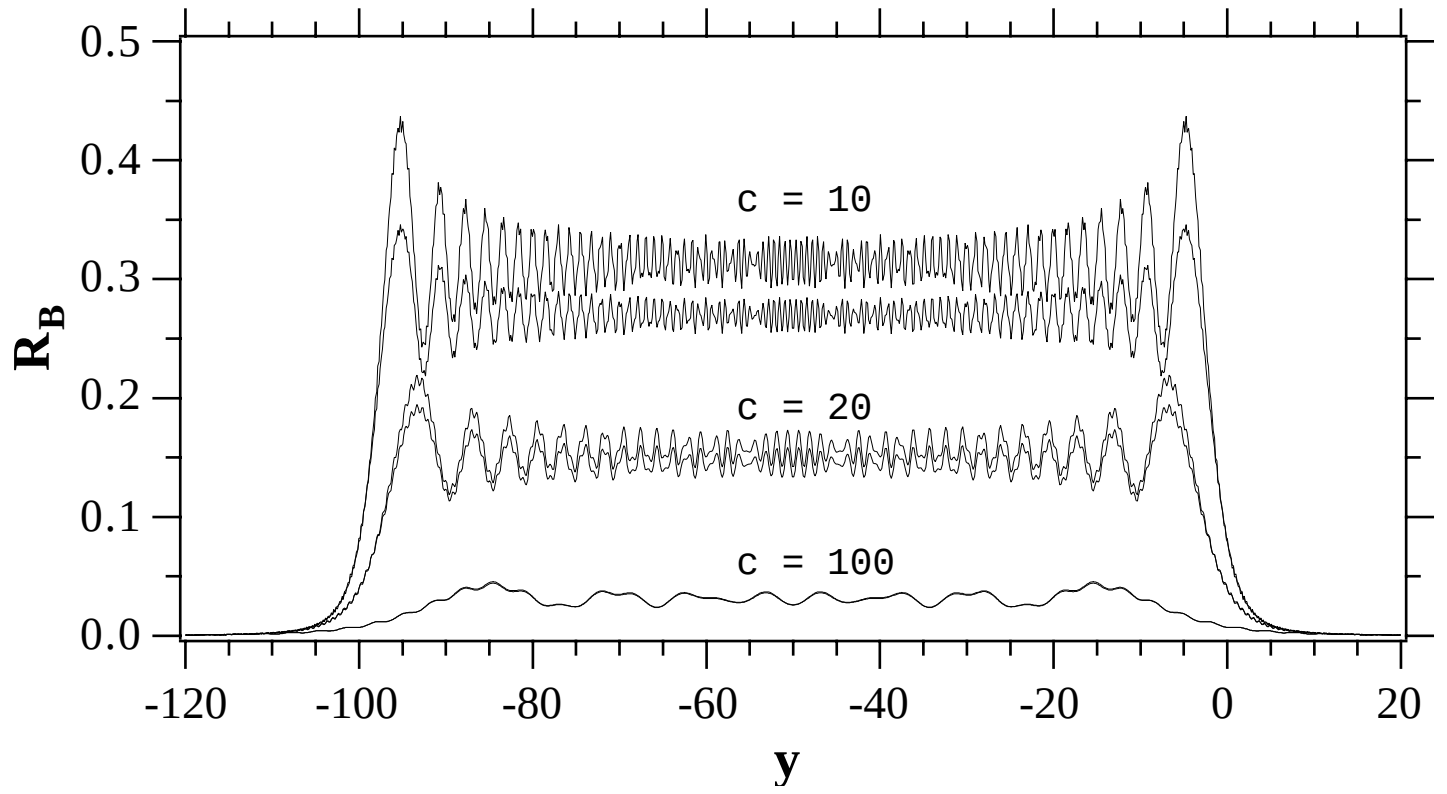

*Abbildung (28):*
*Vergleich der berechneten Reflektionskurven für dicke Kristalle nach der kinematischen und der dynamischen Methode für c·A = 100 und die angegebenen Gradienten. Bei kleinen Gradienten tritt partielle Extinktion ein, weshalb die kinematische Theorie erhöhte Werte liefert.*

Bei der Betrachtung von Gradientenkristallen können auch dicke Proben in der kinematischen Theorie beschrieben werden, solange der Gradient groß und die damit verbundene effektive Dicke, inmitten derer die Braggbedingung innerhalb ihrer natürlichen Breite erfüllt wird, klein ist. Derartige Ergebnisse für c·A = 100 sind in Abbildung (28) graphisch dargestellt. Für c = 10 liefert die kinematische Theorie beträchtlich höhere Reflektivitäten als die Transfermatrizenmethode. Dies beruht ausschließlich auf dem Effekt der partiellen Extinktion, wie sie weiter oben behandelt wurde. Bei c = 20 ist immer noch ein Unterschied zu erkennen, während für c = 100 die beiden Kurven bereits, zumindest auf dieser Skala, völlig übereinanderfallen. Bemerkenswert ist, daß unabhängig von diesem Extinktionseffekt, die schnellen Oszillationen an den richtigen Stellen auftreten und somit, wie in Kapitel 4.1. diskutiert, in ihrer Periodizität nur vom Gradienten und der Kristalldicke abhängen.

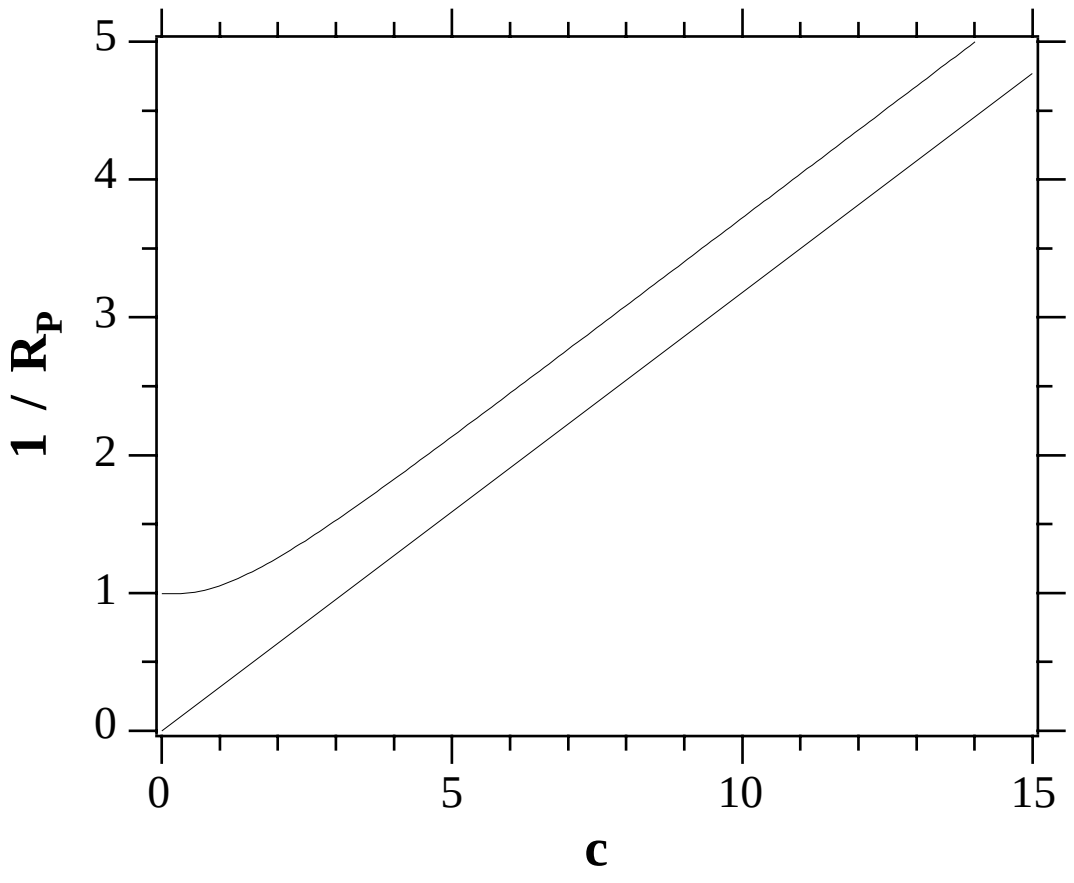

*Abbildung (29):*
*Reziproke Auftragung der Plateaumittelwerte $R_P$ in Abhängigkeit des Gradienten c. Die untere Kurve entspricht der kinematischen Näherung, während die obere dynamisch berechnet wurde und den Extinktionseffekt bei kleinen c mitberücksichtigt.*

Etwas analytischere Kriterien zum Unterschied der beiden Rechenmethoden erhält man, wenn, wie in Abbildung (29) reziprok dargestellt, die lokal über die schnellen Oszillationen gemittelten Plateauwerte $R_P$ als Funktion des Gradienten betrachtet werden. Aus Kapitel 4.1. kennen wir das Verhalten (47)

$$R_P^k = \frac{\pi}{c} \tag{183}$$

für den kinematischen Fall. Wir erinnern uns, daß die Plateaureflektivität für halbunendliche Kristalle hergeleitet wurde. Dabei steigt für $c \to 0$ das Kristallvolumen, innerhalb dessen die Braggbedingung für einen festen Wellenvektor erfüllt ist, gegen unendlich. Proportional mit diesem Volumen steigt auch die Reflektivität, und wir befinden uns in dem Bereich, in dem die kinematische Theorie versagt. Die Kurve für die dynamische Lösung wurde numerisch erstellt, geht für $c \to 0$ in Sättigung, also gegen 1 und folgt für große Gradienten der Hyperbel mit dem angepaßten Parameter m

$$R_P^d = \frac{\pi}{c+m} \;;\; m = 1{,}69\,. \tag{184}$$

Daraus ergibt sich eine Fehlerabschätzung der Ordnung 1 / c beim Gebrauch der kinematischen Streutheorie an dicken Kristallen.

### 4.3.2. Transfermatrizen und Takagi-Taupin

Während der sechziger Jahre wurde von den Autoren Satio Takagi [19] und Daniel Taupin [18] unabhängigerweise ein Formalismus zur Beschreibung leicht gestörter Kristalle im Rahmen der dynamischen Streutheorie hergeleitet, deren grundlegende Differentialgleichungen die nach ihnen benannten Takagi-Taupin-Gleichungen darstellen. Im vorliegenden Kapitel soll die Übereinkunft der daraus hervorgehenden Lösungen für absorptionsfreie, lineare Gradientenkristalle mit denen unserer Transfermatrizenmethode, und damit noch einmal die Richtigkeit letzterer an Beispielen demonstriert werden.
Die Anpassung der von Taupin geforderten, absorptionsbedingten Randbedingungen an den absorptionsschwachen Fall der Neutronenstreuung wurde in einem Artikel von B. Klar und F. Rustichelli durchgeführt [17]. Darin werden jedoch noch die von Taupin eingeführte bzw. auf Zachariasen [14] zurückgehende Notation der Elektrodynamik mit ihren Vektorfeldgleichungen verwendet und erst zum Schluß die entsprechenden Größen für Neutronen eingesetzt, während wir uns hier zwar sehr stark an diesen Artikel lehnen, aber dennoch die durch Rauch und Petrascheck [12, 13] eingeführten Formelzeichen der vorliegenden Arbeit verwenden.

Taupin geht von einer Darstellung der Wellenfunktion

$$\psi(\vec{r}) = u(\vec{r}) \exp\left(i\, \Phi(\vec{r})\right) \tag{185}$$

mit Amplitude $u(\vec{r})$ und Phase $\Phi(\vec{r})$ aus, die im Vakuum durch die reellen Größen

$$u(\vec{r}) = \text{const.} \quad \text{und} \quad \Phi(\vec{r}) = \vec{k}\,\vec{r} \tag{186}$$

gegeben sind. Innerhalb des Kristallmediums wird angenommen, daß die Welle mit der selben Phase $\Phi(\vec{r})$ fortschreitet, so daß $u(\vec{r})$ die Störungen durch das Kristallpotential enthalten muß, und dadurch im allgemeinen komplex wird. Dem entspricht im Idealkristall der Blochwellenansatz. Nun entwickelt man

$$u(\vec{r}) = \sum_{\vec{G}} u_G(\vec{r}) \exp\left(i\, \Phi_G(\vec{r})\right) \tag{187}$$

in eine Fourierreihe mit den Phasen

$$\Phi_G(\vec{r}) = \Phi(\vec{r}) + 2\,\pi\, n_G(\vec{r})\,. \tag{188}$$

Das Indexfeld $n_G(\vec{r})$ ist ein Skalarfeld und beschreibt die Lage der Gitterebenen. Dabei werden letztere der Reihe nach durchnumeriert. $n_G(\vec{r})$ steigt zwischen benachbarten Ebenen um 1 an und ist an den Ortskoordinaten jeder Ebene ganzzahlig. Z. B. wächst bei einem ungestörten Kristallgitter $n_G(\vec{r})$ linear in Richtung des Normalenvektors zu den Ebenen. Abbildung (30) verdeutlicht diesen Sachverhalt für einen Gradientenkristall.

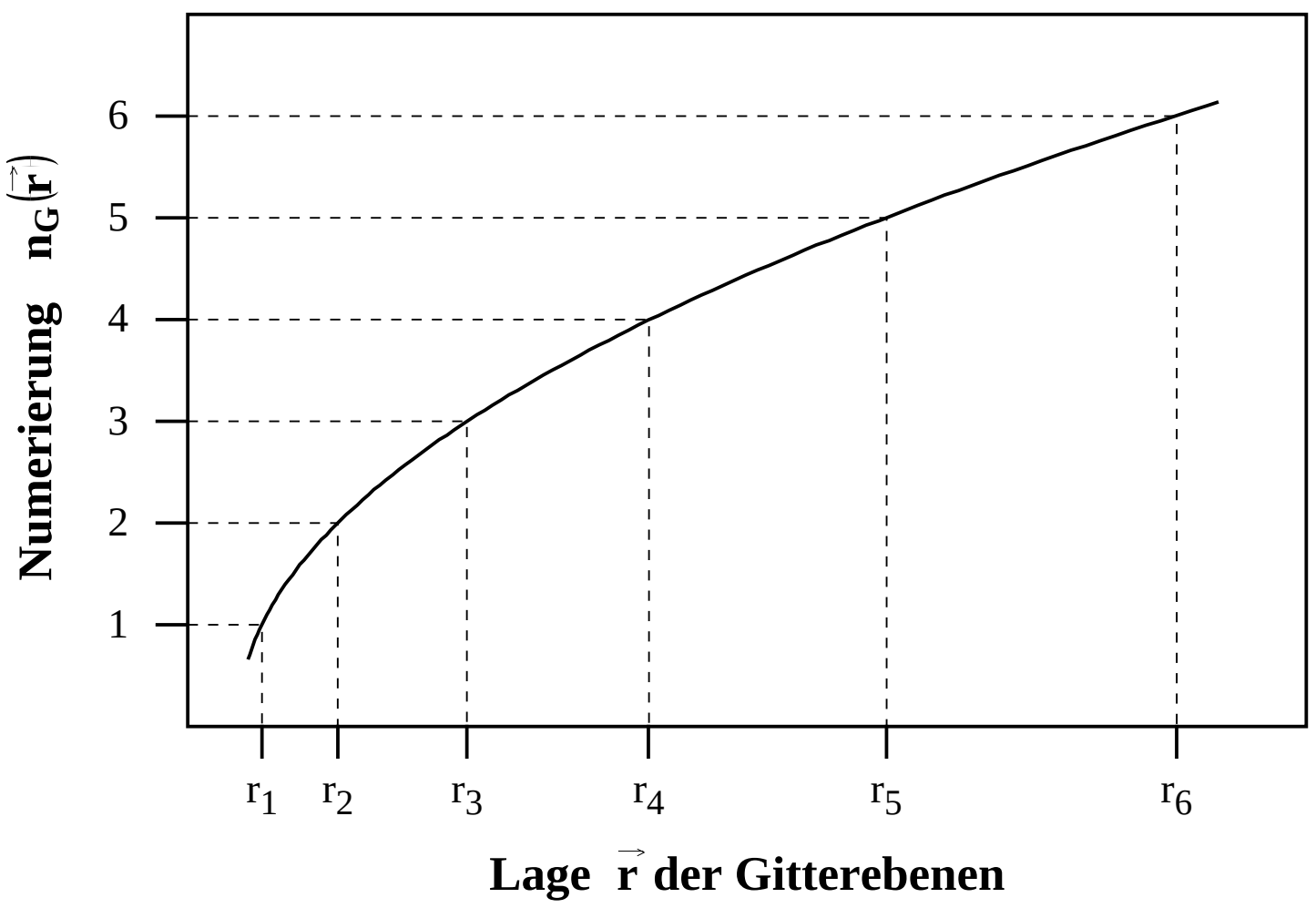

*Abbildung (30):*
*Das von Taupin eingeführte Indexfeld $n_G(\vec{r})$ ändert sich örtlich so, daß es an den Gitterebenen ganzzahlig ist und von Ebene zu Ebene anwächst.*

Entwickelt man das Kristallpotential

$$V(\vec{r}) = \sum_{\vec{G}} V(\vec{G}) \exp\left(i\, 2\, \pi\, n_G(\vec{r})\right) \qquad (189)$$

mit den ungestörten Fourierkomponenten (57) nach dieser Periodizität, setzt es zusammen mit der Wellenfunktion (185) in die Schrödingergleichung (53) ein, so erhält man das Takagi-Taupin-Grundgleichungssystem

$$-i\frac{2}{k}\frac{\partial u_G}{\partial x_G} = \frac{V(\vec{0})}{E} u_G + \frac{V(\vec{G})}{E} u_0 - \alpha\, u_G + \frac{i}{k^2} u_G \left(\Delta \Phi_G\right) \qquad (190)$$

und

$$-i\frac{2}{k}\frac{\partial u_0}{\partial x_0} = \frac{V(\vec{0})}{E} u_0 + \frac{V(\vec{G})}{E} u_G + \frac{i}{k^2} u_0 \left(\Delta \Phi_0\right) \qquad (191)$$

worin $x_0$ und $x_G$ die Koordinaten von $\vec{r}$ in dem aus vorwärtsgebeugter Richtung $\vec{S}_0$ und abgebeugter Richtung $\vec{S}_G$ aufgespannten, schiefwinkligen Koordinatensystem darstellen:

$$\vec{r} = x_0\, \vec{S}_0 + x_G\, \vec{S}_G \qquad (192)$$

Die Abweichung $\alpha$ von der geometrischen Braggposition ist wie gehabt durch (68) definiert. In der Arbeit von Taupin wurde gezeigt, daß die Ausdrücke mit den Laplaceoperatoren in (190) und (191) bei kleinen Gittervariationen gegen den Rest vernachlässigbar sind. Das Differentialgleichungssystem läßt sich, hier für den absorptionsfreien Fall, mit A (65), y (66), b (67), X (63) und

$$\hat{X}(A) = \frac{1}{\sqrt{|b|}} \frac{u_G}{u_0} = \frac{1}{\sqrt{|b|}} X(A) \tag{193}$$

in

$$-i \frac{\partial \hat{X}}{\partial A} = \hat{X}^2 - 2 \hat{X} y + 1 \tag{194}$$

umwandeln, wobei sowohl $\hat{X}$ als auch y von A abhängen. Für einen linearen Gittergradienten setzen wir gemäß (166)

$$y = y_0 + c A \tag{195}$$

und erhalten den für uns endgültigen Ausdruck

$$-i \frac{\partial \hat{X}}{\partial A} = \hat{X}^2 - 2 \hat{X} (y_0 + c A) + 1 \, . \tag{196}$$

Mit der Randbedingung im Braggfall

$$\hat{X}(A_{max}) = 0 \tag{197}$$

wird $\hat{X}$ eindeutig bestimmt, so daß man mit

$$R_B(y_0) = \left|\hat{X}(0)\right|^2 \tag{198}$$

die Reflektivität des abgebeugten Strahls berechnen kann. Die Integration erfolgt numerisch und punktweise für jedes $y_0$ der Reflektionskurve.

Wie wir im oberen Teil der Abbildung (31) sehen, stimmen die aufgeführten Beispiele für die Berechnung mittels Taupin oder Transfermatrizen für große und für kleine Gradienten genau überein. Die Absolutbeträge der Differenzen beider Ergebnisse sind im unteren Teil wiedergegeben. Für kleine Gradienten sind vor allem Abweichungen in den Flanken, bei großen c im Plateaubereich festzustellen. Bemerkenswert ist die Asymmetrie des Fehlers beim kleinen Gradienten, der wohl auf numerische Probleme zurückzuführen ist.

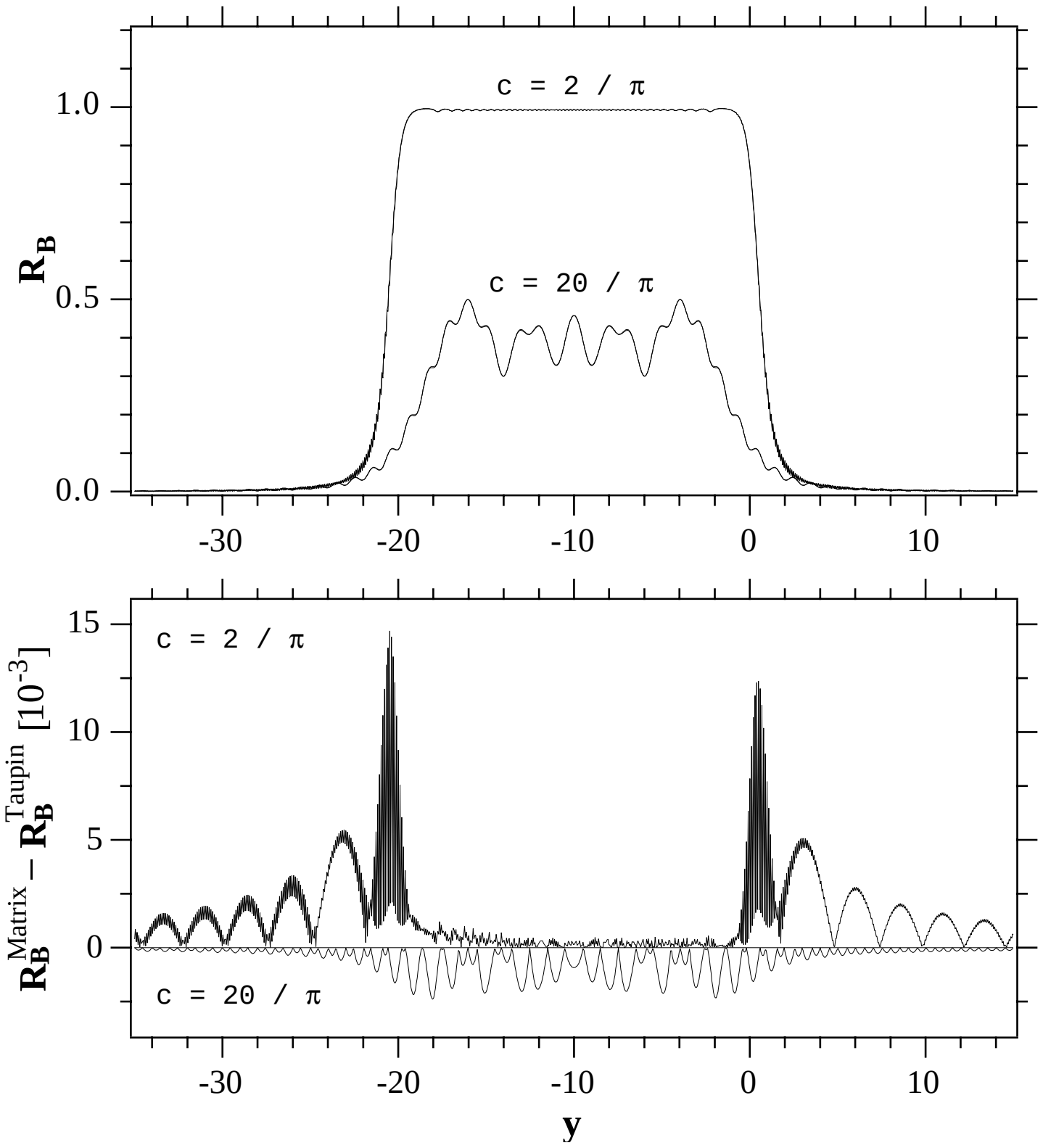

*Abbildung (31):*
*Zum Vergleich übereinandergezeichnete Reflektionskurven im oberen Teilbild, die einmal nach der Taupin- und zum anderen nach der Transfermatrizenmethode berechnet wurden. Da auf dieser Skala kein Unterschied festzustellen ist, wurden im unteren Teilbild deren Differenzen dem Betrag nach wiedergegeben, und zwar nach oben für kleine, nach unten für große Gradienten. Die Asymmetrie für den kleinen Gradienten kann durch numerische Rechenfehler gedeutet werden.*

# 5. Die Kristallzucht

Die Idee eines Gradientenmonochromators auf der Basis von $Si_{1-x}Ge_x$ Legierungsvariationen geht mit der Anmeldung eines Patents durch Maier-Leibnitz und Rustichelli [10] bis ins Ende der sechziger Jahre zurück. Sie beruht auf der gleichbleibenden Gitterstruktur sowie der vollständigen Mischbarkeit der beiden neutronenfreundlichen Elemente. Nicht zuletzt baut man auch auf einen, aus der Halbleiterindustrie entstandenen, außerordentlich umfangreichen Erfahrungsbereich.

## 5.1. Das Materialsystem

Silizium und Germanium kristallisieren beide in der Diamantstruktur. Um Gradientenkristalle zu erhalten, werden bei der Legierung ausgehend von Silizium statistisch mehr und mehr Siliziumatome durch Germaniumatome ersetzt. Nach dem Vegard'schen Gesetz erwartet man einen linearen Anstieg der Gitterkonstanten a mit der Germaniumkonzentration x. Dies ist in Abbildung (32) durch die strichpunktierte Linie wiedergegeben. Strenggenommen weicht dieser Zusammenhang jedoch von der Linearität ab. Meßwerte dazu sind in dem Artikel von Dismukes [20] oder in Tabellenwerken [21] zusammengestellt. Die durchgezogene Linie zeigt eine quadratische Ausgleichsrechnung zu diesen Werten und ist durch das Polynom

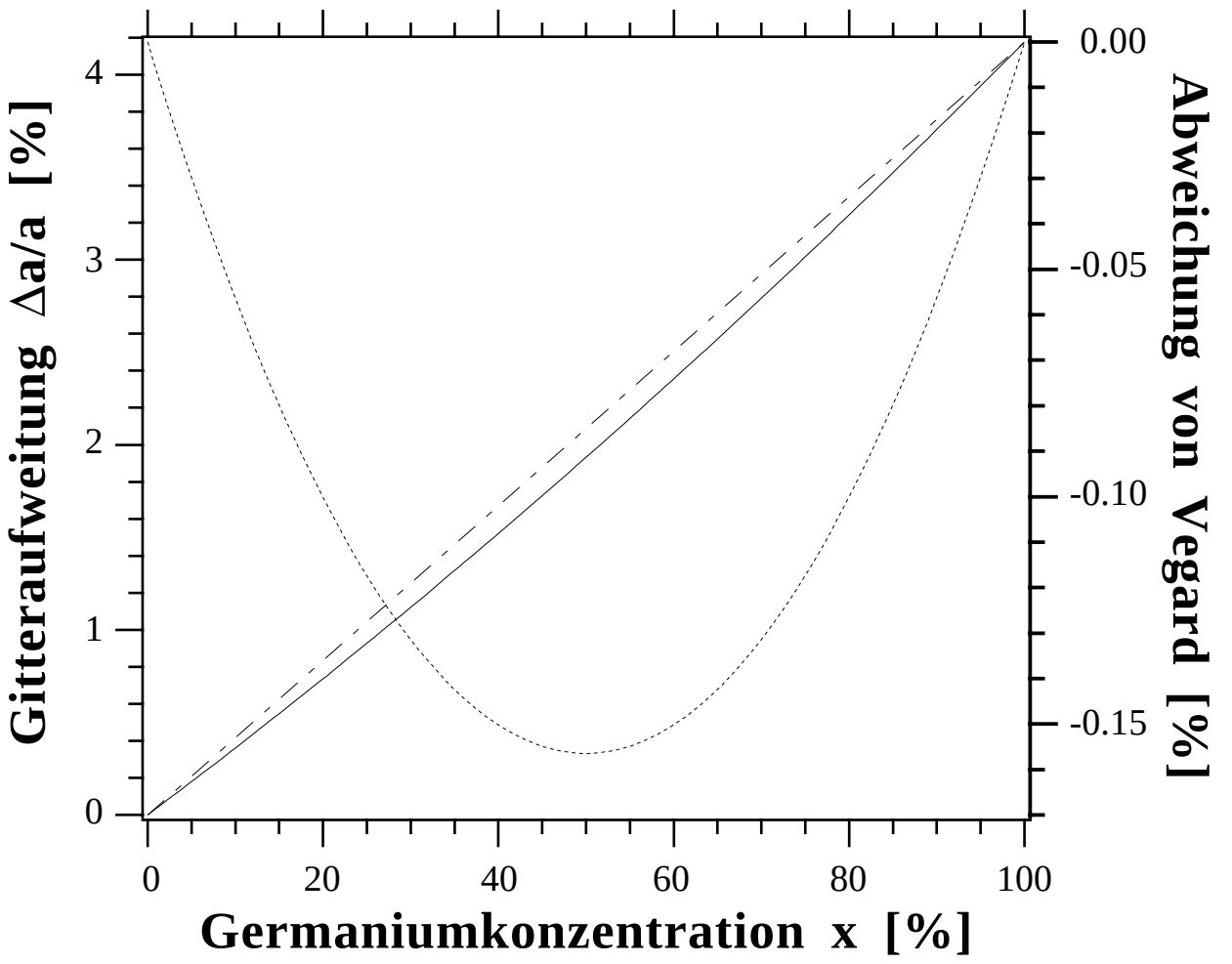

*Abbildung (32):*
*Gitteraufweitung einer $Si_{1-x}Ge_x$ Legierung in Bezug auf reines Silizium: Vegard'sches Gesetz (strichpunktiert), tatsächlich beobachtbare (durchgezogen) und deren Differenz (punktiert) [20, 21]. Die tatsächliche Gitterkonstante liegt immer etwas unterhalb der durch das Vegard'-sche Gesetz idealisierten. Die Gesamtaufweitung beträgt 4,18 %.*

$$\frac{\Delta a}{a} = 0{,}03552\ x + 0{,}006258\ x^2 \qquad (199)$$

bestimmt. Die Abweichung zum Vegard'schen Gesetz ist durch die punktierte Linie mit der rechten Skala wiedergegeben.

Das zugehörige Phasendiagramm für den Übergang zwischen Festkörper und Flüssigkeit ist in Abbildung (33) wiedergegeben. Alle Versuche, die oben genannten Kristalle aus der Schmelze zu ziehen waren bislang erfolglos. Dies kann wohl damit begründet werden, daß einerseits die Erstarrungslinie im Phasendiagramm sehr weit von der Schmelzlinie entfernt ist und die von diesen beiden Linien aufgespannte Linse sehr flach liegt, so daß die Konzentrationen bei vorgegebener Temperatur in Festkörper und Schmelze sehr unterschiedlich sind. Um eine vorgegebene Festkörperkonzentration zu erhalten muß diesem bei der Erstarrung ein erheblicher Materialtransport nahe der Grenzfläche entgegenwirken. Man kann sich vorstellen, daß bei deren Variation, wie sie bei einem Gradientenkristall stufenlos auftreten soll, das System sehr leicht unkontrollierbar wird. Andererseits ist das Kristallgitter des Silizium nahe der Schmelztemperatur so weich, daß die, wie hier vom Germanium dargestellt, durch eingelagerte Fremdatome auftretenden Spannungen nicht mehr vom Gitter aufgenommen werden, sondern relaxieren. In diesem Fall entstehen leicht Instabilitäten, die zu polykristallinem Wachstum führt.

In den letzten Jahren haben sich weitere Kristallzuchtmethoden, die Epitaxien entwickelt und bedeutende Stellungen, insbesonders bei der Herstellung dünner Schichten, eingenommen. Bei der chemischen Gasphasenepitaxie (CVD), die sich durch große Wachstumsgeschwindigkeiten auszeichnet, werden Trägermoleküle nahe an einem Substrat aufgespalten und dann auf diesem, je nach Prozeßbedingungen kristallin, abgelagert. Das Kristallwachstum geschieht hier weit unterhalb des Schmelzpunktes, wo das Siliziumgitter mechanische Spannungen aufnehmen kann und somit der bei der Zucht aus der Schmelze auftretende Störprozeß entfällt. Außerdem kann die gewünschte Festkörperzusammensetzung in sehr einfacher Weise durch die Mischungsverhältnisse der Trägergase gesteuert und im Laufe des Wachstums beliebig variiert werden. Diese Methode wurde von Magerl und Holm zunächst im Rahmen eines abgeschlossenen Systems, jedoch für eine feste Legierungszusammensetzung [22] erfolgreich auf das System $Si_{1-x}Ge_x$ angewendet. Um nun auch Konzentrationsgradienten zu verwirklichen, schlugen Magerl et al. eine CVD-Herstellungsmethode auf Silan- und Germanbasis ($SiH_4$ bzw. $GeH_4$) vor und führten 1990 eine der vorliegenden Arbeit zugrundeliegenden Machbarkeitsstudie mit Erfolg durch [23, 24]. Da für die Anwendung der Gradientenkristalle in der Neutronenstreuung erhebliche Dicken benötigt werden, arbeiten die Autoren mittels kleinflächigen Substraten in einem induktionsgeheizten, röhrenförmigen Ofen hauptsächlich auf die Wachstumsgeschwindigkeit hin. Es wurden Wachstumsgeschwindigkeiten von bis zu 0,7 µm/min erreicht, die, wie die Schichtdicken auch, gegenüber herkömmlichen CVD-Methoden um einen Faktor 100 bis 1000 mal größer sind.

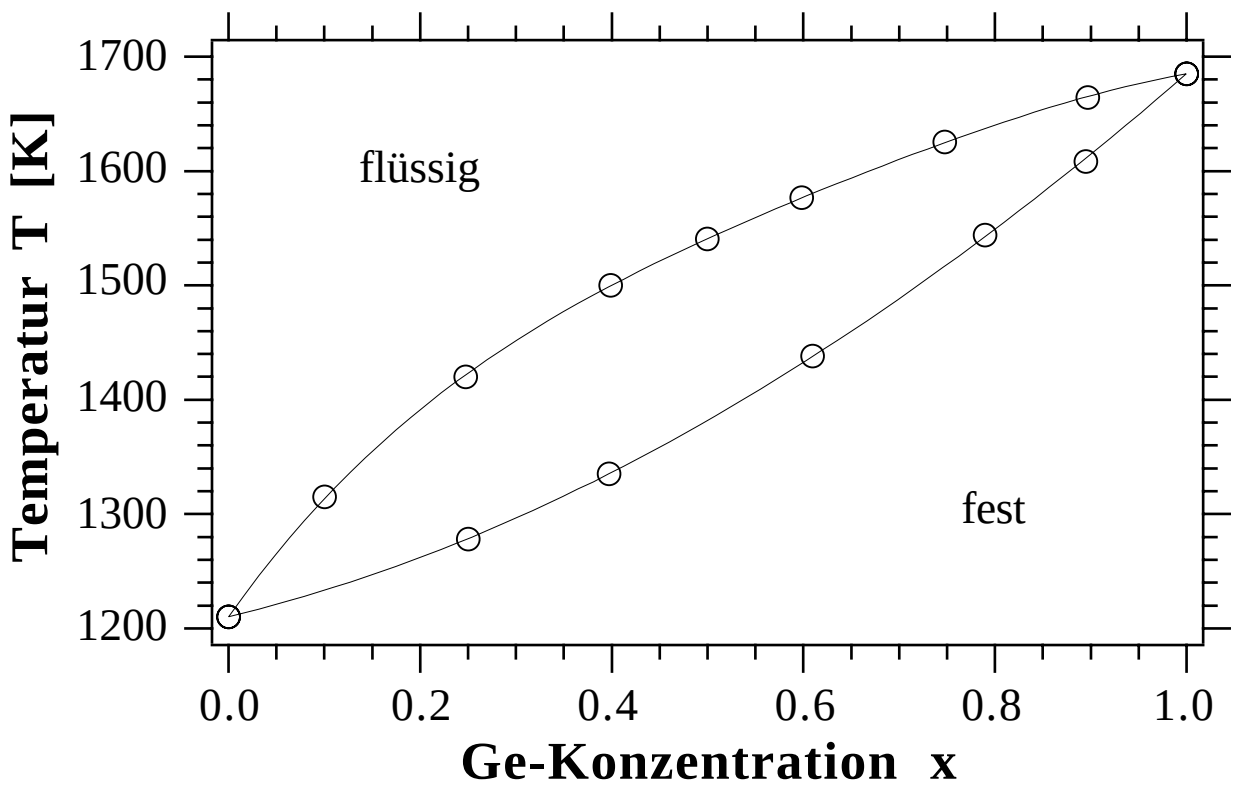

*Abbildung (33):*
*Phasendiagramm des Systems $Si_{1-x}Ge_x$ zwischen Schmelze und Festkörper [21].*

Damit wurde es erstmals möglich, auf diese Art und Weise millimeterdicke Kristalle innerhalb von vernünftigen Zeiträumen, nämlich einigen Tagen herzustellen.

Die Aufgabe in diesem Teil der vorliegenden Arbeit bestand darin, einen neuen Reaktionsofen für großflächige Substrate mit Durchmessern von 10 cm zu bauen und die Kristalle für die anschließenden Experimente zu ziehen. Die Hauptschwierigkeiten bestanden vor allem in der Entwicklung eines geeigneten Heizelements sowie in der Führung der Gase, um ein homogenes, kristallines Wachstum über den benötigten Zeitraum zu gewährleisten.

## 5.2. Die Kristallzuchtanlage

Die Anlage zur Gasphasenabscheidung besteht im wesentlichen aus dem Gassystem und einem Zuchtofen, der sich wiederum in ein Gasströmungssystem und ein Heizelement im inneren eines Vakuumtopfes unterteilen läßt. Eine Übersicht ist in Abbildung (34) dargestellt. Im folgenden sollen diese Komponenten beschrieben werden.

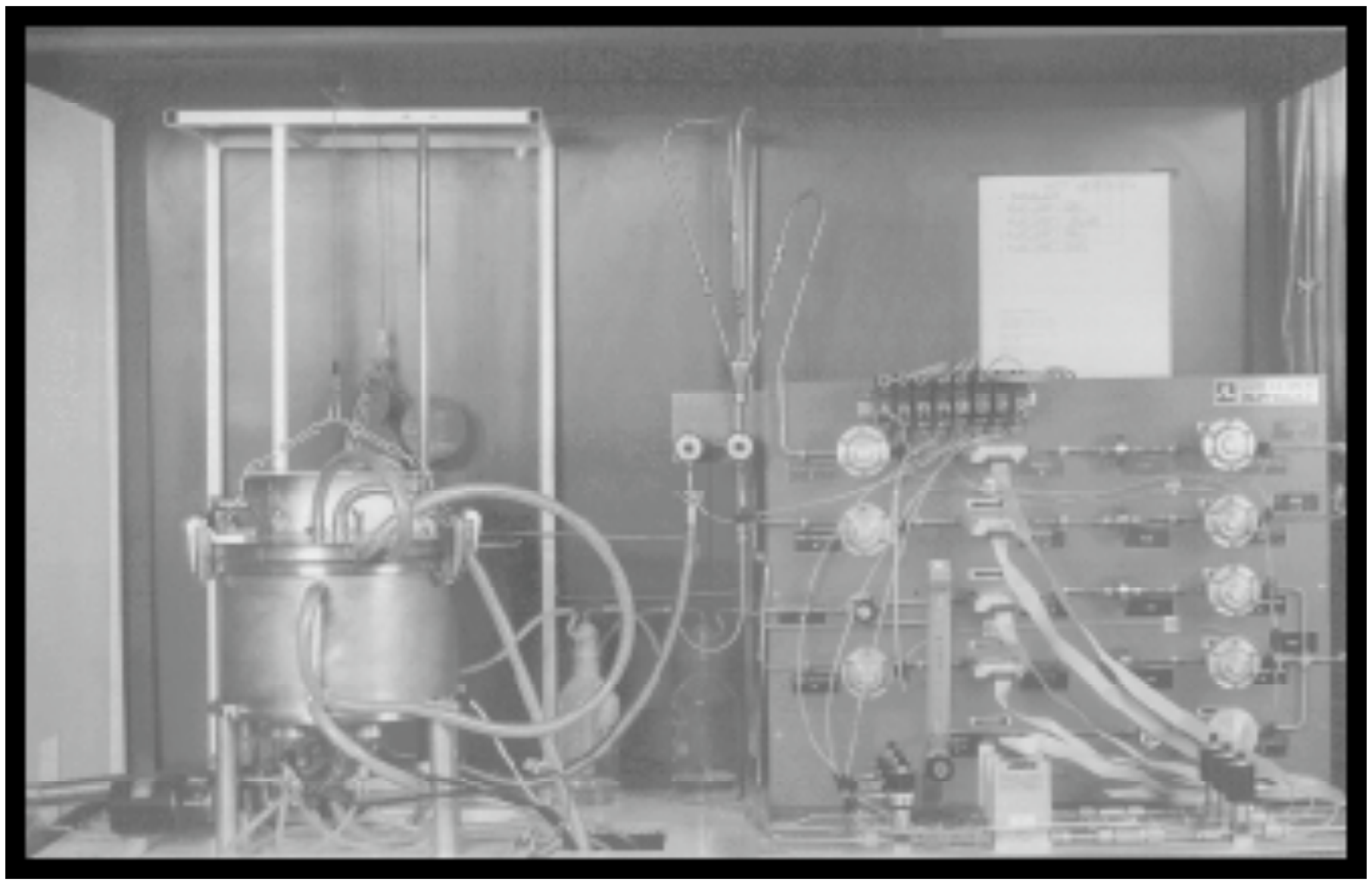

*Abbildung (34):*
*Die Versuchsanlage zur Kristallzucht. Im linken Teil der kesselförmige Reaktionsofen mit den Schläuchen für die Wasserkühlung. Rechts im Bild das Gaszuführsystem mit den preßluftgesteuerten Ventilen und den Massenflußreglern. Im Bildzentum sehen wir die trichterförmige Mischkammer, ganz links unten den Vakuumschlauch zu den Pumpen.*

### 5.2.1. Das Gassystem

Im Zuchtofen werden wohldefinierte Bedingungen der Gase erwartet, die vom Gassystem kontrolliert und geregelt werden. Prinzipiell kann man hier wieder zwischen den beiden Komponenten vor und hinter dem Zuchtofen, dem Gaszuführ- und dem Gasabführsystem unterscheiden. Das erste übernimmt die Regelung von Zusammensetzung und Durchfluß, während das zweite den Druck steuert.

#### 5.2.1.1 Das Gaszuführsystem

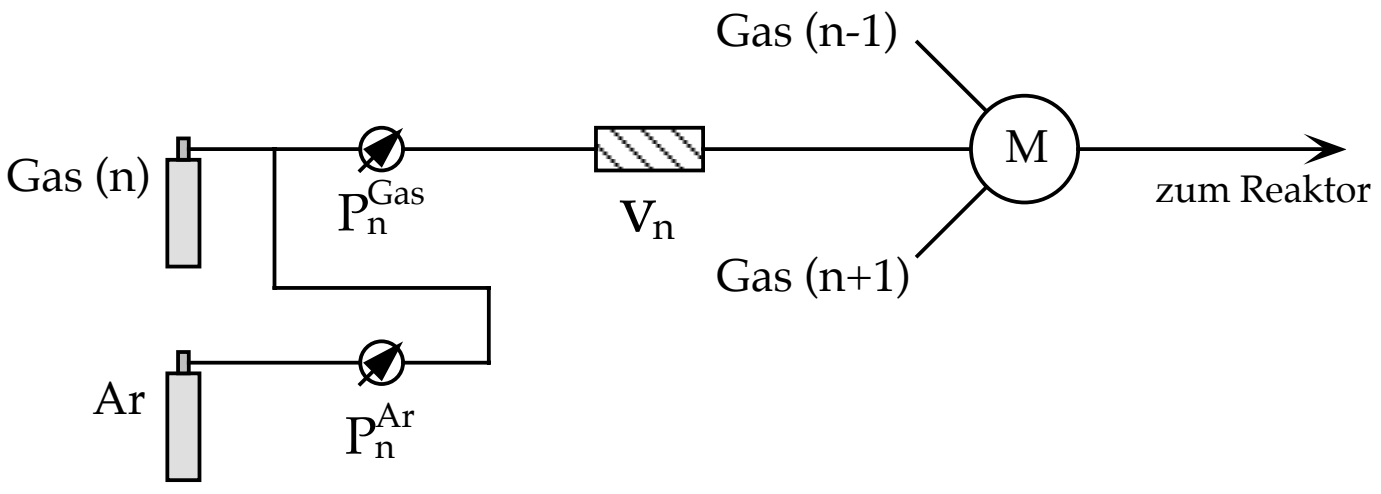

*Abbildung (35):*
*Eine Linie im Gassystem mit Einrichtung zum Argonspülen. Gas (n), Ar: Flaschen der entsprechenden Gase; $P_n$: Druckminderer, $V_n$: Massenflußregler; M: Mischkammer mit weiteren Gaslinien.*

Das Gaszuführsystem besteht aus vier verschiedenen, je wie in Abbildung (35) skizzierten Linien für jedes der in Tabelle (4) aufgelisteten Reaktionsgase. Als Quelle dienen handelsübliche Druckflaschen. Nach je einem Druckminderer auf einige Bar werden die Gase zu je einem elektronisch einstellbaren Massenflußregler geleitet. Mit diesem läßt sich der Mischungsanteil des Reaktionsgases einstellen und konstant halten bzw. gezielt zeitlich verändern. Nach einer Mischkammer, die die Gase vermengt, werden sie dem Zuchtofen zugeleitet. Jede Leitung, mit Ausnahme der von $H_2$ kann aus Sicherheits- und Sauberkeitsgründen mit Argon geflutet werden. Dafür muß manuell je ein System von Hähnen geöffnet bzw. geschlossen werden.

Ein wesentlicher Teil des Gaszuführsystems sind die elektronisch gesteuerten Massenflußregler, die den Durchfluß und damit das genaue Mischungsverhältnis im Reaktionsofen bestimmen. Eine schematische Darstellung ihres Aufbaus ist in Abbildung (36) wiedergegeben. Der zentrale Teil besteht aus einer Strömungsvorrichtung, die so gestaltet ist, daß die Gase sie laminar durchfließen. Ein kleiner Teil der Gase wird, ebenfalls laminar, durch eine Sensorkanüle geleitet. Wegen der Laminarität besteht eine Proportionalität der Ströme durch beide Kanäle, so daß es genügt, den Sensorfluß zu messen um den Gesamtfluß durch ein Ventil zu regeln. Die Signalaufnahme besteht aus zwei Temperaturmessungen $T_1$ und $T_2$ an der Kanüle, zwischen denen ein Heizelement H geschaltet ist. Wird durch H pro Zeiteinheit eine Wärmemenge $\Delta\dot{Q}$ zugeführt, die einen Temperaturunterschied $\Delta T$ verursacht, so erhält man mittels der molaren Wärmekapazität $C_M$ den molaren Fluß

$$\Phi_M = \frac{1}{C_M}\,\frac{\Delta\dot{Q}}{\Delta T}\ . \qquad (200)$$

| Gas | chemische Formel | Mischung | Maximalfluß |
|---|---|---|---|
| Silan | $SiH_4$ | 1% – 3 % in Ar | 2 l / min |
| German | $GeH_4$ | 1% in Ar | 0,1 oder 2 l / min |
| Wasserstoff | $H_2$ | 100 % | 2 l / min |
| Chlorwasserstoff | HCl | 1% in $H_2$ | 1 l / min |
| Argon | Ar | 100%, 3 Flaschen | |

*Tabelle (4):*
*Verfügbarkeit der Gase in unserer Epitaxieanlage.*

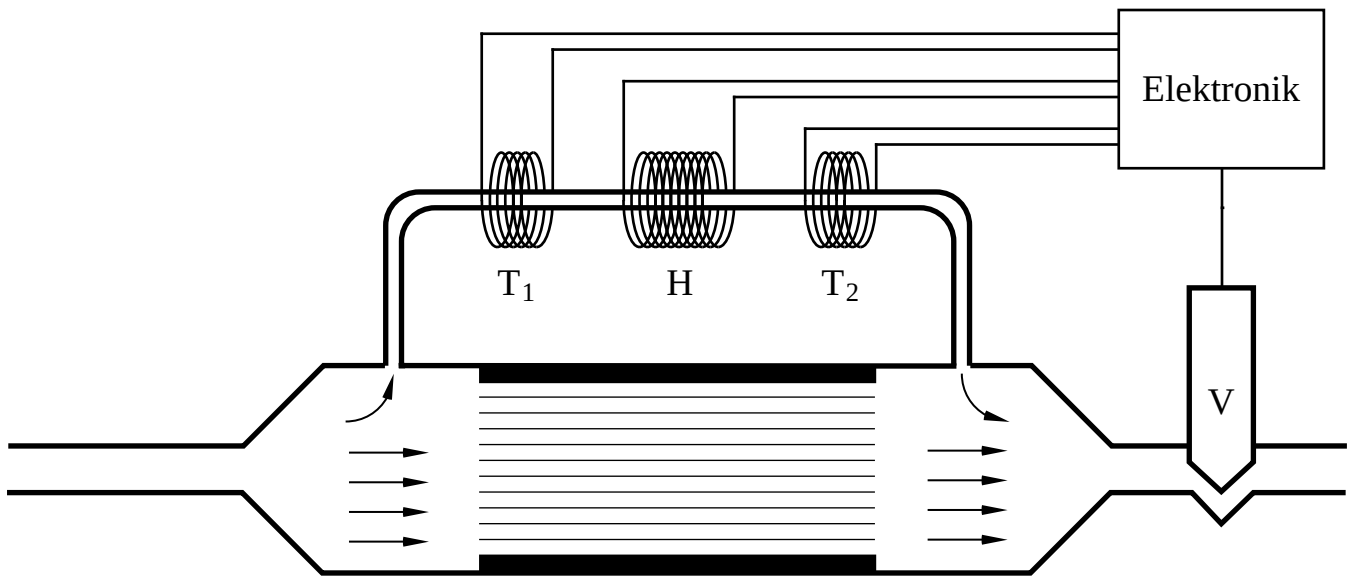

*Abbildung (36):*
*Schematische Darstellung eines Massenflußreglers.*

Diese Meßgröße wird durch eine Elektronik mit der Sollgröße verglichen, die dann das Steuerventil V und somit den Gasfluß regelt.

| **Gaseigenschaften** | **N** |
|---|---|
| einatomig (Ar, He) | 1,03 |
| zweiatomig ($N_2$, $O_2$, CO) | 1,00 |
| dreiatomig ($CO_2$, $SO_2$) | 0,94 |
| mehratomig ($NH_3$, $CH_4$) | 0,88 |

*Tabelle (5):*
*Gasformfaktoren N für verschiedenartige Gase.*

Normalerweise werden die Massenflußregler von der Herstellerfirma auf eine bestimmte Gaszusammensetzung geeicht. Dennoch kann es vorkommen, daß verschiedene Gase dieselbe Leitung benutzen, wofür ein Konversationsfaktor

$$\gamma = \frac{\Phi_M^2}{\Phi_M^1} = \frac{C_M^1}{C_M^2} = \frac{C_{M,0}^1\ N^2}{C_{M,0}^2\ N^1} \qquad (201)$$

aus den Flußverhältnissen bei gleichbleibenden $\Delta\dot{Q}$ und $\Delta T$ berücksichtigt werden muß. Die Molekülformfaktoren $N^1$ und $N^2$ berücksichtigen die Änderung der bei 20 ˚C tabellierten molaren Wärmekapazität zur Arbeitstemperatur $T_A$ von etwa 50 ˚C:

$$C_M(T_A) = C_{M,0}(T = 20\ ˚C)\,\frac{1}{N}\ . \qquad (202)$$

Die gebrauchten Größen werden von der Herstellerfirma im Datenblatt mitgeliefert und sind für die bei uns benötigten Gase in den Tabellen (5) und (6) aufgeführt. Dabei wird häufig schon ein Konversationsfaktor $\gamma$ angegeben, der durch (201) mit Stickstoff als Vergleichsgas 1 hervorgeht.

Wird ein Gemisch aus n Gasen verwendet, so addieren sich die Konversationsfaktoren, mit ihren Volumenanteilen $V_n$ gewichtet, reziprok:

$$\frac{1}{\gamma} = \sum_{\nu=1}^{n} \frac{V_\nu}{\gamma_\nu} . \tag{203}$$

| Gas | chemische Formel | M [g / mol] | ρ [g / l] | $c_M$ [J K / mol] | N | γ |
|---|---|---|---|---|---|---|
| Silan | $SiH_4$ | 32,118 | 1,438 | 42,870 | 0,88 | 0,596 |
| German | $GeH_4$ | 76,622 | 3,423 | 44,271 | 0,88 | 0,580 |
| Wasserstoff | $H_2$ | 2,016 | 0,090 | 28,698 | 1,00 | 1,016 |
| Chlorwasserstoff | HCl | 36,461 | 1,635 | 29,569 | 1,00 | 0,981 |
| Argon | Ar | 39,948 | 1,784 | 20,840 | 1,04 | 1,453 |
| Stickstoff | $N_2$ | 28,014 | 1,250 | 29,135 | 1,00 | 1,000 |

*Tabelle (6):*
*Materialkonstanten der verwendeten Gase: Molmasse M, Dichte ρ, molare Wärmekapazität $c_M$, Molekülformfaktor N und Konversationsfaktor γ zum Umeichen der Massenflußregler in Bezug auf Stickstoff.*

### 5.2.1.2. Das Gasabführsystem

Während das Gaszuführsystem dem Zuchtofen die Flüsse und Mischungsverhältnisse zur Verfügung stellt, regelt das Gasabführsystem seinen Druck. Es besteht im wesentlichen aus einem Rootspumpensystem, einem Druckmesser am Ofenausgang sowie einem Regelventil. Die einzelnen Komponenten sind durch Vakuumschläuche großen Durchmessers miteinander verbunden. Abgase werden durch ein Kupferrohr aus dem Gebäude geleitet, an dessen Ende sich auch schon mal durch die Silanreaktion mit dem Luftsauerstoff eine Flamme entzünden kann.
Eine Elektronik sorgt wieder für den Vergleich des Druckes mit dem Sollwert um das Ventil zu regeln. Damit ist ein Druckbereich zwischen 0,1 und 10 Torr einstellbar, dessen untere Grenze jedoch sehr von dem durchströmenden Gasfluß abhängt und üblicherweise dann bei 1 bis 2 Torr liegt.

### 5.2.2. Der Reaktionsofen

Im Gegensatz zu den Vorarbeiten sollte ein neuer Ofen so konzipiert werden, daß $Si_{1-x}Ge_x$ Kristalle mit bis zu 10 cm Substratdurchmesser gezogen werden können. Dazu muß eine handelsübliche Siliziumscheibe auf eine Temperatur zwischen 1100 K und 1500 K geheizt werden, während das Reaktionsgemisch über ihre Oberfläche strömt. Da sich die Trägergase an allen heißen Oberflächen zersetzen, müssen, um einen ungewollten Niederschlag zu vermeiden, sowohl die Ofenwände als auch eventuelle elektrische Zuleitungen gekühlt werden.

#### 5.2.2.1. Versuche mit einem Halogenlampenofen

Zunächst verwendeten wir einen von der Firma JIPELEC® neu entwickelten, srahlengeheizten Ofen, dessen Reaktionskammer in Abbildung (37) schematisch dargestellt ist. Das Substrat liegt auf drei Stützen im unteren Teil des wenige Zentimeter hohen Raumes. Seitlich sind oben Gaseinlaßdüsen und unten ein ringförmiger Gasauslaß vorgesehen. Der Deckel der Kammer besteht aus einer doppelten Quarzglasscheibe, durch deren Zwischenraum, ebenso wie durch Kanäle in den Wänden, Kühlflüssigkeit strömt. Geheizt wird durch ein Feld von Halogenlampen, das sich über dem Fenster befindet und dessen thermische Strahlung dieses durchdringt um vom Substrat mit heizender Wirkung absorbiert zu werden. Die Temperatur kann durch ein Thermoelement oder ein Pyrometer gemessen und geregelt werden.

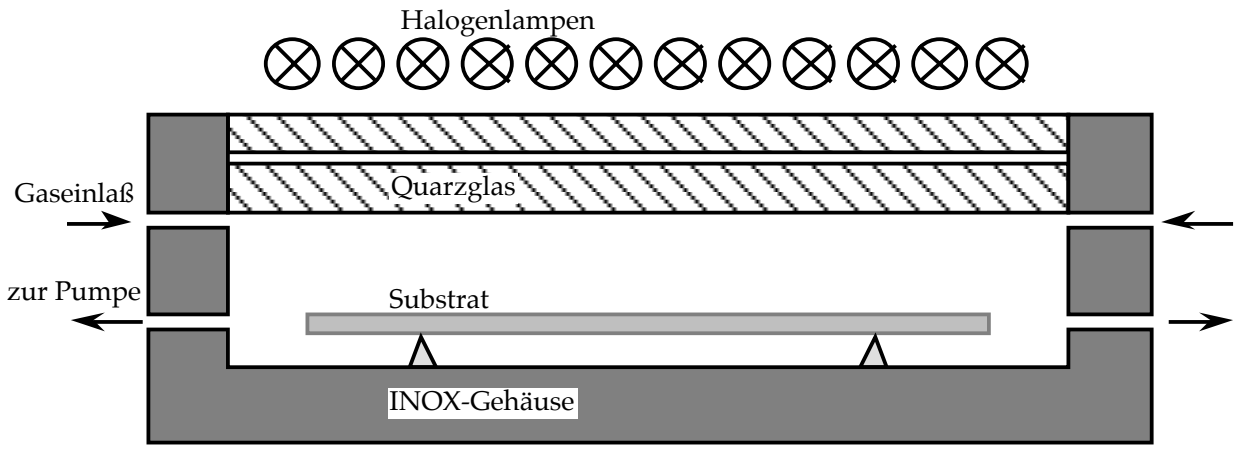

*Abbildung (37):*
*Lampengeheizter Ofen. Es stellte sich sehr bald heraus, daß eine interne Heizung notwendig wird.*

Wie schon erwähnt sind die mit diesem Ofen durchgeführten Experimente von großen Schwierigkeiten begleitet: Zunächst konnten nur relativ tiefe Prozeßtemperaturen von maximal 900 °C erreicht werden, da sonst die Lampen überlastet wurden und durchgebrannt sind. Eine erfolgversprechende Idee war, das im infraroten absorbierende Kühlwasser zwischen den beiden Scheiben des Quarzfensters durch eine weniger absorbierende Flüssigkeit zu ersetzen. Die Herstellerfirma des Ofens entwickelte einen Kreislauf mit Wärmeaustauscher. Dieser wurde installiert und mit VOLTALEF®, einem Öl gefüllt. Wie die in Abbildung (38) wiedergegebenen Messungen zeigen, werden somit bei gleicher Lampenleistung bis um 300 K höhere Temperaturen erreicht.

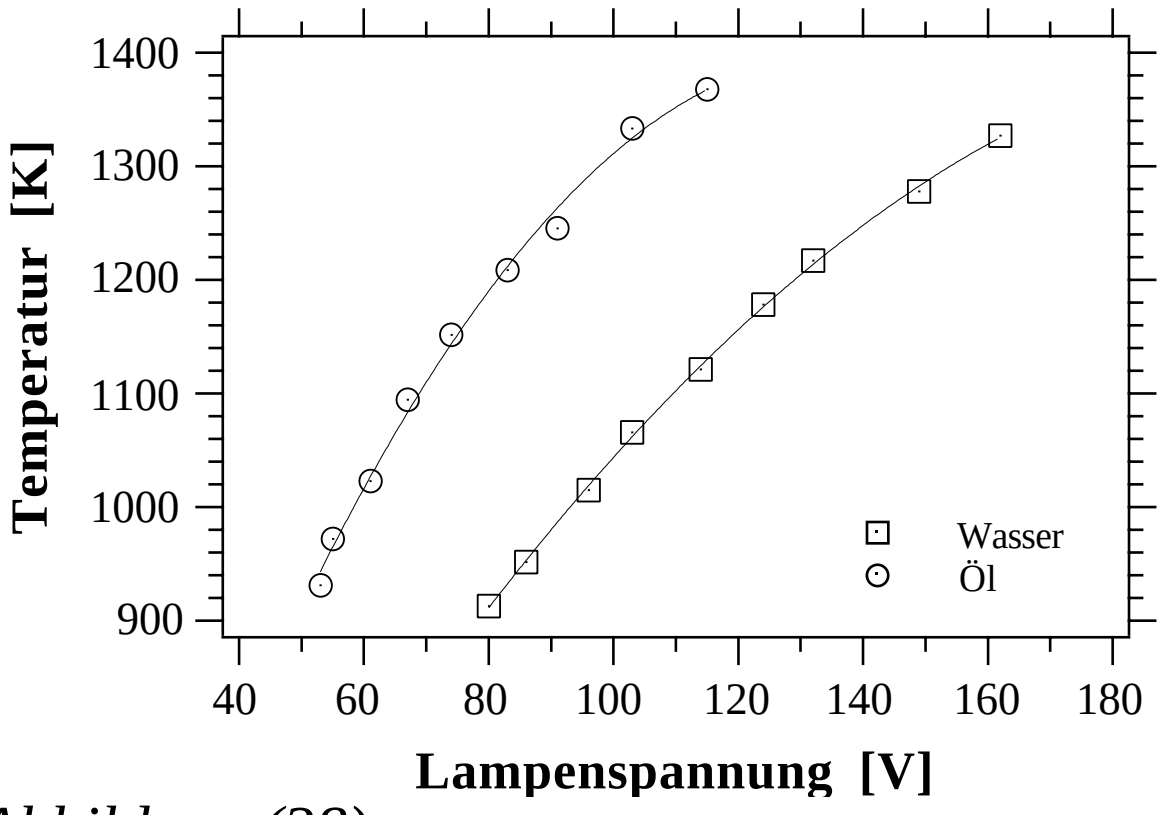

*Abbildung (38):*
*Verbesserung der Heizwirkung durch Wechseln der Kühlflüssigkeit zwischen den Quarzscheiben des Eintrittsfensters. Vierecke: $H_2O$, Kreise: VOLTALEF® - Öl. Das Öl absorbiert weniger im Infraroten, so daß die Prozeßtemperaturen bei gleicher Lampenleistung um etwa 300 K höher liegen.*

Ein wesentlicheres Problem wird durch einen, auf dem Quarzfenster fest haftenden Silizium-Germaniumniederschlag hervorgerufen, der sich ab 1100 K bildet und den Eintritt der thermischen Strahlung in den Reaktionsofen schwächt und schließlich verhindert. Folglich muß die Lampenleistung ständig erhöht werden, um die Probe auf konstanter Temperatur zu halten. Das System erreicht ziemlich schnell seine technischen Grenzen, was Langzeitversuche, wie sie für uns interessant sind, unmöglich macht.
Der Mechanismus des Niederschlags ist folgender: Silan- und Germanmoleküle werden vorwiegend an heißen Oberflächen aufgespalten, wobei sich das frei werdende Silizium (Germanium) auf diesen abscheidet. Ein Teil des Spektrums der von den Halogenlampen ausgehenden thermischen Strahlung wird bereits im Quarzglas der Glühbirne absorbiert. Das durchgelassene Spektrum passiert die Quarzfenster des Reaktionsofens ohne wesentliche Absorption und heizt das Substrat. Dieses wiederum strahlt das gesamte Planck'sche Spektrum ab, also auch die Wellenlängen, die vom Quarzglas des Fensters bereits absorbiert werden, und dieses somit von innen heizen.

#### 5.2.2.2. Der neue Ofen

Wie wir sehr bald gelernt haben, ist die zuvor beschriebene Art von Reaktionsofen für unsere Zwecke ungeeignet. Sie war wohl mit Erfolg für dünne, epitaktische Schichten bei niedrigen Zersetzungstemperaturen und schnellen Aufheizbedingungen konzipiert, wohingegen wir in unserer Anlage dicke Schichten durch hohe Wachstumsraten erzielen wollen und somit hohe Temperaturen bei hohen Gasflüssen über lange Zuchtzeiten beanspruchen. Daher konzentrierten wir uns auf ein internes Heizelement und gaben einen wassergekühlten Kessel in Auftrag, wie er in Abbildung (39) dargestellt ist. Innerhalb diesem konnten verschiedene Typen von Heizelementen und Gasführungssystemen ausprobiert werden.

Im wesentlichen besteht der aus Edelstahl gefertigte Ofen aus einem runden, wassergekühlten Kessel, der nach oben hin durch einen ebenfalls gekühlten Deckel abgeschlossen wird. Durchmesser und Höhe betragen rund 30 cm. Die Reaktionsgase werden durch ein Röhrchen in der Deckelmitte eingeleitet und am Boden abgepumpt.
Im Inneren des Kessels befindet sich, wie in Abbildung (40) dargestellt, eine waagerechte Quarzglasscheibe mit einem Durchmesser von 15 cm, auf die das Siliziumsubstrat gelegt wird. Sie wird von unten durch ein schlangenförmig gestaltetes Graphitband geheizt. Um einen ungewollten Niederschlag zu vermeiden, liegt das Heizelement in einer Vertiefung aus gepreßter Aluminiumoxidkeramikwolle, die nach oben hin direkt durch die Quarzplatte abgedeckt wird. Die Stromzufuhr wird durch, von innen bis unter die Keramik wassergekühlte Leitungen gewährleistet. Die Isolation zu den Gasströmungen muß mit besonderer Sorgfalt vorgenommen werden, da einerseits das Graphitband durch ein ungewolltes Depot sehr spröde wird und andererseits Silizium bei den Prozeßtemperaturen metallisch leitend wird und somit Kurzschlüsse verursacht.
Die Temperatur wird durch ein Thermoelement unter der Quarzglasscheibe bestimmt und geregelt. Zur Versorgung des Heizelements steht ein mit einem Temperaturregelkreis versehenes Netzteil von 75 A und 110 V zur Verfügung.

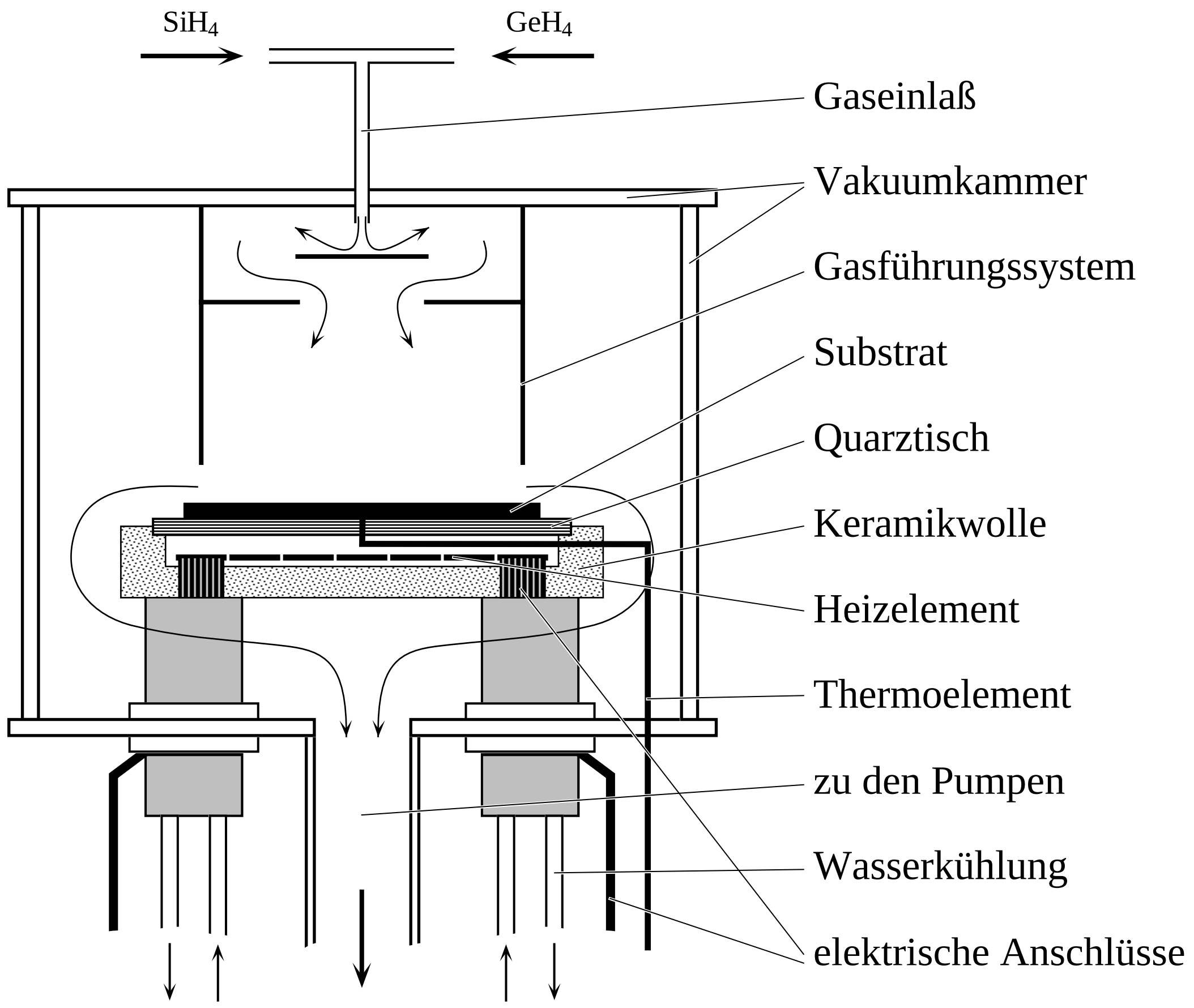

*Abbildung (39):*
*Schematische Darstellung des für die Gasphasenepitaxie entwickelten Zuchtofens.*

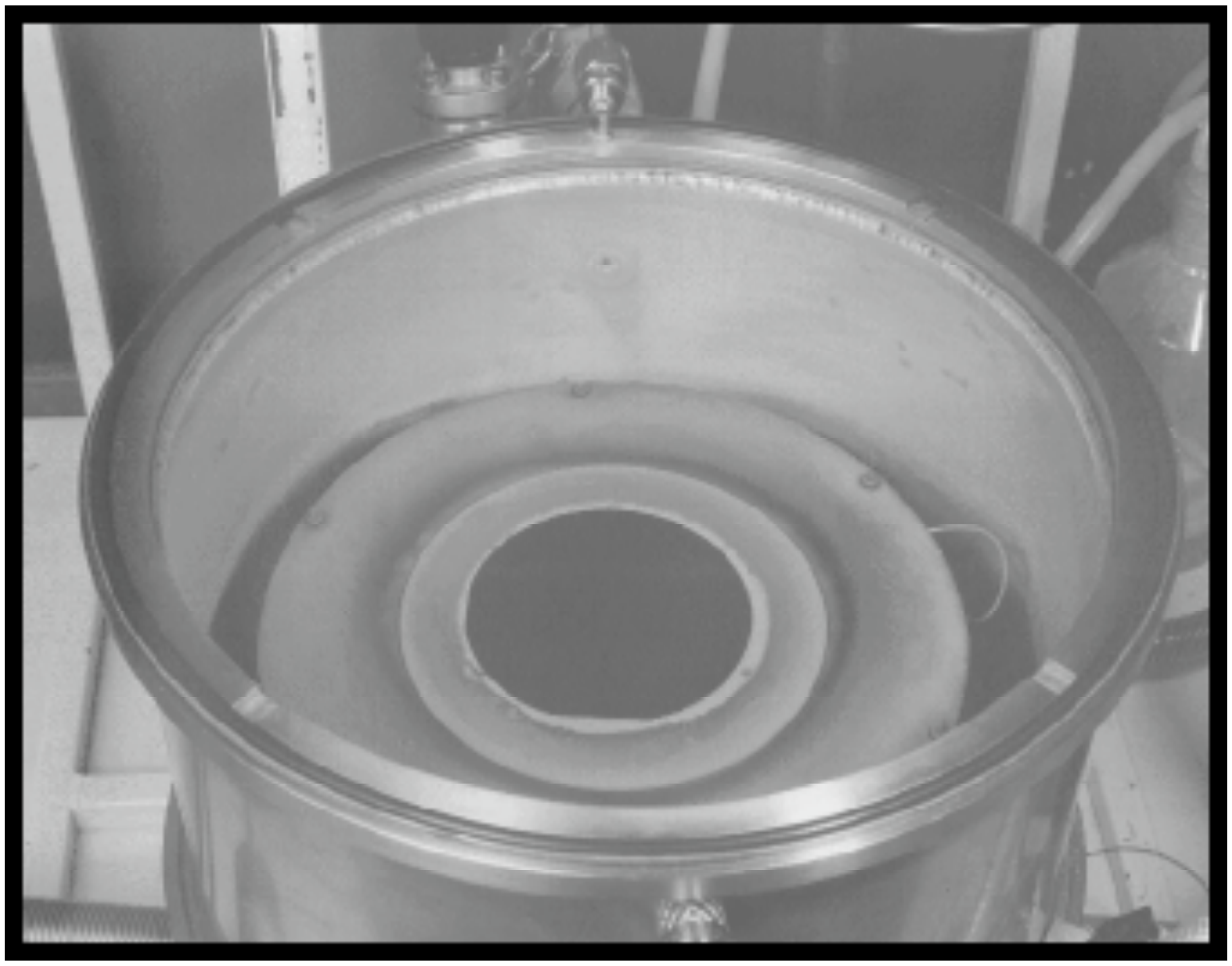

*Abbildung (40):*
*Blick in den Reaktionsofen bei heruntergenommenem Deckel und Gasführungssystem. Die dunkle Scheibe im Zentrum ist das Siliziumsubstrat auf der Quarzplatte.*

#### 5.2.2.2.1. Das Heizelement

Die relativ hohe Prozeßtemperatur in der Silanatmosphäre fordert besondere Ansprüche an das Heizelement, zumal sich, wie erwähnt, bei ungenügender Isolation millimeterdicke Siliziumniederschläge bilden können, die schon bei einem Bruchteil dieser Stärke zu elektrischen Kurzschlüssen führen können. Da seine Lebensdauer daher nicht abschätzbar war, wurde von vornherein ausgeschlossen, teure, industrielle Heizelemente für Epitaxieanlagen anzuschaf-fen. Vielmehr sollte es kostengünstig und leicht herstellbar und damit ersetzbar werden. Um den Abmessungen des Substrats und der Ofengeometrie zu entsprechen, sollte es flach sein und eine möglichst homogene, runde Heizfläche bilden. Außerdem darf es nicht allzusehr mit der Atmosphäre reagieren. Die meisten Metalle bilden mit Silan spröde, nichtleitende Silizide, die dann ein mechanisches und elektrisches Problem hervorrufen. Daher wählten wir eine kartonartige, 0,5 mm dicke Graphitfolie als Widerstandsmaterial, die auch den Vorteil seiner leichten Bearbeitung mit dem Skalpell sowie seine mechanische Flexibilität in sich birgt.

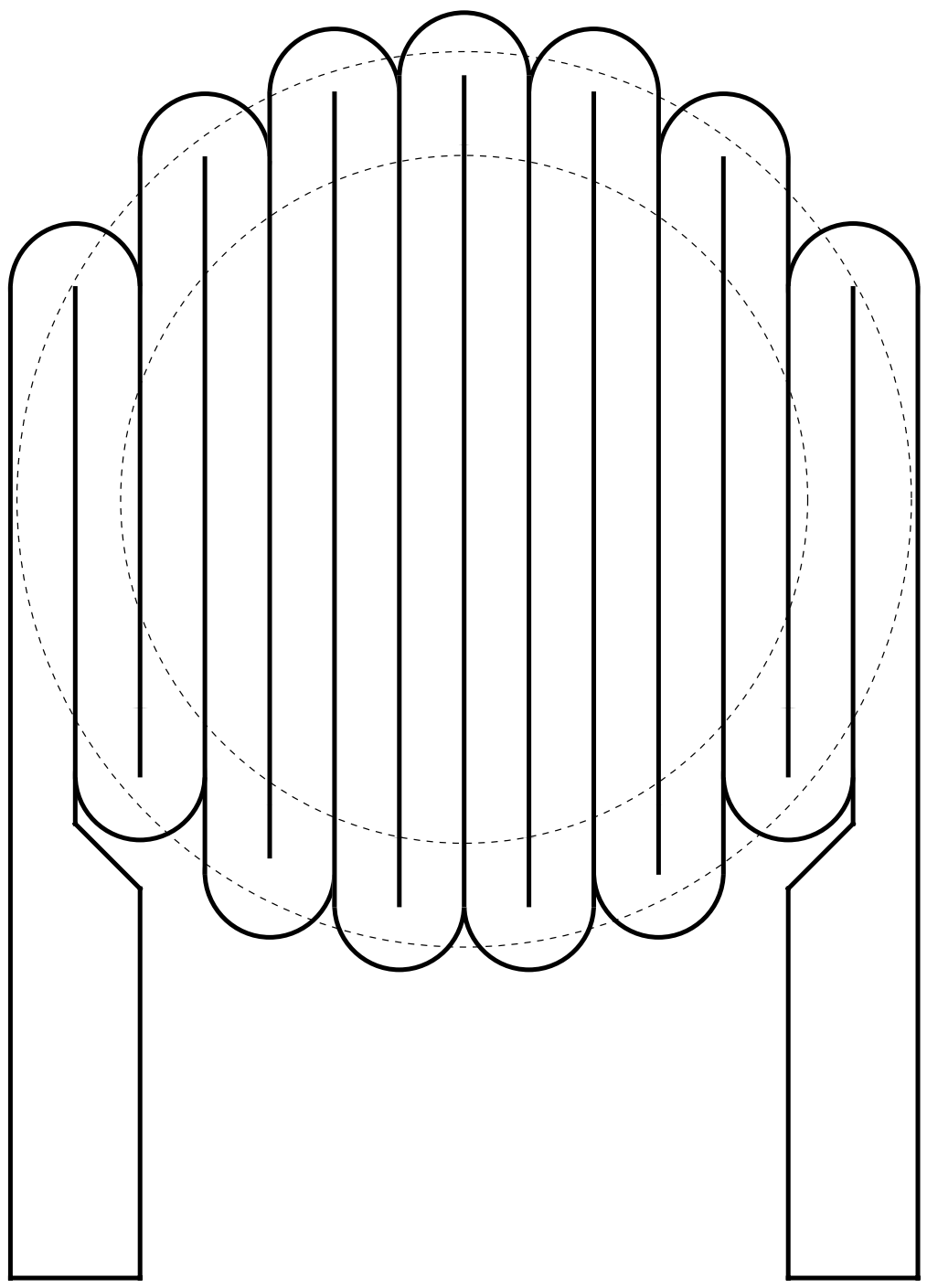

*Abbildung (42):*
*Gestaltung des Heizelements. Der innere Kreis gibt die Substratlage (Ø 10 cm), der äußere die der Quarzträgerscheibe wieder.*

Der spezifische Widerstand (Abbildung (41)) dieses Materials ist bei Raumtemperatur $\rho_0 = 10\ \mu\Omega$ m und sinkt bis etwa 1200 K auf die Hälfte ab, um dann bei weiterer Temperaturerhöhung ziemlich konstant zu bleiben. Der Widerstand des Resistors mit der Länge $\ell$ und des Querschnitts A ergibt sich somit aus

$$R = \rho \frac{\ell}{A} . \qquad (204)$$

Drücken wir A = b d durch Foliendicke d und Breite b der Leiterbahn aus und setzen

$$b\,\ell = \frac{\pi D^2}{4} \qquad (205)$$

der gesamten Heizfläche mit Durchmesser D, lösen (205) nach b auf, so erhalten wir für die Leiterbahnbreite

$$b = \frac{D}{2}\sqrt{\frac{\pi \rho}{R d}} . \qquad (206)$$

Das verwendete Netzteil hat mit den Nennwerten U = 110 V und I = 75 A einen inneren Widerstand von ca. $R_i = 1{,}5\ \Omega$. Da sich sein spezifischer Widerstand bei der Prozeßtemperatur halbiert, sollte das Heizelement im Kalten für R = 3 Ω ausgelegt werden. Mit d = 0,05 cm, D = 15,2 cm resultiert eine Bahnbreite von 1,1 cm. Ein derartiges Element wurde gemäß der Gestaltung in Abbil-dung (42) aus der Graphitfolie ausgeschnitten und installiert. Die beiden Kreise zeigen einmal den Durchmesser des Quarztisches sowie den des Siliziumsubstrats.

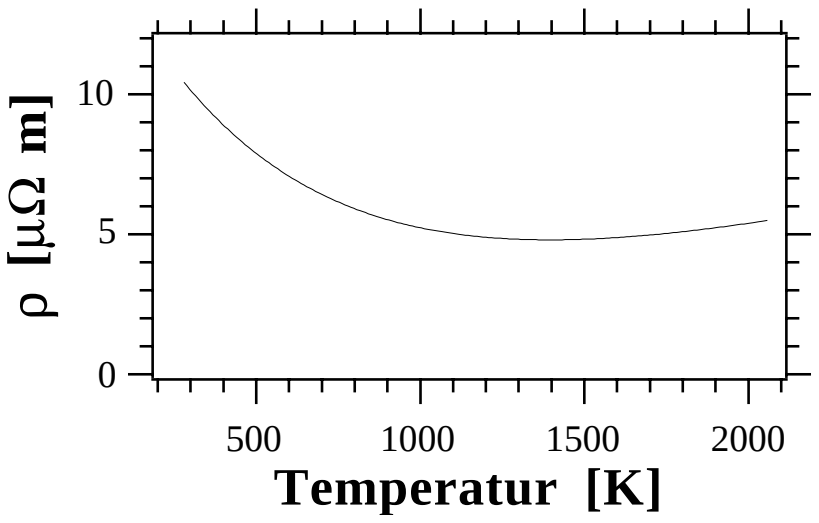

*Abbildung (41):*
*Spezifischer Widerstand für die verwendete Graphitfolie.*

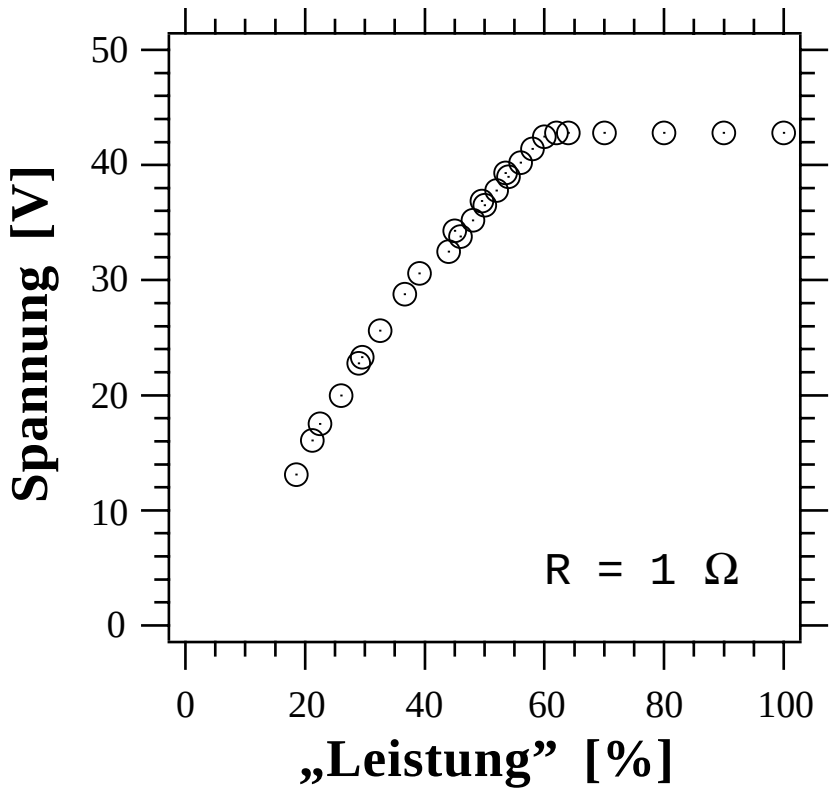

*Abbildung (43): Spannungskennlinie gegen einen Regelparameter des Netzteiles der ein Maß für die Leistung ist. Wichtig ist der Knick, bei der das Netzteil an seine Grenzen stößt und trotz höherer Sollwerte die Spannung am Heizwiderstand in Sättigung geht. Diese Maximalspannung hängt vom Heizwiderstand ab und ist optimiert, wenn dieser mit dem Innenwiderstand des Netzteils identisch ist.*

Im folgenden wurden Heizelemente verschiedenen Widerstands hergestellt und zur Optimierung ausprobiert. Abbildung (43) zeigt das Verhalten der am Heizelement abgegriffenen Spannung gegen einen, am Regelkreis einstellbaren Parameter, der zwar ein Maß für die Leistung ist, jedoch aufgrund der Wechselstromzerhackung nicht linear von ihr abhängt. Wichtig in der Meßkurve ist nur das Abknicken, hier bei etwa 60 %, bei der eine, dem Widerstand entsprechende Maximalspannung erreicht wird. An diesem Knickpunkt stößt das Netzteil an seine Stromgrenze von 75 A, so daß der Sollwert nicht mehr erreicht werden kann. Der Widerstand ist also schlecht an den Innenwiderstand angepaßt.

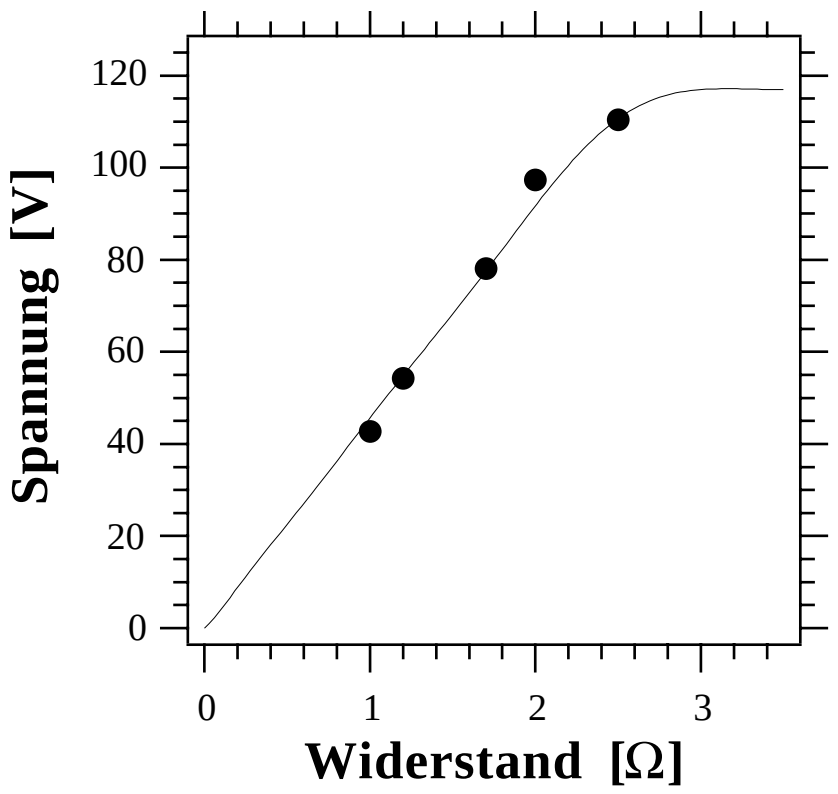

*Abbildung (44): Vom Netzteil gelieferte Maximalspannung als Funktion des Heizwiderstands im kalten Zustand. Links der Abknickstelle ist das System strombegrenzt, rechts von ihr spannungsbegrenzt. Der optimale Widerstand liegt zwischen 2,5 und 3 Ω.*

Die so erhaltenen Spannungsmaximalwerte sind für verschiedene Resistorwiderstände in Abbildung (44) angegeben. Solange man an die Strombegrenzung stößt, steigt sie linear mit dem Widerstand bis zu dem Punkt, an dem Innen- und Heizwiderstand gleich sind, an. Eine weitere Erhöhung des Widerstandes verursacht keinen Spannungsanstieg mehr, da die vom Netzteil lieferbare Maximalspannung bereits erreicht wurde.

#### 5.2.2.2.2. Das Gasführungssystem

Neben einem homogen geheizten Substrat spielen auch die Gasströmungs- und Druckverhältnisse im Kessel eine wesentliche Rolle. Inhomogene Strömungen können einerseits aufgrund des hohen Flusses das Substrat inhomogen kühlen und andererseits verschiedene Materialtransporte bewirken, die dann lokal in Sättigung gehen. Beide Mechanismen können das Wachstum derart beeinträchtigen, daß teils einkristalline, teils polykristalline Wachstumsbedingungen vorliegen.
Läßt man die Gase aus dem Einlaßröhrchen (siehe Abbildung (39)) direkt auf die Probe treffen, so entwickelt sich am Auftreffpunkt des Primärstrahls ein polykristalliner Fleck, während sich auf dem Rest des Substrats sehr geringe, aber einkristalline Niederschläge bilden. Es wurden verschiedene Geometrien und duschkopfähnliche Verteiler ausprobiert woraus wir gelernt haben, daß Gitter- und Lochraster immer auf die Probe abgebildet werden. Vielmehr muß der primäre Strahl gebrochen, aber dennoch frei auf die Probe zuströmen gelassen werden, was durch die Einbauten in Abbildung (39) verwirklicht wurde. Ein vertikaler Zylinder führt die Gase bis in die Nähe der Oberfläche. Ein fünf Mark großes Plättchen unterhalb des Gaseinlasses bricht den Primärstrahl. Damit die Gase nicht an den Zylinderwänden nach unten strömen, werden sie durch einen Ring in die Mitte zurückgeleitet, von wo aus sie sich bis zum Substrat homogen ausbreiten können. Der Zylinder wurde leider nur mit 5 cm im Durchmesser angefertigt, weshalb die kristallinen Abscheidungen auch kaum diesen Wert überschreiten.

## 5.3. Der Arbeitsvorgang bei der Kristallzucht

Im folgenden soll kurz der Ablauf eines normalen Versuches geschildert werden.

Das Substrat, normalerweise eine Si [111] - oder Si [100] - Scheibe wird gewogen und anschließend zwei Minuten lang in ein Bad verdünnter Flußsäure gegeben, um die Oberfläche vom Oxid zu befreien. Nachdem das Substrat mit hochreinem Wasser gespült und mit $N_2$ trockengeblasen wurde, wird es in den Reaktionsofen gelegt. Um die neue Oxidation möglichst gering zu halten, wird dieser schnell, aber sorgfältig geschlossen und sofort abgepumpt. Sobald das Vakuum unter 0,1 Torr gesunken ist, kann mit der Heizphase begonnen werden. Die Temperatur steigt innerhalb weniger Minuten auf den Sollwert zwischen 1200 K und 1500 K. Um das Substrat vollständig von Oxydresten zu befreien, kann man im unteren Temperaturbereich eine Viertelstunde Wasserstoff und fünf Minuten lang das Chlorwasserstoffgemisch bei einem Torr über die Probe strömen lassen. Damit soll die Oberfläche leicht angeätzt und abgetragen werden. Bei den oberen Temperaturen wird dieser Ätzprozeß zu stark, so daß die Oberfläche bereits rauh wird und es als besser erscheint, gasförmige Si-O Verbindungen im bestmöglichen Vakuum zwanzig Minuten lang abdampfen zu lassen.
Zum Beginn des Wachstumsprozesses wird eine Zeit lang kein German beigemischt, damit sich ein möglichst störungsfreier Übergang zwischen Substrat und Depot bildet. Je nach Versuchsreihe wird nun das Silan-German-Mischungsverhältnis zeitlich verändert. Am Ende des Wachstumsprozesses wird die Gaszufuhr abgestellt, somit das Vakuum erhöht und die Heizung abgeschaltet. Nach dem Abkühlen wird der Druck entweder durch Luft, Wasserstoff oder Argon langsam ansteigen gelassen. Entnahme der Probe und Wägung. Durch den Massenzuwachs kann auf die mittlere Schichtdicke der Probe geschlossen werden. Optische Begutachtungen mittels Auge und Mikroskop geben erste Aufschlüsse über Homogenität und Kristallinität der Schicht.

## 5.4. Zuchtergebnisse

Mit der oben beschriebenen Versuchsanlage wurden Experimente an etwa 40 verschiedenen Siliziumsubstraten durchgeführt, die meistens zur Optimierung der Prozeßparameter nach Veränderungen, etwa am Heizelement oder am Gasführungssystem galten. Solche Testreihen wurden meistens ohne Germanzusatz in relativ kurzen Aufdampfzeiten von einer halben bis ganzen Stunde abgewickelt. Wurden damit zufriedenstellende Ergebnisse in Bezug auf Wachstumsgeschwindigkeit, Homogenität und vor allem Kristallinität erreicht, so wurde die Reihe mit verschiedenen Germaniumkonzentrationsprofilen und längeren Aufdampfzeiten fortgeführt.

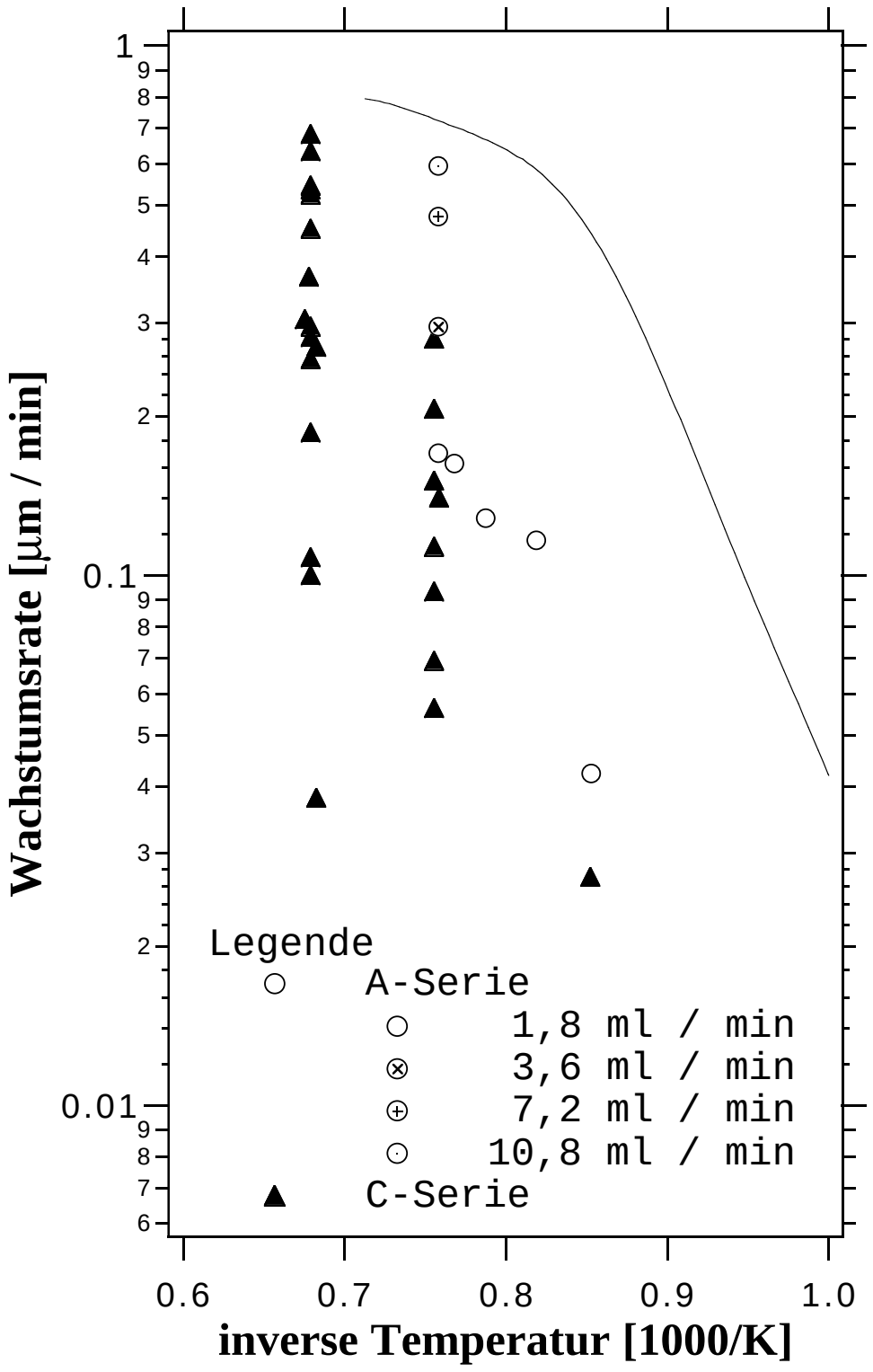

*Abbildung (45):*
*Die Wachstumsraten der einzelnen Proben als Funktion der reziproken Temperatur. Die durchgezogene Linie gibt die Werte bei 1 Torr nach [25] wieder. Ein ähnliches Verhalten zeigen die Ergebnisse der A Serie, wobei die Wachstumsraten auch sehr vom Germandurchsatz abhängen. Die Punkte der C Serie dienen zur Orientierung. Hier wurden Gasflüsse, Geometrien und Konzentrationen variiert, so daß sich die Werte über eine ganze Größenordnung verteilen.*

Tabelle (7) stellt die für die weiteren Analysen wichtigen Proben sowie einige Herstellungsparameter zusammen. Die auf C endenden Proben sind in dem neu entwickelten Reaktionsofen hergestellt worden. Zusätzlich sind noch einige, auf A endende Proben aufgeführt, die aus der Machbarkeitsstudie von M. Hedrich et al. [23] hervorgegangen sind und im Rahmen der vorliegenden Arbeit charakterisiert wurden.

Die Probendicken wurden, vor allem bei dünnen Abscheidungen aus dem Massenzuwachs bestimmt. Zu diesen ist noch je eine Substratdicke reinen Siliziums von 290 µm zu addieren. Aus ihnen ergeben sich sofort die Wachstumsraten, die in Abbildung (45) graphisch gegen die inverse Temperatur aufgetragen sind. Die durchgezogene Linie ist aus der Literatur übernommen und zeigt den Verlauf bei einer Atmosphäre Druck [25]. Sie zeigt deutlich zwei Bereiche, nämlich einen bei niedrigeren Temperaturen, der durch die Oberflächenreaktion begrenzt wird und einen bei höheren Temperaturen, dessen Wachstumsrate sich durch den limitierten Gastransport ergibt. Im ersten Fall werden die Silanmoleküle gemäß eines Arrheniusgesetzes an der Oberfläche aufgespalten, so daß die Temperatur der begrenzende Faktor ist. Wird letztere ständig erhöht, so kommt man in einen Sättigungsbereich, bei dem sich die Siliziumatome schneller ablagern möchten als sie nachgeführt werden. Dies hat ein Abknicken nach unten von der Arrheniuslinie zur Folge.

| Probe | German / Silan [%] | Zeit [min] | Temp, [K] | Dicke [µm] | Rate [µm/min] |
|---|---|---|---|---|---|
| 78 A | 4,31 - 8,26 | 2 · 60 | 1323 | 60 | 0,50 |
| 82 A | 2,20 - 4,31 - 6,32 - 8,26 | 4 · 20 | 1323 | 57 | 0,71 |
| 83 A | 0 —> 11,1 | 60 | 1323 | 41 | 0,69 |
| 84 A | 4,31 | 60 | 1323 | 46 | 0,76 |
| 85 A | 0 —> 8,26 | 180 | 1323 | 107 | 0,60 |
| 86 A | 0 —> 2,86 | 180 | 1323 | 98 | 0,54 |
| 89 A | 0 —> 2,86 | 1440 | 1323 | 787 | 0,55 |
| 94 A | 0 —> ≈ 4 | | 1323 | 850 | |
| 13 C | 6 | 115 | 1473 | 10 | 0,09 |
| 17 C | 1,6 - 3,2 | 2 · 120 | 1473 | 70 | 0,29 |
| 18 C | 1,6 - 3,2 | 2 · 244 | 1473 | 136 | 0,28 |
| 19 C | 0 - 0 —> 3,2 | 5 + 210 | 1480 | 65 | 0,30 |
| 20 C | 25 | 9 | 1465 | | |
| 28 C | 0 - 5,9 - 10,7 - 15,2 - 19,3 - 23,0 - 26,4 | 29 + 6 · 60 | 1473 | 100 | 0,26 |
| 29 C | 0 - 25 | 30 + 1 | 1473 | 20 | 0,65 |
| 30 C | 0 - 6,3 | 30 + 1 | 1473 | 21 | 0,68 |
| 42 C | 0 —> 10,2 | 1920 | 1475 | 850 | 0,44 |

*Tabelle (7):*
*Zusammenstellung einiger Zuchtbedingungen und -Ergebnisse für die wichtigsten, germaniumhaltigen Proben.*

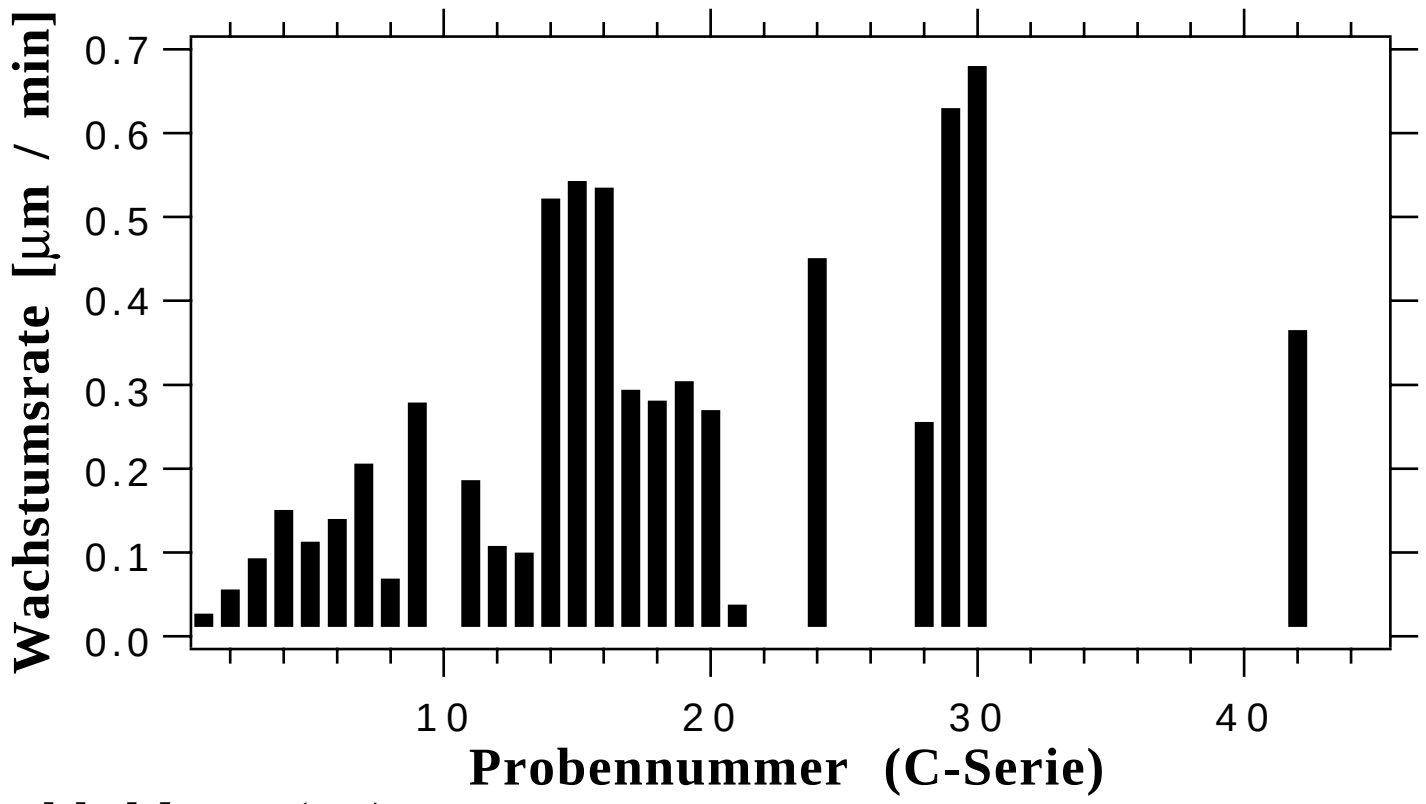

*Abbildung (46):*
*Zusammenstellung der Wachstumsraten für die verschiedenen Proben der C-Serie.*

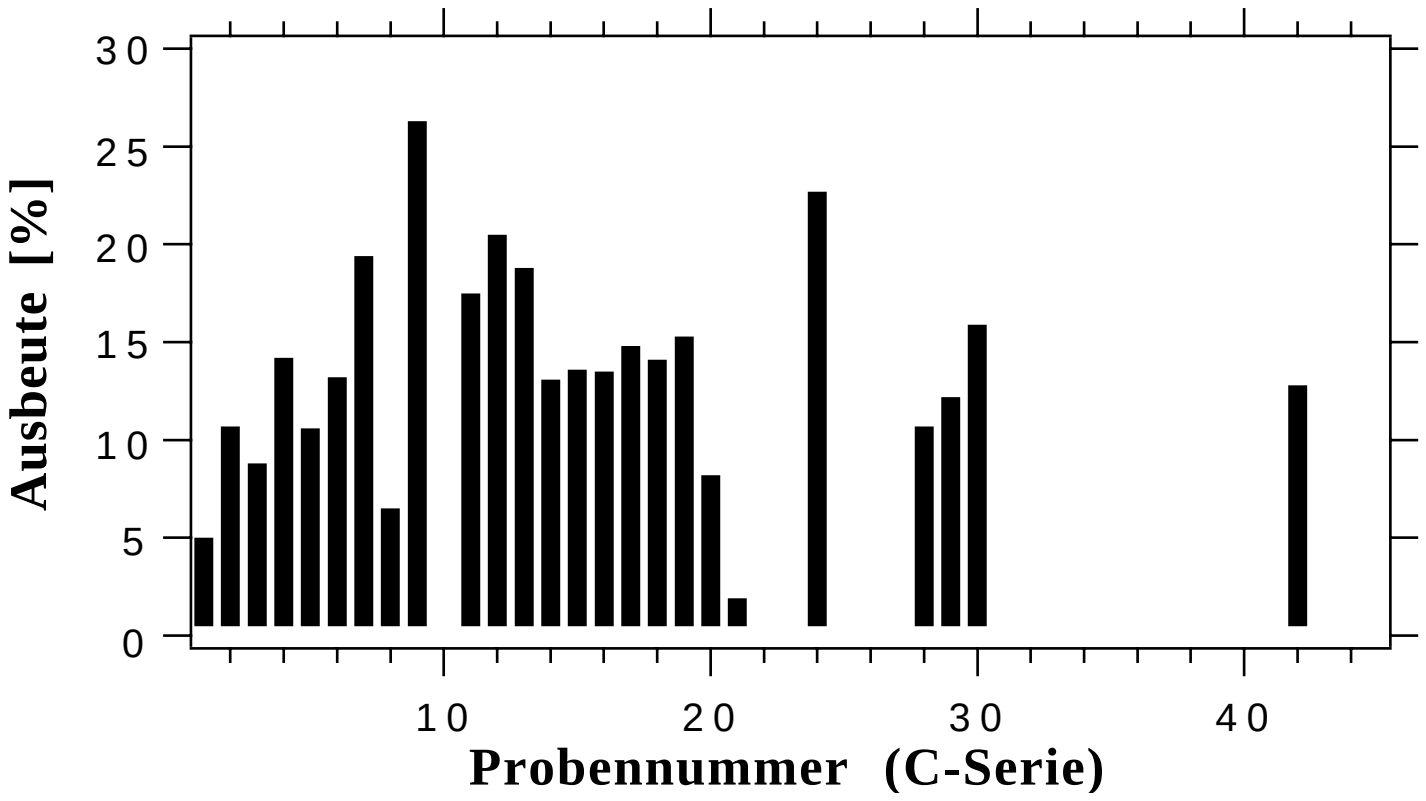

*Abbildung (47):*
*Zusammenstellung der Ausbeuten für die verschiedenen Proben der C-Serie.*

Die durchgezogene Linie stellt eine anerkannte Grenze dar, die zwar überschritten werden kann, wovon die Autoren jedoch erfahrungsgemäß zugunsten der Kristallinität abraten.

Wir befinden uns mit sämtlichen Zuchtergebnissen unterhalb dieser Linie. Dies mag unter anderem mit dem erheblich niedrigeren Arbeitsdruck von nur einigen Torr zusammenhängen. Die runden Symbole der A-Serie in Abbildung (45) zeigen ein ähnliches Sättigungsverhalten wie zuvor beschrieben. Bei Erhöhung des Silanflusses kann jedoch die Wachstumsrate noch erheblich gesteigert werden. Die ausgefüllten Dreiecke zeigen einige Ergebnisse der C-Serie und dienen zur Orientierung. Hier wurde keine explizite Temperaturabhäbgigkeit bei konstantem Gasfluß mehr aufgenommen. Die Wachstumsraten sind nochmals in Abbildung (46) histographisch dargestellt. Bei den Proben 1 C – 10 C wurde das Heizelement optimiert und auf eine möglichst hohe Wachstumsrate bei festem Fluß hingearbeitet. Ab 10 C – 13 C wurden verschiedene Gasführungssysteme zur Optimierung der Homogenität ausprobiert. Die Wachstumsraten sind eher gering. Bei Probe 14 C wurde die Silankonzentration im Argongemisch der Druckflaschen von 1 % auf 3 % erhöht, was sich deutlich in der Wachstumsrate niederschlägt. Die Proben 17 C – 20 C, 28 C und 42 C sind mit Hinblick auf Beugungsexperimente gezogen worden, weshalb zugunsten der Kristallinität nur mit dem halben oder bei 42 C mit dreiviertel Maximalfluß gearbeitet wurde.
Wie wir in Abbildung (45) sehen, hängt die Wachstumsrate nicht nur von der Temperatur, sondern von einer Vielzahl von Parametern ab. Daher ist ein weiterer, wichtiger Gesichtspunkt die Ausbeute, nämlich der auf dem Substrat niedergeschlagene Anteil Siliziummasse zur gasförmig zugeführten, also

$$\eta = \frac{m^{Si}_{Depot}}{m^{Si}_{Gas}} = \frac{m^{Si}_{Depot}}{\rho^{Si}_{Gas}\, t\, \Phi} \,. \tag{207}$$

Darin sind $m^{Si}_{Depot}$ der gemessene Probenmassenzuwachs, $\rho^{Si}_{Gas} = 1{,}25\ \frac{mg}{ml}$ die partielle Gasdichte für Silizium in Silan, $\Phi$ der Silanfluß und t die Aufwachszeit. Abbildung (47) zeigt die Ausbeute für die einzelnen Proben. Vor allem sieht man hier, daß bei Halbierung der Flüsse zwischen 16 C und 17 C praktisch kein Unterschied auftritt.

### 5.4.1. Zusammenfassende Ergebnisse des Zuchtvorgangs

Die maximalen Wachstumsraten liegen bei 0,6 µm / min, wobei anwendungsnahe Proben zugunsten der Kristallinität bei dem halben Wert gezogen wurden. Damit lassen sich innerhalb einiger Stunden Kristalle für die Beugungsanalyse im 20 µm – 200 µm Bereich ziehen. Für dickere Kristalle bis zu 1 mm braucht man bei der Maximalrate etwas länger als einen Tag. Die Ausbeute liegt mit ca. 15 % über dem Vierfachen der Werte um die 3,5 % des alten Ofens und ist, insbesondere bei der gewählten Ofengeometrie äußerst zufriedenstellend.

# 6. Die Probenanalyse

Die Analyse der mit obiger Methode gezüchteten $Si_{1-x}Ge_x$ Kristalle spielt einerseits, indem sie durch die erhaltenen Resultate auf die Eingangsparameter rückkoppelt, eine wichtige Rolle für die Zucht selbst, und führt andererseits über eine schlichte Qualitätskontrolle hinweg zu neuen Erkenntnissen. Angefangen bei der visuellen Begutachtung, über die Mikroskopie, die Mikrosondenanalyse bis hin zur Röntgen-, Gamma- und Neutronendiffraktometrie wurden verschiedene Techniken zu verschiedenen Gesichtspunkten angewandt.

## 6.1. Die optische Mikroskopie

Der erste Eindruck unmittelbar nach der Kristallzucht ergibt sich durch eine optische Begutachtung der Probenoberfläche. Gegen eine Lichtquelle betrachtet, kann sie entweder glänzend oder matt erscheinen, was einem kristallinen oder polykristallinen Wachstum zuzuschreiben ist. Im ersten Fall ist die Oberfläche größtenteils entlang bevorzugten Gitterebenen ausgerichtet, so daß sie das Licht mit einem wohl definierten Ausfallswinkel reflektiert. Häufig ist die Oberfläche entlang weiteren Kristallrichtungen mit Stufen oder regelmäßigen Strukturen übersät, was zu einem etwas matteren Erscheinungsbild führt. Ist die Oberfläche hingegen grau, ohne jegliches Spiegelbild, so handelt es sich um polykristallines Wachstum, dessen wahllos orientierte Oberflächenelemente das Licht diffus streuen. Dies konnte durch Röntgenstreuung an verschiedenen Bereichen ein und derselben Probe bestätigt werden. Abbildung (48) zeigt die Reflektionskurven der nur in das Depot eindringenden Cu $K_\alpha$ Strahlung am 111-Reflex, einmal einen ausgeprägten Braggreflex im spiegelnden, das andere mal konstanten Untergrund im mattgrauen Bereich der Oberfläche.

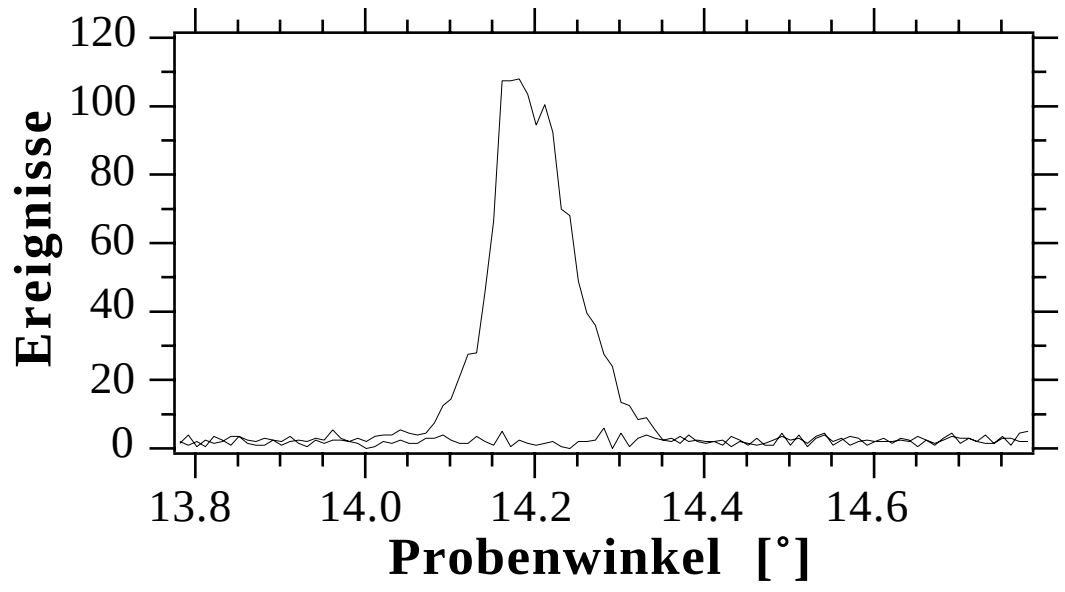

*Abbildung (48): Röntgendoppelkristalldiffraktometrie in der aufgedampften Kristallschicht. Im optisch glänzenden Bereich findet man einen Braggreflex, also Einkristallinität, im mattgrauen nur Untergrund, also Polykristallinität. (Cu $K_\alpha$ an Probe 17 C).*

Nimmt man dem Auge ein Mikroskop zu Hilfe, so kann man schon weitaus bessere Aussagen über die Kristallinität machen. Die Abbildungen (49) bis (54) zeigen einige, typische Erscheinungsformen, nämlich Polykristallinität im mattgrauen Bereich und regelmäßige, die Kristallorientierung widerspiegelnde Strukturen im kristallinen Gebiet, nämlich Dreiecke auf einer (111)- und Vierecke auf einer (100)-Oberfläche.

Bild (49) zeigt eine Aufnahme im polykristallinen Bereich, wie er bei relativ niedrigen Temperaturen um die 1100 K und zu hohen Wachstumsraten entsteht. Man erkennt die einzelnen Kristallite, die in sich regelmäßige Körper, häufig quadratische Grundformen mit dreieckigen Dächern darstellen. Benachbarte Kristallite sind jedoch willkürlich gegeneinander verkippt, so daß keine Kohärenz zwischen ihnen besteht. In diesem Bild handelt es sich um relativ große Kristallite, die je nach Zuchtbedingung, vor allem bei zu niedrigen Temperaturen auch wesentlich feiner ausfallen können. Natürlich kann auch eine oxidierte bzw. unsauber behandelte Oberfläche Anlaß zu polykristallinem Wachstum sein.

Bei höheren Temperaturen erhält man einkristallines Wachstum auf der Substratmatrix. Da bei unserer hohen Wachstumsrate die Oberfläche keine Zeit zu relaxieren findet, tritt Stufen- und Inselwachstum auf. An größeren Defekten entstehen Figuren, die die Symmetrie der Oberfläche widerspiegeln. Einige Aufnahmen sind in den Abbildungen (50) bis (54) gezeigt. Sie erscheinen als gleichseitige Dreiecke in 111 oder Quadrate in 100 Orientierung. Auf ihre genaue Charakterisierung wird im folgenden Kapitel eingegangen.

Die Oberflächen außerhalb der Dreiecke erscheinen, wie in den Bildern (52) und (53), obwohl mit Stufen durchsetzt, relativ glatt, während die in Abbildung (54) eine ausgeprägte Rauhigkeit oder Welligkeit im Mikrometerbereich aufweist. Dabei handelt es sich im letzten Fall um eine nur 0,5 µm dünne $Si_{0,75}Ge_{0,25}$ Schicht, die auf einem reinen Siliziumdepot aufgedampft wurde und daher eine große Fehlanpassung erleidet. Die Welligkeit entsteht durch einen Abbau dieser Spannungen, während sie bei kleinen Fehlanpassungen und Gradientenkristallen verschwindet. Sie wurde vor kurzem von anderen Gruppen ausgiebig theoretisch [26, 27] und experimentell [28] charakterisiert.

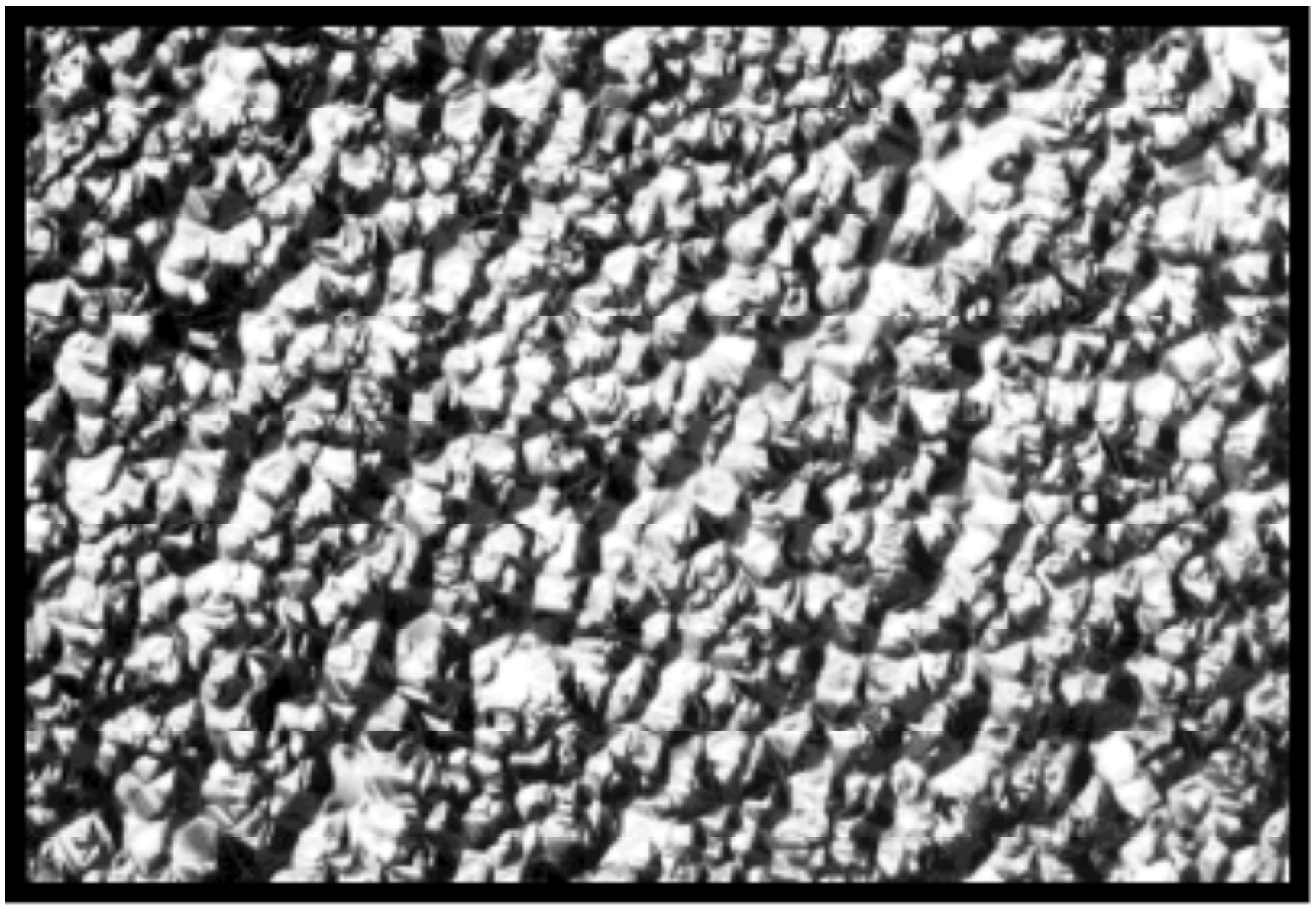

*Abbildung (49):*
*Polykristallin gewachsenes Kristallmaterial, meistens im Randbereich der Scheibe. Die Bilddiagonale entspricht ca. 500 µm.*

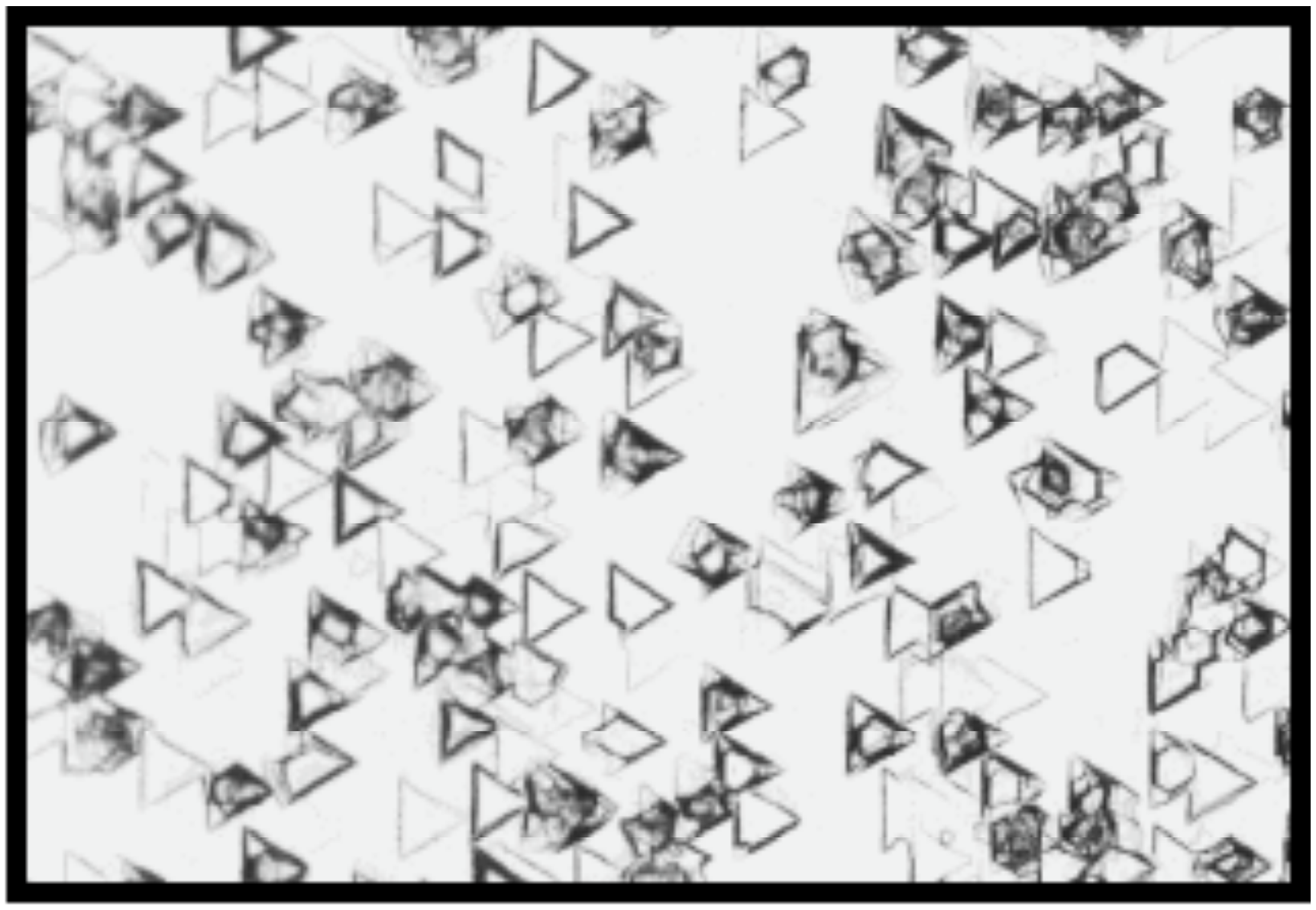

*Abbildung (50):*
*Regelmäßige Dreieckstrukturen auf einer (111) Oberfläche im kristallinen Bereich. Obwohl manche vertieft, andere flach erscheinen, sind ihre Kantenlängen gleich lang. Dies deutet auf eine gemeinsame Entstehung, nämlich an der unsauberen Substratoberfläche hin. Die Bilddiagonale entspricht ca. 500 µm.*

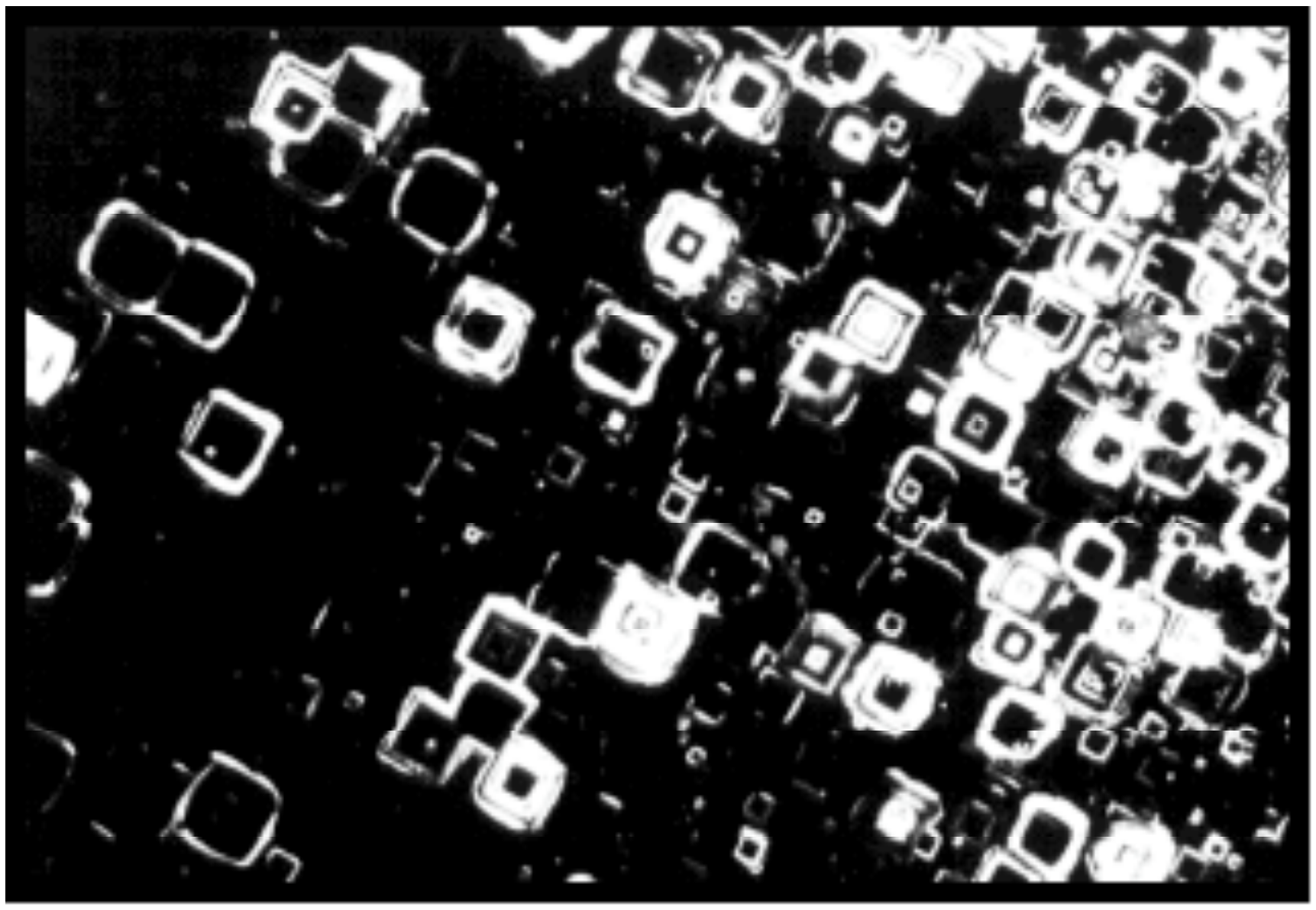

*Abbildung (51):*
*Viereckstrukturen auf einer (100) Oberfläche von 78 A im kristallinen Bereich. Auch hier weisen die häufigsten Quadrate eine maximale Größe auf. Aus dieser kann auf die Schichtdicke geschlossen werden. Die Bilddiagonale entspricht ca. 2 mm.*

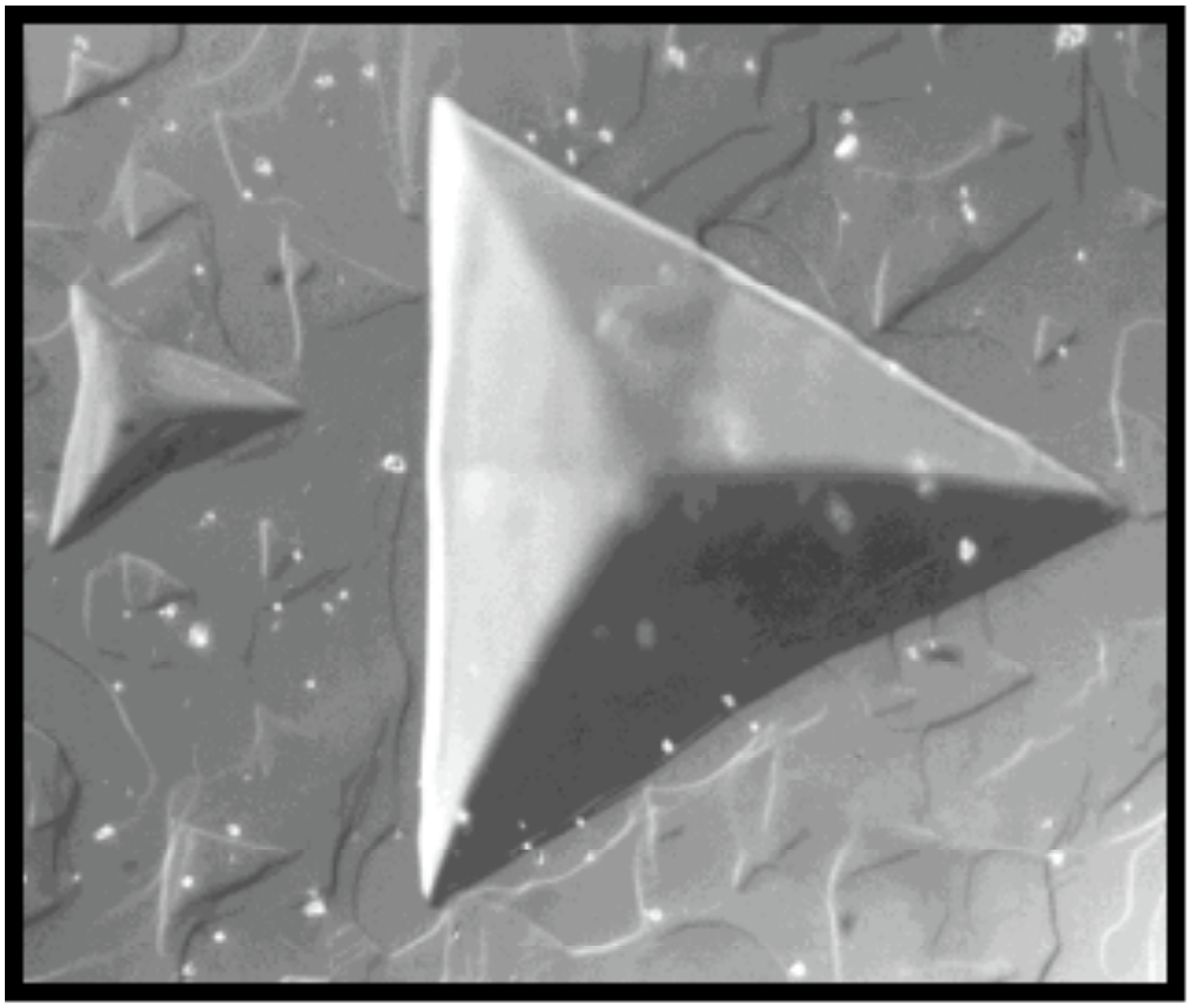

*Abbildung (52):*
*Dreieckstruktur in der sonst recht glatten Oberfläche von 17 C, die noch mit feinen Stufen überzogen ist. Sie stellt sich als eine kraterförmige Vertiefung heraus. Die Bilddiagonale entspricht ca. 100 $\mu$m.*

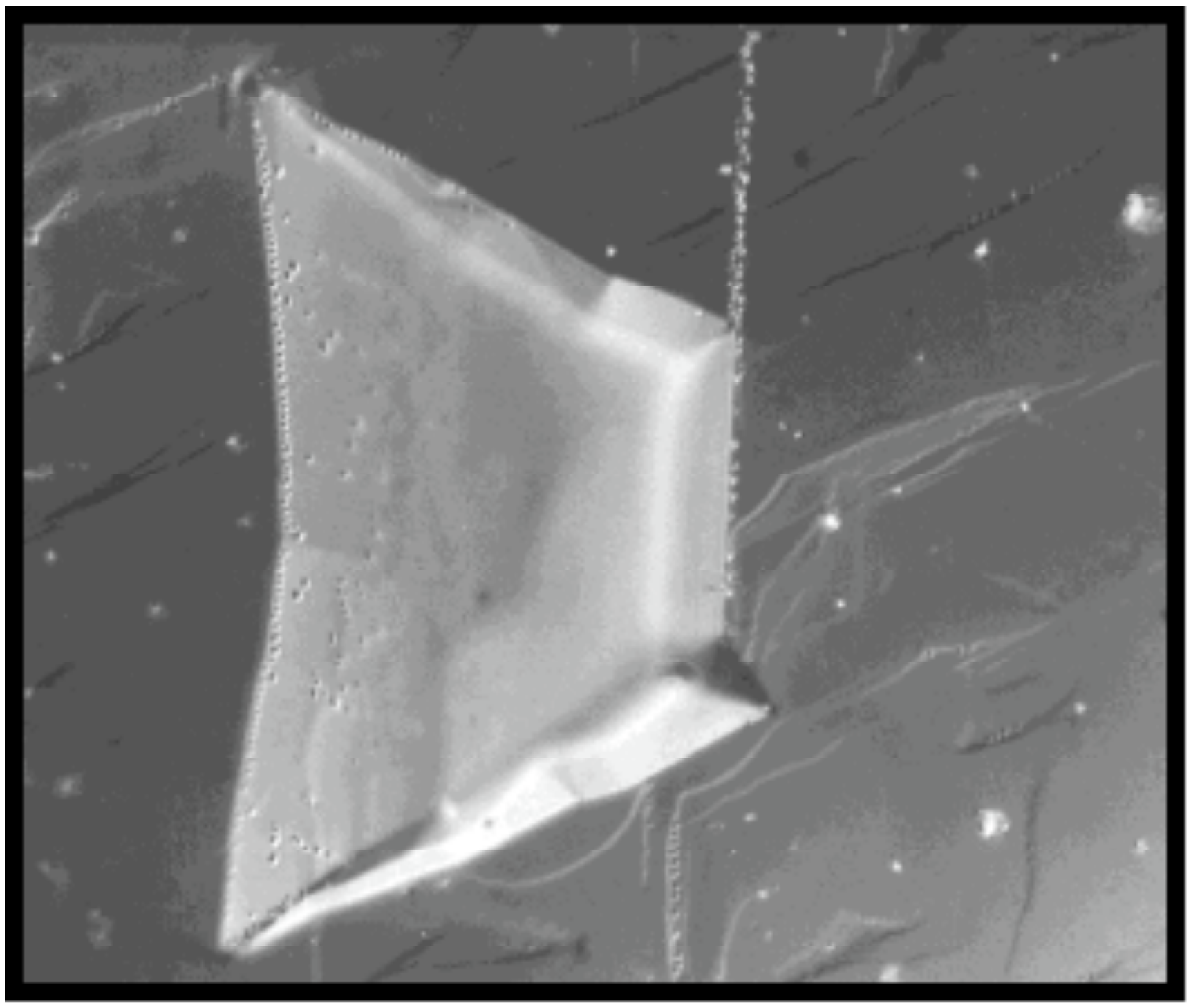

*Abbildung (53):*
*Teilweise aufgefüllte, kraterförmige Vertiefung in der Oberfläche von 18 C. An manchen Stellen treten perlschnurartige Gebilde auf, die auch den linken Kraterrand besetzen. Rechts fehlt die Spitze des großen Dreiecks, die anscheinend durch so eine Konfiguration abgeschnitten wurde. Die Bilddiagonale entspricht ca. 100 $\mu$m.*

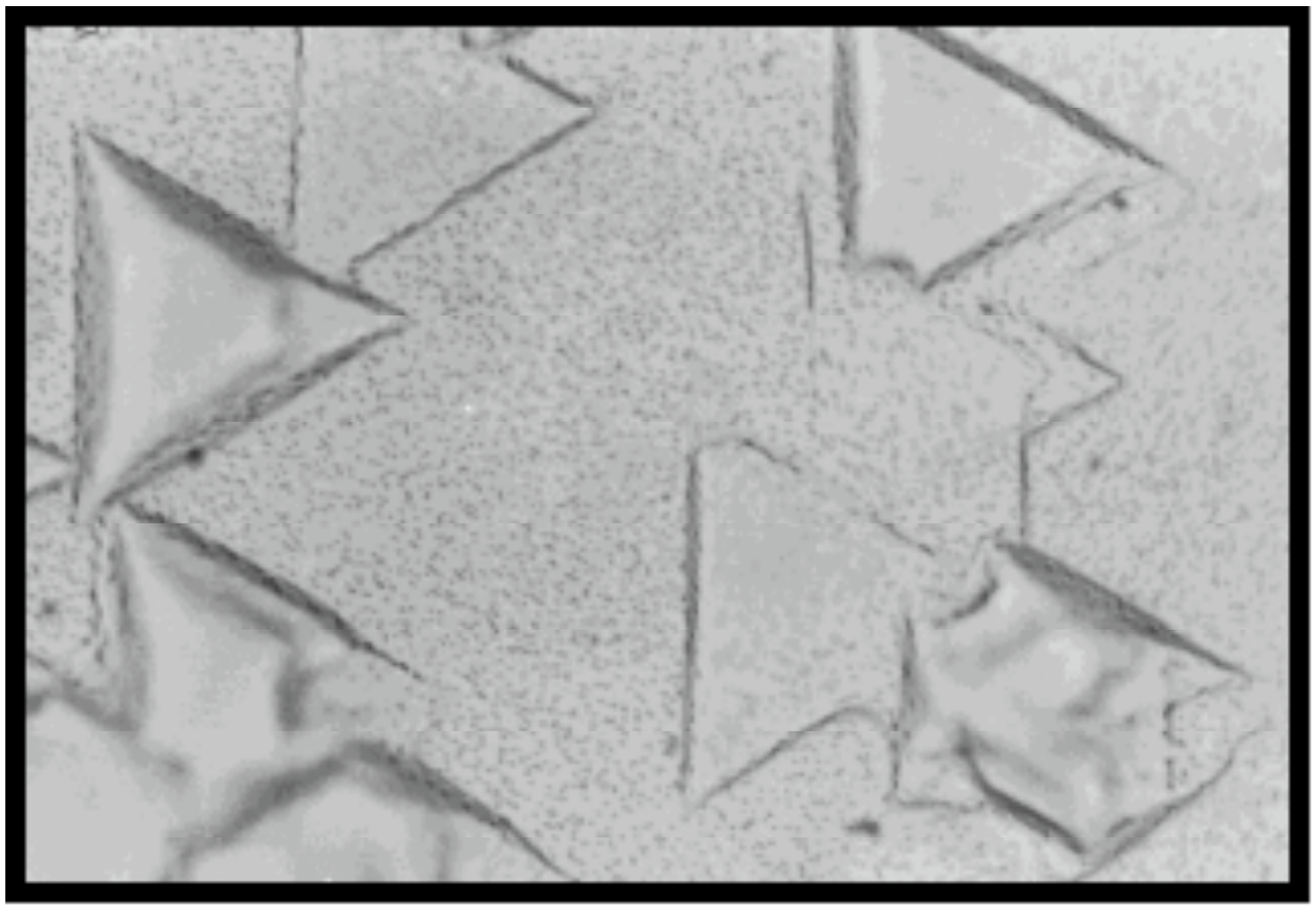

*Abbildung (54):*
*Mehr oder weniger aufgefüllte Dreiecke in einer kristallinen Oberfläche. Die Fläche zwischen den Dreiecken ist wellig. Die Bilddiagonale entspricht ca. 100 $\mu$m.*

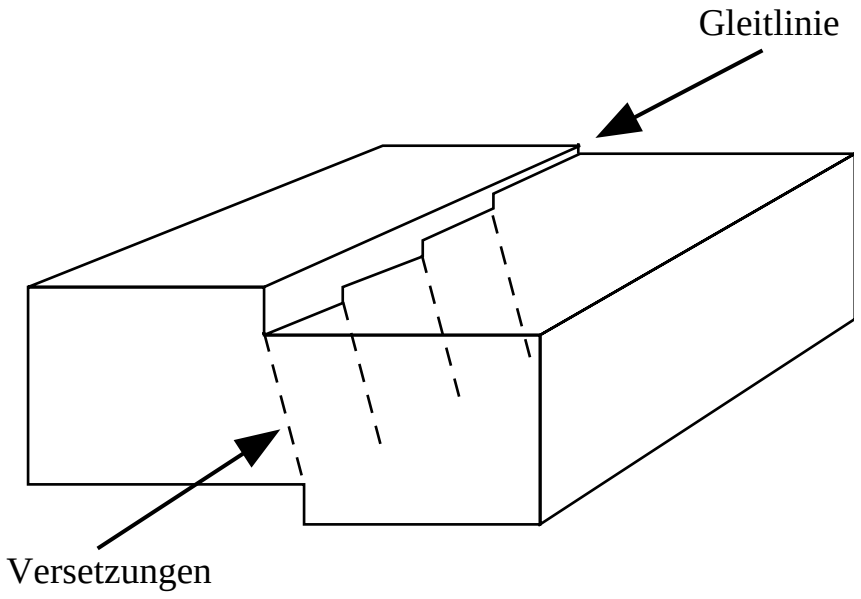

*Abbildung (55):*
*Entstehung von Gleitversetzungen nach McD. Robinson [25].*

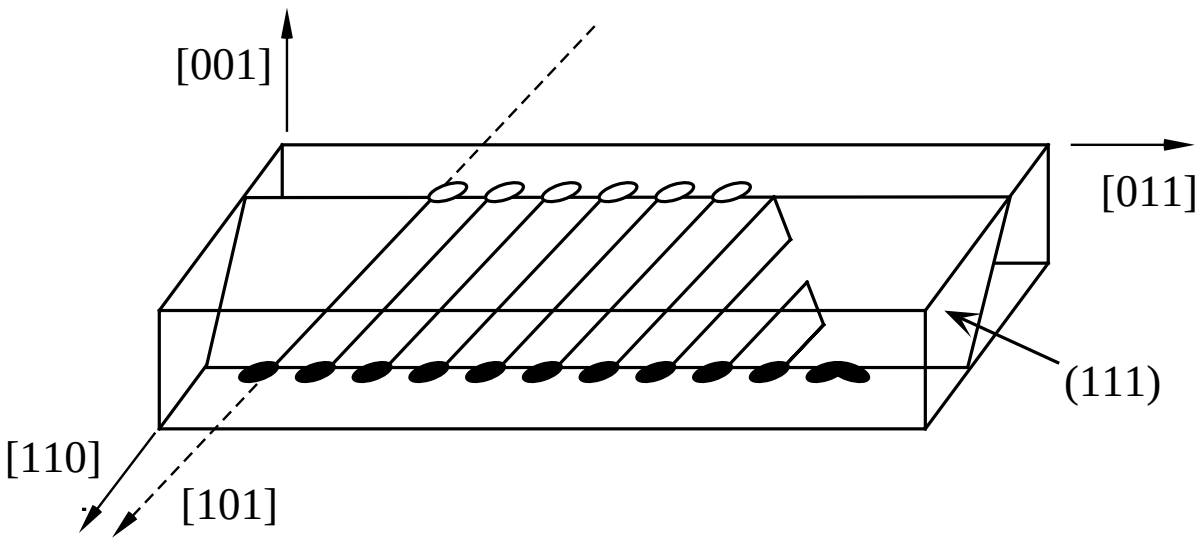

*Abbildung (56):*
*Verlauf der durch Gleiten entstandenen Versetzungslinien entlang den [101] Richtungen am Beispiel einer (100) Oberfläche nach S. M. Hu [29]. Die Schar aller, zum selben Gleitvorgang gehörenden Versetzungslinien liegt in einer oder einigen, sehr eng benachbarten (111) Ebenen, so daß an der Oberfläche eine perlschnurartige Zeile in [011] Richtung entsteht.*

Abbildung (53) zeigt deutlich linear in den [110] Richtungen angeordnete, perlschnurartige Gebilde, die häufig am Rand der Dreiecke aber auch in der glatten Oberfläche vorkommen. Hierbei handelt es sich womöglich um Gleitversetzungen, die wie in Abbildung (55) skizziert, durch Abgleiten im Einfluß mechanischer, etwa durch thermische oder, wie hier durch Legierungsgradienten hervorgerufene Spannungsfelder entstehen. Solche Schraubenversetzungslinen laufen, wie in Abbildung (56) nach S. M. Hu dargestellt entlang einer [101] Richtung in das Volumen hinein [25, 29]. Die gesamte Schar spannt somit eine {111} Ebene auf. Die Versetzungswände können, wie rechts im Bild (56) dargestellt, an einer anderen Versetzungslinie hängen bleiben, wobei dann die perlschnurartige Linie auf der Oberfläche zu erliegen kommt.

### 6.1.1. Die Dreieck- und Viereckstrukturen

Wie wir in den Bildern des vorangegangenen Abschnitts erkennen, werden die auffälligsten Oberflächenstrukturen durch regelmäßige Dreiecke auf 111 oder Quadrate auf 100 Oberflächen gebildet. Ihre Geometrien und ihr Entstehen wurden im Rahmen dieser Arbeit charakterisiert und als mögliche Schichtdickenbestimmung eingesetzt. Außerdem wurden auch schon von McD. Robinson auf sie hingedeutet [25].

Die Strukturen stellen sich meist als Krater heraus, deren Zähligkeit die Kristallinität aufweisen. Ihre Wände erscheinen glatt ansteigend. Im Zentrum einiger weniger sieht man unter dem Mikroskop eine Verunreinigung, etwa ein Staubkorn. Andere erscheinen weniger tief, oder wie in Bild (57) sogar bis zu ihrem Rand aufgefüllt, bis hin zu komplizierteren, stufenförmigen Wänden verschiedener Neigung. Betrachtet man die Gesamtheit der über eine Probe verteilten Krater, so findet man wie z. B. in den Bildern (50) und (51) eine häufig vorkommende Maximalgröße, die unabhängig von der Eigenschaft ist, ob der Krater aufgefüllt ist oder nicht.

Die Orientierungen des Kristalls wurden mit Hilfe der Lauekamera festgelegt womit sich die Kraterränder in beiden Fällen als [110] Richtungen herausstellen. Mit der Erfahrung, daß sich Oberflächen und Stapelfehler bevorzugt parallel zu den (111) Ebenen ausrichten, nehmen wir zunächst an, daß es sich auch bei den Kraterwänden um derartige Orientierungen handelt. Damit erhält man eine Pyramide quadratischer Grundfläche gemäß Abbildung (58) für das

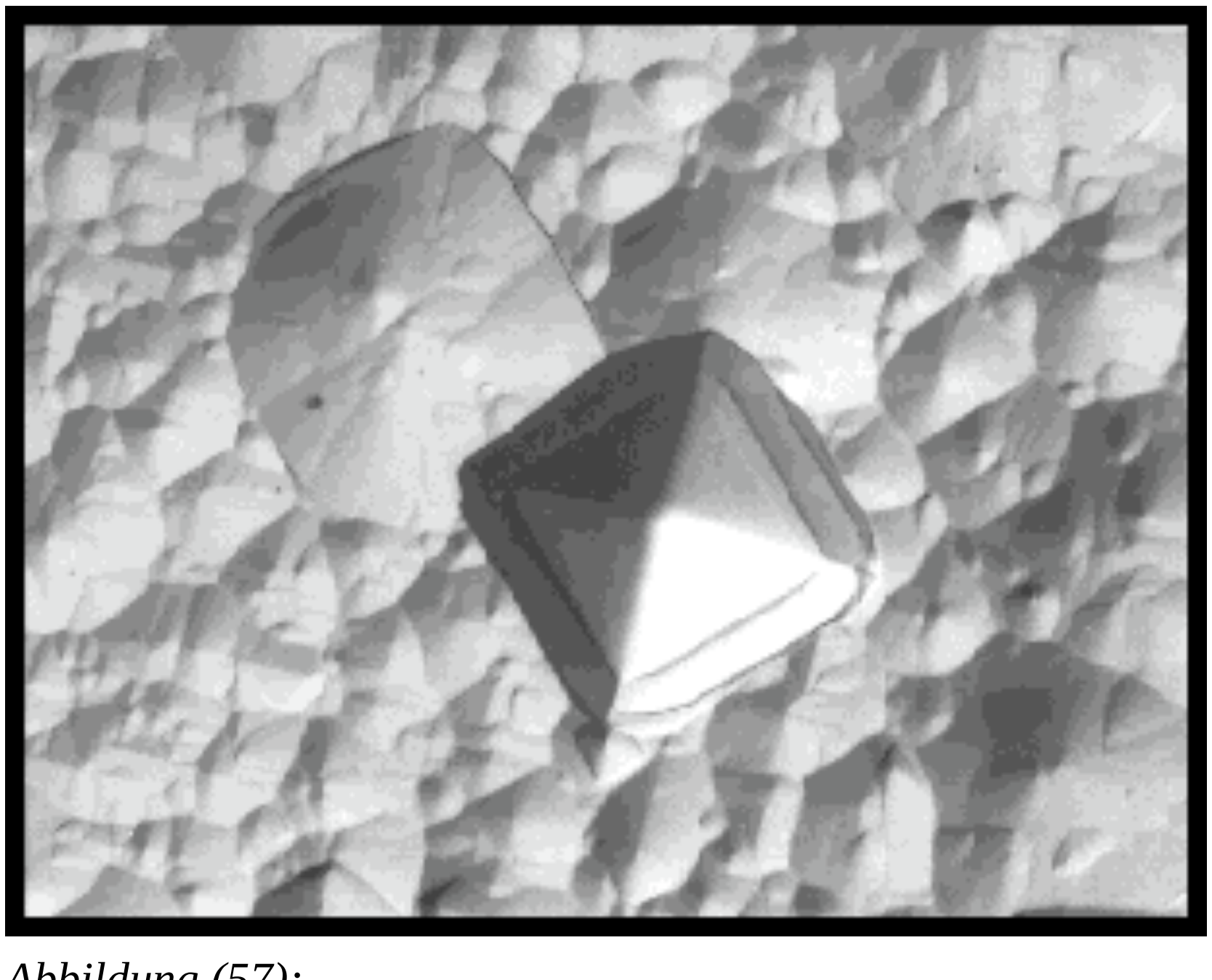

*Abbildung (57):*
*Krater in der (100) Oberfläche von 78 A. Obwohl der rechte bis zur Spitze leer und der linke bis zum Rand aufgefüllt ist, haben sie beide dieselbe Größe und sind somit gleichzeitigen Ursprungs. Die Bilddiagonale entspricht ca. 500 µm.*

100 und einen regelmäßigen Tetraeder nach Abbildung (59) für das 111 Substrat.

Mit diesen Modellen kann man durch trigonometrische Überlegungen die Kratertiefe h als Funktion der Seitenlänge a errechnen, was zu

$$h = \frac{1}{\sqrt{2}}\,a \qquad \text{für das 100 Substrat} \qquad (208)$$

und

$$h = \frac{\sqrt{2}}{\sqrt{3}}\,a \qquad \text{für das 111 Substrat} \qquad (209)$$

führt. Tabelle (8) stellt die so für die maximalgroßen Krater verschiedener Proben erhaltenen Ergebnisse zusammen.

Die, mit dieser Methode zurückgerechneten Schichtdicken stimmen fabelhaft mit den aus anderen Meßmethoden gewonnenen Formen überein. Da diese Tatsache mit der Annahme anderer, niederindizierter Gitterebenen für die Kraterwände nicht erfüllt ist, kann die vorgeschlagene Geometrie dieser Gebilde als bestätigt angesehen werden.

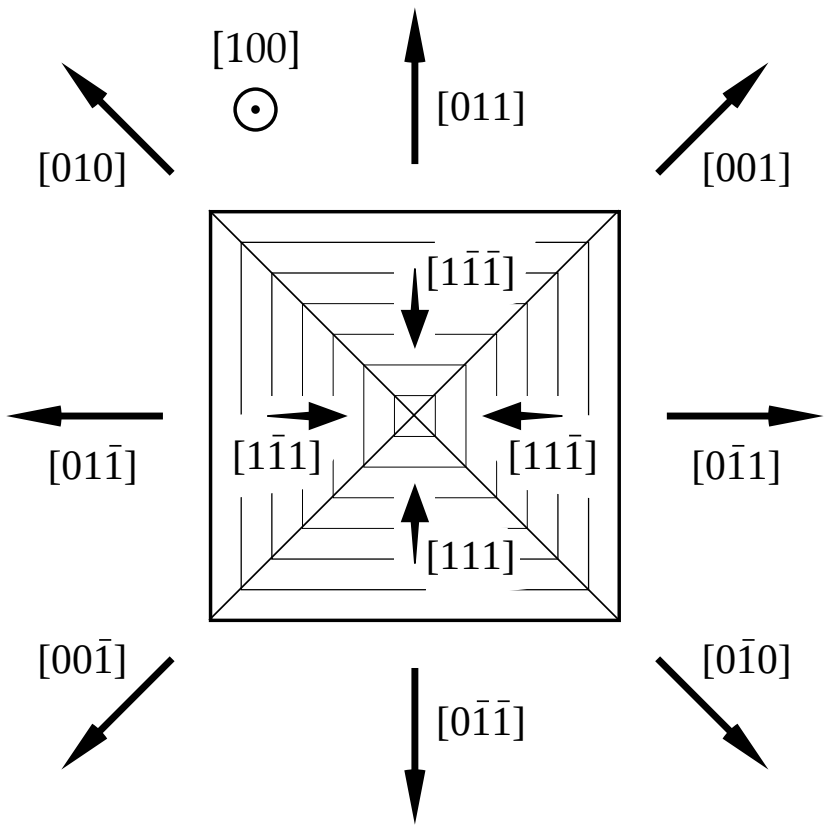

*Abbildung (58):*
*Orientierung der Vertiefungen in einer 100 Oberfläche.*

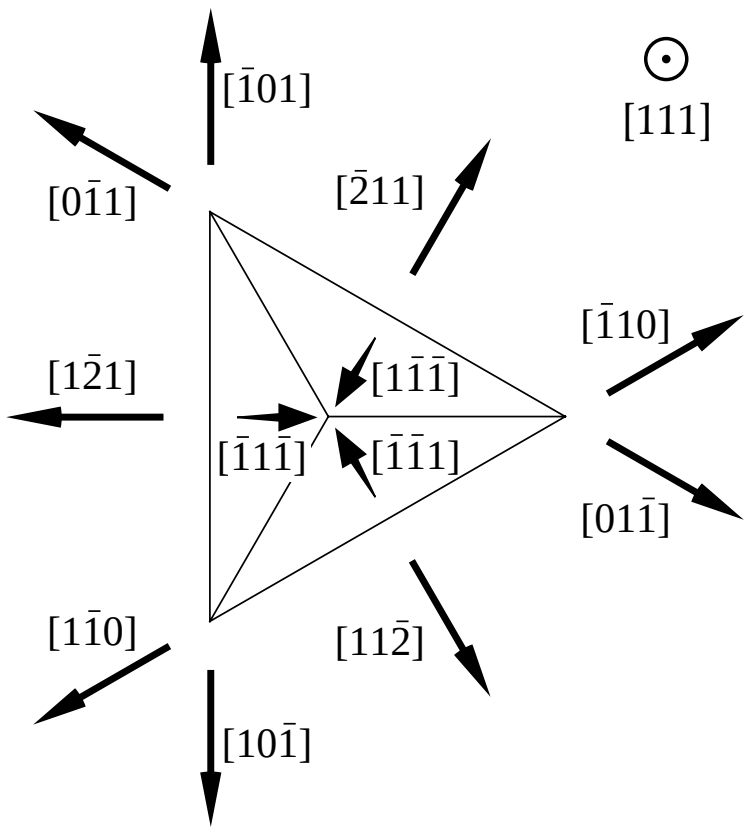

*Abbildung (59):*
*Charakterisierung der tetraederförmigen Krater in einer 111 Oberfläche.*

Durch die häufig vorkommende, maximale Größe der Krater und den vorangegangenen Betrachtungen geht hervor, daß ihre Entstehung auf der Substratoberfläche ihren Ursprung findet. Abbildung (60) zeigt schematisch den Entstehungsprozeß in drei verschiedenen Wachstumsstadien. Etwa ein, bereits erwähntes Staubkorn, Oxidreste oder Kratzer im ersten Teilbild können Anlaß zur Wachstumshemmung sein oder Stapelfehler hervorrufen. Die so entstandenen Defekte pflanzen sich bei weiterem Wachstum in den (111) Ebenen fort. Selbst wenn zu späterem Zeitpunkt auch Wachstum im Kraterboden eintritt, wächst immer noch ein Stapelfehler oder eine Stufe in gleicher Richtung weiter, so daß die am Ende gemeinsame Kantenlänge der Entstehungstiefe entspricht. Die Kraterverteilung zeugt also von der Unsauberkeit im Zuchtprozeß während der Entstehungsgeschichte des Kristalls.

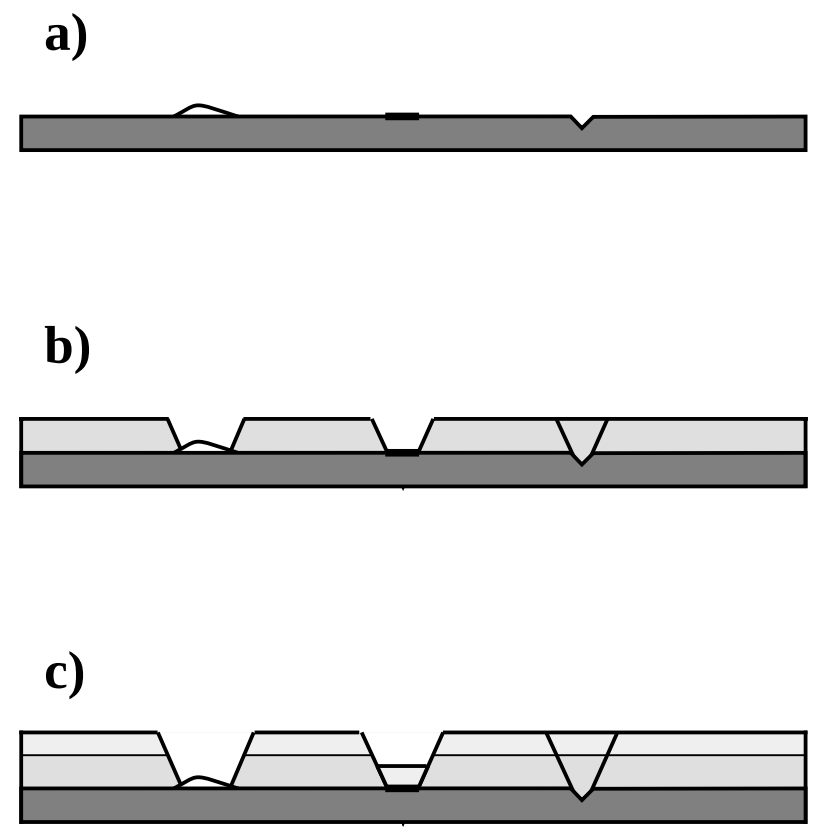

*Abbildung (60):*
*Entstehungsgeschichte der Krater. Staubkörner, Oxidreste oder Kratzer auf der Substratoberfläche in a) bilden Wachstumshemmungen oder Stapelfehler in [111] Richtungen, b), die bei weiterem Wachstum auch teilweise aufgefüllt werden können, c). in jedem Fall sieht man eine Struktur gleicher Orientierung und Seitenlänge auf der Oberfläche.*

| Probe | Orientierung | Seitenlänge | Kratertiefe | Schichtdicke |
|---|---|---|---|---|
| 78 A | [100] | 95 µm | 67 µm | 63 µm (µ-Sonde) |
| 16 C | [111] | 22 µm | 18 µm | 16 µm (Wägung) |
| 17 C | [111] | 65 µm | 53 µm | 56 µm (µ-Sonde) |
| 30 C | [111] | 33 µm | 27 µm | 21 µm (Wägung) |

*Tabelle (8):*
*Zusammenstellung der Seitenlängen der regelmäßigen Wachstumsfiguren verschiedener Proben und die daraus berechnete Kratertiefe im Vergleich mit der durch Wägung oder Mikrosonde bestimmten Schichtdicke.*

## 6.2. Krümmung der Kristalle

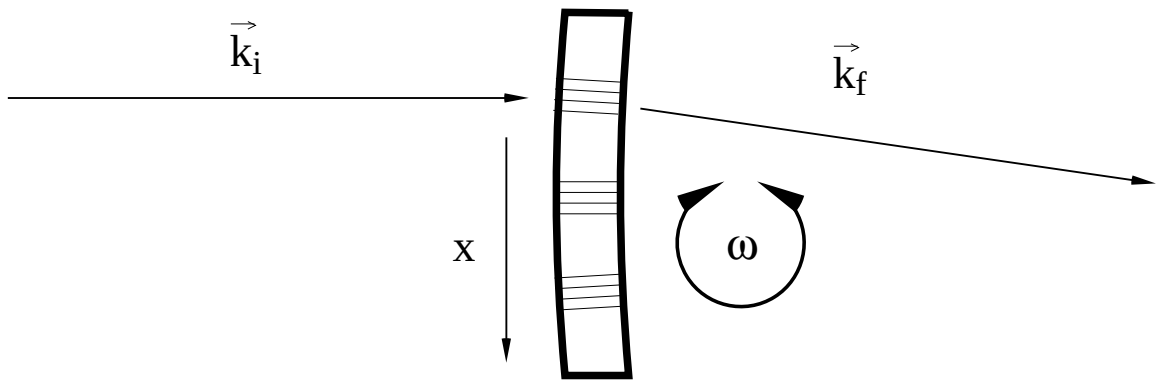

*Abbildung (61):*
*Messung der makroskopischen Krümmung eines Kristalls. Bei Verschiebung der Probe um $\Delta x$ kann aus der Verschiebung des Braggmaximums um $\Delta\omega$ auf den Krümmungsradius geschlossen werden.*

Durch die thermische Abkühlung der Probe von der Zuchttemperatur auf die Umgebung erwartet man aufgrund verschiedener Ausdehnungskoeffizienten im Substrat und der aufgedampften Schicht Gitterverzerrungen. Dabei gehen wir von dem Modell aus, daß das bei der hohen Temperatur sehr weiche Kristallgitter mit den durch Germaniumbeimischungen hervorgerufenen Fehlanpassungen durch Einbau von Versetzungen möglichst relaxiert ist. Durch verschiedene thermische Ausdehnung in den einzelnen Lagen kann beim Abkühlen ein Bimetalleffekt auftreten und somit eine makroskopische Krümmung erscheinen. Dies wurde vor allem bei den dicken Kristallen beobachtet.

Die Krümmung der gezogenen Kristalle kann mit Hilfe der Gammadiffraktometrie untersucht werden. Diese Methode ist wegen der kleinen Streuwinkel auf den Gradienten unempfindlich und eignet sich besonders gut zur Untersuchung der Verkippung von Gitterebenen. Die Meßanordnung ist schematisch in Bild (61) wiedergegeben.
Auf den fünf Instrumenten GAD1 - GAD5 am ILL werden eine gemeinsame $^{137}$Cs Quelle mit einer Wellenlänge von $\lambda_{Cs}$ = 0,01874 Å, sowie eine gemeinsame $^{198}$Au Quelle mit $\lambda_{Au}$ = 0,03011 Å angeboten. Die Auflösungen sind geometrisch durch die Quellengröße und einer Blende vor der Probe bestimmt und betragen z. B. auf dem GAD1 Spektrometer $4{,}4\cdot10^{-4}$ rad für die Cäsium- bzw. $1{,}2\cdot10^{-4}$ rad für die Goldquelle. Die Darwinbreiten für ideales Si [111] betragen bei obigen Wellenlängen etwa $4\cdot10^{-7}$ rad und sind damit gegen die geometrische Auflösung völlig vernachlässigbar.

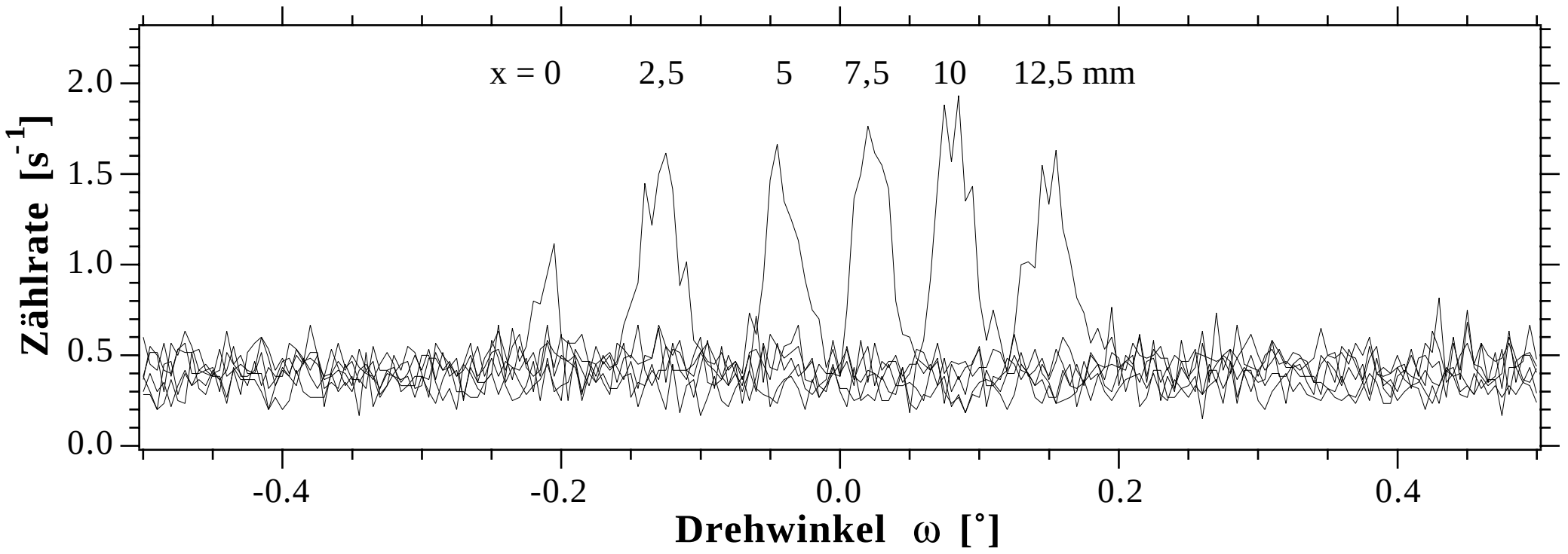

*Abbildung (62):*
*Aus dem Untergrund herausragende Braggreflexe bei verschiedenen Translationsstellungen x der Probe. Die Verschiebung des Maximums beschreibt die Krümmung des Kristalls.*

Außer einer ω-Abtastung, bei der die Probe gedreht wird, kann dieselbe auch in den beiden Dimensionen x und y senkrecht zur Strahlachse verschoben werden. Damit lassen sich durch einen auf wenige $mm^2$ beschränkten Strahl Mosaizität und Kristallorientierung an verschiedenen Positionen der Probe messen.
Speziell erhält man die Krümmung eines Kristalls, wenn man die Änderung des Maximums eines Braggreflexes als Funktion des Ortes auswertet. Verschiebt sich bei einer Translation des Kristalls um $\Delta x$ die Braggposition um $\Delta\omega$, so ergibt sich der Krümmungsradius $\rho_K$ zu

$$\rho_K = \frac{\Delta x}{\Delta\omega} \quad . \tag{210}$$

Abbildung (62) zeigt eine Meßreihe an dem 850 µm dicken Gradientenkristall 94 A. Die Translationsrichtung ist parallel zum Streuvektor $(0\bar{2}2)$, der in der Kristalloberfläche liegt. Der Krümmungsradius dieser Meßreihe sowie in der Richtung senkrecht dazu ist in Abbildung (63) ausgewertet. Dieser Kristall zeigt eine erhebliche Krümmung, die sich später auch auf die weiteren Diffraktionsergebnisse auswirken wird. Typische Werte, auch für dünnere Proben sind Radien von zwei bis zehn Metern. Wir sehen auch, daß die Proben nicht gleichmäßig und isotrop gebogen sind.

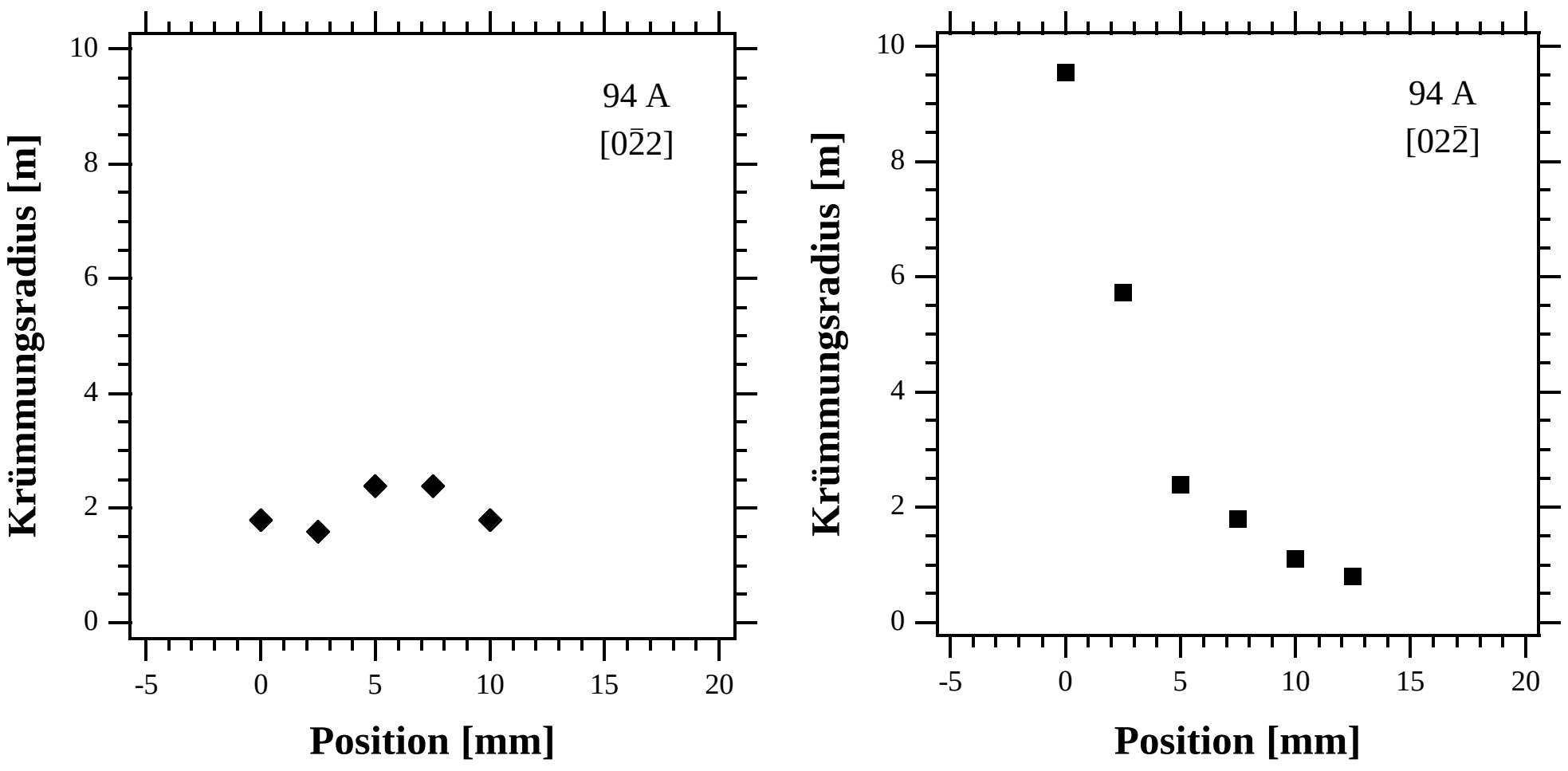

*Abbildung (63):*
*Krümmungsradien des Gradientenkristalls 94A als Funktion zweier zueinander senkrechter Richtungen in der Kristalloberfläche.*

## 6.3. Die Konzentrationsprofile und Einzelschichtdicken

Im Vorangegangenen haben wir die geometrischen und kristallinen Eigenschaften der $Si_{1-x}Ge_x$ beschrieben und den Legierungsaspekt vernachlässigt. Ein Hauptpunkt der Kristallzucht konzentriert sich aber auch auf die Germaniumverteilung in der Siliziummatrix, etwa ihren Anstieg als Funktion der Kristalldicke, die absoluten Konzentrationen des Festkörpers im Vergleich zur Gasphase, die Wachstumsgeschwindigkeit bei verschiedenen Zusammensetzungen.

Zur Beantwortung dieser Fragen ist es notwendig, die Konzentrationsprofile über den Kristallquerschnitt zu bestimmen. Neben komplizierteren, zerstörungsfreien Methoden, wie z. B. mit Rutherfordrückstreuung von A. Magerl durchgeführt [30], ist es naheliegend, die Probe zu brechen und die Bruchkante mikroskopisch zu analysieren.

Mit dem lichtoptischen Mikroskop läßt sich dabei höchstens die Gesamtschicht erkennen, da das Depot wegen der vielen Fehlstellen etwas rauher bricht als das Substrat. Abbildung (64) stellt eine so gewonnene Dickenabhängigkeit als Funktion des Probenortes zusammen. Insbesondere bestätigt sie wieder die Tendenz zu polykristallinem Wachstum bei überhöhter Rate.

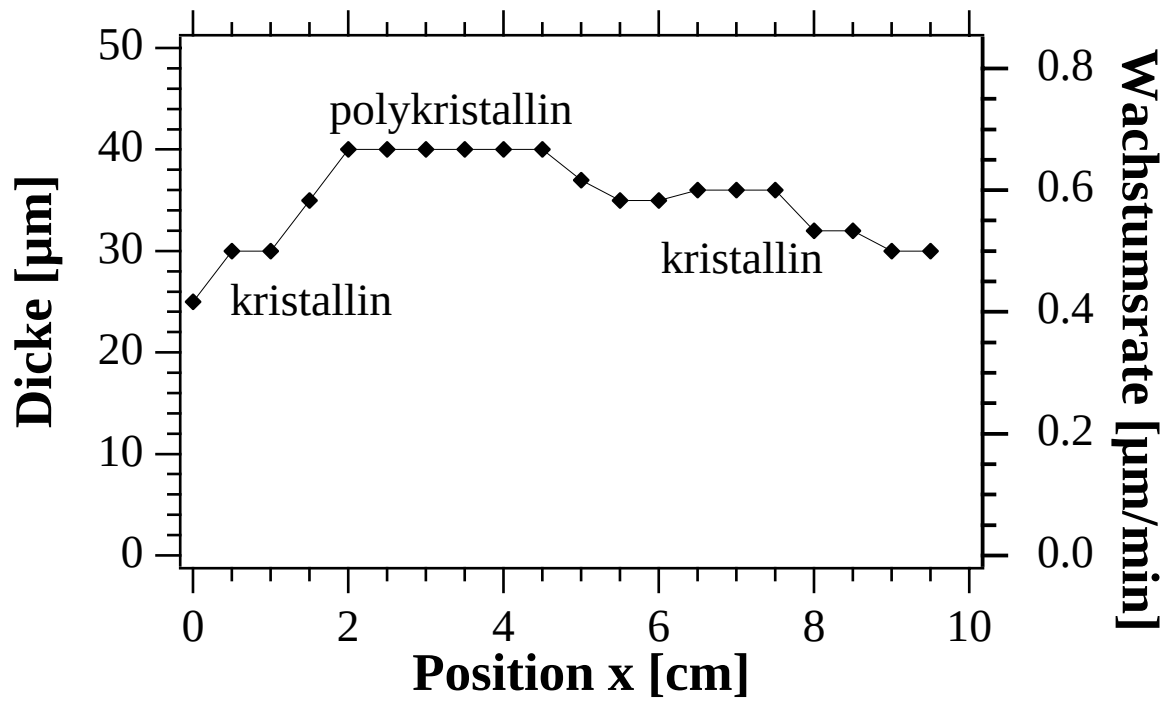

*Abbildung (64):*
*Lichtmikroskopisch an einer Bruchkante gemessene Dickenabhängigkeit des Siliziumdepots 14 C. Teile der Oberfläche sind einkristallin, andere polykristallin gewachsen. Man erkennt die Tendenz polykristallinen Wachstums bei erhöhter Rate.*

Ein Kontrast zwischen Schichten verschiedener Germaniumkonzentrationen läßt sich mit dem Lichtmikroskop jedoch nicht erkennen. Nimmt man sich ein Elektronenmikroskop mit Mikrosondenanalyse zu Hilfe, so kann man, wie im folgenden beschrieben, genaue Aussagen über die Konzentrationsprofile erhalten. Dazu sollen zunächst die Grundzüge dieser Methode erläutert werden, bevor wir die Daten einiger Gradienten- und Stufenkristalle zusammenstellen. Derartige Geräte standen uns in der Ruhruniversität Bochum sowie durch Auftragserteilung an der Universität in Grenoble zur Verfügung.

### 6.3.1. Elektronenmikroskopie und Mikrosondenanalyse

Poliert man die Kante eines gebrochenen Kristalls und sieht sie sich unter dem Elektronenmikroskop an, so erhält man bereits bei kleinen Konzentrationsunterschieden im Prozentbereich gute Kontraste auf der Photoplatte. Abbildung (65) zeigt eine derartige Mikrophotographie des Stufenkristalls 28 C. Links, ganz dunkel, befindet sich das Siliziumsubstrat, während nach rechts hin immer dünnere Schichten, der Reihe nach durchnumeriert, mit mehr und mehr Germaniumgehalt liegen. Die Schicht 1 kann man durch andere Belichtungen, die mehr auf das dunkle Substrat abgestimmt sind, hervorheben.

Um die absolute Germaniumkonzentrationen der einzelnen Lagen zu bestimmen, macht man sich die Wechselwirkung des Elektronenstrahls mit der Probe zu Nutze: Trifft der unter einen Mikrometer fokussierte Strahl auf die Probe, so wird dieser teilweise gestreut oder unter Aussendung verschiedener Sekundärstrahlungen absorbiert. Wie in der klassischen Röntgenröhre entsteht hier auch eine elementspezifische charakteristische Strahlung, die zur Konzentrationsanalyse der Legierung verwendet werden kann. Ein energiedispersiver Halbleiterdetektor oder ein nachgeschaltetes Kristalldiffraktometer zerlegen die Röntgenintensitäten spektral, wobei man typischerweise ein Spektrum gemäß Abbildung (66) erhält. Mit einer angeschlossenen Datenverarbeitungsanlage kann man sofort die einzelnen Maxima identifizieren und aus den gemessenen Intensitätsverteilungen die Konzentrationsverhältnisse ermitteln.

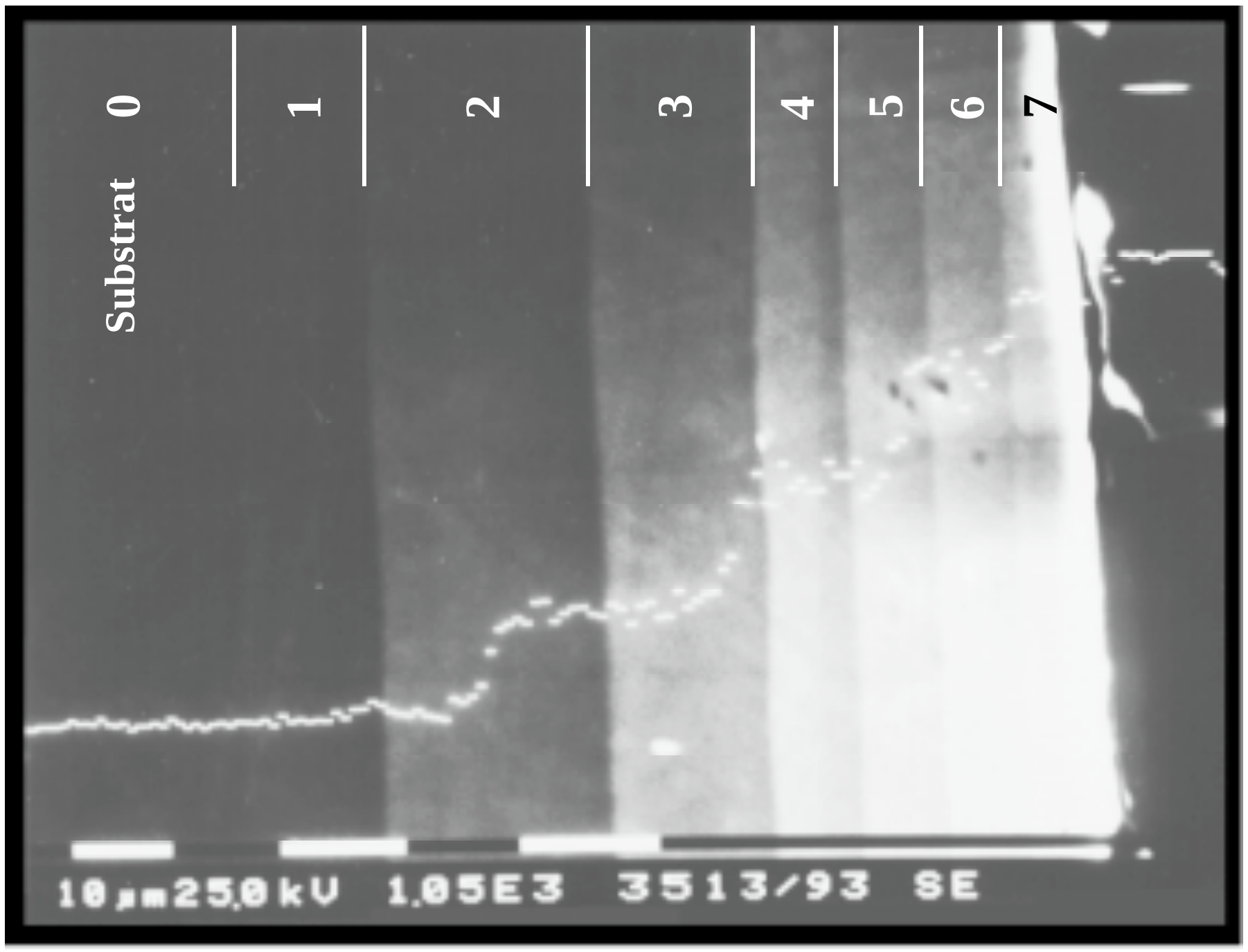

*Abbildung (65):*
*Elektronenmikrograph des Stufenkristalls 28 C. Die Oberfläche liegt links im Bild senkrecht und erscheint wegen der ungenügenden Ladungsabführung sehr hell. Das Substrat besteht aus reinem Silizium. Schicht 0 enthält nur etwa 1 % Germanium und fällt daher sehr dunkel aus, während in den anderen Schichten bis zu 35 % mehr und mehr Germanium beigemischt wurde.*

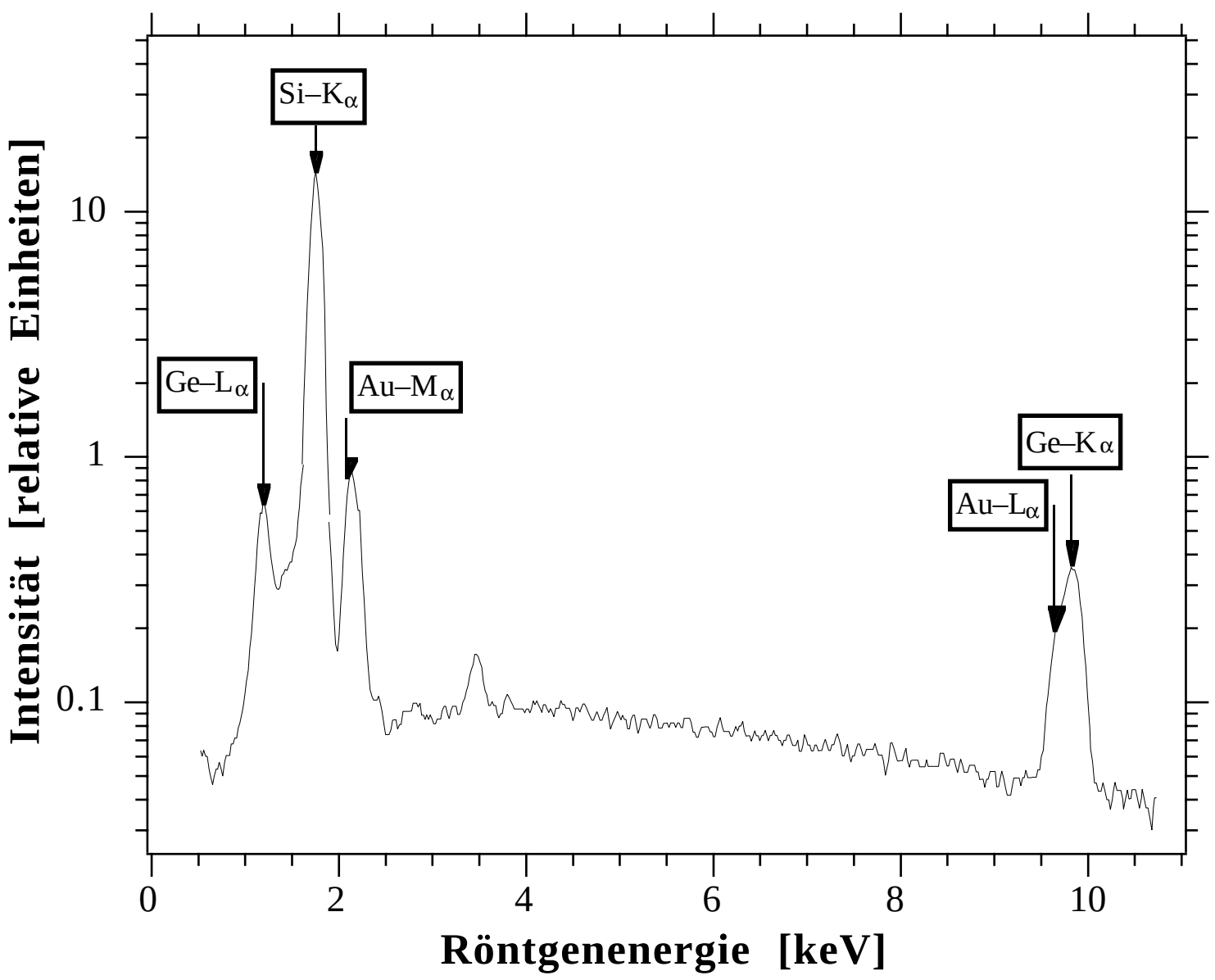

*Abbildung (66):*
*Charakteristisches Röntgenspektrum bei der Mikrosondenanalyse von $Si_{1-x}Ge_x$ Kristallen. Die Goldlinien rühren von einer Bedampfung der Oberfläche zugunsten ihrer Leitfähigkeit her.*

Läßt man nun den Elektronenstrahl langsam über den Probenquerschnitt wandern und taktet die Analyse in zeitliche Intervalle, so erhält man die Konzentration in Abhängigkeit des Ortes. Eine derartige Meßkurve ist, jedoch um 10 µm nach rechts verschoben, der Mikrographie in Abbildung (65) überlagert.

Die Ortsauflösung wird, obwohl eine feinere Fokussierung möglich ist, auf etwa 1 µm beschränkt, da durch Vielfachstreuung der Elektronen in der Probe ein Volumen entsprechenden Durchmessers ausgeleuchtet wird.

### 6.3.2. Mikrosondenergebnisse

Einige der wichtigsten und typischen Resultate der Mikrosondenanalyse sind in Abbildung (67) zu sehen. Die beiden oberen Messungen wurden an der Universität Bochum durchgeführt, die restlichen, mit wesentlich besserer Statistik, sind an der Universität Grenoble in Auftrag gegangen.

In der linken Spalte sind die Konzentrationsprofile einiger Stufenkristalle, rechts die von Gradientenkristallen gezeigt. Die sprunghaften Anstiege bei 78 A und 82 A im Bereich der instrumentellen Auflösung von 1 µm bis 2 µm sowie die linear ansteigenden Profile der Gradientenkristalle selbst über große Dicken hinweg zeugen von der perfekt beherrschbaren Zusammensetzung im Festkörper. Insbesonders an den Stufenkristallen ersichtlich, tritt keine meßbare Interdiffusion benachbarter Schichten auf.

Zusammen mit den Gasmischungsanteilen aus Tabelle (7) kann man das Konzentrationsverhältnis

$$\gamma = \frac{x_{fest}}{x_{gas}} \qquad (211)$$

zwischen Festkörper und Gasphase erhalten. Abbildung (68) zeigt diesen Zusammenhang für eine Reihe von Ergebnissen verschiedener Kristalle. Dabei häufen sich die Meßpunkte entlang den Geraden $\gamma = 0{,}73$ für die A- und $\gamma = 1{,}4$ bzw. 2,3 für die C-Serie. Über die Deutung dieser krassen Unterschiede soll hier nur spekuliert werden: Wie bereits erwähnt, wird bei den relativ niedrigen Temperaturen der A-Serie um die 1200 K die Wachstumsrate überwiegend durch die Oberflächenreaktion beschränkt. Dabei diffundiert Silan aufgrund des Massenunterschieds schneller an die Oberfläche heran als German, so daß sich im Abscheideprozeß überwiegend Silizium gegenüber Germanium anbietet und sich dies wiederum auf eine zu niedrige Germaniumkonzentration im Festkörper auswirkt. Bei den hohen Zuchttemperaturen um knapp 1500 K der C-Serie wird die Wachstumsgeschwindigkeit vorwiegend durch die Gasnachführung beschränkt. Hierbei bildet sich oberhalb des Substrats eine verarmte Silanzone, in der dann auch die verweilenden Germanmoleküle genügend Zeit finden an die Oberfläche zu gelangen. Zudem kommt noch ein weiterer Aspekt, daß sich German mit kleinerem Energieaufwand, also schneller zersetzt als Silan, wodurch sich im Festkörper eine höhere Germaniumkonzentration gegenüber der Gasphase einstellt. Sicherlich spielen auch andere Parameter und eine wesentlich kompliziertere Chemie über Zwischenprodukte in den Reaktionsvorgang mit ein. So wurde in der A-Serie mit [100]- und in der C-Serie mit den schneller wachsenden [111] Substraten sowie mit völlig unterschiedlichen Ofengeometrien gearbeitet.

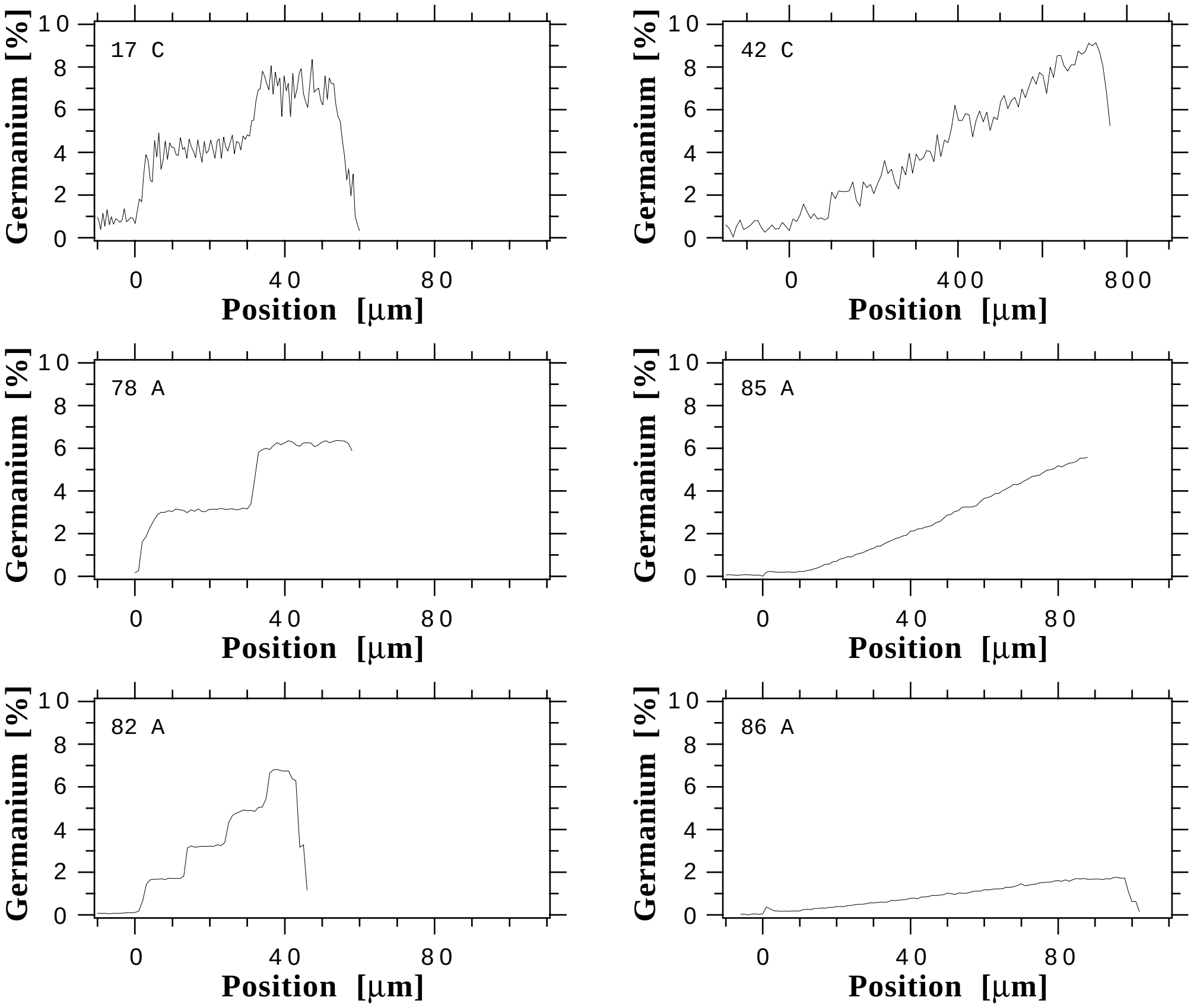

*Abbildung (67):*
*Konzentrationsprofile verschiedener Proben mit stufenförmigem oder linear kontinuierlichem Anstieg des Germaniumanteils. Deren Verlauf läßt sich beliebig durch die zeitliche Änderung der Gasmischungsverhältnisse während des Wachstums einstellen. Man beachte die unterschiedliche Skala bei 42 C. Die Kurve zu 85 A wurde nicht bis zur Oberfläche hinausgemessen. Die beiden oberen Kurven sind wegen der schlechten Statistik verwackelt.*

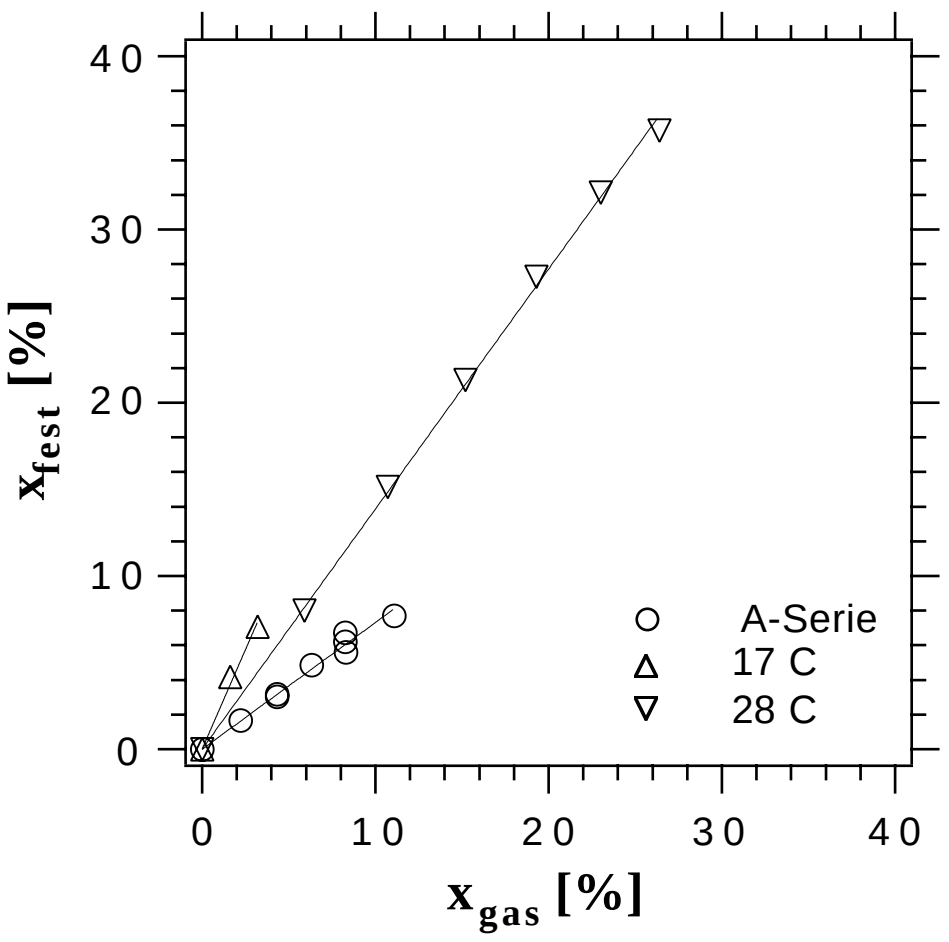

*Abbildung (68):*
*Die Germaniumkonzentration $x_{fest}$ im Festkörper ist nicht notwendigerweise identisch mit der in der Gasphase $x_{gas}$. Das Verhältnis hängt stark von den Zuchtbedingungen ab. Numerische Werte sind in Tabelle (9) zusammengestellt.*

| Kristall | $x_{gas}$ | $x_{fest}$ | $\gamma$ |
|---|---|---|---|
| 82 A | 2,20 | 1,65 | 0,75 |
| 82 A | 4,31 | 3,17 | 0,74 |
| 82 A | 6,32 | 4,87 | 0,77 |
| 82 A | 8,26 | 6,73 | 0,82 |
| 78 A | 4,31 | 3,03 | 0,70 |
| 78 A | 8,26 | 6,20 | 0,75 |
| 83 A | 11,1 | 7,70 | 0,69 |
| 85 A | 8,30 | 5,60 | 0,67 |
| 17 C | 1,60 | 4,18 | 2,61 |
| 17 C | 3,20 | 7,08 | 2,21 |
| 28 C | 5,90 | 8,0 | 1,36 |
| 28 C | 10,7 | 15,1 | 1,41 |
| 28 C | 15,2 | 21,3 | 1,40 |
| 28 C | 19,3 | 27,3 | 1,41 |
| 28 C | 23,0 | *32,2* | 1,40 |
| 28 C | 26,4 | *35,7* | 1,35 |

*Tabelle (9):*
*Zusammenstellung der Konzentrationsverhältnisse zwischen Festkörper und Gasgemisch. Die beiden kursiv gedruckten Werte sind aus Neutronenstreudaten vervollständigt.*

Wie wir in den folgenden Kapiteln sehen werden, ist die Mikrosondenanalyse ein wesentlicher Eingangsparameter für die Interpretation der Röntgen- und Neutronenbeugungsergebnisse. Andererseits können letztere wiederum zur Bestimmung mittlerer Konzentrationsprofile eingesetzt werden.

# 7. Experimentelle Beugungseigenschaften der $Si_{1-x}Ge_x$ Kristalle

Um dem angestrebten Ziel des vorliegenden Projekts, dem Einsatz von Gradientenkristallen als Neutronenmonochromator näherzukommen, ist die Studie der Neutronen- und Röntgenbeugungseigenschaften speziell gezogener Testkristalle ein wesentlicher Punkt. Durch derartig gewonnene Ergebnisse lassen sich wiederum auf die Kristallstruktur und die Zusammensetzung Rückschlüsse ziehen, so daß diesem Kapitel die Doppelrolle zwischen Überprüfung und Analyse zukommt.

Um sich später auf den direkten Vergleich und die Ergebnisse der durchgeführten Experimente konzentrieren zu können, sollen zunächst die für die Beugungsexperimente relevanten Meßapparate in einem untergeordneten Kapitel zusammengestellt werden.

Der Vollständigkeit halber sei erwähnt, daß auch auf Röntgenbeugung beruhende Geräte, wie die Lauekamera zur Kristallorientierung oder die Gammadiffraktometrie zur Krümmungsmessung verwendet wurden, deren Resultate bereits an geeigneter Stelle aufgeführt sind und nicht direkt mit der Beugungsanalyse der Legierungsvariation in Zusammenhang stehen.

## 7.1. Die verwendeten Diffraktometer

Die Experimente wurden an verschiedenen Diffraktometern, sowohl mit Neutronen- als auch mit Röntgenstrahlen, durchgeführt. Wichtige Kriterien für die Wahl einer Maschine sind ihre Auflösung in der durch die Legierungsvariation verursachten, longitudinalen Änderung des reziproken Gittervektors sowie eine ausreichende Transmission der Strahlung durch millimeterdicke $Si_{1-x}Ge_x$ Proben. Während letzteres für Neutronen kein Problem ist, haben wir im Fall der Röntgenbeugung hochenergetische Strahlung im 45 keV bis 100 keV Bereich gewählt, deren Absorptionslänge wie im Kapitel 2. geschildert, bereits schon einige Zentimeter beträgt.

### 7.1.1. Das Neutronenrückstreuspektrometer

Die Auflösung $\Delta k/k$ eines Kristalldiffraktometers wird mit Betrachtung der logarithmischen Ableitung des Braggesetzes

$$\frac{\Delta k}{k} = \frac{\Delta G}{G} + \cot(\theta)\,\Delta\theta \qquad (212)$$

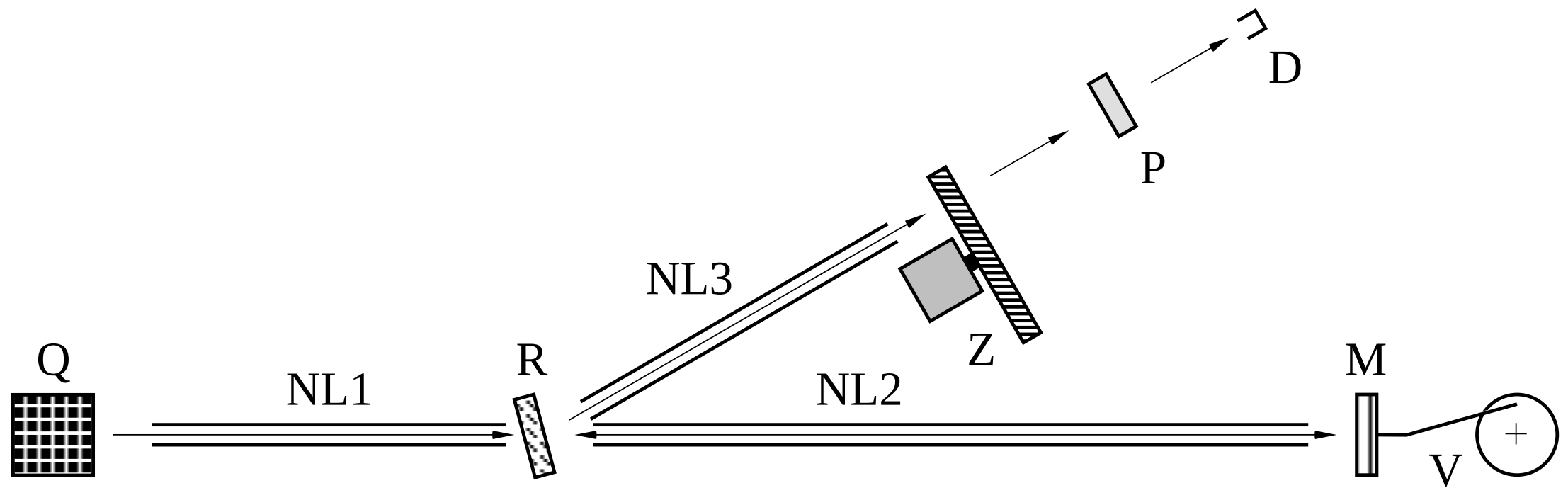

*Abbildung (69):*
*Schematische Darstellung des IN10 am ILL bei Transmissionsstellung der Probe. Neutronen aus der kalten Quelle Q werden vom Monochromator M zurückreflektiert, durch den Reflektor R aus dem Primärleiter NL1 ausgeblendet und durch NL3 zur Probe P hin geführt. Hinter letzterer steht der Detektor D. Die Wellenzahl wird durch den Dopplereffekt V am Monochromator durchgestimmt. Ein Zerhacker Z unterdrückt den Untergrund.*

durch die Definition ΔG des Streuvektors G am Monochromator und einer Winkelvariation Δθ, etwa der Strahldivergenz oder der Monochromatormosaizität bestimmt. Während ΔG/G bis zur natürlichen Linienbreite von $3{,}7 \cdot 10^{-5}$ sehr wohl definiert sein kann, liefert der zweite Summand in (212) bei normalen Streugeometrien Werte von $10^{-2}$ bis $10^{-3}$. Wählt man jedoch die spezielle Rückstreugeometrie von $\theta = \pi/2$, so verschwindet der Winkelterm in erster Ordnung und man kann Auflösungen im Bereich der natürlichen Linienbreite erreichen [31, 32].
In Abbildung (69) ist die am IN10 des ILL verwendete Rückstreuanordnung in ihren wesentlichen Gesichtspunkten wiedergegeben. Ein allgemeinerer Abriß mit den technischen Daten findet sich in der Intstrumentenbeschreibung des ILL [33].

Die Neutronen der kalten Quelle Q werden durch die Leiter NL1 und NL2 zum Si [111] Monochromator M geführt, der die wohl definierte Wellenzahl von $k_0 = 1{,}00$ Å$^{-1}$ zurückwirft. Diese Neutronen treffen auf den, leicht oberhalb des Primärstrahls angebrachten Reflektor R, der sie in den Leiter NL3 und damit zum Probenort P umlenkt. Hinter der Probe steht ein Detektor D. Um den monochromatischen Strahl durchzustimmen, nutzt man den durch Bewegung des Monochromators entstehenden Dopplereffekt. Damit kann ein dynamischer Bereich von $\Delta k = \pm 1{,}4 \cdot 10^{-2}$ Å$^{-1}$ abgetastet werden. Die Intensitäten werden in einem Vielkanalanalysator, dessen Kanäle den verschiedenen Dopplergeschwindigkeiten und damit Wellenlängen zugeordnet sind, gespeichert und können somit über lange Zeit integriert werden. Detaillierte Angaben hierzu

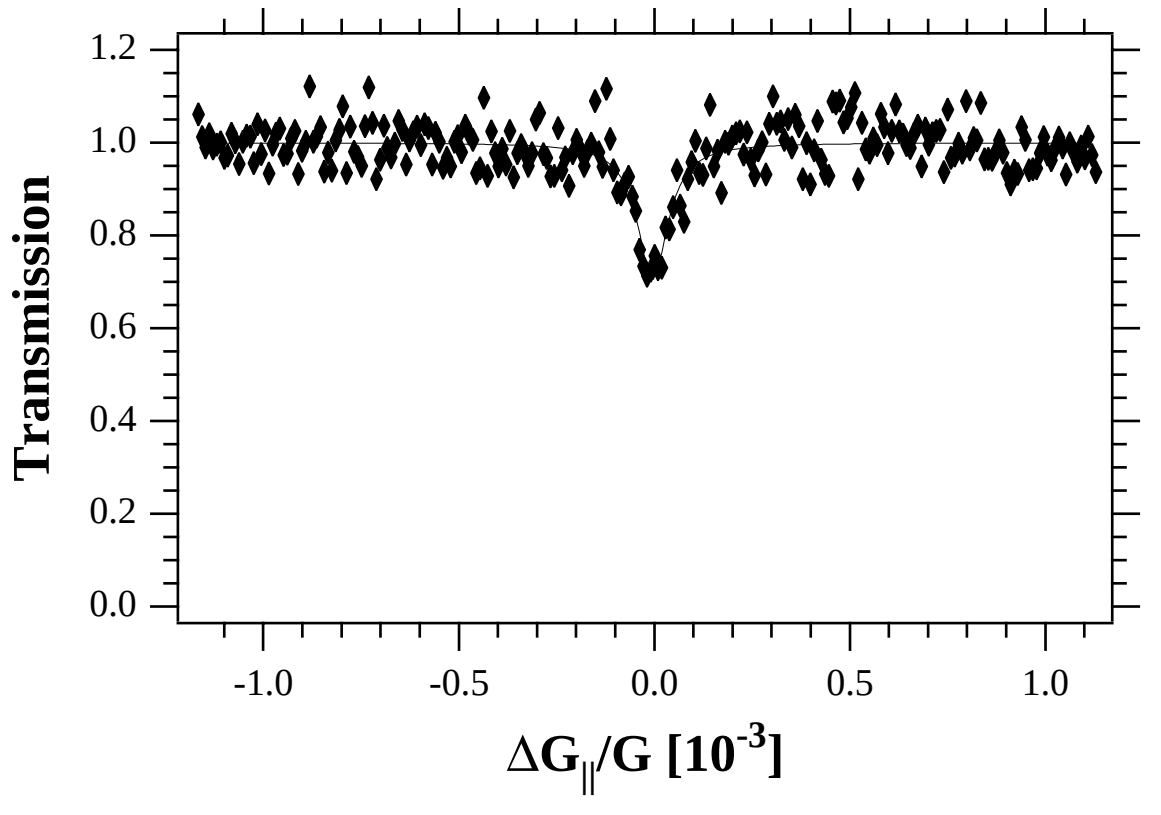

*Abbildung (70):*
*Mittels Idealkristall gemessene Auflösungsfunktion des IN10 in Transmissionsgeometrie.*

sowie zur Standardauswertung und Normierung finden sich in meiner Diplomarbeit [34].

Nimmt man ein Spektrum ohne Probe als Funktion der Wellenzahl auf, so trifft der gesamte monochromatische Strahl mit voller Intensität in den Detektor und wir erhalten den Normierungswert 1. Stellt man nun eine Si [111] Probe an die Position P und richtet ihre Gitterebenen zur Rückstreuung aus, dann tritt bei $k_0$ Braggreflektion ein. Es werden also Neutronen aus dem Strahl zurückreflektiert, die die Intensität im Detektor schwächen. Verstimmt man jetzt durch den Dopplereffekt die einfallende Wellenzahl, so gerät sie außer Braggbedingung, woraufhin die Intensität am Detektor wieder auf den Wert 1 maximaler Transmission ansteigt. Die Kurve in Abbildung (70) zeigt die mit einer perfekten Siliziumprobe gewonnene Auflösungsfunktion des Instruments. Die Linie erscheint wegen des unperfekt zusammengesetzten Monochromators um 3 gegenüber der natürlichen verbreitert. Dies geht auf Kosten der Linientiefe, und vergrößert zudem um 1,5 den theoretischen Wert der integrierten Reflektivität von $\mathfrak{R}_y = \pi$.

Mißt man $Si_{1-x}Ge_x$ Kristalle, so haben sie, im Vergleich zum Monochromator leicht unterschiedliche reziproke Gittervektoren, die sich in einer Verschiebung des Braggreflexes aus dem Zentrum des Meßbereichs ausdrückt. Gewissermaßen tastet man mit der monochromatisierten Wellenzahl die reziproke Gittervektorverteilung ab. Da Winkelabweichungen erst in zweiter Ordnung eingehen, liegt die Stärke des Rückstreuspektrometers in der Messung der longitudinalen Richtung, also der Längenvariation des Gittervektors.

### 7.1.2. Das Flugzeitrückstreuspektrometer

Am Forschungsreaktor München steht eine Variante des linearen Rückstreuspektrometers, bei dem die Wellenzahl des Primärstrahls durch eine Flugzeitanalyse ermittelt wird.

Der Aufbau des Instruments ist in Abbildung (71) dargestellt: Ein Zerhackersystem Z1, Z2, Z3 sorgt für einen eindeutigen, polychromatischen Startimpuls, dessen verschiedene Wellenzahlen über die Flugstrecke von ca. 145 m zu den Detektoren 0 - 8 zeitlich auseinanderlaufen. Dadurch können die Wellenzahlen zeitlich nacheinander den Kanälen eines Vielkanalanalysators zugewiesen werden [35]. Die Probe P kann, je nach Orientierung der Netzebenen, in räumlich verschiedene Detektoren reflektieren, wobei gleichzeitig auch die Transmission gemessen wird. Die Auflösung des Instruments ist durch die Öffnungszeit des ersten Zerhackers Z1 sowie die fest vorgegebene Laufstrecke gegeben und beträgt bei unseren Messungen $5 \cdot 10^{-4}$.

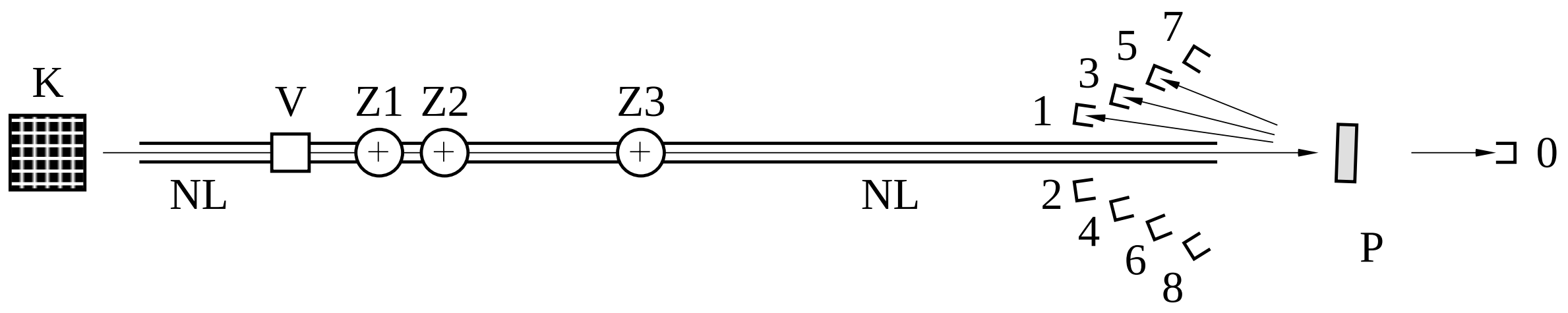

*Abbildung (71):*
*Schema des Neutronenflugzeitspektrometers in naher Rückstreuung am Forschungsreaktor München. Reaktorneutronen starten im Kern K und werden vom Zerhackersystem Z1, Z2, Z3 in wohldefinierte Pakete unterteilt, deren Wellenzahlen über eine 145 m lange Flugstrecke im Leiter NL auseinanderlaufen. Die Probe streut, je nach Orientierung in einige der Detektoren 1 - 8. Auch die Transmission kann durch den Detektor 0 gemessen werden. V ist ein Strahlverschluß.*

Typische Transmissions- und Reflektionskurven sind in Abbildung (72) gezeigt und später in Abbildung (77) ausgewertet: Das Transmissionsspektrum zeigt im wesentlichen die Wellenlängenverteilung des gepulsten Strahls. Aus diesem Kontinuum sind wieder die Braggreflexe der Probe herausgeschnitten. Direkt sieht man die Reflexe im Reflektionsspektrum, jedoch aufgrund unterschiedlicher Detektorabstände zyklisch in den Kanalnummern vertauscht. Die Normierung findet nach zyklischem Rückvertauschen der Kanalnummern durch ein parallel gemessenes Monitorspektrum statt.

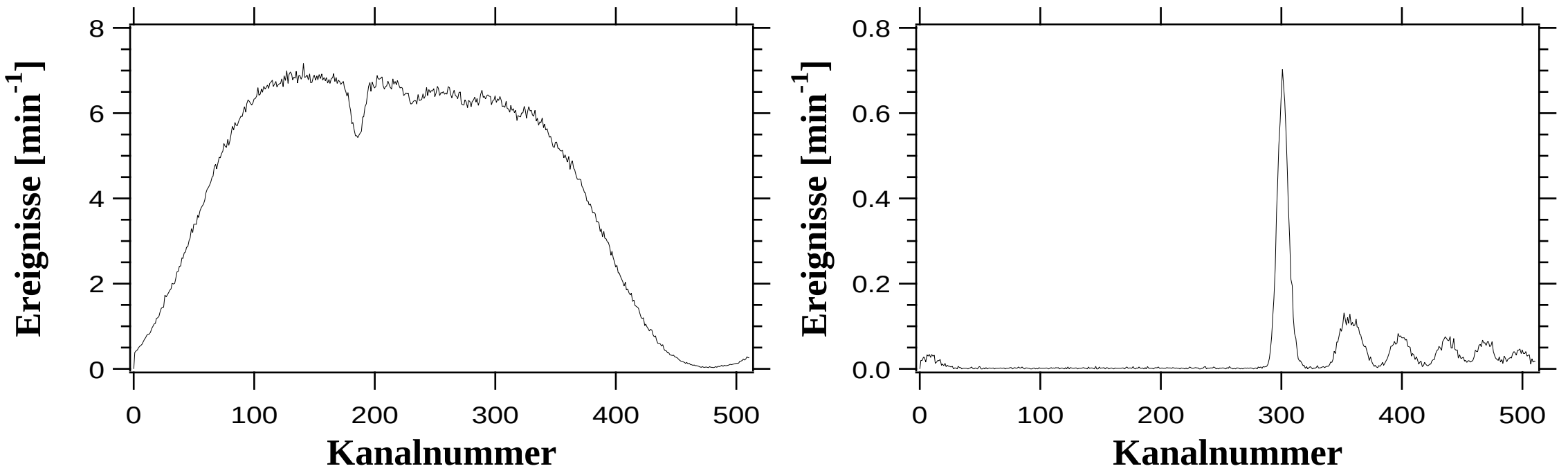

*Abbildung (72):*
*Typische, unnormierte Flugzeitspektren, wie sie direkt in Transmission, links, oder Reflektion, rechts, gemessen werden.*

### 7.1.3. Das Röntgendreikristalldiffraktometer

Neben Neutronenstrahlen wurden auch Beugungsexperimente mit geeigneten Röntgenstrahlen durchgeführt, die bei entsprechend hoher Energie ähnlich geringe Absorption erleiden. Dazu stand uns Meßzeit an einem neuentwickelten Dreikristalldiffraktometer mit 100 keV Strahlung am HASYLAB des DESY in Hamburg sowie die Dreikristallversion des X27C am NSLS des BNL in Brookhaven mit 45 keV zur Verfügung. Beide Instrumente sind vom gleichen Typ, während der angebotene Energiebereich von der Synchrotronstrahlungsquelle abhängig war. Eine schematische Darstellung, die bis auf Kleinigkeiten für beide Instrumente gültig ist, ist in Abbildung (73) wiedergegeben. Ein von der Synchrotronquelle Q ausgehender Röntgenstrahl trifft auf einen perfekten Siliziummonochromator M, der in Lauestellung, eine wohldefinierte Wellenzahl in Betrag und Richtung selektiert. In Braggbedingung, streut die Probe P diese Welle auf den perfekten Siliziumanalysatorkristall A, der seinerseits, ebenfalls abgestimmt, in den Detektor D reflektiert. Um eine dispersionsfreie Anordnung zu erhalten wählt man einander angepaßte Streuvektoren gleichen Betrags an Monochromator, Probe und Analysator und betreibt die Anlage in der (+–+) Geometrie. Somit wird die Richtungswellenzahlkorrelation bei der sukzessiven Reflektion an den Kristallen bestmöglich ausgenutzt und die Auflösung nur durch die natürliche Linienbreite des Idealkristalls beschränkt. Eine sehr ausführliche Studie hierzu wurde von dem Mitarbeiter H.-B. Neumann aus der HASYLAB-Gruppe erstellt [36].

Das Streudiagramm ist in Abbildung (74) dargestellt: Monochromator und Analysator halten den auf die Probe einfallenden Wellenvektor $\vec{k}_i$ bzw. den zu beobachtenden Wellenvektor $\vec{k}_f$ eindeutig fest. Enthält die Probe, etwa in Braggbedingung, einen Streuvektor $\vec{G} = \vec{k}_f - \vec{k}_i$, so wird Intensität in den Detektor D gestreut. Die Verteilung der reziproken Gittervektoren kann nun durch zwei Arten linear unabhängiger Variationen abgetastet werden: Zum einen dreht man die Probe um den Winkel ω, womit man transversal zum Streuvektor fährt. Offensichtlich gilt

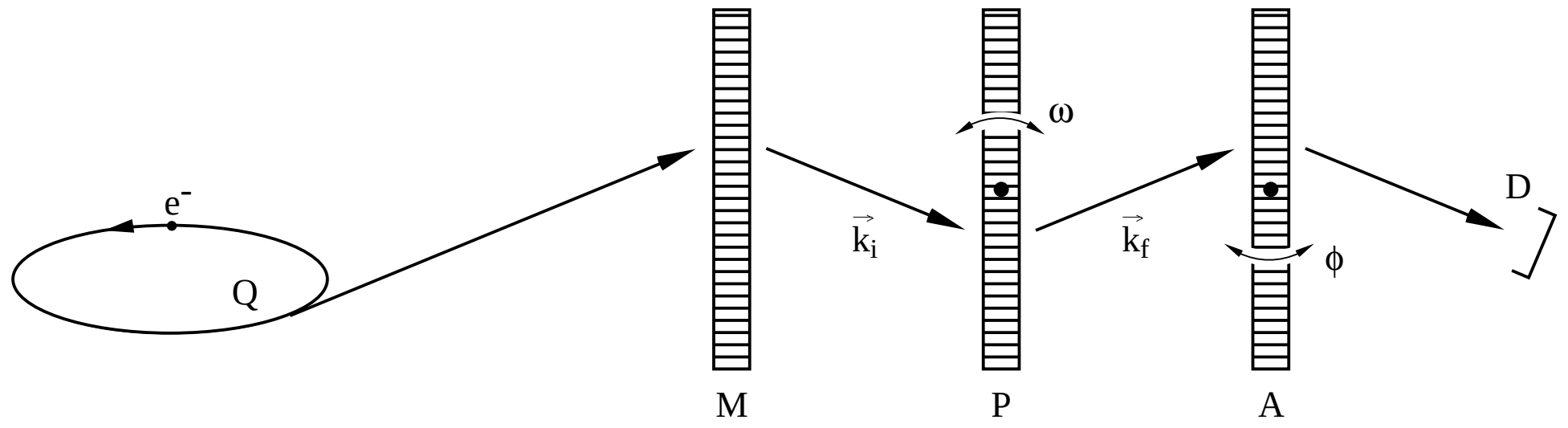

*Abbildung (73):*
*Schematische Darstellung des Röntgendreikristalldiffraktometers. Strahlen von der Synchrotronquelle Q werden durch M monochromatisiert, in P gestreut, in A analysiert und mit dem weitgeöffneten Detektor D registriert. Die Spektren können als Funktion der Drehwinkel ω und ϕ gemessen werden, wobei unterschiedliche Richtungen im reziproken Raum abgetastet werden.*

$$\frac{\Delta G_\perp}{G} = \omega\,. \qquad (213)$$

Dies liefert die Mosaikverteilung bei einem festen, longitudinalen Wert. Andererseits kann man den analysierenden Wellenvektor $\vec{k}_f$ durch Rotation des Analysators um $\phi$ verdrehen, wobei man sich auf einer um den Braggwinkel $\theta$ zur longitudinalen Richtung geneigten Linie bewegt. Hierbei variiert man Longitudinal- als auch Transversalkomponente, und wir erhalten bei festgehaltenem $\omega$:

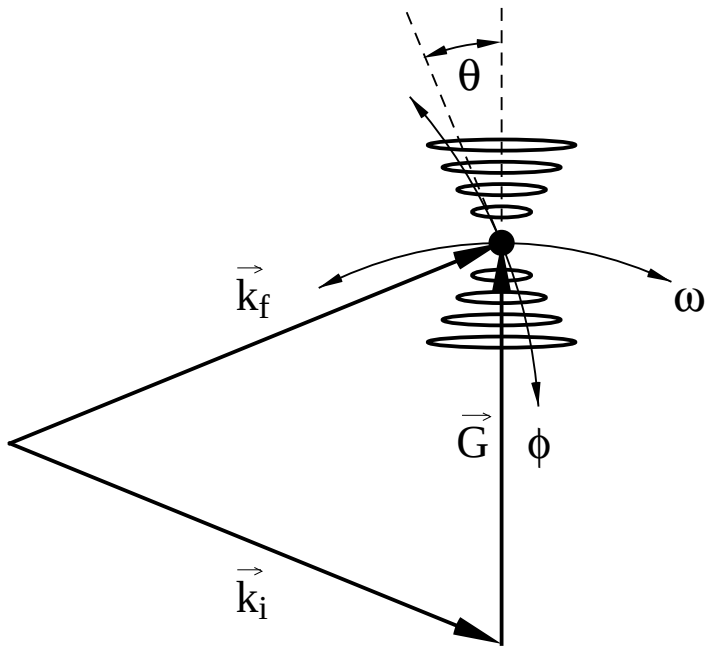

*Abbildung (74):*
*Streudreieck im reziproken Raum. Durch $\vec{k}_i$ und $\vec{k}_f$ wird der Streuvektor $\vec{G}$ analysiert. Je nach Rotation der Probe ($\omega$) oder des Analysators ($\phi$) bewegt man sich auf verschiedenen Linien durch die Streuvektorverteilung. Durch Kombination beider kann die zweidimensionale Streuebene abgetastet werden.*

$$\frac{\Delta G_\parallel}{G} = \frac{k}{G}\cos(\theta)\,\phi = \frac{\phi}{2\tan(\theta)} \qquad (214)$$

und

$$\frac{\Delta G_\perp}{G} = \frac{k}{G}\sin(\theta)\,\phi = \frac{\phi}{2}\,. \qquad (215)$$

Für das Verhältnis ergibt sich

$$\frac{\Delta G_\perp}{\Delta G_\parallel} = \tan(\theta)\,, \qquad (216)$$

wobei, um ein schnelles Bild zu erhalten, bei kleinen Streuwinkeln, wie hier um die 2° und 4°, die transversale Komponente vernachlässigt werden kann.

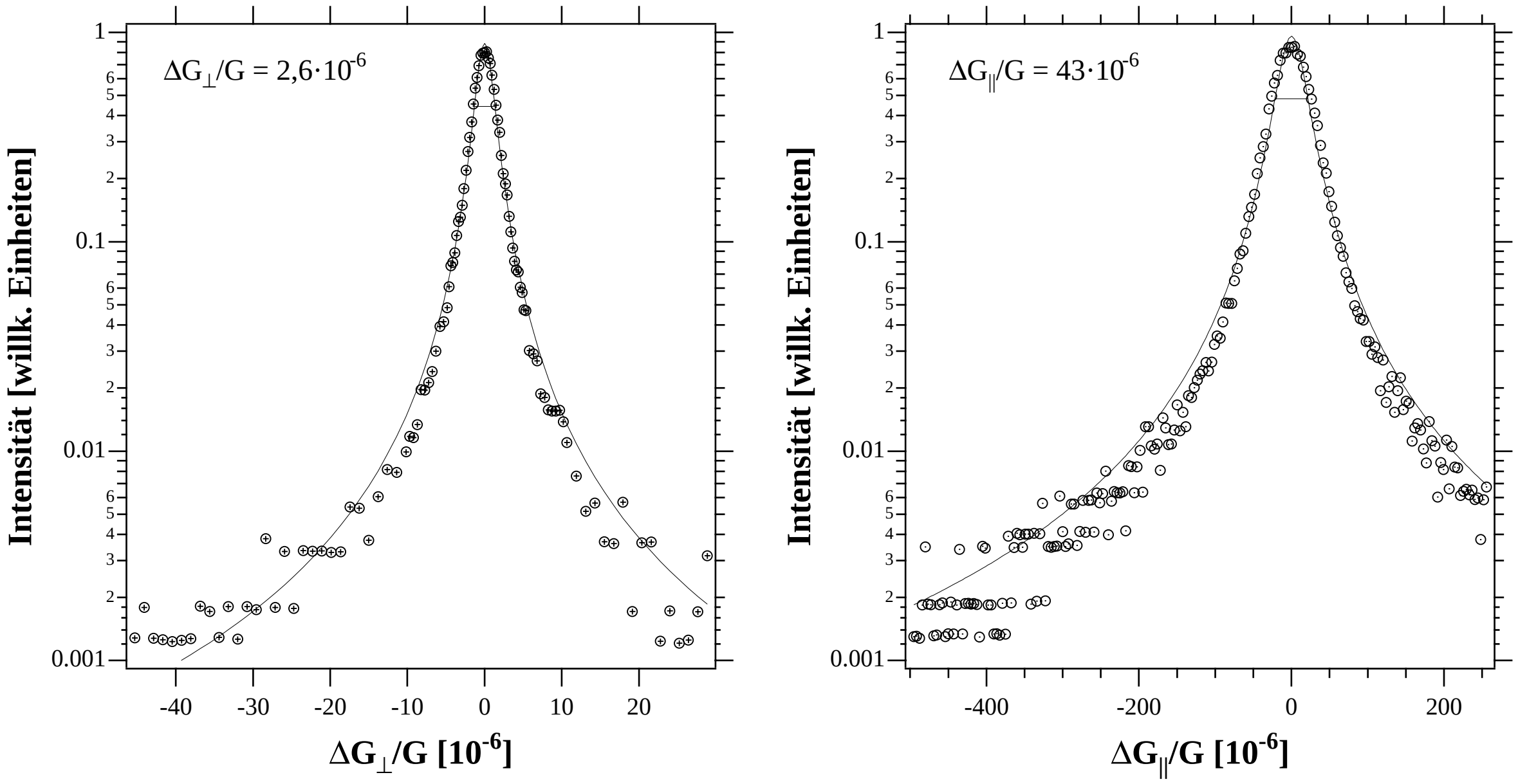

*Abbildung (75):*
*Transversal und longitudinal gemessene Auflösungsfunktionen des Dreikristalldiffraktometers am HASYLAB*

Die Auflösungen werden durch die natürlichen Linienbreiten der Braggreflexe im Rahmen der dynamischen Streutheorie von Monochromator und Analysator gegeben und sind zusammenfassend in Tabelle (10) aufgelistet. Die gemessenen Werte mit einem Idealkristall an der Probenposition sind in Abbildung (75) gezeigt. Sie unterscheiden sich von den theoretischen Größen nur um einen Faltungsfaktor.

### 7.1.4. Zusammenstellung der Diffraktometerdaten

Eine Zusammenstellung der verwendeten Streudaten und Auflösungen ist in Tabelle (10) wiedergegeben. Die aufgrund der Darwinbreite berechneten Halbwertsbreiten sind ungefaltete Werte, das heißt, um sie mit den Meßwerten zu vergleichen, müssen sie mit einem Faltungsfaktor ≥ 1,4 multipliziert werden. Die experimentellen Auflösungen wurden mit Idealkristallen am Probenort bestimmt. Das HASYLAB-Spektrometer entspricht somit genau den theoretischen Größen. Am NSLS war die Transversalkomponente durch die grobe Schrittweite von 0,001° des Goniometers limitiert. Dies liefert einen vergrößerten Wert für die transversale Auflösung, der nicht durch eine etwaige Mosaizität des Idealkristalls interpretiert werden darf. Die Rückstreuspektrometer haben theoretisch eine sehr schöne, jedoch nur longitudinale Auflösung, die am IN10 zugunsten des Flusses und am FRM aufgrund des Flugzeitmonochromators beschränkt waren. Rein transversal messen die Gammadiffraktometer, deren Auflösungen noch nicht durch ihre natürliche Linienbreite, sondern durch die Quellengröße und das Blendensystem gegeben sind.

Während Rückstreu- und Gammadiffraktometer in komplementären Richtungen messen und jeweils über die andere Komponente integrieren, erhält man auf dem Dreikristalldiffraktometer beide Komponenten gleichzeitig, und das mit ähnlicher oder besserer Auflösung. Zudem wird der hohe Aufwand des Abrasterns des reziproken Raumes aufgrund der hohen Intensität der Synchrotronquelle wieder wettgemacht, so daß sich die Gesamtmeßzeit nicht verlängert.

Die Dreikristalldiffraktometrie gibt guten Aufschluß über die Streuvektorverteilung aufgrund Mosaizität, Germaniumkonzentrationen oder innerer Spannungen. Mit Hinblick auf einen Neutronenmonochromator spiegelt jedoch nur die Neutronenstreuung die richtigen Intensitätsverteilungen wider.

Nicht zuletzt können auch die unterschiedlichen Strahlquerschnitte zum Vergleich von punktförmigen mit flächendeckenden Analysen herangezogen werden.

| | NSLS | HASYLAB | γ-Au | γ-Cs | IN10 | FRM |
|---|---|---|---|---|---|---|
| Strahlung | Röntgen | Röntgen | Röntgen | Röntgen | Neutronen | Neutronen |
| E | 45 keV | 100 keV | 411,8 keV | 661,7 keV | 2080 μeV | 2080 μeV |
| Si-Reflex | 220 | 220 | 220 | 220 | 111 | 111 |
| G [$Å^{-1}$] | 3,272 | 3,272 | 3,272 | 3,272 | 2,004 | 2,004 |
| k [$Å^{-1}$] | 22,80 | 50,68 | 208,7 | 335,3 | 1,002 | 1,002 |
| θ [°] | 4,114 | 1,850 | 0,4492 | 0,2796 | 90 | 90 |
| F | 8 | 8 | 8 | 8 | $4\sqrt{2}$ | $4\sqrt{2}$ |
| f | 8,706 | 8,706 | 8,706 | 8,706 | 1 | 1 |
| $b_c$ [fm] | 2,818 | 2,818 | 2,818 | 2,818 | 4,149 | 4,149 |
| \|V(G)/E\| | $2,961\cdot10^{-7}$ | $5,995\cdot10^{-8}$ | $3,535\cdot10^{-9}$ | $1,369\cdot10^{-9}$ | $1,834\cdot10^{-5}$ | $1,834\cdot10^{-5}$ |
| Δ [μm] | 92,83 | 206,7 | 851,6 | 1368 | 34,19 | 34,19 |
| Δθ [“] | $8,533\cdot10^{-1}$ | $3,832\cdot10^{-1}$ | $9,301\cdot10^{-2}$ | $5,788\cdot10^{-2}$ | | |
| $\Delta G_{\perp}/G$ theo | $4,137\cdot10^{-6}$ | $1,858\cdot10^{-6}$ | $4,509\cdot10^{-7}$ | $2,806\cdot10^{-7}$ | | |
| $\Delta G_{\parallel}/G$ theo | $2,876\cdot10^{-5}$ | $2,876\cdot10^{-5}$ | | | $2,118\cdot10^{-5}$ | $2,118\cdot10^{-5}$ |
| $\Delta G_{\perp}/G$ exp | $1,7\cdot10^{-5}$ | $2,6\cdot10^{-6}$ | $1,2\cdot10^{-4}$ | $4,4\cdot10^{-4}$ | | |
| $\Delta G_{\parallel}/G$ exp | $4,3\cdot10^{-5}$ | $4,3\cdot10^{-5}$ | | | $7,5\cdot10^{-5}$ | $5\cdot10^{-4}$ |
| exp/theo ⊥ | 4,2 | 1,4 | 270 | 1600 | | |
| exp/theo \|\| | 1,5 | 1,5 | | | 3,5 | 24 |
| B [$mm^2$] | 0,8·1,5 | 5·5 | 1·10 | 1·10 | 10·10 | 30·30 |

*Tabelle (10):*
*Zusammenstellung der Streudaten und Auflösungen der verwendeten Diffraktometer. NSLS, HASYLAB: Dreikristalldiffraktometer, γ-Au, γ-Cs: Gammadiffraktometer am ILL, IN10: Rückstreuspektrometer am ILL, FRM: Flugzeitrückstreuspektrometer.*
*E: Energie, G: Streuvektor, k: Wellenzahl, θ: Braggwinkel, F: Strukturfaktor, f: Formfaktor, $b_c$: kohärente Streulänge, V(G)/E: Kristallpotential (4π el. Suszeptibilität), Δ: Pendellösungsperiode, ΔG/G: ungefaltete, theoretische Auflösungen, aufgrund der Darwinbreiten, experimentell gemessene Auflösungen und Verhältnisse, B: Strahldimensionen parallel und senkrecht zur Streuebene.*

## 7.2. Diffraktionsergebnisse

Durch die dispersionslose, hochenergetische Röntgendreikristalldiffraktometrie können hochaufgelöste, zweidimensionale Ausschnitte aus der Streuebene im reziproken Raum abgetastet werden. Abbildung (76) zeigt derartige Ergebnisse in der Umgebung des Si $[02\bar{2}]$ Reflexes verschiedener Proben im gleichen Maßstab. Die Bildkoordinaten wurden direkt aus dem Probenwinkel $\omega$ horizontal und dem Analysatorwinkel $\phi$ vertikal zusammengestellt. Dadurch ergibt sich mit den Gleichungen (214) und (215) ein schiefwinkliges Koordinatensystem des reziproken Raumes, dessen Achsen durch gestrichelte Linien eingezeichnet sind. Die Streuintensitäten sind logarithmisch durch den angegebenen Schlüssel kodiert und nicht absolut normiert. Die Teilbilder 28 C und 78 A zeigen die Ergebnisse an Stufenkristallen, die anderen beiden, 19 C und 94 A, die von Gradientenkristallen. Allen Teilbildern gemeinsam ist das vom dicken Substrat herrührende, sehr intensive Hauptmaximum bei (0, 0). Bei den Stufenkristallen treten weitere Braggmaxima der einzelnen Schichten und bei den Gradientenkristallen je ein ausgeprägtes Plateau in longitudinaler Richtung auf. Die zugehörenden Schichtdicken sind dünn gegen die Pendellösungsperiode, so daß ihre Intensitäten entsprechend schwach ausfallen. Die gegenüber dem Substrat verkürzten Streuvektoren dieser Nebenreflexe entsprechen, wie erwartet, im realen Raum durch die Germaniumbeimischung vergrößerten Gitterkonstanten. In der transversalen Richtung findet mit zunehmender, longitudinaler Abweichung vom Substratreflex eine zunehmende Verbreiterung der Reflexe aufgrund anwachsender Mosaizität statt. Die Bilder zeigen in der Vertikalen, also um den Braggwinkel gegen die longitudinale Richtung geneigt, einen durchgehenden Streifen erhöhter Intensität, der sich als Artefakt der Auflösungsfunktion, nämlich als lorentzförmigen Ausläufer des intensiven Substratreflexes, herausstellt.

Im folgenden sollen anhand von Projektionen der Intensität auf die longitudinale Richtung und ausgewählten Schnitten parallel zu den Koordinatenachsen die wesentlichen Resultate vorgestellt werden.

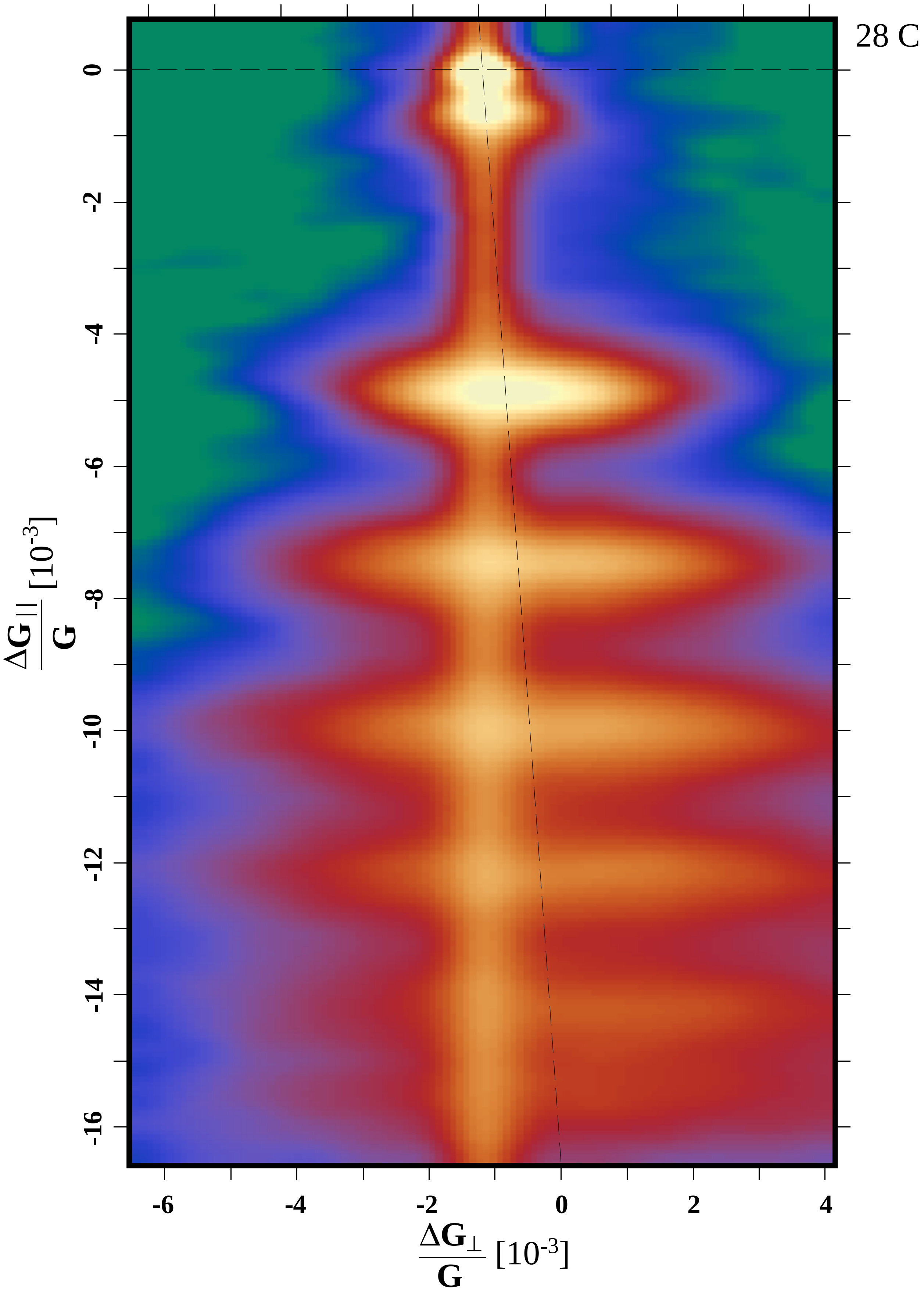

*Abbildung (76a):*
*Siehe nächste Seite.*

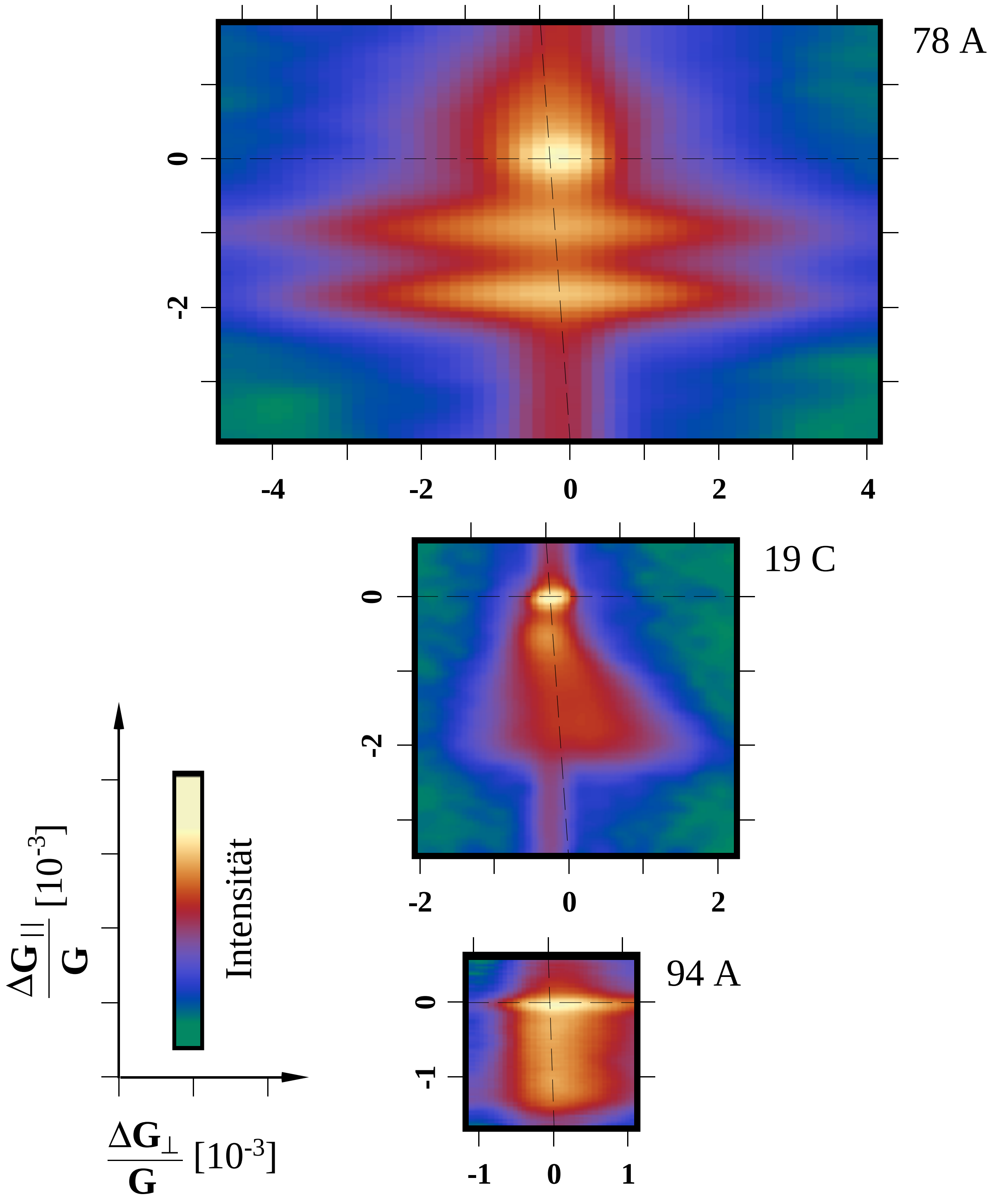

*Abbildung (76b):*
*Zweidimensionale Abrasterung der reziproken Streuebene mit den Röntgendreikristalldiffraktometern. Die Ausschnitte liegen um die* $02\bar{2}$ *Reflexe einiger Stufen und Gradientenkristalle. Der Maßstab ist für alle vier Teilbilder identisch. Die gestrichelten Linien geben das um den Braggwinkel verscherte Koordinatensystem der longitudinalen und transversalen Richtungen wieder. Am Ursprung liegt jeweils der intensive Substratreflex, während sich zu kürzeren Streuvektoren die Beiträge der* $Si_{1-x}Ge_x$ *Legierungen erstrecken. Die transversale Verbreiterung erleidet hauptsächlich in den ersten Schichten einen kräftigen Sprung, während sie in zunehmender Entfernung vom Substrat bedeutend weniger zunimmt. Die Intensitätsverteilungen können transversal integriert werden und sind in den nachfolgenden Abbildungen zusammengestellt.*

### 7.2.1. Die Meßergebnisse an Stufenkristallen

Die Stufenkristalle haben den Vorteil einer quantitativen Auswertung der aus den Zuchtparametern hervorgegangenen kristallinen Eigenschaften. Jede Schicht unterschiedlicher Konzentration weist einen individuellen Braggreflex auf, deren Positionen, longitudinale und transversale Breiten und integrierte Reflektionsvermögen insgesamt mehr Meßparameter liefern als bei Gradientenkristallen.

Bild (77) zeigt im oberen Teilbild die über die Transversalkomponente integrierte Intensität des 45 keV Spektrums zu 28 C aus Bild (76). Man erkennt unter den acht Maxima den intensiven Substratreflex, dicht daneben einen schmalen, also durch gute Kristallqualität ausgezeichneten Legierungsreflex sowie sechs weitere $Si_{1-x}Ge_x$ Reflexe in größerer Entfernung. Die Meßkurve kann gut durch die durchgezogene Linie, die Summe von acht Lorentzkurven beschrieben werden. Aus den numerisch angepaßten Parametern gehen die Maximalpositionen sowie die integrierten Reflektivitäten der individuellen Reflexe einer jeden Schicht hervor und lassen sich in den Tabellen (11) und (12) mit weiteren Daten vergleichen.

Das untere Teilbild zeigt die mit der Flugzeitanalyse am Si [111] Reflex gewonnene Reflektionskurve desselben Kristalls. Hier lassen sich die Meßwerte eher durch eine Gaußverteilung pro Maximum anpassen, da die Auflösungsfunktion durch den Strahlzerhacker bestimmt ist und einer Dreiecksfunktion ähnelt. Das erste Nebenmaximum wird nicht mehr aufgelöst, während die anderen deutlich voneinander getrennt sind. Das Spektrum wurde auch in Transmission gemessen, womit die Ergebnisse absolut normiert werden konnten. Die erhaltenen Werte sind ebenfalls in den Tabellen (11) und (12) eingetragen.

Aus den durch Mikrosondenanalyse gewonnenen Schichtdicken D wurden durch Gleichung (65) die auf das Kristallpotential normierten Dicken A der dynamischen Beugungstheorie sowie durch (95) die integrierten Reflektivitäten $\Re_y$ berechnet. Da die Einzelschichtdicken zu den Nebenreflexen wesentlich kleiner als die Pendellösungsperiode sind, wurden die experimentellen Werte $\Re_y^{exp}$ durch die kinematische Beziehung $\Re_y = \pi\ A_{kin}$ auf die Dicken $A_{kin}$ und $D_{kin}$ beider Einheitensysteme zurückgerechnet. Die experimentellen Ergebnisse für die Nebenmaxima liegen in derselben Größenordnung, aber systematisch etwas unterhalb den theoretischen. Da der reflektierte Strahl über drei Detektoren summiert wurde, mag dieser Intensitätsverlust auf die nicht alles erfassende Detektoranordnung zurückzuführen sein. Die Intensitätsauswertung erscheint somit äußerst zufriedenstellend und spiegelt die erwarteten Werte der Theorie wider. Der Substratreflex hingegen fällt aufgrund von Spannungs- und Verzerrungsfeldern dreimal stärker aus als beim Idealkristall. Die Röntgenmessungen konnten nicht absolut ausgewertet werden, weshalb diese Ergebnisse nur relativ mit den Dicken zu vergleichen sind.

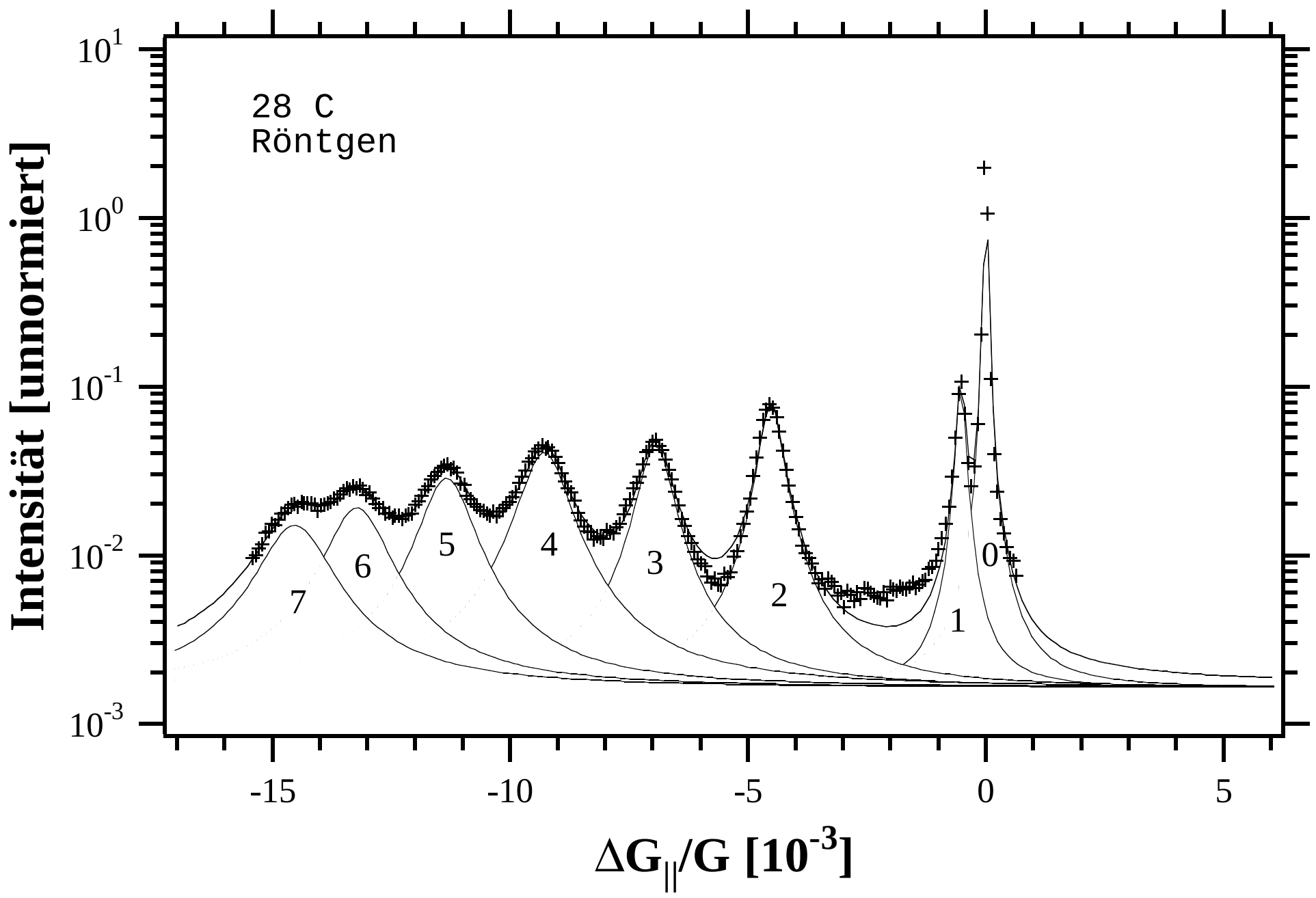

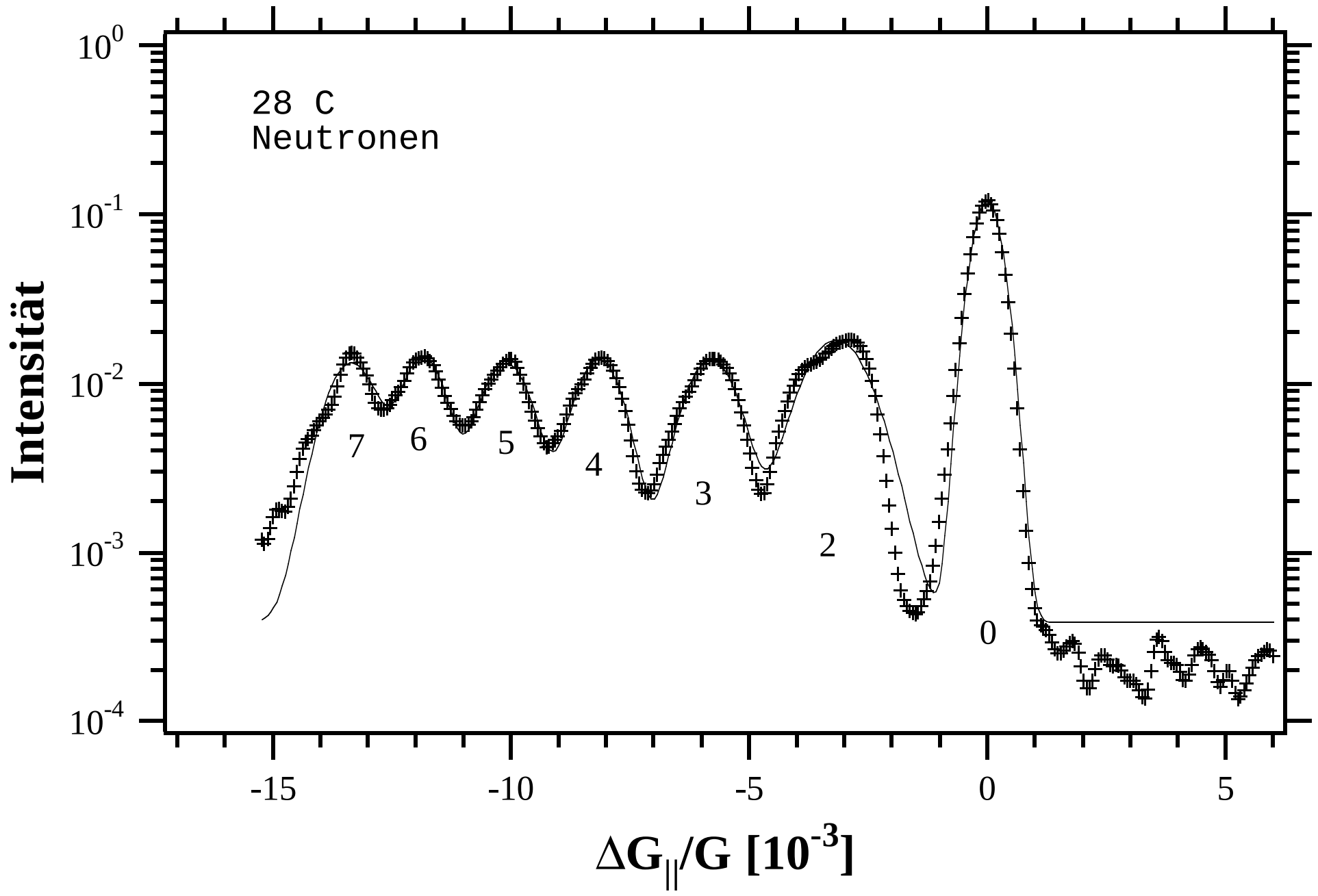

*Abbildung (77):*
*Das obere Teilbild zeigt das longitudinale Spektrum nach transversaler Integration der in Abbildung (76) zu 28 C dargestellten Intensitäten. Die acht Reflexe sind durchgehend vom Substrat mit 0 beginnend durchnumeriert und entsprechen den verschiedenen Germaniumkonzentrationen der $Si_{1-x}Ge_x$ Schichten. Die Neutronendaten im unteren Teilbild wurden durch Flugzeitanalyse in Rückstreuung aufgenommen und sind absolut normiert. Die Reflektivitäten der Einzelreflexe entsprechen absolut den theoretischen Werten. Allerdings sind die Legierungsreflexe der Neutronendaten näher am Substratmaximum als die der Röntgenmessung. Da hier orthogonale Streuvektoren benutzt wurden, deutet dies auf eine tetragonale Verzerrung hin. Bemerkenswert ist auch die hohe Gitteraufweitung von 1,4 %!*

| Nr. | D [µm] | A | $\Re_y$ | $\Re_y^{exp}$ | $A_{kin}$ | $D_{kin}$ [µm] | $\Re_X$ [willk.] |
|---|---|---|---|---|---|---|---|
| 0 | 280 | 25,7 | 3,14 | 9,1 | 2,9 | 32 | 8,5 |
| 1 | 11,6 | 1,07 | 2,48 | | | | 3,0 |
| 2 | 19,6 | 1,80 | 2,97 | 3,2 | 1,0 | 11 | 6,0 |
| 3 | 15,1 | 1,39 | 2,77 | 2,0 | 0,65 | 7,0 | 4,9 |
| 4 | 7,50 | 0,689 | 1,88 | 1,8 | 0,58 | 6,3 | 6,3 |
| 5 | 7,10 | 0,652 | 1,80 | 1,7 | 0,55 | 6,0 | 4,6 |
| 6 | 7,00 | 0,643 | 1,78 | 1,9 | 0,59 | 6,4 | 3,6 |
| 7 | 7,00 | 0,643 | 1,78 | 1,5 | 0,49 | 5,3 | 3,2 |

*Tabelle (11):*
*Auswertung der integrierten Reflektivitäten. D: Schichtdicke, $A = \pi D / \Delta$: Schichtdicke in Einheiten der Pendellösungsperiode $\Delta$, $\Re_y$, $\Re_y^{exp}$: theoretische und experimentelle integrierte Neutronenreflektivitäten, $A_{kin}$, $D_{kin}$ aus $\Re_y^{exp}$ zurückgerechnete kinematische Dicken, $\Re_X$: integrierte, experimentelle Röntgenreflektivitäten.*

| Nr. | x [%] | $x_n$ [%] | $\Delta G/G_x$ | $\Delta G/G_n$ | $\Delta G/G_X$ | $\Delta G/G_n - \Delta G/G_X$ | $\Delta h/h_T$ |
|---|---|---|---|---|---|---|---|
| 0 | 0,00 | 0,00 | 0,00 | 0,00 | 0,00 | 0,00 | 0,00 |
| 1 | | | | | $5{,}31\cdot10^{-4}$ | | |
| 2 | 8,00 | 9,44 | $2{,}88\cdot10^{-3}$ | $3{,}41\cdot10^{-3}$ | $4{,}53\cdot10^{-3}$ | $1{,}12\cdot10^{-3}$ | $2{,}61\cdot10^{-4}$ |
| 3 | 15,1 | 16,4 | $5{,}52\cdot10^{-3}$ | $5{,}98\cdot10^{-3}$ | $6{,}96\cdot10^{-3}$ | $9{,}79\cdot10^{-4}$ | $4{,}93\cdot10^{-4}$ |
| 4 | 21,3 | 22,6 | $7{,}85\cdot10^{-3}$ | $8{,}33\cdot10^{-3}$ | $9{,}30\cdot10^{-3}$ | $9{,}68\cdot10^{-4}$ | $6{,}94\cdot10^{-4}$ |
| 5 | 27,3 | 27,6 | $1{,}02\cdot10^{-2}$ | $1{,}03\cdot10^{-2}$ | $1{,}14\cdot10^{-2}$ | $1{,}07\cdot10^{-3}$ | $8{,}88\cdot10^{-4}$ |
| 6 | | 32,0 | | $1{,}20\cdot10^{-2}$ | $1{,}32\cdot10^{-2}$ | $1{,}25\cdot10^{-3}$ | $1{,}04\cdot10^{-3}$ |
| 7 | | 35,8 | | $1{,}35\cdot10^{-2}$ | $1{,}45\cdot10^{-2}$ | $1{,}03\cdot10^{-3}$ | $1{,}17\cdot10^{-3}$ |

*Tabelle (12):*
*Auswertung der Linienpositionen und tetragonalen Verzerrung. x: Ge Konzentration, $x_n$: aus den Neutronendaten zurückgerechnete Ge Konzentration, $\Delta G/G$: Gitteraufweitung aus Konzentration x, Neutronenmessung n, Röntgenmessung X, Differenz: gemessene Querkontraktion, $\Delta h/h_T$: Querkontraktion aus thermischer Ausdehnung.*

Die Linienpositionen sollen den Germaniumkonzentrationen entsprechen. Für die Neutronendaten trifft dies relativ gut zu, so daß die fehlenden Mikrosondenergebnisse aus diesen Streudaten vervollständigt wurden und bereits in Tabelle (9) und Abbildung (68) für die Interpretation der Konzentrationsverhältnisse eingegangen sind. Im Gegensatz dazu sind die Nebenreflexabstände vom Substratreflex der Röntgenmessung systematisch um $1\cdot10^{-3}$ gegenüber den Neutronendaten vergrößert. Hierbei handelt es sich jedoch nicht um einen Fehler, sondern die Tatsache einer tetragonalen Verformung des einst kubischen $Si_{1-x}Ge_x$ Gitters. Während die Röntgenmessungen an einem $02\bar{2}$ Reflex senkrecht zum Schichtennormalenvektor $\vec{n}$ stattgefunden haben, wurden die Neutronendaten am 111 Reflex parallel zu $\vec{n}$ durchgeführt.

Der gleiche Sachverhalt ist auch an den Spektren von 78 A zu erkennen, die unmittelbar nacheinander auf derselben Maschine studiert wurden: Abbildung (78) zeigt in der oberen Kurve die Intensitätsverteilung, wie sie bei einem longitudinalen Schnitt aus Abbildung (76) hervorgeht. Dabei wurde der senkrecht zum Normalenvektor $\vec{n}$ in der (100) Oberfläche liegende Streuvektor $02\bar{2}$ verwendet. Im Abstand von $\Delta G/G = 0{,}9\cdot 10^{-3}$ und $\Delta G/G = 1{,}8\cdot 10^{-3}$ vom Substratreflex finden sich die zu $Si_{0,97}Ge_{0,03}$ und $Si_{0,94}Ge_{0,06}$ gehörenden Maxima. Die darunter liegende Kurve zeigt eine entsprechende Messung an dem um 45° zu $\vec{n}$ geneigten, aus der Kristalloberfläche herausragenden 202 Reflex. Hier sind die $Si_{1-x}Ge_x$ Positionen näher am Substrat als zuvor. Auf die Oberflächennormale projiziert bedeutet dies, daß die relativen Netzebenenabstände parallel zu $\vec{n}$ gegenüber den dazu senkrechten um $0{,}1\cdot 10^{-4}$ bzw. $0{,}3\cdot 10^{-4}$ verkleinert sind, das Gitter also tetragonal verzerrt ist. Die genauen Werte hierzu lassen sich in Tabelle (13) vergleichen.

Bei der $Si_{1-x}Ge_x$ Epitaxie kann bei sehr dünnen Schichten eine tetragonale Verformung aufgrund der Fehlanpassung zwischen dem Substrat- und dem Legierungsgitter auftreten. Dabei bliebe, in Abbildung (79a) dargestellt, für die Gitterkonstanten parallel zur Oberfläche der Substratwert erhalten, während das Gitter in der Normalenrichtung frei relaxierte und damit die durch das Germanium bedingte Volumenaufweitung aufnähme und den Abstand in dieser Richtung vergrößerte. Wir beobachten jedoch eine entgegengesetzte Verzerrung. Zudem betrachten wir Schichtdicken, die um drei Größenordnungen über den kritischen Werten liegen, bei denen die genannte Verzerrung den Spannungen nicht mehr standhält und durch Einbau von Fehlanpassungsversetzungen relaxiert. Die hohe Zuchttemperatur macht das Gitter weich, so daß diese Relaxation noch begünstigt wird. Wir gehen also davon aus, daß das Gitter bei der Zuchttemperatur, wie in Bild (79b) idealisiert dargestellt, vollkommen relaxiert und die Einheitszelle bei genügender Entfernung von der Grenzschicht, wie im Substrat, kubisch ist. Natürlich treten in der Nähe der Grenzschicht Fehlanpassungsversetzungen auf, die auch, hier nicht dargestellt, aus ihr heraus ins Volumen fortschreiten können.

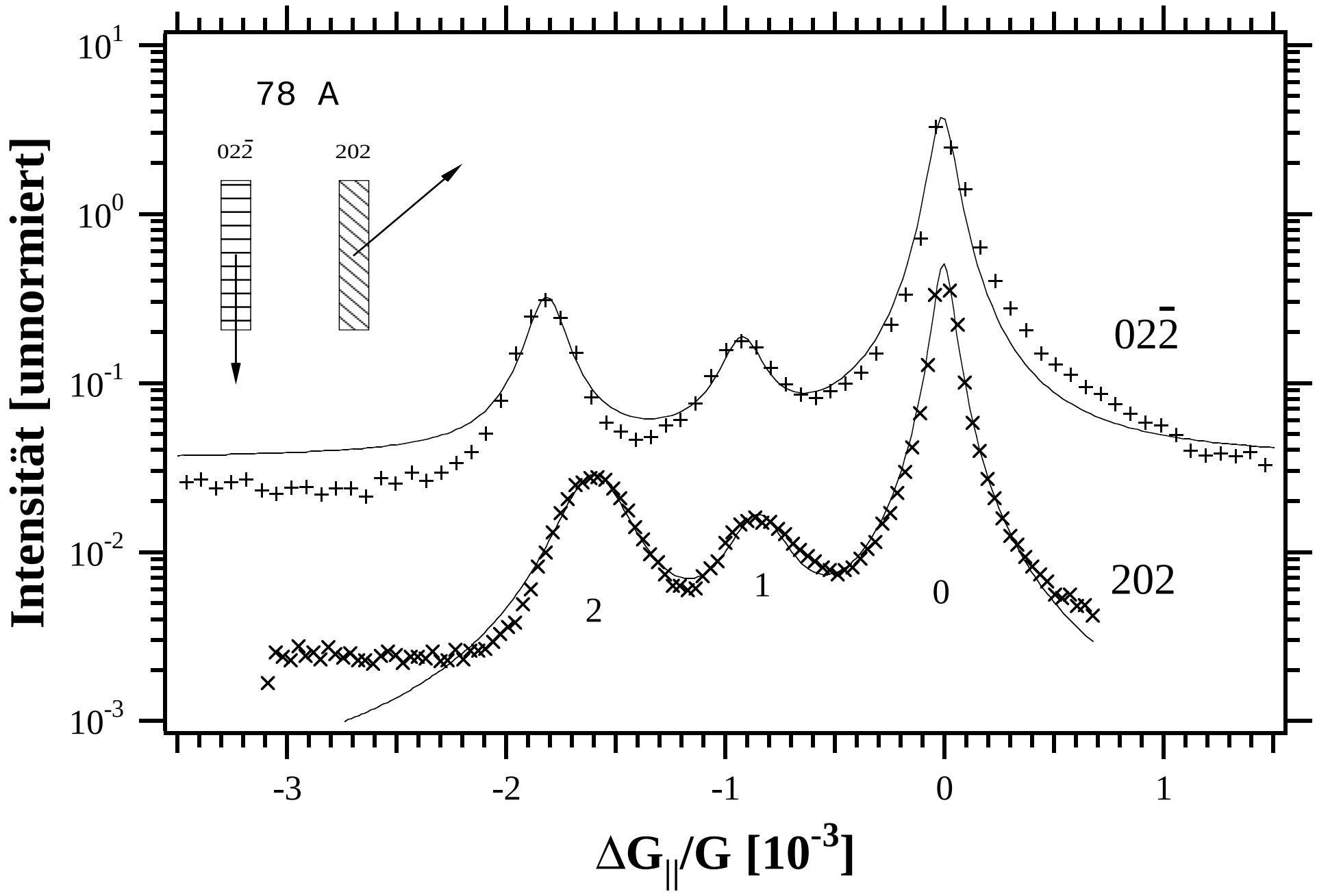

*Abbildung (78):*
*Longitudinale Reflektionskurven zweier an sich symmetrischer Reflexe des Kristalls 78 A. Die Nebenmaxima des 202 Reflexes befinden sich näher am Substratmaximum als die des $02\bar{2}$. Dies deutet auf eine tetragonale Gitterverzerrung hin und kann eindeutig durch die thermisch unterschiedliche Zusammenziehung beim Abkühlen von der Zuchttemperatur beschrieben werden. Links oben ist die Lage der verwendeten Streuvektoren relativ zur makroskopischen Kristallscheibe angedeutet.*

| Nr. | x [%] | $\Delta G/G_x$ | $\Delta G/G$ $02\bar{2}$ | $\Delta G/G$ 202 | $\Delta(\Delta G/G)$ 100 | $\Delta L/L$ | $\Delta h/h$ |
|---|---|---|---|---|---|---|---|
| 1 | 3.03 | $1.08\cdot10^{-3}$ | $9.30\cdot10^{-4}$ | $8.51\cdot10^{-4}$ | $1.11\cdot10^{-4}$ | $9.76\cdot10^{-5}$ | $8.37\cdot10^{-5}$ |
| 2 | 6.20 | $2.20\cdot10^{-3}$ | $1.83\cdot10^{-3}$ | $1.60\cdot10^{-3}$ | $3.20\cdot10^{-4}$ | $2.00\cdot10^{-4}$ | $1.71\cdot10^{-4}$ |

*Tabelle (13):*
*Meßdaten zur tetragonalen Verzerrung am* $02\bar{2}$ *und 202 Reflex, deren Projektion auf die Oberflächennormale [100] sowie thermischer Ausdehnungsunterschied $\Delta L/L$ und resultierende Querkontraktion $\Delta h/h$ in [100] Richtung.*

Kühlt man nun den Kristall von der Zuchttemperatur auf Raumtemperatur ab, so möchte sich das $Si_{1-x}Ge_x$ Gitter aufgrund des thermischen Ausdehnungskoeffizienten stärker zusammenziehen als das Silizium. Da es aber fest auf dem wesentlich dickeren Substrat haftet, zwingt letzteres dem ersten die zur Oberfläche parallele Ausdehnung auf, während die Normalkomponente wieder frei relaxieren und damit weiter schrumpfen kann. Tatsächlich ist das Gitter durch diesen Mechanismus, wie beobachtet und in Abbildung (79c) dargestellt, tetragonal gestaucht, während es bei der Verzerrung durch Fehlanpassung gedehnt würde.

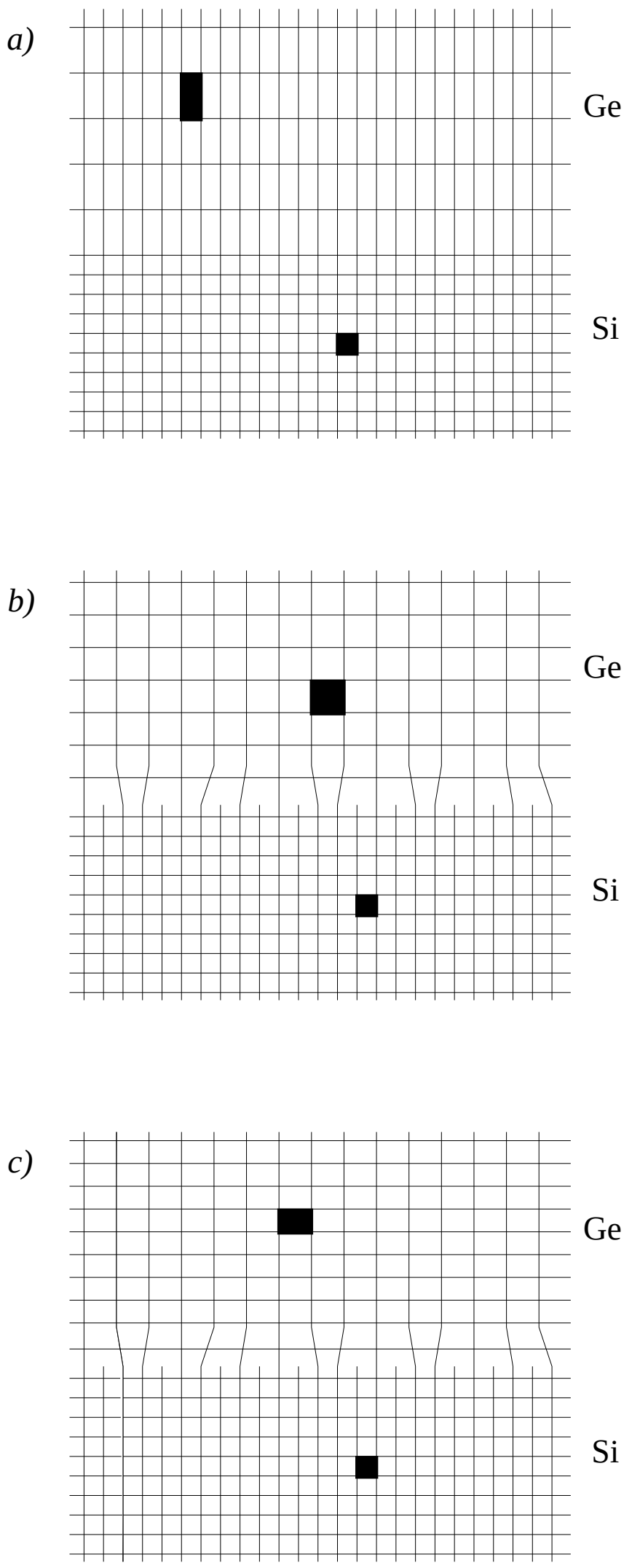

*Abbildung (79):*
*Verschiedene Möglichkeiten der Anpassung eines $Si_{1-x}Ge_x$ Gitters auf einem Siliziumsubstrat: a) epitaktisch verspannt, b) relaxiert, c) thermisch, durch Abkühlung verzerrt.*

Eine Abschätzung dieser Verformung kann durch die Elastizitätstheorie gegeben werden. Seien x, y die Koordinaten der Substratoberfläche und z in Normalenrichtung, so kann man die relativen Dehnungen

$$\varepsilon_n = \frac{\Delta n}{n} \quad ; \quad n = x, y, z \tag{217}$$

in allen drei Raumrichtungen definieren. Der Querkontraktionskoeffizient $\nu$ beschreibt die lineare Stauchung eines Materials in den übrigen Dimensionen, wenn in einer z. B. in x gestreckt wird:

$$\begin{aligned} \varepsilon_z &= -\nu\, \varepsilon_x \\ \varepsilon_y &= -\nu\, \varepsilon_x \end{aligned} \tag{218}$$

Die Dehnung unserer Schicht in der x-Richtung wirkt sich nicht nur auf z aus, sondern koppelt auch an y. Zieht man letztere Komponente wieder zurück in ihre Ausgangsposition, so verringert sich z mit einer weiteren Rückkopplung auf x um weitere $\nu \cdot \nu \cdot \varepsilon_x$ usw., und wir erhalten unter Berücksichtigung eines Faktors 2 durch gleichzeitige Dehnung in x und y Richtung und $\varepsilon = \varepsilon_x = \varepsilon_y$

$$\varepsilon_z = -2\, \nu \left\{ 1 - n + n^2 - n^3 + \cdots \right\} \varepsilon \tag{219}$$

oder, die Reihe im Bronstejn [37] nachgeschlagen

$$\varepsilon_z = -2\, \frac{\nu}{1-\nu}\, \varepsilon \, . \tag{220}$$

Im Fall der thermisch bedingten Spannung ergibt sich

$$\varepsilon_{\Delta T} = \left( \frac{\Delta L}{L} \right)_{SiGe} - \left( \frac{\Delta L}{L} \right)_{Si} \tag{221}$$

durch die unterschiedliche Längenänderungen $\Delta L/L$ bei Abkühlung von der Zuchttemperatur zur Raumtemperatur. Die Werte lassen sich in Tafelwerken [38] für Silizium und Germanium nachschlagen und wurden für die Legierungen linear interpoliert. Der Querkontraktionskoeffizient ist wohl für reines Silizium und Germanium bekannt, erhöht sich aber bei einer Legierung, und wurde mit $\nu = 0{,}3$ abgeschätzt. Die so errechneten Werte sind in den Tabellen (12) und (13) eingetragen und geben hervorragend die beobachtete Verzerrung wieder.

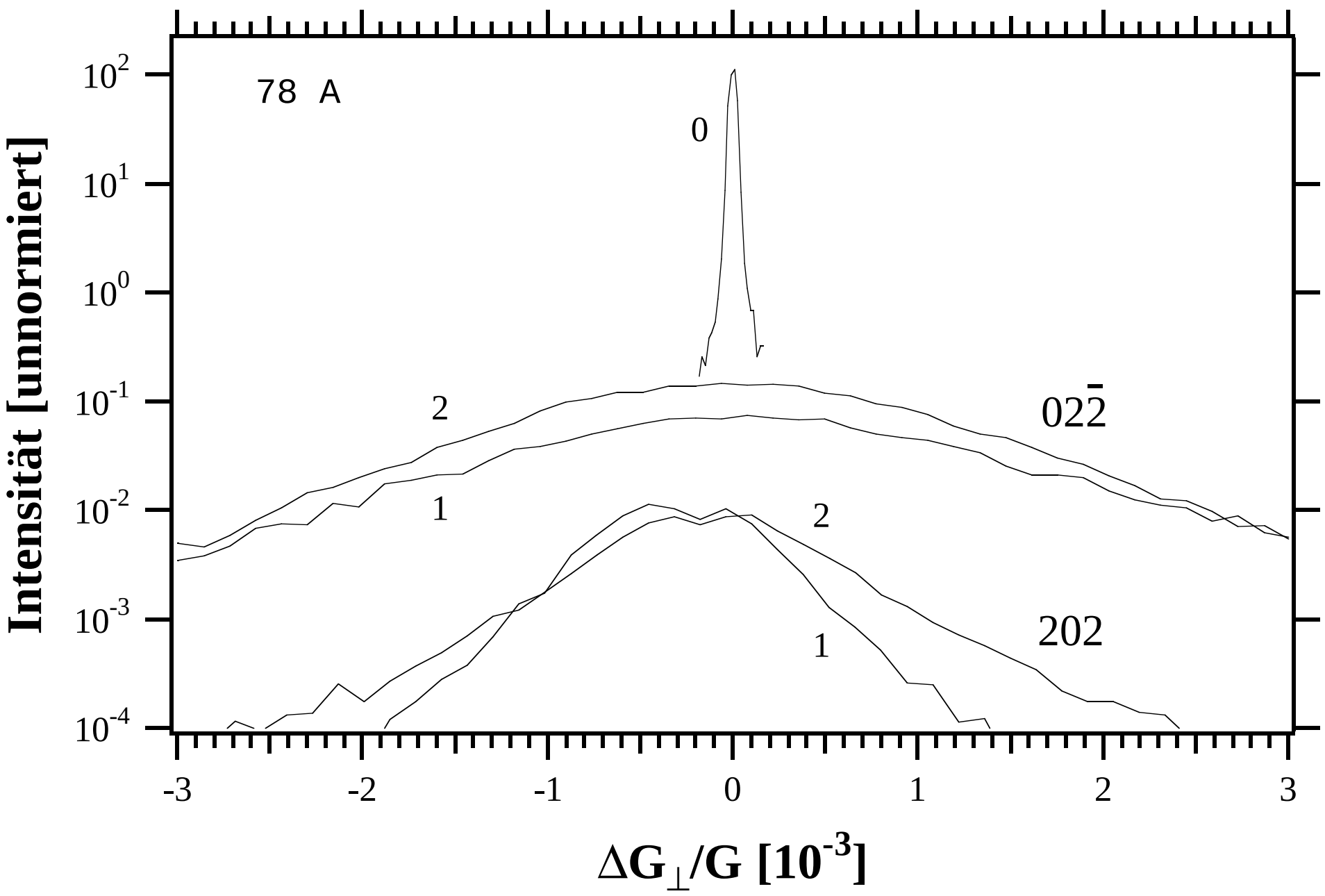

*Abbildung (80):*
*Transversale 45 keV Röntgenmessungen an 78 A. Die Legierungsreflexe sind wesentlich breiter als der Substratreflex. Die an sich symmetrischen 202 und* $02\bar{2}$ *Reflexe zeigen sehr unterschiedliche Breiten. Dies läßt auf eine vorwiegend vertikale Verbiegung der Gitterebenen schließen.*

Den ersten drei Teilbildern in Abbildung (76) ist eine transversale Verbreiterung mit zunehmender Germaniumkonzentration gemeinsam. Bei den Stufenkristallen nimmt diese Verbreiterung besonders zwischen Substrat- und erstem Nebenreflex zu, während sie sich bei großen Gradienten (19 C) linear und bei kleinen Gradienten (94 A) vernachlässigbar gegenüber der makroskopischen Krümmung verhält.

Die transversale Verbreiterung entspricht einer örtlichen Verkippung der Gitterebenen gegenüber der Vorzugsrichtung. Insbesonders werden bei der Bildung von Fehlanpassungsversetzungen, wie aus Bild (79b) oder (79c) an der Grenzschicht ersichtlich, die vertikal verlaufenden Netzebenen verbogen, während die horizontalen glatt bleiben. Genau dieser Sachverhalt wird auch in den Streudaten beobachtet und ist in Abbildung (80) dargestellt: Die Graphen des oberen Teilbildes wurden wiederum am $02\bar{2}$ von 78 A gemessen, während die des unteren den 202 Vektor benutzten. Die Nebenmaxima des $02\bar{2}$ Reflexes sind etwa 50 mal breiter als das fast ideale Substratmaximum, während die des 202 Reflexes sich nur um die Hälfte aufweiten. Aufgrund der geometrischen Betrachtung des um 45° aus der Normalen geneigten Streuvektors hätten wir eine 0,7 fache Verringerung der Mosaizität erwartet. Dies läßt tatsächlich darauf schließen, daß vorwiegend die vertikalen Gitterebenen verbogen werden, während die parallel zur Oberfläche liegenden relativ glatt bleiben.

Betrachtet man die voll ausgeprägten Intensitäten der Verbreiterung, so kann die Verbiegung nicht nur an der Grenzfläche zwischen den Legierungen auftreten, sondern müssen sich wellenförmig, immer wieder durch den Einbau etwa neuer Versetzungen oder Stapelfehler durch die ganze Schicht ausbreiten. Das erklärt auch das Ausbleiben einer weiteren Verbreiterung beim Sprung von der ersten zur zweiten Schicht, da die bereits vorhandene Versetzungsstruktur leichter eine weitere Fehlanpassung aufnehmen kann als ein Idealkristall.

Die Abbildungen (81) und (82) zeigen die longitudinalen und transversalen Intensitätsverteilungen der Streuung am $02\bar{2}$ Vektor der heißgezogenen Kristalle 17 C und 18 C. Mit dem

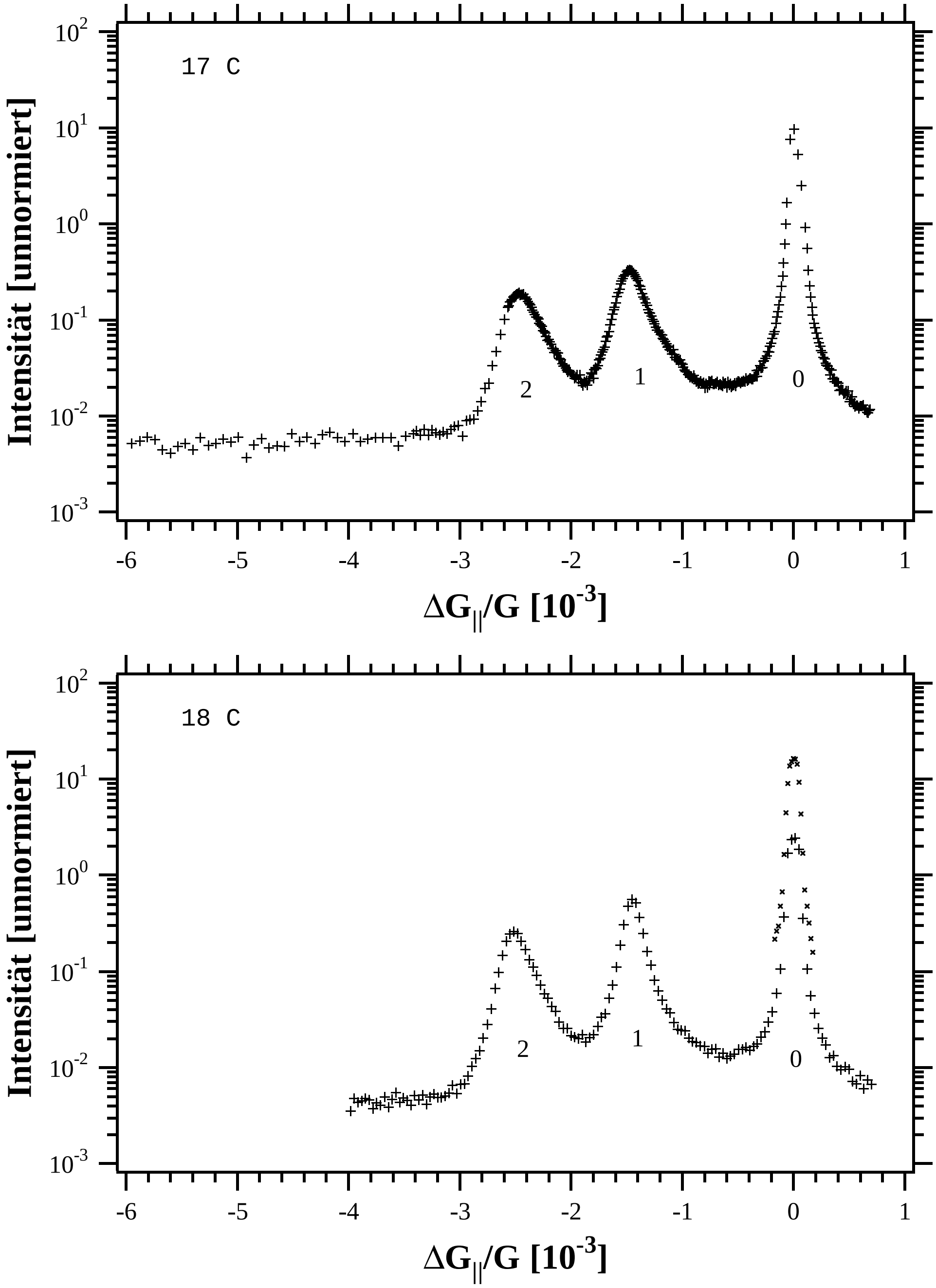

*Abbildung (81):*
*Schön ausgebildete Legierungsreflexe der Kristalle 17 C und 18 C. Die Proben wurden unmittelbar hintereinander mit gleichen Konzentrationen und verdoppelten Schichtdicken gezogen. Letzteres spiegelt sich durch die vergleichsweise erhöhte Intensität bei 18 C wieder. Die starke Asymmetrie der Nebenreflexe läßt ein Nachhinken der Gitterkonstanten nach der Germaniumkonzentration vermuten.*

Auge betrachtet erscheinen ihre Oberflächen äußerst spiegelnd und homogen, obwohl, wie früher diskutiert, unter dem Mikroskop Terrassen und ins Volumen reichende Gleitversetzungen beobachtet werden. Letztere entstehen durch thermische Spannungen im hochgeheizten, sehr weichen Gitter und reichen damit auch ins Substrat hinein. Dies erklärt, daß bereits die transversalen Substratmaxima verbreitert sind und eine nichtgaußförmige Verteilung annehmen, die auch den Nebenmaxima aufgeprägt ist.

In longitudinaler Richtung sind schöne Nebenreflexe ausgebildet. Betrachtet man sie genauer,

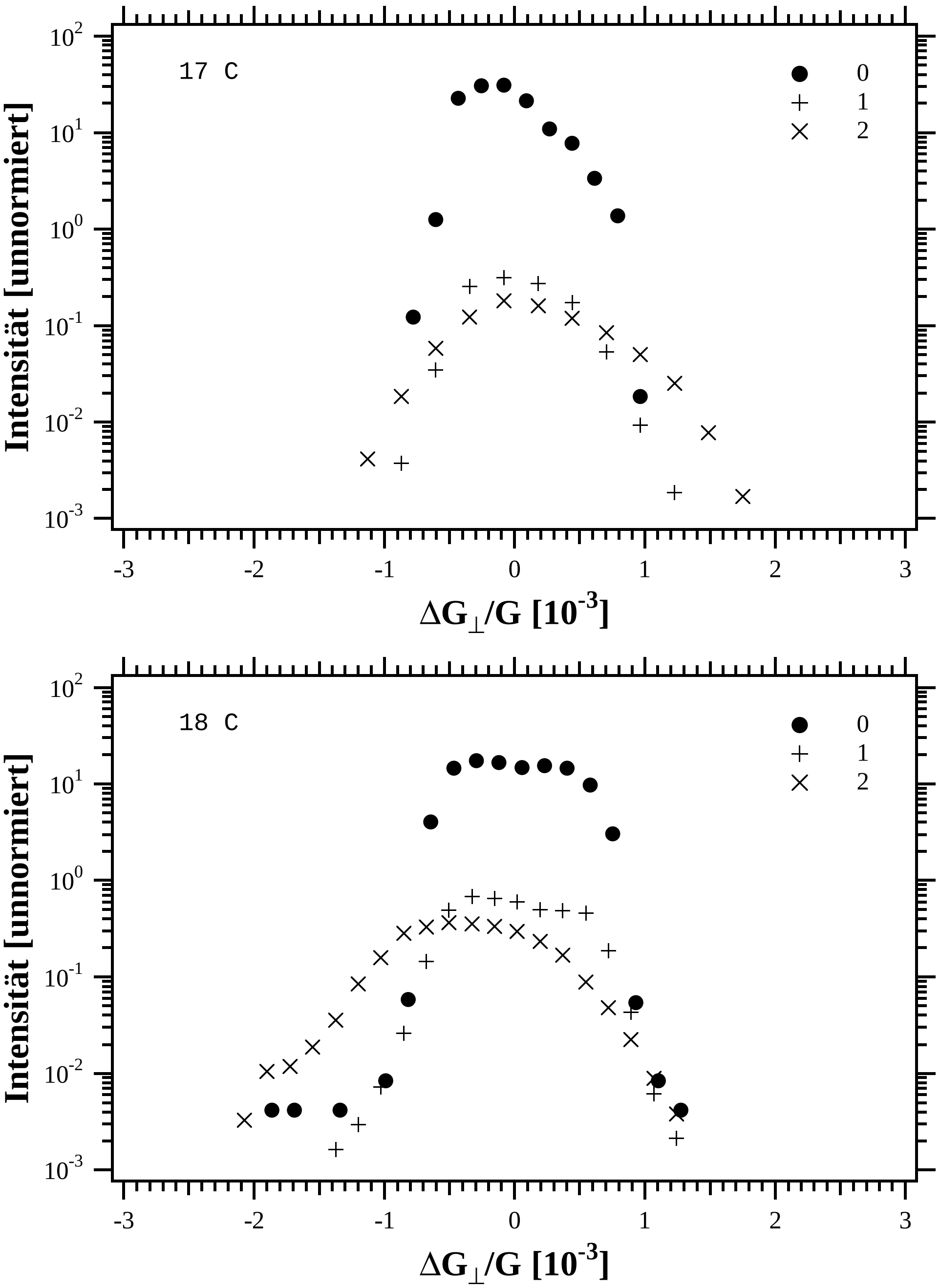

*Abbildung (82):*
*Transversale Intensitätsverteilungen an den Proben 17 C und 18 C. Hier war die Zuchttemperatur so hoch, daß auch das Substrat weich wurde und Versetzungen bekam. Damit ist dessen Reflektionskurve schon sehr breit. Eine zusätzliche Verbreiterung der Nebenmaxima bleibt unwesentlich.*

so besitzen sie eine starke Asymmetrie zugunsten längerer Streuvektoren, also kürzerer Gitterkonstanten. Auch zwischen den Maxima findet sich eine, gegenüber dem Untergrund in den Ausläufern des Spektrums deutlich erhöhte Intensität. All das deutet darauf hin, daß während der Kristallzucht die Anpassung der Gitterkonstanten der Germaniumkonzentration nachhinkt. Wir gehen vom idealen Silizium, dem Substrat aus und schalten, durch den hohen Gasfluß gewährleistet, plötzlich auf eine erhöhte Germaniumkonzentration um. Das bereits aufgebaute Gitter kann ohne schwerfällige Versetzungsbewegungen nicht mehr verändert werden, wodurch die Spannungen der Fehlanpassung erst in den darüberliegenden Schichten, direkt während des Wachstums durch Versetzungen allmählich abgebaut werden und somit das Gitter langsam seinen Netzebenenabstand an den erwarteten Wert anpaßt.

### 7.2.2. Experimentelle Reflektionskurven der Gradientenkristalle

Integriert man in Abbildung (76) das zu 19 C gehörende Spektrum über die Transversalkomponente so erhalten wir die in Abbildung (83) dargestellte longitudinale Intensitätsverteilung. Die vom 280 µm dicken Substrat herrührende Reflektionskurve liegt über der Pendellösungsperiode von 93 µm der 45 keV Strahlung und kann gut durch die angepaßte, durchgezogene Lorentzkurve beschrieben werden. Links davon beobachten wir ein schön ausgebildetes, vom 65 µm dicken Gradientenkristall herstammendes Plateau. Das schwache Maximum bei $-0{,}5 \cdot 10^{-3}$ rührt von einem, durch die Diffraktionsanalyse entdeckten Fehler bei der Kristallzucht, nämlich einem Germanleck zugunsten erhöhter Gitterkonstanten her, der insbesondere bei sehr kleinen Konzentrationen ins Gewicht fällt. Da keine Germaniumkonzentrationsanalyse vorliegt, wurde die maximale Gitteraufweitung mit $\Delta G/G = 1{,}8 \cdot 10^{-3}$ aus der Plateaubreite bzw. $2{,}0 \cdot 10^{-3}$ vom Substratreflex und der daraus resultierenden Endkonzentration von x = 5,6 % aus der Meßkurve herausgelesen. Mit diesen Parametern kann die punktiert eingezeichnete, theoretische Reflektionskurve durch die Transfermatrizenmethode errechnet werden. Da die Meßdaten in ihrer Intensität nicht absolut normierbar sind, wurde die Theoriekurve auf der Ordinate rein willkürlich unterhalb der Meßkurve durch die punktierte Linie eingetragen. Eine Faltung mit der als Auflösungsfunktion betrachteten, angepaßten Lorentzkurve des Substratmaximums liefert das durchgezogene Plateau. Hier sind die schnellsten Oszillationen herausgemittelt. Die durchgeführte Faltung erscheint wesentlich, da die Linienform, z. B. am linken Plateaurand besser mit den Meßdaten zusammenfällt als die der ungefalteten Kurve.

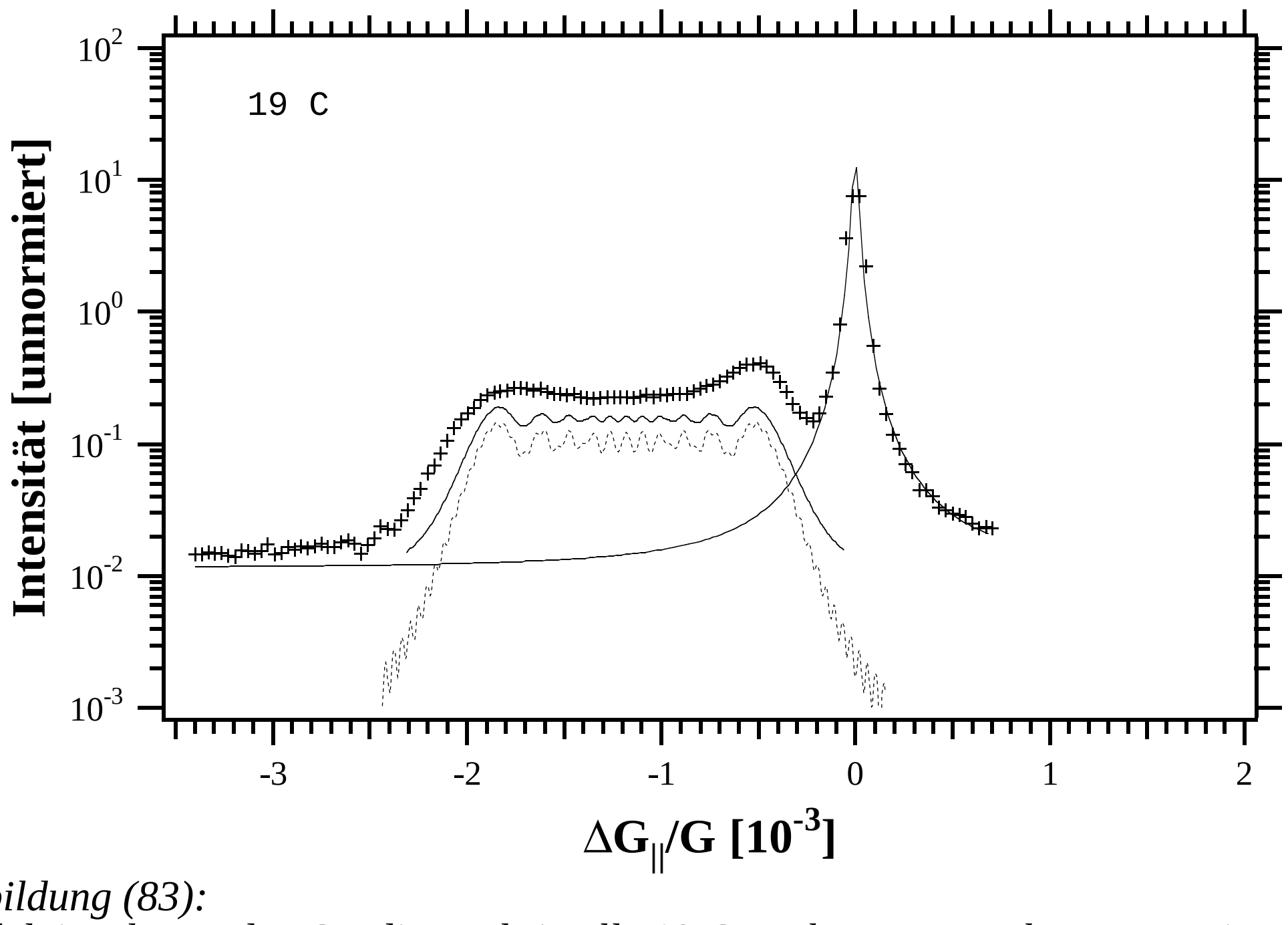

*Abbildung (83):*
*Reflektionskurve des Gradientenkristalls 19 C nach transversaler Integration der Meßdaten in Abbildung (76). Links neben dem lorentzförmig ausgebildeten Substratreflex befindet sich das vom Gradienten herrührende Plateau. Darunter sind die durch Transfermatrizen gerechneten Kurven ungefaltet (punktiert) und gefaltet auf der Ordinate willkürlich eingetragen.*

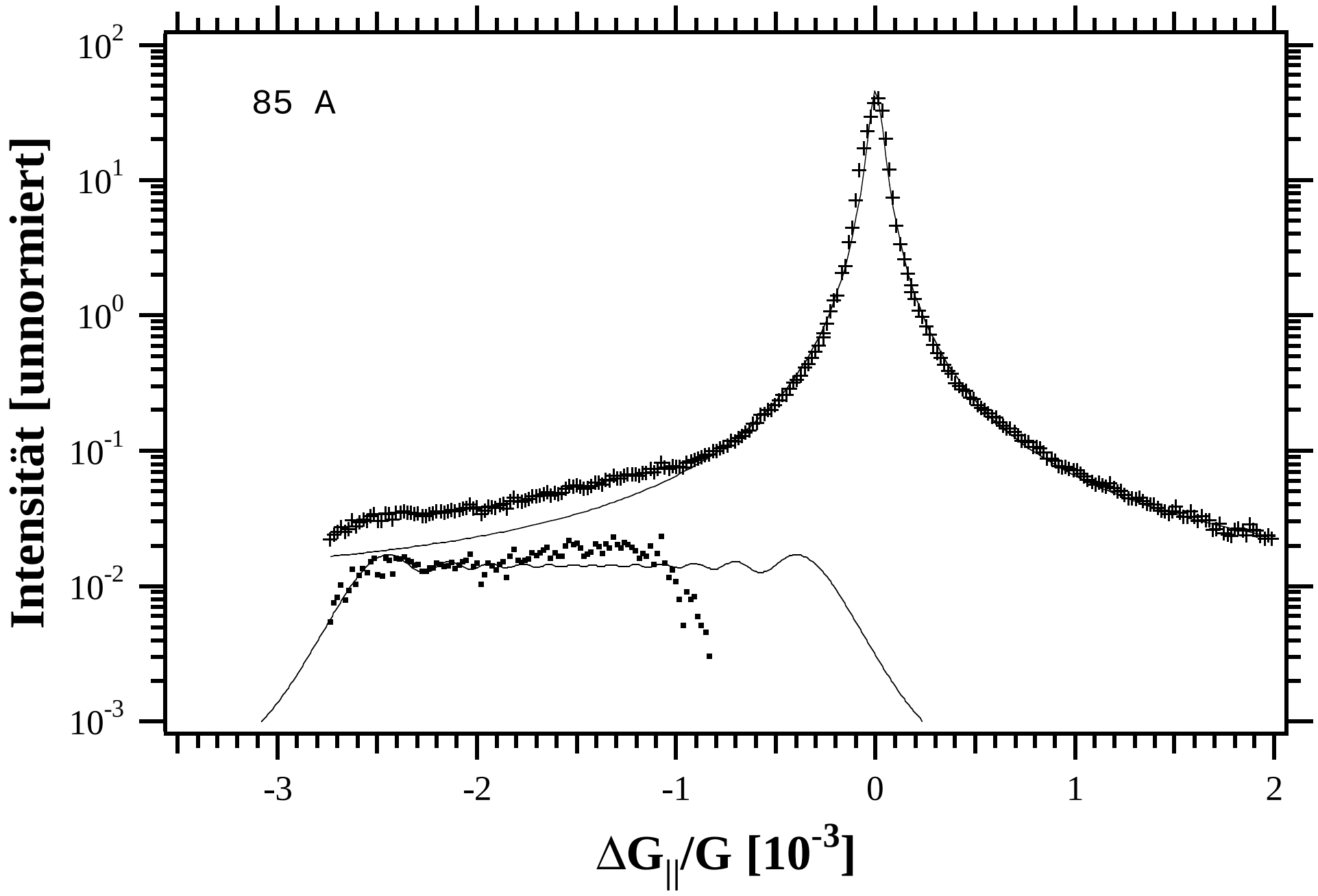

*Abbildung (84):*
*Longitudinaler Schnitt um den* $02\bar{2}$ *Reflex eines steilen Gradientenkristalls bei 100 keV Röntgenstrahlung. Der Gradientenbeitrag ist um mehr als 3 Größenordnungen schwächer als der des Substrats und macht sich nur im linken Teil des Ausläufers bemerkbar. Nach Subtraktion des lorentzförmigen Substratanteils erhält man ein flaches Plateau. Im linken Teil läßt sich die Linienform hervorragend durch die Transfermatrizentheorie voraussagen.*

Die Reflektionskurve von 85 A bei einem longitudinalen Schnitt der reziproken Streuebene durch den Substratreflex zeigt Abbildung (84). Bei der hier verwendeten 100 keV Strahlung liegt die Schichtdicke von 107 µm im Vergleich zur Pendellösungsperiode von 207 µm noch ungünstiger, so daß die vom Gradienten gestreute Intensität noch wesentlich geringer ausfällt. Zudem wurde nicht über die Transversalkomponente integriert. Nach Subtraktion der vom Substrat herrührenden Lorentzkurve erhält man wieder ein ausgeprägtes Plateau, dessen linker Rand bei einem etwas höheren als durch die Zusammensetzung erwarteten Wert von $\Delta G/G = -2{,}7^{-3}$ abfällt. Die rechte Flanke ist ein Artefakt der Subtraktion und enthält keinerlei physikalischen Inhalt. Mit der Dicke von 107 µm läßt sich wieder eine theoretische Reflektionskurve berechnen, falten und auf der Intensitätsskala übereinanderschieben. Hierbei fällt das erste Oszillationsminimum am linken Plateaurand mit einem experimentellen Intensitätsminimum zusammen. Dies ist ein starker Hinweis dafür, daß die hergeleitete, theoretische Beschreibung der Reflektionskurven durch Transfermatrizen experimentelle Ergebnisse richtig beschreibt. Obwohl die Rechensimulationen auf der Intensitätsskala wegen mangelnder Normierung angepaßt wurden, kann mit dem vorgegebenen Gradienten und der Schichtdicke die Linienform vorausgesagt werden.

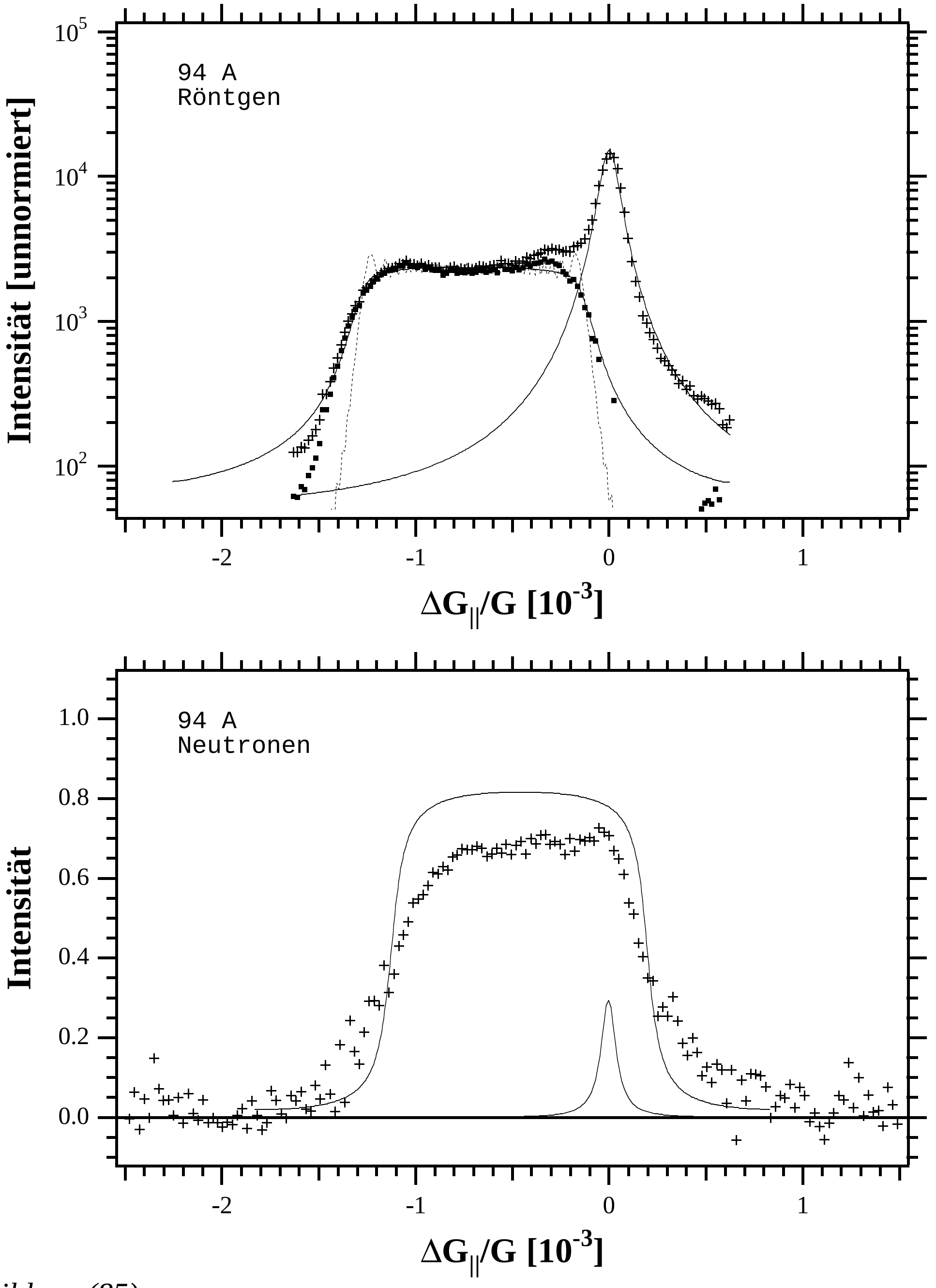

*Abbildung (85):*
*Longitudinale Intensitätsverteilungen von 100 keV Röntgen- und Neutronenstrahlen an ein und derselben Probe, 94 A. Die punktierte, theoretische Kurve im oberen Graph ist zu kastenförmig. Erst nach einer Faltung mit der Auflösungsfunktion werden ihre Kanten rund und passen ausgezeichnet mit den Meßdaten des Gradientenbeitrags zusammen. Im Gegensatz zu den Röntgendaten ist die Neutronenmessung absolut normiert und liegt etwas unterhalb der theoretischen Kurve. Letztere wurde ausschließlich aufgrund des Gradienten und der Kristalldicke berechnet. Die kleine Lorentzlinie bezeichnet die gemessene Auflösungsfunktion.*
*Die Intensitätsgewinne gegenüber den Auflösungsfunktionen betragen 2,0 im Röntgen- und 25 im Neutronenfall.*

Abbildung (85) gibt im oberen Teilbild die mit 100 keV Strahlung gemessene, über die Transversalkomponente integrierte Intensität des zu 94 A gehörenden Spektrums in Bild (76) wieder. Darunter ist die durch Neutronenrückstreuung gewonnene, absolut normierte Aufnahme des 850 µm dicken Gradientenkristalls dargestellt. Die Röntgendaten sind wie in den Beispielen zuvor ausgewertet. Hier ergibt sich der Substratreflex sehr breit, was wohl auf die relativ schlechte Kristallhomogenität, dessen starke Krümmung, sowie einen großen Strahlquerschnitt zurückzuführen ist. Der Gradientenbeitrag läßt sich wieder gut vom Substrat trennen und paßt der Form nach ausgezeichnet zu der gefalteten, auf der Ordinate angepaßten Theoriekurve. Hier wird nochmals demonstriert, wie wichtig die Faltung mit der Auflösungsfunktion erscheint, da die ungefaltete Theoriekurve zu kastenförmig ist und zu schnell abfällt.

Bei den Neutronendaten ist zusätzlich die schon in Abbildung (70) ausgewertete, absolut normierte Auflösungsfunktion des Instruments sowie die mit dieser gefaltete, theoretisch berechnete Reflektionskurve des Gradientenbeitrags eingezeichnet. Obwohl die gemessene Plateaureflektivität zugunsten der Flanken um 20 % unter der berechneten liegt, ist das Ergebnis äußerst zufriedenstellend und demonstriert hervorragend die kastenförmige Intensitätsverteilung, wie sie für die Anwendung eines Neutronenmonochromators von Bedeutung sein wird. Wie früher schon erwähnt, ist die Oberfläche dieses dicken Kristalls sehr inhomogen, speziell mit mehreren 100 µm tiefen Wachstumsstrukturen übersät, so daß der Fehler auch durch die zu optimale Dickenbestimmung und das damit verbundene, mangelnde Streuvolumen interpretiert werden kann.

Der Neutronengewinnfaktor gegenüber der gemessenen Auflösungsfunktion beträgt bei diesem Kristall 25 und entspricht demnach exakt der Vergrößerung des Streuvolumens von der Pendellösungsperiode von 34 µm auf die Dicke des Gradientenkristalls von 850 µm. Generell ist die Plateaureflektivität der Neutronenmessung gegenüber des Röntgenergebnisses wesentlich erhöht, da, wie im theoretischen Teil dieser Arbeit diskutiert, die Dicke des Gradientenkristalls mit der Pendellösungsperiode verglichen werden muß, und diese im ersten Fall sechs mal geringer ausfällt, als im letzten. Damit erklärt sich auch der recht schwache Gewinn von 2,0 gegenüber dem Substratreflex im Fall der 100 keV Röntgenbeugung.

# 8. Zusammenfassung

Im ersten Teil der vorliegenden Arbeit wurden zwei theoretische Modelle zur Beschreibung der Beugungseigenschaften von Gradientenkristallen, eines im Rahmen der kinematischen Theorie, das andere auf einer Transfermatrizenmethode basierend, aufgestellt und hergeleitet. Der erste Ansatz führt zu rein analytischen Ergebnissen, die insbesonders für eine Beschreibung der Breiten der Reflektionskurven und der Plateauoszillationen geeignet erscheint. Die zweite liefert nach numerischer Berechnung die exakten, extinktionsbestimmten Intensitätsverteilungen. Sie beschreibt analytisch durch eine Matrix die Kopplung der vorwärts- und abgebeugten Wellenfunktionen, sowie deren Propagation durch eine planparallele Kristallschicht. So können beliebige Kristallmedien, deren Eigenschaften sich entlang der Grenzflächennormalen ändern, durch sukzessive Multiplikation solcher Matrizen beschrieben werden. Insbesonders ist für die Größe der Unstetigkeiten an den Grenzflächen keinerlei Einschränkung gegeben, so daß die Anwendung, wie hier an Multilagensystemen demonstriert, weit über die Gradientenkristalle hinausgeht. Beide Theorien sind untereinander konsistent und stimmen gleichermaßen mit den Lösungen des Takagi-Taupin Formalismus überein.

Vom Standpunkt der Kristallzucht sei hervorgehoben, daß im Rahmen dieser Arbeit erstmals neuartige $Si_{1-x}Ge_x$ Gradientenkristalle großflächig mit Wachstumsraten bis zu 0,6 µm/min hergestellt wurden. Die Schichtdicken betragen bis zu einigen 100 µm. Damit ist eine Größenordnung erreicht, die grundsätzlich eine Anwendbarkeit als Neutronenmonochromator in den Bereich des Möglichen bringt. Die maximal erreichte Gitteraufweitung liegt im Prozentbereich und kann ohne Schwierigkeiten beliebig bei der Kristallzucht gesteuert werden.

Durch lichtmikroskopische Beobachtungen wurden regelmäßige Oberflächendefekte und Versetzungsstrukturen beobachtet, deren Orientierungen und geometrische Eigenschaften charakterisiert wurden. Mikrosondenanalysen lieferten Konzentrationsprofile, die wiederum mit den Zuchtbedingungen verglichen und durch Beugungsanalyse überprüft wurden. Dabei wurde erstmalig eine tetragonale Gitterverzerrung entdeckt, die auf thermisch bedingte Spannungen zurückgeführt wird. Sie entsteht bei der Abkühlung von der Zuchttemperatur auf Umgebungstemperatur durch verschiedene thermische Ausdehnungskoeffizienten von Epitaxieschicht und Substrat und wurde durch eine Messung der Asymmetrie ansonsten symmetrischer Reflexe charakterisiert. Hierbei deutet eine anisotrope transversale Aufweitung der Braggreflexe auf Netzwerke von Versetzungen hin, wie sie durch die Relaxation der Gitterfehlanpassung entstehen.

Zur Überprüfung der Beugungseigenschaften an Gradientenkristallen wurden Streumethoden sowohl für Neutronen- als auch für Röntgenstrahlen genutzt. Die verwendeten Neutronen-

rückstreu- und Flugzeitspektrometer, sowie die hochenergetische Röntgendreikistalldiffraktometrie zeichnen sich durch ihre hohe Impulsauflösung des reziproken Raumes aus. Letztere Methode erlaubt insbesonders ein präzises, zweidimensionale Abtasten der Umgebung des reziproken Gitterpunktes. Die mit der Transfermatrixmethode berechneten Reflektionskurven schmiegen sich gut an die experimentellen Ergebnisse an. In dieser Arbeit wurden Kristalle durch beide Strahlungsarten untersucht, was im Rahmen der dynamischen Streutheorie unterschiedlichen Gradienten gleichkommt. Da absolute Intensitätsmessungen meist nicht möglich waren, liegt die Stärke der Anpassung in der Beschreibung der Linienform an den Plateaurändern sowie in der Übereinstimmung mit einem beobachteten Plateauoszillationsminimum.

Mit Hinblick auf einen Neutronenmonochromator zeigen die Diffraktionsergebnisse einen bis zu 25-fachen Intensitätsgewinn gegenüber der gemessenen Auflösungsfunktion. Vergleicht man diesen Wert mit der berechneten Reflektivität perfekten Siliziums, so ergibt sich ein Faktor 40. Dabei wurden die Gradienten zugunsten der Analyse und Machbarkeitsstudie zu steil gezogen, so daß die Maximalreflektivität von 100 % noch nicht erreicht ist. Die Aufweitung des reziproken Gittervektors beträgt in diesem Beispiel das 70-fache der natürlichen Linienbreite eines Idealkristalls und wurde mit $\Delta d/d = 1{,}4 \cdot 10^{-2}$ bis auf das 700-fache vorangetrieben.

# 9. Ausblick

Obwohl die gegenwärtigen Ergebnisse vielversprechend sind, können die $Si_{1-x}Ge_x$ Gradientenkristalle noch nicht am Fließband produziert werden. Der Wachstumsprozeß ist sehr kritisch auf geringfügige Veränderungen der Zuchtparameter und sollte noch mit Hinblick auf die Gasströmungsverhältnisse und Sauberkeit des eingebrachten Substrats verbessert werden. Zudem können kostengünstigere Flüssigquellen als Grundlage für die Epitaxiematerialien zum Einsatz kommen.

Um optimale Reflektionseigenschaften zu erlangen, müssen Kristalle mit großen Gitteraufweitungen wesentlich mehr Streuvolumen anbieten. Dicken bis in den Zentimeterbereich wären wünschenswert, damit die Intensitätsgewinne den Größenordnungen oben genannter Gitteraufweitungen gleich kommen. Dabei scheint es günstiger, von einem Germaniumsubstrat aus zu wachsen, da dieses Element eine größere Streukraft als Silizium besitzt und somit kleinere Dicken ausreichten. Außerdem sollten andere Systeme, wie $Si_{1-x}C_x$ oder das bereits bekannte $Cu_{1-x}Ge_x$ für diese Epitaxiemethode ins Auge gefaßt werden.

Die theoretischen Ergebnisse können noch in Hinblick verschiedener Einzelheiten ausgemessen werden. Eine Beobachtung der Plateauoszillationen kann an hochindizierten Reflexen sehr steiler Gradienten oder stark gekrümmten, idealen Kristallen durchgeführt werden. Dazu wird die ständig verbesserte Auflösung von Dreikristalldiffraktometern an Synchrotronstrahlungsquellen eine wesentliche Hilfe sein. Aus der Feinstruktur von Reflektionskurven lassen sich die Details der Streuvektorverteilungen erkennen. Nicht zuletzt kann die Transfermatrizenmethode zur Beschreibung komplizierter Schichtenstrukturen der Halbleiterindustrie erweitert werden und in der Auswertung der zerstörungsfreien Materialanalyse Einzug finden.

# 10. Literaturverzeichnis

# 11. Danksagung

An dieser Stelle möchte ich mich bei all denen bedanken, deren Interesse und praktische Unterstützung zum Gelingen dieser Arbeit beigetragen haben.

Herzlichen Dank an meinen Doktorvater Prof. T. Springer für die Betreuung dieser Arbeit, sowie sein oft ins Detail vertiefte Interesse und den daraus hervorgegangenen Diskussionen.

Dem Initiator dieses Themas, Privatdozent Andreas Magerl möchte ich für die tiefgehende Betreuung vor Ort in Grenoble danken. Die täglichen Gespräche und Diskussionen formten wesentlich den Verlauf der Arbeit und entwickelten sich zu einer guten Zusammenarbeit. Sein Organisationstalent ermöglichte die Meß- und Konferenzreisen, an denen er meistens auch selbst mitwirkte. Ohne den Rahmen dieser Danksagung sprengen zu wollen, möchte ich auch die privaten Kontakte und gemeinsamen Ausflüge unserer Familien erwähnen.

Für die Bereitstellung der Doktorandenstelle am Institut Laue-Langevin und dem damit verbundenen Vertrauen möchte ich meinen Dank Prof. P. Armbruster aussprechen.

J’exprime mes remerciements au Prof. Roland Madar, qui a dirigé la croissance des cristaux dans son laboratoire et qui a contribué aux missions pour les expériences et les conférences.

Herrn Prof. P. H. Dederichs herzlichen Dank für die Anstöße im theoretischen Teil dieser Arbeit.

Den Verantwortlichen der Diffraktometer am HASYLAB Prof. J. Schneider und Hansi Neumann, am NSLS Dr. Peter Siddons und Dr. Jerry Hastings, sowie am Forschungsreaktor München Dr. E. Steichele und ihren Mitarbeitern möchte ich für die Einführung und Betreuung während den Meßzeiten an ihren Instrumenten danken.

Meinem Kollegen und Nachfolger Christian Doll sei herzlich für seine tiefgehenden Fragen zum Verständnis der dynamischen Beugungstheorie gedankt, durch die manche Probleme erst erkannt oder geweckt wurden.

Un remerciement aux techniciens du groupe monochromateur, Roland Hehn et Bernard Nief pour leur aide, mais surtout à Alain Escoffier, qui a fortement contribué à l’installation et aux modifications du dispositif de croissance de cristaux.

Neben den Betreuern und Christian Doll möchte ich auch Frau Barbara Standke als Lektoren dieses Textes danken.

Un grand merci à l'artiste Blandine Leclerc, qui a mis la gravure de la dédicace à ma disposition.

Zuletzt, aber nicht weniger gern möchte ich auch der großen Zahl der Vorgesetzten, Kollegen, Mitarbeiter und Freunden danken, die über den Konsens der vorliegenden Arbeit hinaus das Leben am und im Umfeld des ILL geprägt haben.

# 12. Lebenslauf

| | |
|---|---|
| *Name:* | Klaus-Dieter Liß |
| *Geburtstag:* | 10. Juni 1962 |
| *Geburtsort:* | Idar-Oberstein, Rheinland–Pfalz, Deutschland |
| *Staatsangehörigkeit:* | Deutsch |
| *Adresse:* | 3 rue du Champa<br>F-38450 Les Saillants du Gua<br>Tel.: (+33) 76.72.25.79, eMail: liss@esrf.fr |
| *Familienstand:* | verheiratet mit Laure Jacqueline, geb. Payan |
| *Kind:* | Katharina Christel Maria Liß, geboren am 23. August 1994 |
| | |
| *1968-1969* | Erste Klasse Grundschule in Erding |
| *1969-1972* | Grundschule am Niederbronner Weg, Fürstenfeldbruck, Bayern |
| *1972-1973* | Hauptschule Nord, Fürstenfeldbruck |
| *1973-1976* | Graf-Rasso-Gymnasium, Fürstenfeldbruck |
| *1976-1983* | Internatschüler im Gymnasium mit Schülerheim Hohenschwangau, Bayern |
| *1983* | Abiturabschluß am Gymnasium Hohenschwangau |
| | |
| *1983-1990* | Studium der allgemeinen Physik an der Technischen Universität München (TUM), München, Deutschland |
| *1985, 1986* | Zwei zweimonatliche Werkstudentenpraktika in Biophysik an der TUM |
| *1986, 1987, 1987, 1988* | Vier zweimonatliche Werkstudentenpraktika am Institut Laue Langevin (ILL), Grenoble, Frankreich |
| *1989-1990* | Diplomarbeit am ILL: *“Untersuchungen zur Verbesserung der Energieauflösung für Neutronenrückstreuspektrometer unter Verwendung von Idealkristallen mit geringer dynamischer Breite der Reflexionskurve.”* |
| *1990* | Diplom in allgemeiner Physik |
| *8-1990-11-1990* | dreimonatige Tätigkeit in der Neutronenstreuung am AECL, Chalk River, Kanada |
| *seit 12-1990* | Doktorarbeit am ILL, Grenoble, Frankreich und IFF/KFA Jülich: *“Strukturelle Charakterisierung und Optimierung der Beugungseigenschaften von* $Si_{1-x}Ge_x$ *Gradientenkristallen, die aus der Gasphase gezogen wurden”* |
| *27. Oktober 1994* | Mündliche Doktorprüfung |